\documentclass[twoside,12pt]{article}
\usepackage{epsfig}

\def\Journal#1#2#3#4{{#1} {#2} (#4) #3 }
\def\NCA{{\em Nuovo Cimento} A}
\def\PHYS{{\em Physica}}
\def\NPA{{\em Nucl. Phys.} A}
\def\MATH{{\em J. Math. Phys.}}
\def\PRO{{\em Prog. Theor. Phys.}}
\def\NPB{{\em Nucl. Phys.} B}

\def\PLB{{\em Phys. Lett.} B}

\def\PL{{\em Phys. Lett.}}
\def\PRL{\em Phys. Rev. Lett.}
\def\PREV{\em Phys. Rev.}
\def\PREP{\em Phys. Rep.}
\def\PRA{{\em Phys. Rev.} A}
\def\PRD{{\em Phys. Rev.} D}
\def\PRC{{\em Phys. Rev.} C}
\def\PRB{{\em Phys. Rev.} B}

\def\ANNP{\em Ann. Phys. (N.Y.)}
\def\RMP{{\em Rev. Mod. Phys.}}
\def\CHEM{{\em J. Chem. Phys.}}
\def\INT{{\em Int. J. Mod. Phys.} E}

\def\q{\vec q}

\newcommand{\be}{\begin{equation}}
\newcommand{\ee}{\end{equation}}
\newcommand{\bea}{\begin{eqnarray}}
\newcommand{\eea}{\end{eqnarray}}
\newcommand{\nn}{\nonumber}
\begin{document}

\title{
Two-Body Correlations in Nuclear Systems
}

\author{H.\ M\"uther
\\
Institut f\"ur Theoretische Physik,\\ Universit\"at T\"ubingen,
T\"ubingen, Germany
\\
\\
A. Polls\\
Departament d'Estructura i Constituents de la Mat\`eria\\
Universitat de Barcelona,
E-08028 Barcelona, Spain}

\maketitle

\begin{abstract} Correlations in the nuclear wave-function beyond the mean-field 
or Hartree-Fock approximation are very important to describe basic properties of
nuclear structure. Various approaches to account for such correlations are
described and compared to each other. This includes the hole-line expansion, the
coupled cluster or ``exponential S'' approach, the self-consistent evaluation of
Greens functions, variational approaches using correlated basis functions and
recent developments employing quantum Monte-Carlo techniques. Details of these 
correlations are explored and their sensitivity to the underlying 
nucleon-nucleon interaction. Special
attention is paid to the attempts to investigate these correlations in
exclusive nucleon knock-out experiments induced by electron scattering.
Another important issue of nuclear structure physics is the role of relativistic
effects as contained in phenomenological mean field models.  The sensitivity of 
various nuclear structure observables on these
relativistic features are investigated. The report includes the discussion of
nuclear matter as well as finite nuclei.
\end{abstract}

\section{Introduction}
One of the central challenges of theoretical nuclear physics is the attempt to
describe the basic properties of nuclear systems in terms of a realistic
nucleon-nucleon (NN) interaction. Such an attempt typically contains two major
steps. In the first step one has to consider a specific model for the NN
interaction. This could be a model which is inspired by the
quantum-chromo-dynamics\cite{faes0}, a meson-exchange or One-Boson-Exchange
model\cite{rupr0,nijm0} or a purely phenomenological ansatz in terms of
two-body spin-isospin operators multiplied 
by local potential
functions\cite{argo0,urbv14}. Such models are considered as a realistic
description of the NN interaction, if the adjustment of parameters within the
model yields a good fit to the NN scattering data at energies below the
threshold for pion production as well as energy and other observables of the
deuteron.

After the definition of the nuclear hamiltonian,
 the second
step implies the solution of the many-body problem of $A$ nucleons interacting
in terms of such a realistic two-body NN interaction. The simplest approach to
this many-body problem of interacting fermions one could think of would be the
mean field or Hartree-Fock approximation. This procedure yields very good
results for the bulk properties of nuclei, binding energies and radii, 
if one employs simple phenomenological NN forces like e.g.~the Skyrme forces, 
which are adjusted to describe such nuclear structure data\cite{skyrme}.
However, employing realistic NN interactions the Hartree-Fock approximation
fails very badly: it leads to unbound nuclei\cite{art99}. 

This failure of the Hartree-Fock approximation is a consequence of the strong
short-range components of a realistic NN interaction, which are necessary to
describe the NN data. The Hartree-Fock wavefunction describes the nucleus as a
system of nucleons moving independent from each other in a mean field derived
from the average interaction with all other nucleons. This implies that the
wavefunction contains large amplitudes of configurations,
 in which two
nucleons are so close to each other, that they are exposed to the very repulsive
components of the NN interaction at short distances. The hard-core 
potentials\cite{hamada}, which were very popular in the sixties, describe these
components in terms of an infinitely repulsive core for relative distances
smaller than the radius of this hard core of about 0.4 $fm$. In this case a
Hartree-Fock calculation would even yield an infinite repulsive energy. Modern
models for the NN interaction, in particular the meson-exchange models which
lead to non-local NN interactions, contain softer cores. Nevertheless, a
careful treatment of two-body short range correlations beyond the mean field
approximation is indispensable to describe the structure of nuclei in terms of
realistic NN interactions. The same is true for the correlations which are
induced by the strong tensor components in the NN interaction, which mainly
originate from the pion exchange contributions.

Various different tools have been developed to account for such correlation
effects. Below we will describe some of these methods including the Brueckner
hole-line expansion\cite{bruek1,bruek2,bruek3}, the
coupled cluster or ``exponential S'' approach\cite{kuem,bish}, 
the self-consistent evaluation of Greens functions\cite{wim1}, 
variational approaches using correlated basis functions\cite{fhnc1,fhnc2,fhnc3}
and recent developments employing quantum Monte-Carlo
techniques\cite{monc1,monc2}. 

Using such methods one obtains correlated many-body wave functions, which are
rather sensitive to the NN interaction under consideration. Therefore the
question arises which kind of experimental observables might be considered to
investigate details of these correlations. The hope is that different
predictions derived from the various model of the NN interaction will allow to
distinguish between the various model for the NN interaction at short distances.
Therefore, finally such nuclear structure studies will help to explore the
details of the strong baryon baryon interaction at short distances. 

What is 
 the effect of correlations on the nuclear wave function? In order
to discuss this question let us consider for the moment a system of infinite
nuclear matter. The mean field or Hartree-Fock wave function of nuclear matter
corresponds to the Slater determinant of plane wave states, in which all
single-particle states with momenta below the Fermi momentum $k_F$ are occupied.
Correlations on top of this Hartree-Fock state will yield a depletion of
single-particle states with momenta $k$ below $k_F$ and a non-vanishing
occupation of states with high momentum. From this consideration one may think
that the study of high-momentum components in the single-particle wave functions
of nuclear states might be an ideal tool to explore correlation effects in the
nuclear wave function. Therefore it has been suggested to 
 study these
high-momentum components by means of exclusive single nucleon knock-out
experiments like $(e,e'N)$ with the $(A-1)$ nucleus remaining in the ground state
or other well defined state\cite{nikhef,mainz}. However, a detailed analysis of
the spectral function for such knock-out experiments shows that the expected
high-momentum components in the nuclear wavefunction could only be observed at
high missing energies, i.e.~with an excitation energy of the residual 
target nucleus well above the threshold for the emission of a secondary
nucleon\cite{artur1,wim2,poll1}. Therefore one expects exclusive two-nucleon 
knock-out experiments like $(\gamma ,NN)$ or ($e,e'NN$)\cite{onder,rosner} to be
more sensitive to the effects of correlations. Details for the analysis of such
nucleon knock-out experiments induced by inelastic electron scattering will be
given below. 

Up to this point we merely discussed the nuclear system within the framework of
the non-relativistic quantum many-body theory. At first sight this seems to be
well justified since the binding energies of nuclei with single-particle
potentials of -50 MeV or so, are very small on the scale of the nucleon rest
mass. It has already been discussed, however, that the short-range components of
the NN interaction are very strong with large repulsive components, which are
compensated by strong attractive ones. Within the meson-exchange model for the
NN interaction the repulsive components are generated by the exchange of the
$\omega$, a vector meson, while the attractive parts are described in terms of
the exchange of a scalar particle, the $\sigma$ meson\cite{rupr0}. This $\sigma$
meson does not represent a meson in the usual sense but its exchange is used to
describe the correlated two-pion exchange without and with intermediate
excitation of the interacting nucleons\cite{paris,durso}. Calculating the nucleon
self energies from such a meson exchange model within a Hartree approximation,
one finds that the $\omega$ exchange yields a component $\Sigma_0$, which 
transforms under a Lorentz transformation like the time-like component of a
vector, while the scalar meson exchange yields a contribution $\Sigma_s$, which
transforms like a scalar. Inserting this self energy into the Dirac equation for
a nucleon in the medium of nuclear matter leads to binding effects which are as
small as the -50 MeV discussed above. This small binding effect, however,
results from a strong cancellation between the repulsive $\Sigma_0$ and the
attractive $\Sigma_s$ component. The attractive scalar component $\Sigma_s$
leads to Dirac spinors for the nucleons in the nuclear medium, which contain a
small component significantly enhanced as compared to the Dirac spinor of a free
nucleon. This effect is often described in terms of an effective Dirac mass
$m^*$ for the nucleon, which can be of the order of 600 MeV in nuclear matter
around saturation density.

This modification of the Dirac spinors in the nuclear medium requires a
self-consistent evaluation of the matrix elements of the meson exchange
interaction. Within the phenomenological $\sigma - \omega$ or Walecka model of 
nuclear physics\cite{wal1,serot} it is this self-consistency, which provides the
saturation of nuclear matter. This relativistic effect, however, is also
observed in Dirac-Brueckner-Hartree-Fock studies, which are based on realistic
NN interactions\cite{anast,brock,malf1,weigel}. Including these relativistic
features improves the predictions for the saturation point of nuclear matter as
well as the bulk properties of finite nuclei
significantly\cite{broc1,fri93,fri94,boersma}. The question is: Are there other
observables, which are sensitive to the enhancement of the small component of
the nucleon Dirac spinors in the nuclear medium? 

\begin{figure}[tb]
\begin{center}
\begin{minipage}[t]{8 cm}
\epsfig{file=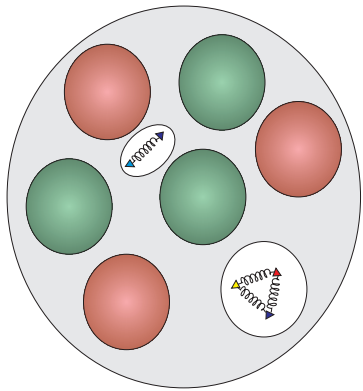,scale=0.5}
\end{minipage}
\begin{minipage}[t]{16.5 cm}
\caption{Cartoon of a nucleus, displaying the size of the nucleons as compared
to the typical distance to nearest neighbors. Also indicated are the internal
structure of nucleons and mesons.\label{fig1}}
\end{minipage}
\end{center}
\end{figure}

The calculation scheme discussed so far, determine the interaction of two
nucleons in the vacuum in a first step and then solve the many-body problem of
nucleons interacting by such realistic potentials in a second step, is of course
based on the picture that nucleons are elementary particles with properties,
which are not affected by the presence of other nucleons in the nuclear medium. 
One knows, of course, that this is a rather simplified picture: nucleons are
built out of quarks and their properties might very well be influenced by the
surrounding medium. A cartoon of this feature is displayed in Fig.~\ref{fig1}.
So the question is, how important are the sub-nucleonic degrees of freedom, must
we expect a change of the nucleon properties in the nuclear medium? At low
energies it might be sufficient to account for the internal structure of the
nucleons by considering the possibility that the nucleon may get excited to the
$\Delta (3,3)$ excitation at 1232 MeV. It has been demonstrated\cite{elster}
that the process of two interacting nucleons, which polarize each other to such
isobar excitations are an important ingredient to the medium range attraction of
the NN interaction. Attempts have been made to account for such isobar
excitations also in microscopic studies of nuclear structure\cite{oldrep,franz}.

Furthermore,
not only the basic properties of the nucleons, their mass or electromagnetic
form-factor might be affected by the nuclear medium, also the masses and thereby
the propagation of the mesons might be modified as well\cite{rhom,pimas1,pimas2}. 
This will have consequences for the meson-exchange model of the NN interaction.
For example, a lowering of the meson masses in the nuclear medium as compared to
the vacuum would lead to a larger range of the corresponding meson exchange
interaction terms, which should lead to different results in the nuclear
structure calculation.

\begin{figure}[tb]
\begin{center}
\begin{minipage}[t]{8 cm}
\epsfig{file=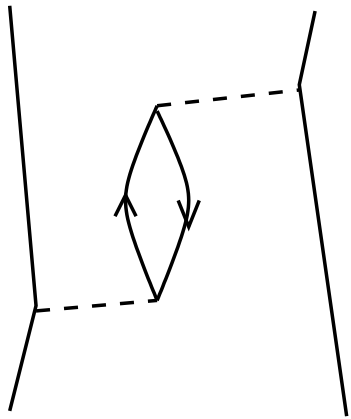,scale=0.7}
\end{minipage}
\begin{minipage}[t]{16.5 cm}
\caption{Two nucleons interacting in a nuclear medium, see discussion in the
text. \label{fig2}}
\end{minipage}
\end{center}
\end{figure}

In studying such features of sub-nucleonic degrees of freedom, however, one
should be careful to avoid the mixing of different points of view on the same
process. As an example, let us consider a process like the one indicated by the
diagram in Fig.~\ref{fig2}: Two nucleons, represented by the two upward going
lines, interact with each other by the exchange of e.g.~a pion, which is
represented by the dashed line. Propagating in a nuclear medium between 
these two nucleons, the pion may interact with a third nucleon leading to an
intermediate particle-hole excitation of the nuclear system. Such processes 
might be considered as a modification of the pion propagator in the nuclear 
medium, which could be characterized by a modification of the effective pion 
mass inside a nucleus. The process displayed in Fig.~\ref{fig2}, however, will
also be included in any many-body calculation based on two-nucleon interactions,
which accounts for three-nucleon correlations or includes effects of ring
diagrams. Therefore effects which one may identify with a modification of 
meson propagators are accounted for in an approach, which does not even mention
any mesonic degrees of freedom.

The pion re-scattering process displayed in  Fig.~\ref{fig2} may also lead to an
excitation of the intermediate nucleon to e.g.~the $\Delta$ resonance. Again
such terms might 
be
 considered as a modification of the meson propagator.
Such terms would also be accounted for in a many-body theory based on baryon -
baryon interactions, like e.g.~the Argonne V28\cite{v28}, which includes the
$\Delta$ degrees of freedom explicitly. Within the framework of a many-body
theory which does not account for such $\Delta$ excitations, such a process
might be taken into account by means of a three-nucleon interaction. This
example has been given to demonstrate the model dependence of three-body forces
or statements about the relevance of mesonic degrees of freedom. The approaches
which we will discuss throughout this review, are all based on realistic NN
interactions and try to solve the many-body problem as good as possible.
It is the aim to study
how far such an approach, which ignores all sub-nucleonic degrees of
freedom, can predict the properties of nuclear systems at zero temperature and
without external pressure.  Remaining discrepancies between theoretical
prediction and empirical data can then bet attributed to the necessity to
account for sub-nucleonic degrees of freedom, explicitly.

Such many-body calculations can then also be used to obtain predictions for
nuclear matter under extreme conditions like in neutron stars, supernova
explosions\cite{ruder} or in central heavy ion collisions. They yield a
prediction for the equation of state of nuclear matter at high densities and/or
high temperatures which is free of any parameters. The success or failure of 
such theoretical calculations in predicting the properties of normal nuclear 
matter from the NN interactions of two nucleons in the vacuum can be used to
judge the reliability of these predictions for nuclear matter at extreme
conditions.

Nuclear systems provide an intriguing and challenging object for the 
development of quantum many-body theories. The underlying NN interaction is
non-trivial: It contains a rich operator structure with strong tensor
components and is in general non-local. The interacting particles must be
considered as quasiparticle and the inclusion of internal degrees of freedom
might get important. An extension of the many-body theory to account for
relativistic effects might be necessary. Empirical data are available and should
be reproduced for finite systems with particle numbers ranging from $A=2$ to
infinite nuclear matter. For other systems of condensed matter or Fermi liquids,
like electron gas or liquid Helium 3, reliable data on finite samples are 
difficult to obtain but of high interest. Therefore nuclei are an ideal testing
ground for many-body theory for finite systems in particular. The experience
collected here should be rather useful in the study of other systems.

After this introduction, we will present an outline of some many-body 
approaches which are used in nuclear physics. This includes the Brueckner -
Bethe hole-line expansion, the coupled cluster or ``exponential S'' approach, 
the self-consistent evaluation of Greens functions, variational approaches 
using correlated Jastrow-type basis functions and
variational and quantum Monte Carlo techniques. A short review on
the status of realistic NN interactions and on the differences between those
models will be given in the first part of section 3. This section 3 also
contains the discussion of results for bulk properties of infinite nuclear 
matter and finite nuclei. As an example on experimental efforts to explore the
effects of correlations in detail, we will report on the analysis of inelastic
electron scattering in section 4. The section 5 describes attempts to extend the
many-body theory to account for relativistic features as well as the sensitivity
of some observables to these relativistic effects.

\section{Many-Body Approaches}
\subsection{\it Hole - Line Expansion \label{sec:holeline}}
\begin{figure}[tb]
\begin{center}
\begin{minipage}[t]{8 cm}
\epsfig{file=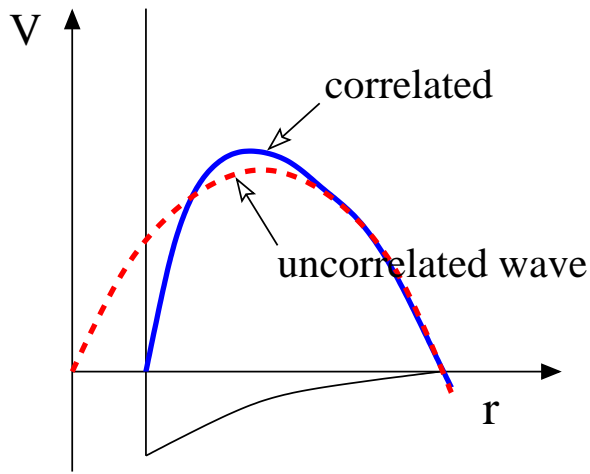,scale=0.7}
\end{minipage}
\begin{minipage}[t]{16.5 cm}
\caption{Schematic picture of a NN interaction with hard core and its effect on
the correlated NN wave function $\Psi(r)$. \label{fig3}}
\end{minipage}
\end{center}
\end{figure}

As it has been discussed already above one problem of nuclear structure
calculations based on realistic NN interactions is to deal with the strong
short-range components contained in all such interactions. This problem is 
evident in particular when
so-called hard-core potentials are employed, which are infinite for relative
distances smaller than the radius of the hard core $r_c$. The matrix elements
of such a potential $V$ evaluated for an uncorrelated two-body wave function
$\Phi (r)$ diverges since $\Phi (r)$ is different from zero also for relative
distances $r$ smaller than the hard-core radius $r_c$ (see the schematic picture
in Fig.~\ref{fig3}. A way out of this problem is to account for the two-body
correlations induced by the NN interaction in the correlated wave function
$\Psi (r)$ or by defining an effective operator, which acting on the
uncorrelated wave function $\Phi (r)$ yields the same result as the bare
interaction $V$ acting on $\Psi (r)$. This concept is well known for example in
dealing with the scattering matrix $T$, which is defined by
\be
<\Phi \vert T \vert \Phi > = <\Phi \vert V \vert \Psi > \; . \label{eq:tmat}
\ee
As it is indicated in the schematic Fig.~\ref{fig3}, the correlations tend to 
enhance the amplitude
of the correlated wave function $\Psi$ relative to the uncorrelated one
at distances $r$ for which the interaction is attractive. A reduction of
the amplitude is to be expected for small distances for which $V(r)$ is
repulsive. From this discussion we see that the
correlation effects tend to make the matrix elements of $T$ more attractive
than those of the bare potential $V$.  For two nucleons in the vacuum the $T$
matrix can be determined by solving a Lippmann-Schwinger equation
\bea
T \vert \Phi > &= &V \left\{ \vert \Phi > + \frac{1}{\omega  - H_0 +
i\epsilon } V \vert \Psi >\right\}\nonumber \\
& = &  \left\{ V + V \frac{1 }{\omega  - H_0 +i\epsilon } T\right\} \vert 
\Phi >\, . \label{eq:lipschw}
\eea

In these equations the starting energy $\omega$ stands for the energy of the two
interacting nucleons and $H_0$ represents the operator for the nucleons in
the intermediate state without residual interaction, i.e.~the kinetic energy.

This concept of an effective operator accounting for correlation effects in the
wave function is one way to present the Brueckner-Bethe-Goldstone (BBG)
approach to the many-body problem. To show the features of the BBG approach,
we  consider the nuclear hamiltonian, a sum of the kinetic energy
$t_{kin}$ and the two-body interaction $V$, and introduce an appropriate
single-particle potential $U$ 
\be
 H=H_0 + H_1, \qquad H_0=t_{kin}+U, \qquad H_1=V-U\,. \label{eq:h01}
\ee
The eigenstates of $H_0$ for $A$ particles, i.e.~Slater determinants build from
the corresponding single-particle wavefunctions, provide a basis of the
Hilbert-space for the $A$ nucleon problem
\be 
 H_0 \Phi_i (1,\dots ,A) = E_i^0 \Phi_i (1,\dots ,A),  \label{eq:schh0}
\ee
with $E_i^0$ the sum of the single particle energies.
This basis $\Phi_i$ is used to split the Hilbert-space into a model space and the
rest. If one is interested in the
ground-state properties of closed shell nuclei like $^{16}O$ this model space
could be defined by a single Slater determinant. If one is interested in 
the states of open shell nuclei at low energies, this model space could consist 
out of those functions $\Phi_i$ which represent a typical set of basis states 
for a shell-model configuration mixing calculation\cite{glaude}. The concept of
a model space, however, can also be considered for the system of infinite
nuclear matter. One may choose the model space to contain all Slater
determinants, which can be constructed by considering occupied states with
momenta below a model space momentum $k_M$. If this momentum $k_M$ is equal to
the Fermi momentum of a free Fermi gas $k_F$ of the density considered, the
model space reduces just to this model wave function\cite{tomnm}. With this
choice for a model space one can define projection operators $\cal P$ and 
$\cal Q$, which
project onto the model space and on the rest of the A-nucleon Hilbert space,
respectively. It is the aim to derive an effective hamiltonian $H_{eff}$, which
is defined within the model space, with eigenvalues $E_i$ identical to
the
eigenvalues of the original hamiltonian $H$ and eigenvectors, which are just the
projection of the exact eigenvectors $\Psi_i$ of $H$ on the model space. This
implies that
\be
{\cal P} H_{eff} {\cal P} \Psi_i = E_i {\cal P} \Psi_i \,. \label{eq:heffmod}
\ee
With this separation of the Hilbert-space into a model space and a rest also the
treatment of correlations is separated: There are correlations which can be
represented by 
 degrees of freedom within this model space. They are
treated explicitly and accounted for by the diagonalization of the effective
hamiltonian $H_{eff}$ (see Eq.~(\ref{eq:heffmod})) within the model-space. 
This will
typically be correlations which are described in terms of single-particle
excitations around the Fermi energy with low excitation energy, small momentum
transfers and therefore of longer range. These long range correlations are quite
sensitive to the shell structure of finite systems and will therefore be
different for nuclear matter and specific finite nuclei. The correlations, which
cannot be represented by the mixing of configurations within the model space are
treated by means of the effective operator. Such correlations include
excitations up to states high above the Fermi surface. This implies large
excitation energies and high momentum transfers, which means correlations of
short range. 

Several expansions have been formulated for this effective
hamiltonian. There are the energy-dependent expansions of Feshbach \cite{fesh}
and of Bloch and Horowitz \cite{bloho}. They correspond to the Brillouin
- Wigner perturbation expansion and therefore yield an effective hamiltonian
which depends on the exact energy to be calculated. To get rid of this
energy-dependence in the case of multi-dimensional degenerate model-spaces
one has to consider the so-called folded diagram expansion, which has been 
formulated e.g.~by Brandow \cite{bran} and Kuo, Lee and Ratcliff \cite{kulera}. 
Using this folded-diagram formulation, the effective hamiltonian could be 
written as
\be
H_{eff} = H_0 + H_1 + \left\{ H_1 {{\cal Q} \over E^0 - H_0 } H_{eff}
+ \hbox{folded diagrams} \right\}_{\hbox{linked}}, \label{eq:folded}
\ee
with $E^0$ referring to the eigenvalue of $H_0$ for the state in the model
space considered.

\begin{figure}[tb]
\begin{center}
\begin{minipage}[t]{8 cm}
\epsfig{file=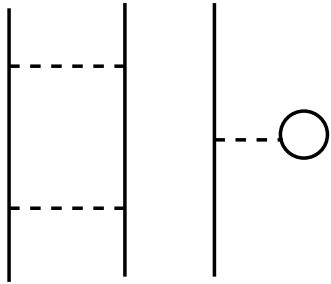,scale=0.7}
\end{minipage}
\begin{minipage}[t]{16.5 cm}
\caption{Example of an unlinked diagram, which should be ignored in the
expansion of $H_{eff}$ in Eq.~(\protect{\ref{eq:folded}}) \label{fig4}}
\end{minipage}
\end{center}
\end{figure}

Note that in this expansion one only needs to consider the contribution of the
so-called linked diagrams. This implies that contributions like the one
characterized by the diagram in Fig.~\ref{fig4} can be ignored. Such unlinked
contributions contain products of matrix elements of $H_1$ (the two-body parts
are represented by the dashed lines in Fig.~\ref{fig4}), which are completely
unlinked in the sense that the summation over single-particle quantum numbers of
the states attached to the matrix elements of $H_1$ are completely disconnected.
Translated into the representation of diagrams this means that the corresponding
diagram can be separated into two pieces without cutting a solid or dashed line,
representing the propagation of a nucleon and an interaction, respectively. This
restriction to linked diagram contribution is of particular importance for
systems with many particles, like e.g.~infinite nuclear matter. Unlinked
contributions, which represent a combination of processes which are independent
from each other, diverge in the limit of particle number to infinity.
Details on the definition and on techniques to evaluate the contribution of
folded diagrams can be found in \cite{bran,kulera,MKO,leesuz,artufold}.
Folded diagrams do not occur if the model space is of dimension one, an
assumption, which is most commonly made in calculating ground-state properties 
of closed shell nuclei or infinite nuclear matter. 

\begin{figure}[tb]
\begin{center}
\begin{minipage}[t]{12 cm}
\epsfig{file=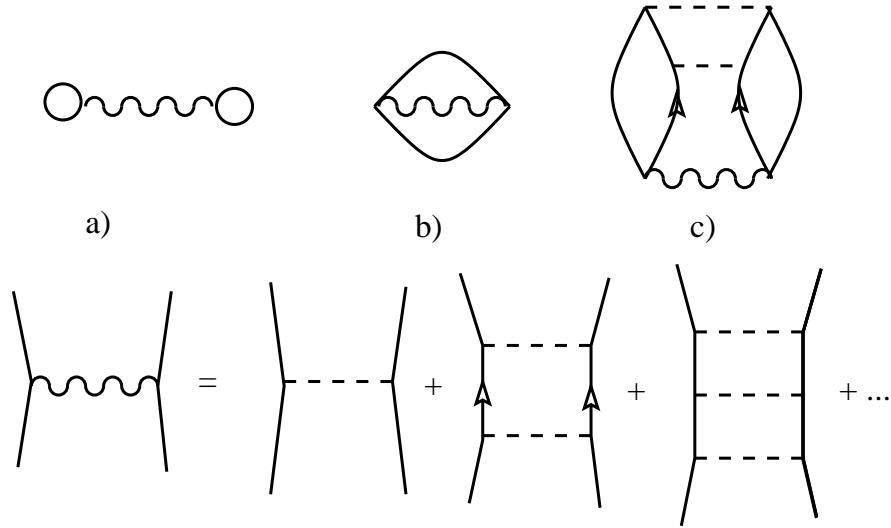,scale=0.6}
\end{minipage}
\begin{minipage}[t]{16.5 cm}
\caption{The upper half of this figure displays Goldstone diagrams, which are
are included in the calculation of the ground-state energy within the BHF
approximation (\protect\ref{eq:ebhf}). Diagram a) corresponds to the direct, b)
to the Fock-exchange contribution. The wiggly lines represent the $G$ matrix and
the lower half of the figure visualizes the Bethe-Goldstone
Eq.~(\protect\ref{eq:betheg}) demonstrating that the $G$ interaction line
represents the sum of all particle-particle ladder diagrams. This implies that
e.g.~the Goldstone diagram displayed in c) is redundant as it is contained
already in a). \label{fig5}}
\end{minipage}
\end{center}
\end{figure}

From the expansion in Eq.~(\ref{eq:folded}) it is obvious that $H_{eff}$ 
will not only contain one-body and two-body operators but, in general, also 
operators involving three and more nucleons. If, however, we restrict the
discussion to the evaluation of ground-state properties and consider only the
effective two-body operators in $H_{eff}$, the expansion is reduced to
the summation of ladder diagrams and one obtains the Bethe-Goldstone equation
\be
G(\omega ) = V + V \frac{Q}{\omega - h_{12}} G \label{eq:betheg}\, ,
\ee
which defines an effective NN interactions of two nucleons. Note that the
projection operator on many-body states ${\cal Q}$ outside the model space in
(\ref{eq:folded}) has
been replaced by the Pauli operator $Q$, which is a two-body operator defined by
\be
Q\vert ij> = \cases{ \vert ij> & if i and j are single-particle unoccupied in
$\Phi$ \cr
0 & else\cr}, \label{eq:pauli}
\ee
with $\Phi$ referring to the Slater-determinant defining the model-space.
The energy denominator in (\ref{eq:betheg}) corresponds to the excitation 
energy of the intermediate
two-particle two-hole state, i.e.~it is the sum of the single-particle energies
for the interacting nucleons in states below the Fermi energy, expressed by the
starting energy $\omega$ minus the energy of the two nucleons in the
intermediate states above the Fermi energy, denoted by the operator $h_{12}$.
(See also the graphical representation in Fig.~\ref{fig5})

The Bethe-Goldstone equation (\ref{eq:betheg}) is quite similar to the 
Lippmann-Schwinger Eq.~(\ref{eq:lipschw}) for the scattering matrix $T$. The only
differences are the Pauli operator $Q$ and the energy denominator which in
(\ref{eq:betheg}) is defined in terms of single-particle energies of the
many-body system instead of the kinetic energies in Eq.~(\ref{eq:lipschw}).
Therefore the solution of the Bethe-Goldstone equation, the $G$-matrix,
corresponds to an effective interaction between two nucleons, which accounts for
correlation effects in the nuclear medium. Similar to Eq.~(\ref{eq:tmat}) one can
define for each product wave function of two uncorrelated single-particle wave
functions $\vert \phi_{\alpha \beta} >$ a correlated wave function $\vert
\psi_{\alpha \beta} >$ by
\bea
G \vert \phi_{\alpha \beta} >  & = & V \vert\psi_{\alpha \beta} > \nonumber\\
& = & V \left\{1 + \frac{Q}{\omega - h_{12}}G \right\} \vert
\phi_{\alpha \beta} > \,,
\label{eq:gcorel1}
\eea 
which implies that the correlated wave function $\vert \psi_{\alpha \beta} >$
can be identified with 
\be
\vert\psi_{\alpha \beta} > = \vert \phi_{\alpha \beta} > + \frac{Q}{\omega - 
h_{12}}G \vert \phi_{\alpha \beta} > \label{eq:gcorel2}
\ee
The second term on the right hand side of this equation, the difference between
the correlated and uncorrelated wave function, is called the defect function.
If the uncorrelated state refers to two single-particle states $\alpha$ and
$\beta$ below the Fermi surface, the difference between the starting energy
$\omega = \epsilon_\alpha + \epsilon_\beta$ ($\epsilon_\alpha$ denoting the
single-particle energies of $H_0$)
 and the eigenvalue of $ h_{12}$ is negative
for all intermediate states, which are restricted to two-particle states above
the Fermi surface. This means that the summation or integration over
intermediate particle states does not meet any pole in the propagators of
(\ref{eq:gcorel1}) or (\ref{eq:gcorel2}). This implies that the matrix elements
of $G$ are real. There exist not phase shifts between the uncorrelated and the
correlated wave function, the defect function vanishes for large relative
distances. This vanishing of the defect function at large $r$ is called the
healing property, the correlated wave functions ``heals'' to the uncorrelated
one at large $r$.

One of the main points of the Brueckner-Bethe-Goldstone approach to the
many-body system is to
evaluate the contributions to the effective hamiltonian 
$H_{eff}$ in Eq.~(\ref{eq:folded}) not in a perturbation expansion in which the
contributions are ordered with respect to the numbers of bare interaction $V$
terms, but in terms of the $G$-matrix.

Up to this point we have not specified yet, how to choose the unperturbed
hamiltonian $H_0$ or the single-particle potential $U$ in Eq.~(\ref{eq:h01}).
The aim is of course to choose $H_0$ such that the Slater determinant $\Phi$
defining the model space is close to the exact wavefunction for the ground-state.
Therefore it seems quite natural to define the single-particle potential $U$ in
analogy to the Hartree-Fock definition with the bare interaction $V$ replaced by
the corresponding $G$-matrix. To be more precise, the Brueckner-Hartree-Fock
(BHF) definition of $U$ is given by 
\be
<\alpha \vert U \vert \beta> = \cases{ \sum_{\nu \le F} <\alpha \nu \vert 
\frac{1} {2} \left(
G(\omega_{\alpha \nu}) + G(\omega_{\beta \nu}) \right) \vert
\beta \nu >, & if $\alpha$ and $\beta$ $\le F$ \cr 
\sum_{\nu \le F} <\alpha \nu \vert
G(\omega_{\alpha \nu}) \vert \beta \nu >, & if $\alpha\le F$ and $\beta > F$
\cr 0 & if $\alpha$ and $\beta$ $>F$, \cr}\, . \label{eq:ubhf}
\ee
In this definition $\alpha \le F$ refers to single-particle states $\alpha$
below the Fermi surface and the starting energies in calculating $G$ are defined
by $\omega_{\alpha h} = \epsilon_\alpha + \epsilon_h$ using the single-particle
energies
\be 
\epsilon_\alpha = <\alpha \vert t_{kin} + U\vert \alpha > \,.\label{eq:epsbhf}
\ee

For matrix elements of $U$ involving hole states states, 
i.e. $<\alpha \vert U 
\vert \beta>$ with $\alpha$ and/or $\beta$ $\le F$, it can be shown by a
theorem of Bethe, Brandow and Petschek (BBP)\cite{BBP}, that the 
on-shell definition of the starting energy in the BHF choice for $U$ 
(\ref{eq:ubhf}) yields an exact cancellation of many diagrams
of higher order in $G$. The BBP theorem cannot be applied to the 
particle-particle matrix elements of $U$ ($\alpha$ and $\beta\ >F$). Therefore
the choice $U=0$ for particle states, the so-called conventional choice, 
has been favored in many BHF calculations.

Looking at the Eqs.~(\ref{eq:betheg}), (\ref{eq:ubhf}) and (\ref{eq:epsbhf}),
which define the BHF approach, one can see that these equations request the
solution of a self-consistency problem: in order to solve the Bethe-Goldstone
equation (\ref{eq:betheg}) one has to know the single-particle states $\vert
\alpha>$ and single-particle energies $\epsilon_\alpha$, to define the Pauli
operator $Q$ and starting energies, respectively. On the other hand, one should
know the $G$-matrix already to determine the single-particle states and energies
by diagonalizing $t_{kin}+U$. This self-consistency requirement goes beyond the
usual self-consistency requirement, which request the knowledge of the hole
states to define $U$. A self-consistent solution of the BHF equations can be
obtained by solving the Bethe-Goldstone equation and the Hartree-Fock equation
in an iteration scheme. After the self-consistency has been achieved, it is easy
to calculate the total energy by
\be
E_{BHF} = \frac{1}{2} \sum_{\nu \le F} t_\nu + \epsilon_\nu \; , \label{eq:ebhf}
\ee
the sum of single-particle energies $\epsilon_\nu$ and the expectation value of
the kinetic energy $t_\nu$ for the single-particle wave functions of the hole
states. Goldstone diagrams representing this approximation for the energy are
displayed in Fig.~\ref{fig5}. Results for other observables like e.g.~the 
radius of the mass or charge-distribution are usually determined by 
simply calculating
the expectation value of the corresponding operator for the model space
Slater determinant $\Phi$. Strictly speaking one should derive an effective
operator also for such observables, which account for the restriction of the
complete Hilbert space to the model space.

The optimal choice of $H_0$ to be used for the particle states in the BHF
definition of $U$, and consequently also the $h_{12}$ in the Bethe-Goldstone 
equation, has widely been discussed in particular for the case of nuclear
matter calculations \cite{maha1,tomnm}.  Good arguments have been presented to
favor a single-particle spectrum which is continuous at the Fermi surface and
therefore contains an attractive potential for the low-lying particle states
\cite{tomnm} or all particle states \cite{maha1}. The answer to the questions,
which is the optimal choice, can only be obtained by evaluating the
contributions of higher order correction and their sensitivity to the choice of
$U$.

So before we can answer the question, how to choose $U$ in an optimal way, we
should discuss, how to go beyond the BHF approximation, which is also often
called the lowest order Brueckner theory. The total energy is calculated in the
BHF approximation (see Eq.~(\ref{eq:ebhf})) by taking into account the contribution
of all ladder diagrams with any number of intermediate two-particle states,
which are summed up in the $G$-matrix. This means that all diagrams with two
hole lines are taken into account. The first step beyond this two-hole line 
approach would be to include the contributions of all linked diagrams with three
hole lines. This ordering with respect to the number of hole lines, which is the
basic assumption of the hole line expansion, can be justified with the following
argument: Linked diagrams including $n$ hole lines describe processes, in which
$n$ nucleons are interacting in a coherent way. This means that all $n$
nucleons should be within a volume that they can all interact with each other.
If the range of the strong interaction $r_V$ is smaller than the average
distance to the next neighbor $d_N$, the probability that $n+1$ nucleons are
found within a volume of mutual interaction is smaller than the corresponding
probability for $n$ nucleons by a factor $(r_V/d_N)^3$. In nuclear matter around
saturation density, the average  distance to the next neighbor $d_N$ is around
1.8 fm, which is larger than range of the strong short-range components of the
NN interaction (the radius of a hard core is around 0.4 fm, the range for the
exchange of an $\omega$ meson, $\hbar /\mu c$, corresponds to 0.26 fm), 
however, comparable to the range of the pion exchange (1.45 fm). It is obvious
that this hole line expansion should work at small densities, but may fail at
large densities, at which the assumption, $r_V$ small compared to $d_N$, is not
justified. The convergence of the hole-line expansion will be discussed in
section 3 below.

\begin{figure}[tb]
\begin{center}
\begin{minipage}[t]{13 cm}
\epsfig{file=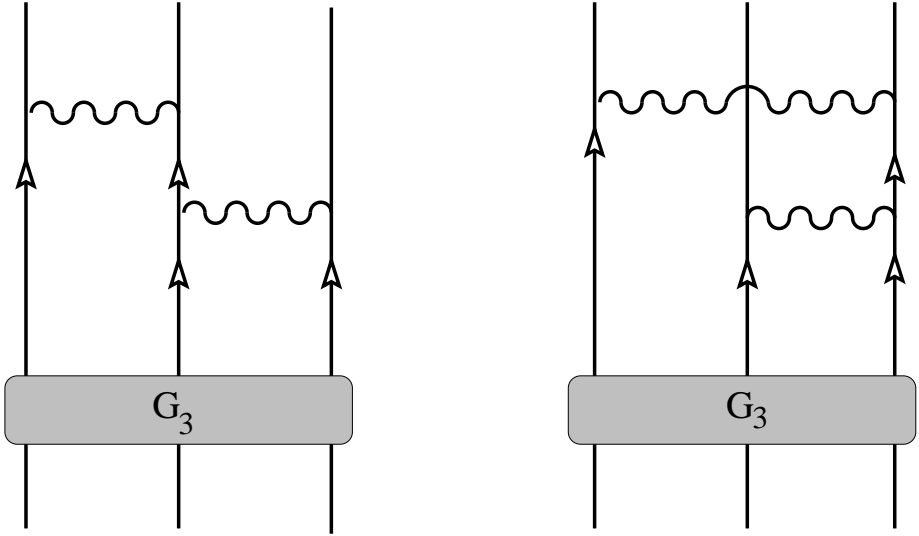,scale=0.7}
\end{minipage}
\begin{minipage}[t]{16.5 cm}
\caption{Three-particle ladder diagrams to be included in the Bethe-Fadeev
equation. \label{fig6}}
\end{minipage}
\end{center}
\end{figure}

The inclusion of all two-hole line contributions requires the solution of the
two-body problem in the nuclear medium, as expressed by the Bethe-Goldstone
equation. For the inclusion of all three-hole lines one has to solve the
three-body problem in the nuclear medium, which corresponds to solving the
Bethe-Fadeev equations\cite{rajar} (see also diagrams in Fig.~\ref{fig6}). 
Techniques and details how to solve the
Bethe-Fadeev equations have been described by Day\cite{day81}. Numerical
solutions for the three-hole line contributions in nuclear matter have recently
been presented by Song et al.\cite{song1,song2}.

Various techniques have been developed to solve the BHF equations, i.e.~the
Bethe-Goldstone equation in particular for nuclear matter and finite nuclei. As
an example for a standard solution in infinite nuclear matter we would like to
refer to the work of Haftel and Tabakin\cite{haftab}. A description, how to
solve the Bethe-Goldstone equation for finite systems, including a FORTRAN
program, can be found in \cite{sauer}.  

\subsection{\it Coupled Cluster Method\label{sec:ccm}}
A very detailed description of the coupled cluster Method (CCM), which has been 
introduced more than 40 years ago by Coester and K\"ummel\cite{coes0,coes1} 
can be found in \cite{kuem}.
More recently a review on this approach has been given by Bishop \cite{bish}.
Also in the CCM or ``Exponential S'' approach one starts assuming 
an appropriate Slater determinant $\Phi$ as a first approximation for the exact
eigenstate of $\Psi$ for the $A$-particle system. If the overlap $<\Psi \vert
\Phi>$ is different from zero, one can always consider the  ``Exponential S''
ansatz  
\be
\Psi = e^S \Phi \label{eq:esansatz}
\ee
with $S$ an operator of the form
\be 
S = \sum_{n=1}^A S_n \label{eq:sumsn}
\ee      
where $S_n$ is an $n$-particle operator which can be written
\be
S_n = \frac{1}{(n!)^2}\sum_{\nu_1\dots\nu_n\atop \rho_1 \dots\rho_n}
<\rho_1 \dots\rho_n \vert S_n \vert \nu_1\dots\nu_n > a^\dagger_{\rho_1} \dots
a^\dagger_{\rho_n} a_{\nu_n} \dots a_{\nu_1} \label{eq:sndef}
\ee
Here and in the following $a^\dagger_\alpha$ ($a_\alpha$) stand for creation
(annihilation) operators for nucleons in a state $\alpha$ with $\alpha$ a
single-particle state of the basis, which also defines the Slater determinant
$\Phi$. We will use labels $\nu_1\dots\nu_n$ to refer to hole states,
i.e.~single particle states which are occupied in the model state $\Phi$ and
indices $\rho_1 \dots\rho_n$ to denote particle states, i.e.~states above the
Fermi surface of $\Phi$. Expanding the exponential in (\ref{eq:esansatz}) one
obtains
\be
\Psi = \left( 1 + S_1 + \frac{1}{2} S_1^2 + S_2 + \frac{1}{6} S_1^3 + S_2 S_1 +
  S_3 \dots \right) \Phi \, . 
\label{eq:esexpand}
\ee
So exact eigenstates of the hamiltonian $\Psi $ is written as a sum of the
reference state $\Phi$, one-particle on-hole ($S_1$) excitations relative to
$\Phi$, two-particle two-hole ($1/2 S_1^2 + S_2$) and so on up to $A$-particle
$A$-hole excitations. Therefore it is obvious that the ansatz
(\ref{eq:esansatz}) is sufficient, but 
one may ask the question, why we are using
the exponential form of the ansatz and not simply write
\bea
\Psi & = &\left( 1 + F_1 + F_2 + F_3 \dots \right) \Phi \nonumber\\
& = & \left( 1 + F\right) \Phi
\label{eq:cim}\,,
\eea
where $F_n$ contains all $n$-particle $n$-hole contributions, which implies
\bea
F_1 & = & S_1 \nonumber \\
F_2 & = & \frac{1}{2} S_1^2 + S_2 \nonumber \\
F_3 & = &\frac{1}{6} S_1^3 + S_2 S_1 + S_3 \label{eq:fands}
\eea   
If the two expansions are treated including all terms up to $n=A$ both
expansions lead to the exact result. The question is, which approach is more
appropriate if one has to truncate the expansion and consider terms $F_i$ or
$S_i$ only up to an order $i\le n$ with $n$ smaller than the total particle
number $A$.

A first answer on the question why the ``exponential S'' ansatz
(\ref{eq:esansatz}) is preferable to the parametrisation (\ref{eq:cim}) 
can be given by Thouless theorem\cite{thoules}: If we restrict the 
``exponential S'' ansatz including only terms up to $n=1$, the Thouless theorem
says that the ansatz (\ref{eq:esansatz}) includes all Slater determinants which
are not orthogonal to $\Phi$. Therefore the solution of the CCM method in the
$S_1$ approximation would correspond to the Hartree-Fock approach. On the other
hand, however, the $F_1$ approach, i.e.~ignore all contributions of $F_i$ with
$i=2\dots A$ in  (\ref{eq:cim}), is more restrictive and more sensitive to the
choice of the initial state $\Phi$.

A more convincing argument, however, can be found by considering the fact that
e.g.~$F_2$ will contain two-particle two-hole excitation which are unrelated and
completely independent from each other. Just from statistical arguments it is
clear that such unlinked contributions will become more and more important if
the size of the system, or the particle number gets large. This problem can be
seen by inspecting the set of equations, which determine the amplitudes $F_i$. 
To derive these equations one considers the Schr\"odinger equation
\be 
H \vert \Psi > = E \vert\Psi> \label{eq:schroed}
\ee
assumes the ansatz (\ref{eq:cim}) and projects from the left with $<\Phi \vert$,
$<\Phi \vert a^\dagger_\nu a_\rho $, $<\Phi \vert a^\dagger_{\nu_1}
a^\dagger_{\nu_2} a_{\rho_2} a_{\rho_1}$ and so on. This yields
\bea
E & = & <\Phi \vert H (1+F) \vert \Phi >\nonumber \\
E <\rho \vert F_1 \vert \nu > & = & <\Phi \vert a^\dagger_\nu a_\rho H (1+F) 
\vert \Phi >\nonumber \\
E <\rho_1 \rho_2\vert F_2 \vert \nu_1 \nu_2> & = & <\Phi \vert a^\dagger_{\nu_2} 
a_{\rho_2} a_{\rho_1}H (1+F)\vert \Phi > \label{eq:fampl}
\eea
One finds that all left-hand sides of these equations contain the energy, which
is an extensive quantity, i.e.~it grows proportional to the particle number $A$.
Therefore also the right-hand side of these equations must be extensive, which
is a reflection of the fact that these equations contain unlinked terms.

The ``exponential S'' ansatz provides a way out of this problem. This can be
seen already from Eqs.~(\ref{eq:fands}), in which e.g.~$F_2$ is rewritten in
terms of unlinked contributions $S_1^2$ and a linked term $S_2$. To demonstrate
this feature we consider the equations which determine the amplitudes $S_n$. To
do this we consider the Schr\"odinger equation in the form
\be
e^{-S} H e^S \vert \Phi > = e^{-S} E \vert \Psi > = E \vert \Phi > 
\label{eq:schroed1}
\ee 
and project it from the left with $<\Phi \vert$, which yields
\bea
E & = & <\Phi \vert e^{-S} H e^S \vert \Phi > = <\Phi \vert H e^S \vert \Phi >
\nonumber \\
& = & <\Phi \vert H\left( 1+S_1+\frac{1}{2} S_1^2 + S_2\right) \vert \Phi >
\label{eq:senerg}
\eea
Note that in the first line we have used the fact $S^\dagger \vert \Phi>=0$,
which is obvious from the definitions of the $S_n$ in (\ref{eq:sndef}). The other
equations are obtained by multiplying (\ref{eq:schroed1}) from the left with 
$<\Phi \vert a^\dagger_\nu a_\rho $, $<\Phi \vert a^\dagger_{\nu_1}
a^\dagger_{\nu_2} a_{\rho_2} a_{\rho_1}$ etc. which yields
\bea
<\Phi \vert a^\dagger_\nu a_\rho e^{-S} H e^S \vert \Phi > & = & 0 \nonumber \\
<\Phi \vert a^\dagger_{\nu_1}a^\dagger_{\nu_2} a_{\rho_2} a_{\rho_1}
e^{-S} H e^S \vert \Phi > & = & 0  \label{eq:expseq0}
\eea
and so on. Note that these equations do not contain the extensive quantity $E$
as the corresponding Eqs.~(\ref{eq:fampl}) for the $F_l$.

The fact, that the Eqs.~(\ref{eq:expseq0}) do not contain extensive quantities,
growing with $A$, is only a hint that the matrix elements contained in these
equations do not contain terms, which correspond to unlinked diagrams. In order
to proof this feature, one has to consider the evaluation of the matrix elements
more in detail. For that purpose we consider the expansion
\be
e^{-S} H e^S = H + \left[H,S\right] +
\frac{1}{2!}\left[\left[H,S\right],S\right] + 
\frac{1}{3!}\left[\left[\left[H,S\right],S\right] ,S\right]+ \dots
\label{eq:bakhaus}
\ee
As an example for such commutator expression let us consider the commutator
between the operator of kinetic energy with $S_1$
\bea
\left[t_{kin}, S_1\right] & = & \sum_{\alpha\beta\rho\nu} <\alpha \vert
t_{kin}\vert \beta> <\rho \vert S_1\vert \nu > \left[a^\dagger_\alpha a_\beta ,
a^\dagger_\rho a_\nu\right] \label{eq:commu1}\\
& = &
\sum_{\alpha\nu}\left\{\sum_\rho <\alpha \vert t_{kin}\vert \rho > <\rho \vert 
S_1\vert \nu > a^\dagger_\alpha a_\nu\right\} - \sum_{\beta\rho} \left\{\sum_\nu
<\nu \vert t_{kin}\vert \beta > <\rho \vert S_1\vert \nu > a^\dagger_\rho
a_\beta \right\} \nonumber
\eea
Each commutator removes one $a^\dagger$ and one $a$ operator from the operator
product, and links single-particle labels of two amplitudes together. Translated
into the language of diagrams such a link means that there is an internal line,
a single-particle propagator, connecting the operator symbols of $t_{kin}$ and
$S_1$. 
The components of $S$ commute with each other since they all contain creation
operator for particle states and annihilation operators for hole states in
$\Phi$. Therefore links produced by the commutators in (\ref{eq:bakhaus}) can
only occur between the hamiltonian $H$ and $S$. This implies that the expansion
in (\ref{eq:bakhaus}) is finite: if e.g.~the hamiltonian contains one- and
two-body operators only, there will only be up to four nested commutators on the
right hand side.  Furthermore we see that the left-hand sides of 
Eqs.~{\ref{eq:expseq0}) correspond to matrix elements of a linked operator
product between the unperturbed model ground-state $\Phi$ as ket- and 
$n$-particle $n$-hole states
 relative to $\Phi$ as  bra-state. 

In order to obtain explicit equations for the amplitudes $S_n$ one introduces
the so-called n-particle subsystem amplitudes $\Psi_n$ defined by
\be
<x_1 \dots x_n \vert \Psi_n \vert \nu_1 \dots \nu_n > = <\Phi \vert
a^\dagger_{\nu_1} \dots a^\dagger_{\nu_n} a_{x_n} \dots a_{x_1} \vert \Psi >\, .
\label{eq:defpsin}
\ee
So this subsystem amplitudes correspond to the overlap of the exact $A$-particle
wave function $\Psi$ with an $n$-particle $n$-hole excitation of the reference
state $\Phi$. Note, however, that the label $x_i$ does not necessarily refer to
quantum numbers of a single-particle state in $\Phi$, it may also refer to
coordinates of a nucleon in the usual space or in momentum space. As an example
we consider explicitly
\bea 
<x_1\vert \Psi_1 \vert \nu_1 > & = & <x_1\vert \nu_1 > + <x_1\vert S_1 \vert 
\nu_1 > \nonumber \\
<x_1x_2\vert \Psi_2 \vert \nu_1 \nu_2 > & = & {\cal A}\left\{<x_1\vert \Psi_1 
\vert \nu_1 > <x_2\vert \Psi_1 \vert \nu_2 >\right\} + <x_1x_2\vert S_2 \vert
\nu_1 \nu_2 > \label{eq:examp12}
\eea
If we consider $x_1$ to represent coordinate, we see that the single-particle 
amplitude contains the uncorrelated wave function of the hole state $<x_1\vert 
\nu_1 >$  and possible corrections defined by $S_1$. In a similar way $\Psi_2$
contains the anti-symmetrized (${\cal A}$ represents the operator of
antisymmetrisation of the indices $\nu_i$) product of the $\Psi_1$ plus
corrections due to  $S_2$ This would imply that $<x_1x_2\vert S_2 \vert\nu_1 
\nu_2 >$ plays a role similar to the defect function in (\ref{eq:gcorel2}).
For the three- and more-particle amplitudes it is convenient to introduce
also\cite{zab1}
\bea
< x_1x_2\rho \vert \chi_3^{(12)} \vert \nu_1 \nu_2 \nu_3 > & = & <x_1x_2\rho 
\vert \Psi_3 \vert \nu_1 \nu_2 \nu_3 > - {\cal A}\left\{<\rho \vert \Psi_1 
\vert \nu_3 > <x_1x_2\vert S_2 \vert\nu_1 \nu_2 >\right\}\nonumber \\
& = & {\cal A}\left\{<x_2 \vert \Psi_1 \vert \nu_2 > <x_1\rho\vert S_2
\vert\nu_1 \nu_3 >\right\} \nonumber \\
&& \qquad + {\cal A}\left\{<x_1 \vert \Psi_1 \vert \nu_1 > 
<x_2\rho\vert S_2\vert\nu_2 \nu_3 >\right\}\nonumber \\
&& \qquad + < x_1x_2\rho \vert S_3 \vert \nu_1 \nu_2 \nu_3 >\, .
\eea
Furthermore we can define a single-particle potential for hole states $\nu$ by
\be
<x \vert U \vert \nu > = \sum_{\nu ' \le F} <x \nu' \vert V \Psi_2 \vert \nu\nu '
>\, .\label{eq:uexps}
\ee
If we identify $\Psi_2$ with a correlated two-particle wave function like in
(\ref{eq:gcorel2}), the product $V\Psi_2$ plays the same role than the  $G$
matrix in the hole-line expansion and the single-particle potential of
(\ref{eq:uexps}) can directly be compared with the definition of the
single-particle potential  (\ref{eq:ubhf}) in the BHF approach.

Using all these definitions the one-particle, Hartree-Fock like equation,
determining $S_1$ or $\Psi_1$ can be written
\bea
<x_1 \vert t_{kin,1} \Psi_1 \vert \nu_1> + \sum_{\nu ' \le F}<x_1 \nu ' \vert
t_{kin,2}  S_2 \vert \nu_1\nu '> &&\nonumber \\ +  <x_1 \vert U \vert \nu_1 >  + 
\sum_{\nu\nu '\le F} <x_1 \nu \nu ' \vert V_{23} \chi_3^{(23)} \vert \nu_1  \nu
\nu ' >  & = &\sum_{\nu \le F} <x_1 \vert \Psi_1 \vert \nu > 
h_{\nu\nu_1}
\label{eq:s1gl}
\eea 
with $h_{\nu\nu_1}$ the matrix elements for single-particle energy between hole
states
\be
h_{\nu\nu_1} = <\nu \vert t_{kin,1} \Psi_1 \nu_1> + <\nu \vert U \vert \nu_1 >
\label{eq:hnunu1}
\ee
The indices $1,2$ in these equation shall denote that operators of kinetic
energy and the two-body interaction $V$ act on the corresponding single-particle
states in the ket-vector.

The corresponding equation for the two-body amplitudes, determining $S_2$ or
$\Psi_2$ can be written (omitting some corrections proportional to $S_1$)
\bea
<x_1 x_2 \vert Q(t_{kin,1}+t_{kin,2}) S_2 \vert \nu_1 \nu_2 > & - & \sum_{\nu\le
F} \left(<x_1 x_2 \vert S_2 \vert \nu \nu_2 > h_{\nu\nu_1}  + 
<x_1 x_2 \vert S_2 \vert \nu_1 \nu > h_{\nu\nu_2}\right)\nonumber \\
\quad +  <x_1 x_2 \vert Q 
V_{12} S_2\vert \nu_1 \nu_2 > 
& = & - <x_1 x_2 \vert V_{12} \vert \nu_1 \nu_2 > - <x_1 x_2 \vert S_2 P V_{12}
\Psi_2 \vert \nu_1 \nu_2 > \nonumber \\
&& \qquad - \sum_{\nu \le F} <x_1 x_2 \nu\vert Q V_{13}\chi_3^{(13)} + Q V_{23} 
\chi_3^{(23)}\vert \nu_1 \nu_2 \nu >\nonumber \\
&& \qquad - \frac{1}{2}\sum_{\nu\nu ' \le F}<x_1 x_2 \nu\nu '\vert Q V_{34}
\chi_4^{(34)}\vert \nu_1 \nu_2 \nu \nu '>\,. \label{eq:s2gl}
\eea
Here $Q$ refers again to the Pauli operator for two-particle states, which we
have defined already in (\ref{eq:pauli}). From these two equations
(\ref{eq:s1gl}) and (\ref{eq:s2gl}) we can observe some general features of the
problem to determine the amplitudes $S_n$. The different equations are coupled.
If we restrict our considerations to two-body interaction terms only, the n-body
equation, which should provide $S_n$ requires the knowledge of $S_{n+1}$ and
$S_{n+2}$. If the hamiltonian would also contain a three-body interaction,
$S_{n+3}$ would be needed as well. In order to solve the one-body equation 
(\ref{eq:s1gl}) which will determine
$S_1$ or $\Psi_1$, one should know the amplitudes $S_2$ and $S_3$, which are
hidden in $U$, $\Psi_2$ and $\chi_3$, respectively. For the two-body equation 
(\ref{eq:s2gl}) the information on $S_3$ and $S_4$ (contained in $\chi_4$) is
needed.

This implies that one has to truncate the hierarchy of this set of equations by
assuming in the so-called ``SUBn'' approximation that all amplitudes $S_i$ with
$i > n$ are assumed to be zero. As a first example we will consider the SUB2
approximation and compare it to the BHF approach discussed in the preceeding 
subsection. The SUB2 assumption implies that we ignore the last term on the
left-hand side of Eq.~(\ref{eq:s1gl}) ($\chi_3$) and the last two terms in 
Eq.~(\ref{eq:s2gl}). If for the moment we furthermore assume that 
\be
h_{\nu\mu} = \epsilon_\nu \delta_{\nu\mu}
\ee
is diagonal and neglect the matrix elements of $S_2 P V_{12}\Psi_2$ we can
rewrite Eq.~(\ref{eq:s2gl}) into
\be
Q \left(\epsilon_{\nu_1} + \epsilon_{\nu_2} - \left(t_{kin,1}+ t_{kin,2} \right)
\right) S_2 \vert \nu_1 \nu_2 > = Q V \Psi_2 \vert \nu_1 \nu_2 > 
\ee
which corresponds to 
\be
S_2 \vert \nu_1 \nu_2 > = \frac{Q}{\epsilon_{\nu_1} + \epsilon_{\nu_2} -
\left(t_{kin,1}+ t_{kin,2} \right)} V \Psi_2 \vert \nu_1 \nu_2 >\, .
\ee 
Comparing with Eq.~(\ref{eq:gcorel2}) we may identify
\bea
G \vert \phi_{\nu_1 \nu_2} > & \longleftrightarrow & V \Psi_2\vert \nu_1 \nu_2 >
\nonumber\\
\frac{Q}{\omega - h_{12}}G \vert \phi_{\nu_1 \nu_2} >  &\longleftrightarrow &
S_2 \vert \nu_1 \nu_2 >\,, \label{eq:comsbhf}
\eea
where the last line identifies the so-called defect functions in the two
approaches. With this identification  we also see that the single-particle
equation (\ref{eq:s1gl}) corresponds to the single-particle equation if we
ignore the term $t_{kin,2}  S_2$ in that equation.

\begin{figure}[tb]
\begin{center}
\begin{minipage}[t]{9 cm}
\epsfig{file=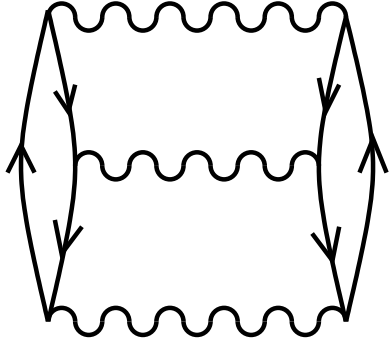,scale=0.9}
\end{minipage}
\begin{minipage}[t]{16.5 cm}
\caption{Hole-hole ladder diagram included in CCM SUB2 approximation
 \label{fig7}}
\end{minipage}
\end{center}
\end{figure}

The main difference between the Coupled Cluster Method (CCM) SUB2 approach and
the BHF approximation is the inclusion of the $S_2 P V_{12}\Psi_2$ in the
two-body Eq.~(\ref{eq:s2gl}) as compared to the Bethe-Goldstone equation
(\ref{eq:betheg}). In terms of diagrams this means that the energy calculated in
the CCM SUB2 approach also includes hole-hole scattering diagrams like the one
displayed in Fig.~\ref{fig7}. This means that the hole-hole ladders are treated
on the same level as the particle-particle ladders. This should be of particular
importance if the phase space of hole-hole states, which can be reached by the
two-body interaction is as large as the corresponding space of particle-particle
configurations. As we will see in the discussion of results in chapter 3, for
realistic interactions in nuclear physics, the particle-particle ladders are
dominant as compared to the hole-hole ladders. Therefore the differences between
the results of CCM SUB2 and BHF calculations are usually small. Note that
particle-particle and hole-hole ladders are also often referred to as
particle-particle hole-hole ring diagrams, which are included in a  
pphh RPA calculation\cite{song,jiang,ellis}

Within the framework of the CCM approach one does not have any choice for the
auxiliary potential $U$ as it was the case in the BHF approach. From the
comparison of the two approaches discussed above (see Eq.~(\ref{eq:comsbhf})) one
finds that the CCM SUB2 approach is rather close to the BHF assuming the
conventional choice for the particle state-spectrum in the Bethe-Goldstone
equation, which means that $h_{12}$ is replace by the sum of the kinetic 
energies. All corrections to the CCM SUB2 approach occur due to the inclusion of
the three-body and higher order terms. 

In order to discuss the effects of the three-body amplitude on the calculated
energies, we write the equation for $\chi_3$\cite{zab2}
\bea 
<x_1 x_2 x_3 \vert \chi_3^{(12)} \vert \nu_1 \nu_2 \nu_3 > & = & {\cal A}\bigl\{
<x_3 x_1 \vert S_2 \vert \nu_2 \nu_1 > <x_2 \vert \Psi_1 \vert \nu_2 >
 \nonumber \\ &&\quad
+ <x_2 x_3 \vert S_2 \vert \nu_2 \nu_3 > <x_1 \vert \Psi_1 \vert \nu_1 >
\bigr\} \nonumber \\
&& \quad + \sum_{y_i}<x_1 x_2 x_3 \vert \frac{Q_3}{e_3} \vert y_2 y_3 y_1 ><
y_1 y_2 y_3 \vert G_{12} \chi_3^{(12)}\vert \nu_1 \nu_2 \nu_3 > \nonumber \\
&& \quad + \sum_{y_i}<x_1 x_2 x_3 \vert \frac{Q_3}{e_3} \vert y_2 y_3 y_1 ><
y_3 y_1 y_2 \vert G_{12} \chi_3^{(12)}\vert \nu_1 \nu_2 \nu_3 > \, .
\label{eq:s3gl}
\eea 
In this equation we have identified $V\Psi_2$ with $G$ according to 
(\ref{eq:comsbhf}). $Q_3$ denotes the Pauli operator for three-particle states
and the energy denominator $e_3$ is defined by
\be
e_3 =\epsilon_{\nu_1} +   \epsilon_{\nu_2} + \epsilon_{\nu_3} - \left( t_{kin1}
+ t_{kin2} + t_{kin3}\right)\, .
\ee
The resulting amplitudes $\chi_3$ multiplied with $V$ enter into the one-body
(\ref{eq:s1gl}) as well as two-body Eq.~(\ref{eq:s2gl}). The contribution to the
binding energy $E$ originating from these contributions is of third and higher
order in $G$. The Goldstone diagrams representing the contributions of third
order (omitting exchange diagrams) are displayed in Fig.~\ref{fig8}. The diagram
of Fig.~\ref{fig8}a is a particle-hole ring diagram. The summation of all ph
ring diagram contributions can be obtained by calculating the correlation energy
arising from particle-hole RPA calculations. The diagram displayed in 
Fig.~\ref{fig8}b is a contribution, which within the BHF approach one would try
to cancel by an appropriate definition of the single-particle spectrum $U$ for
particle states. This indicates again that $U$ should be chosen to minimize the
effects of three- and four-body terms.

\begin{figure}[tb]
\begin{center}
\begin{minipage}[t]{12 cm}
\epsfig{file=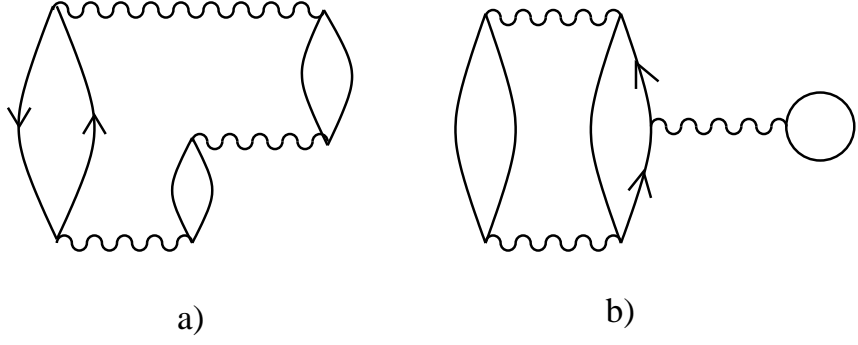,scale=0.5}
\end{minipage}
\begin{minipage}[t]{16.5 cm}
\caption{Lowest order contributions originating from the inclusion of three-body
terms in the CCM. The diagram shown in the left part (a) is a particle-hole 
ring  diagram, while the one displayed in (b) contains a bubble insertion into a
particle line, which one would like to compensate by an appropriate choice of the
single-particle potential for particle states.\label{fig8}}
\end{minipage}
\end{center}
\end{figure}

If one would like to evaluate the amplitudes $<x_1 \dots x_n\vert S_n \vert
\nu_1 \dots \nu_n >$ in the real space representation, which means that $x_i$
refer to the usual coordinates of the nucleons, the basic equations
(\ref{eq:s1gl}) and (\ref{eq:s2gl}) must be considered as differential 
equations. This is particularly useful if the NN interaction $V$ is described
in terms of local potential terms\cite{zab1,zab2}. For non-local potentials it
is me more convenient to use momentum space representation. In this case the
equations lead to inhomogeneous integral equations\cite{stauf}. The CCM equations
have also been solved in a Hilbert space, which is spanned by a basis of
appropriate oscillator functions\cite{heis1}, which leads to a solution of
a coupled system of nonlinear equations.

If the amplitude $S_i$ have been determined, it is easy to evaluate the energy
according to (\ref{eq:senerg}). Knowing $S_i$ we also have the information on
the exact wave function (not normalized). An efficient evaluation of observables
different from the energy can be done using the following considerations. As an
example we consider the single-particle density matrix  
\be
d_{\beta\alpha} = \frac{< \Psi \vert a^\dagger_\beta a_\alpha \vert \Psi >}{<\Psi 
 \vert \Psi >} \label{eq:defsrho1}
\ee
One can rewrite this expression into
\bea
d_{\beta\alpha} & = & \frac{< \Phi \vert e^{S^\dagger}
a^\dagger_\beta a_\alpha e^S\vert 
\Phi >}{<\Phi \vert e^{S^\dagger}e^S \vert \Phi >} \nonumber \\
& = & \frac{< \Phi \vert 
e^{S^\dagger}e^S e^{-S} a^\dagger_\beta a_\alpha e^S\vert 
\Phi >}{<\Phi \vert e^{S^\dagger}e^S \vert \Phi >} \nonumber \\
& = & <\Phi \vert e^{-S} a^\dagger_\beta a_\alpha e^S\vert \Phi > + \sum_n 
\frac{< \Phi \vert e^{S^\dagger}e^S X_n^\dagger \vert\Phi > < \Phi\vert X_ne^{-S} 
a^\dagger_\beta a_\alpha e^S\vert  \Phi >}{<\Phi \vert e^{S^\dagger}e^S \vert 
\Phi >}\label{eq:sd1eq1}
\eea
The last line has been obtained by inserting the unity operator
\be
\vert \Phi ><\Phi \vert + \sum_n X_n^\dagger \vert\Phi > < Phi\vert X_n
\ee
with the n-particle n-hole operator
\be
X_n^\dagger = \frac{1}{(n!)^2} \sum_{\rho_i >F, \nu_i \le F} a_{\rho_1}^\dagger
\dots a_{\rho_n}^\dagger a_{\nu_n} \dots a_{\nu_1}\, . \label{eq:defxn}
\ee
Since $X_n^\dagger$ commutes with $S$, the expression (\ref{eq:sd1eq1}) can be
rewritten into
\be
d_{\beta\alpha} = <\Phi \vert e^{-S} a^\dagger_\beta a_\alpha e^S\vert \Phi > 
+ \sum_n <\Phi\vert X_ne^{-S}a^\dagger_\beta a_\alpha e^S\vert  \Phi >
\frac{<\Psi \vert X_n^\dagger \vert \Psi >}{<\Psi \vert \Psi >}\, .
\label{eq:sd1eq2}
\ee
The matrix elements of operators like $e^{-S} a^\dagger_\beta a_\alpha e^S$ can
be calculated employing an expansion similar to the expansion in 
(\ref{eq:bakhaus}). The factors of the form $<\Psi \vert X_n^\dagger \vert \Psi
>/<\Psi \vert \Psi >$, on the other hand correspond to the matrix elements of
the $n$-body density operator, which brings us back to the starting point in
(\ref{eq:defsrho1}). This means that the final matrix elements can be calculated
in an iterative scheme\cite{heis1,emrich}.

\subsection{\it Many-Body Theory in Terms of Green's Functions
\label{subsec:green}}

The two-body approaches discussed so far, the hole-line expansion as well as the
CCM, are essentially restricted to the evaluation of ground-state properties.
The Green's function approach, which  will 
shortly be introduced in this section
also yields results for dynamic properties like e.g.~the single-particle
spectral function which is closely related to the cross section of particle
knock-out and pick-up reactions. It is based on the time-dependent
perturbation expansion and also assumes a separation of the total hamiltonian
into an single-particle part $H_0$ and a perturbation $H_1$ as introduced in
(\ref{eq:h01}). A more detailed description can be found e.g.~in the textbooks
of Fetter and Walecka\cite{fetwal}, Negele and Orland\cite{negor} or in the book
by Mattuck\cite{mattuk}, which particularly provides a rather intuitive
interpretation of the Feynman diagrams. A very comprehensive description of the
main features has been presented by Mahaux and Sartor\cite{mahau1}. 
Introductions are also given in various
review articles\cite{wim1,ruder}.

The expectation value of any operator $ O$ is calculated  in the so-called 
interaction picture as
\be
< \Psi^{\rm I} (t) \vert O_{\rm I}(t) \vert \Psi^{\rm I} (t)
>\, ,
\ee
where the time-dependent operator $O_{\rm I}(t)$ in the interaction picture is
related to the operator $O$ in the usual Schr\"odinger picture by
 \be
O_{\rm I}(t) = e^{i H_{0}t} O e^{-i H_{0}t} \, .
\label{eq:operi}
\ee
and the time-dependence of the state $\vert \Psi^{\rm I} (t)>$ can be
derived from
\be
i\frac{\partial}{\partial t} \vert \Psi^{\rm I} (t) > = 
H_{1{\rm I}}(t) \vert \Psi^I (t) > \, .\label{eq:moti}  
\ee
If we introduce the evolution operator in the interaction scheme $U_{\rm
I}(t,t_{0})$ by
\be
\vert \Psi^{\rm I} (t) > = U_{\rm I}(t,t_{0})\vert \Psi^{\rm I} (t_0) 
>\,,
\ee
the equation of motion for the state $\vert \Psi^{\rm I}>$ (\ref{eq:moti})
can be rewritten into a differential equation for the evolution operator
\be
i\frac{\partial}{\partial t} U_{\rm I}(t,t_{0}) =  H_{1{\rm I}}(t)
U_{\rm I}(t,t_{0})
\, .
\ee
This equation can be transformed into an integral equation
\be
U_{\rm I}(t,t_{0})  =  U_{\rm I}(t_{0},t_{0}) - i \int_{t_{0}}^t d t_{1} \,
H_{1{\rm I}} (t_{1}) U_{\rm I}(t_{1},t_{0}) \, .
\ee
Using the fact that $U_{\rm I}(t_{0},t_{0})$ is identical to the unit
operator $\hat{1}$, we can iterate this integral equation to a
perturbative expansion in powers of $H_{1{\rm I}}$:
\bea
U_{\rm I}(t,t_{0}) & = & \hat{1} + (-i )\int_{t_{0}}^t d t_{1} \,
H_{1{\rm I}}(t_{1}) \nn\\ &&\quad +
(-i )^2 \int_{t_{0}}^t d t_{1} \, H_{1{\rm I}}(t_{1}) \int_{t_{0}}^{t_{1}} d
t_{2} \,
H_{1{\rm I}}(t_{2}) + \dots
\label{eq:uexpa1}
\eea
The integration variables used on the left-hand side are nested in a
rather inconvenient way. Therefore one rewrites the term of order $n$ in
this expansion using
\be
\int_{t_{0}}^t d t_{1}\,H_{1{\rm I}}(t_{1}) \dots \int_{t_{0}}^{t_{n-1}} d t_{n}
\, H_{1{\rm I}}(t_{n})  =  \frac{1}{n!} \int_{t_{0}}^t d t_{1} \dots
\int_{t_{0}}^{t}
 t_{n} 
 \,{\cal T}\left( H_{1{\rm I}}(t_{1})\dots H_{1{\rm I}}(t_{n}) \right)
\label{eq:timeo}
\ee
with the time ordering or chronological operator ${\cal T}$, which is
defined for two operators by
\be
{\cal T}\left( A(t_{1}) B(t_{2})\right) = \cases{ A(t_{1})B(t_{2}), & if
$t_{1}\ge t_{2}$, \cr (-1)^m B(t_{2}) A(t_{1}), &
otherwise. }
\label{eq:chrono}
\ee
Here $m$ is the number of exchanges of fermion creation and annihilation
operators
contained in $A$ and $B$, which are needed to bring $A$ and $B$ into
chronological order. Note that for our present purpose
(\ref{eq:timeo}) the factor $(-1)^m$ is always equal to $1$, as the number
of fermion operators defining $H_{1{\rm I}}$ is even.
The definition of ${\cal T}$ in
(\ref{eq:chrono}) for two operators is easily extended to $n$
operators. Applying (\ref{eq:timeo}) to the expansion of the time
evolution operator in (\ref{eq:uexpa1}), one gets
\be
U_{\rm I}(t,t_{0}) = \sum_{n=0}^{\infty} \frac{(-i)^n}{n!}
\int_{t_{0}}^t d t_{1} \dots \int_{t_{0}}^{t}
d t_{n} \,{\cal T}\left( H_{1{\rm I}}(t_{1})\dots H_{1{\rm I}}(t_{n}) \right)
\label{eq:uexpa2}   
\ee
In order to arrive at a perturbation expansion for the calculation of
matrix elements, one assumes that the perturbation $H_1$ is ``switched
off'' at times $t=-\infty$ and $t=+\infty$ and can be switched on in an
adiabatic way for times $t\approx 0$. This can be achieved by a
time-dependent hamiltonian of the form
\be
H_{\alpha}(t) = H_{0} + e^{-\alpha \vert t \vert} H_1
\label{eq:halpha}
\ee
where $\alpha$ is a small positive number that becomes infinitesimal
in the adiabatic limit. This procedure implies that the eigenstates of
the hamiltonian $H_{\alpha}$ are identical to the eigenstates of the
unperturbed hamiltonian $H_{0}$ at times $\vert t \vert = \infty$ and
should evolve to the corresponding eigenstates of the total $H$ at $t
\approx 0$ if we use the time evolution operator $U_{{\rm I}\alpha}(t,t')$ for
the hamiltonian $H_{\alpha}$. If we denote the (nondegenerate) ground state
of $H_{0}$ by $\Phi_{0}$, which is independent of time in the
interaction picture, this means that we obtain an eigenstate of the
exact $H$ at time $t=0$ by
\be
\vert \Psi_{0} (t=0) > = \lim_{\alpha \to 0}
U_{{\rm I}\alpha}(0,-\infty) \vert \Phi_{0} > \, .
\ee
It has been shown by Gell-Mann and Low\cite{gellow} that $\Psi_{0}$ is indeed
an exact eigenstate of $H$ if the perturbation expansion converges.
 
In order to calculate matrix elements we now consider the Heisenberg scheme. In
this  representation,
 which corresponds to the interaction scheme with $H_0=H$, the wave 
function is time-independent and identical to $\Psi_{0}(t=0)$ and the 
time-dependence is completely assigned
 to the operators $O_H(t)$ which are
defined as in (\ref{eq:operi}), replacing $H_o$ by $H$. 
This means that a matrix element of an operator $O$ can be calculated as
\bea
\frac{< \Psi_{0}\vert O_H(t) \vert \Psi_{0}>}
{< \Psi_{0}\vert\Psi_{0}>} & = & 
\lim_{\alpha \to 0}
\frac{< \Phi_{0}\vert U_{{\rm I}\alpha}(\infty ,0) O_{\rm H}(t)
U_{{\rm I}\alpha}(0,-\infty)\vert \Phi_{0}>}
{< \Phi_{0}\vert U_{{\rm I}\alpha}(\infty ,0)
U_{{\rm I}\alpha}(0,-\infty)\vert\Phi_{0}>}   \nn \\
& = & 
\lim_{\alpha \to 0}
\frac{< \Phi_{0}\vert U_{{\rm I}\alpha}(\infty ,t) O_{\rm I}(t)
U_{{\rm I}\alpha}(t,-\infty)\vert \Phi_{0}>}
{< \Phi_{0}\vert U_{{\rm I}\alpha}(\infty ,-\infty)\vert\Phi_{0}>}
\label{eq:matel1}\, .
\eea
Using the explicit representation of the time evolution operator in
(\ref{eq:uexpa2}) one can furthermore show (see, e.g.\cite{fetwal}) 
that the matrix element for any time-ordered product of
two Heisenberg operators can be calculated as
\bea
\frac{< \Psi_{0}\vert {\cal T}\left(A_{\rm H}(t)
B_{\rm H}(t')\right)
\vert \Psi_{0}>} {< \Psi_{0}\vert\Psi_{0}>}
&  = &  \lim_{\alpha \to 0}\Biggl[\frac{1}{< \Psi_{0}
\vert\Psi_{0}>}
\sum_{n=0}^\infty \frac{(-i)^n}{n!}\int_{-\infty}^\infty d t_{1}\dots
\int_{-\infty}^\infty d t_{n}\
 e^{-\alpha(\vert t_{1}\vert +
\dots \vert
t_{n}\vert )}\nn\\ &&\quad\times
 < \Phi_{0}\vert {\cal T}\left(V_{\rm I}(t_{1})
\dots V_{\rm I}(t_{n})A_{\rm I}(t)B_{\rm I}(t')\right)\vert\Phi_{0}
> \Biggr]\, ,\label{eq:timatel}
\eea
which means that one has to evaluate matrix elements of time-ordered
products of operators in the interaction scheme for the ground state of  
the unperturbed hamiltonian $\Phi_{0}$.   

The operators in these matrix elements are time-ordered. Therefore we can apply
Wick's
 theorem\cite{wick} to evaluate them.  In the following we will shortly
review this scheme and illustrate
 how the various contributions are visualized in
terms of Feynman diagrams.
 
First we recall that the operator for the residual interaction $H_1$ as well
as  any other operator  can be expressed
in terms of the basic single-particle creation ($a^{\dagger}_{i}$) and
annihilation operators ($a_{i}$).
It is easy to verify that these operators in the interaction scheme are
given by 
\bea
a_{{\rm I}j}(t) & = & a_{j}\exp\left({-i\epsilon_{j}t}\right)\, ,\nn\\
a^\dagger_{{\rm I}j}(t) & = &
a^\dagger_{j}\exp\left({+i\epsilon_{j}t}\right) \, ,
\label{eq:creint}
\eea
with $\epsilon_{j}$ the single-particle energies defining the
unperturbed hamiltonian $H_{0}$  
The ground
state for this unperturbed hamiltonian, $\Phi_{0}$, is
given by a Slater determinant in which all single-particle states $i$
with an energy $\epsilon_{i}$ below the Fermi energy $\epsilon_{\rm F}$ are
occupied. The Fermi energy separates hole-states  ($\epsilon_{\nu} \le
\epsilon_{\rm F}$, occupied in the unperturbed ground state) from particle
states  ($\epsilon_{\rho} > \epsilon_{\rm F}$, unoccupied in
the unperturbed
ground state). If $M,N,O,P,\dots$ represent creation or annihilation
operators, in the Schr\"odinger or in the interaction picture, one
can define the normal product of such operators by
\be
{\cal N}\left(MNOP\dots \right) = (-1)^\gamma OP\dots MN \dots
\label{eq:normal}
\ee
where the sequence of operators on the right-hand side of this equation is
such that all creation operators for particle states ($a^\dagger_{\rho}$)
and annihilation operators for hole states ($a_{\nu}$) are moved to the
left ($O,P$), whereas all creation operators for hole states
($a^\dagger_{\nu}$)
and annihilation operators for particle states ($a_{\rho}$)
are moved to the
right ($M,N$) and $\gamma$ counts the number of exchanges of these
operators required to obtain the normal ordering. This normal ordering
guarantees that the matrix element of such an ordered product
calculated for the unperturbed state $\Phi_{0}$ vanishes:
\be
< \Phi_{0}\vert {\cal N}\left(MNOP\dots \right) \vert
\Phi_{0}> = 0 \, .\label{eq:norm0}
\ee
Furthermore we define a ``contraction'' of two operators, using the
chronological operator of (\ref{eq:chrono}) for two such operators
$M,N$ in the interaction scheme, as
\bea
\underbrace{MN} & = & < \Phi_{0}\vert {\cal T}\left(MN \right)
- {\cal N}\left(MN \right)\vert \Phi_{0}> \nn\\
& = & < \Phi_{0}\vert {\cal T}\left(MN \right) \vert \Phi_{0}>
\, , \label{eq:contr0}
\eea
which is just a complex number given by 
\be
\underbrace{a_{{\rm I}j}(t)a^\dagger_{{\rm I}k}(t')} =
\cases{\delta_{jk}e^{-i\epsilon_{j}(t-t')} &
if $j$ is a particle state and $t> t'$, \cr
-\delta_{jk}e^{-i\epsilon_{j}(t-t')} & if $j$ is a hole state
and $t' > t$, \cr 0 & otherwise,}\label{eq:contr1}
\ee
\be
\underbrace{a^\dagger_{{\rm I}k}(t')a_{{\rm I}j}(t)} = -
\underbrace{a_{{\rm I}j}(t)a^\dagger_{{\rm I}k}(t')}\, . \label{eq:contr2}
\ee 
Using Wick's
 theorem it is easy to show that matrix elements of time-ordered
products calculated for the unperturbed ground state $\Phi_0$ as in
(\ref{eq:timatel}) can be calculated as a sum of all terms in which the
single-particle operators are pairwise contracted, which means that each pair is
replaced by the corresponding value for the contraction (\ref {eq:contr1}).

The use of Feynman
diagrams provides an easy control of all the contributions. To
demonstrate this, we consider as an example the matrix element occurring
in the first-order term of (\ref{eq:timatel})
\be
\frac{1}{2} \sum_{i,j,k,l} < ij \vert V \vert kl >
< \Phi_{0}\vert {\cal T}\left( a^{\dagger}_{{\rm I}i}(t_{1})
a^{\dagger}_{{\rm I}j}(t_{1}) a_{{\rm I}l}(t_{1}) a_{{\rm I}
k}(t_{1}) a_{{\rm I}\alpha}(t)
a^\dagger_{{\rm I}\beta}(t')\right) \vert \Phi_{0}> \, ,
\label{eq:examp}
\ee
where we have made the substitution
$A_{\rm I}\to a_{{\rm I}\alpha}$, $B_{\rm I}\to
a^\dagger_{I\beta}$, and used the explicit representation of $V$. 
For this example we will furthermore
assume that $t' < t_{1} < t$. As before the operator of the residual 
interaction is
represented in the Feynman diagram by a horizontal dashed line, with
an outgoing and incoming
arrow at each end. The outgoing arrows refer to creation operators
contained in $V$ ($a^{\dagger}_{{\rm I}i}$, $a^{\dagger}_{{\rm I}j}$) in our
notation, and the incoming ones refer to annihilation operators. The
external operators are represented by a dot with an ingoing and an
outgoing arrow for the annihilation and creation operator, respectively.
These objects are displayed in Fig.~\ref{fig9}a. With the
assumption that the vertical axis represents a time axis the objects
are ordered according to the
choice of our example: $t' < t_{1} < t$.
 
\begin{figure}[tb]
\begin{center}
\begin{minipage}[t]{10 cm}
\epsfig{file=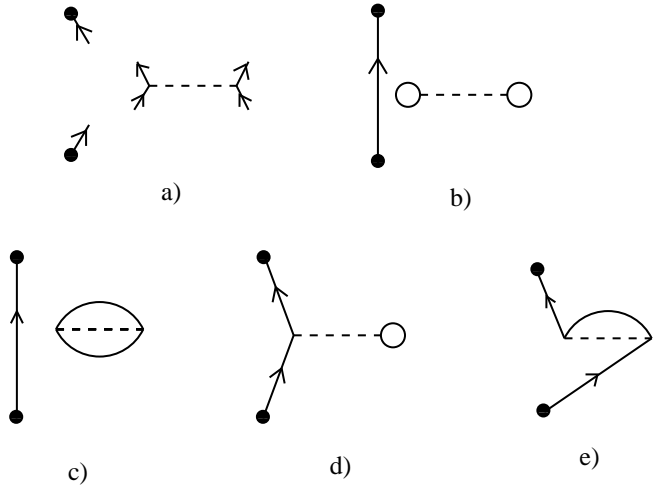,scale=0.4}
\end{minipage}
\begin{minipage}[t]{16.5 cm}
\caption{Building blocks for Feynman diagrams {\bf (a)} and completely
contracted contributions as discussed in the text.\label{fig9}}
\end{minipage}
\end{center}
\end{figure}  

Any contraction implies that two of the lines with arrows must be
paired. In the graphical representation this is achieved by connecting
them. Looking at the contractions which yield results different from
zero (see (\ref{eq:contr1})--(\ref{eq:contr2})), one finds that only
those pairs of lines in which the arrows point into
the same direction must be connected.
Furthermore we can distinguish connected lines with
an arrow pointing upwards, which refers to a particle state, i.e., the
corresponding summation indices in  (\ref{eq:examp}) can be
restricted to particle states, whereas connected lines with an arrow
pointing downwards refer to  hole lines.

All diagrams representing the completely contracted terms of the
expression shown in (\ref{eq:examp}) are displayed in
Fig.~\ref{fig9}b)--e). Note that we show only those diagrams that are
topologically distinct in the sense that a diagram obtained from
another one by just mirroring the ends of an interaction line is not
displayed again. Recalling Wick's theorem, it is evident that all
non-vanishing contributions to the perturbation
calculation of the $n$th-order term in the matrix
element of a time-ordered product of operators in (\ref{eq:timatel})
can be obtained in the following way.
\begin{itemize}
\item Draw $n$ interaction lines (dashed lines in Fig.~\ref{fig9})
and mark on the creation and annihilation operators for the external
operators to be calculated (dots in Fig.~\ref{fig9}).
\item Construct all diagrams by connecting the ``arrows'' linked to
these basic building blocks according to the rules given in
the example of Fig.~\ref{fig9}.
\item Keep in mind that all possible time orderings of the interaction
vertices relative to the external operators must be considered (see time
integrations in (\ref{eq:timatel})).
\item Each of the resulting diagrams represents a non-vanishing
contribution to the evaluation of (\ref{eq:timatel}) and there exist
well-established Feynman rules that translate the contribution of the
diagram into a calculable expression (see, e.g.\cite{mattuk}).
\item Consider only the contribution of linked diagrams. The linked cluster
theorem, which has already been discussed in sections \ref{sec:holeline} and
\ref{sec:ccm} also holds for this case (see e.g.\cite{fetwal}).
For the example displayed in Fig.~\ref{fig9} this implies that only the 
contributions of the linked diagrams displayed in Fig.~\ref{fig9}d) and e) have
to be taken into account.
\end{itemize}

Up to this point the diagrams have been used only as a kind of
book-keeping tool to identify all non-vanishing contributions in the
perturbation expansion. However, we have seen already that each
connecting line in those diagrams represents a contraction, and a line
with an arrow pointing upwards stands for [see (\ref{eq:contr0})]
\be
\underbrace{a_{{\rm I}j}(t)a^\dagger_{{\rm I}k}(t')} =
< \Phi_{0}\vert {\cal T}\left(
a_{{\rm I}j}(t)a^\dagger_{{\rm I}k}(t')\right)\vert \Phi_{0}> \, ,
\ee
with the creation of a particle taking place before the annihilation
($t>t'$). This means that we can ignore the operator ${\cal T}$ and
rewrite this contraction as
\bea
< \Phi_{0}\vert a_{{\rm I}j}(t)a^\dagger_{{\rm I}
k}(t')\vert \Phi_{0}>
& = & \sum_{\beta} < \Phi_{0}\vert a_{{\rm I}j}(t) \vert \beta >
< \beta \vert a^\dagger_{{\rm I}k}(t')\vert \Phi_{0}> \nn\\
& = & \delta_{jk}e^{-i\epsilon_{j}(t-t')} \quad
\mbox{for $j$ a particle
state}.
\eea
In the first line of this equation we have inserted a summation over a
complete set of states $\vert \beta > $ with one particle in
addition to the number of fermions in $\vert \Phi_{0}> $, in
order to show that this contraction describes the product of a
probability amplitude to create a particle at a time $t'$,
producing a state $\beta$ and the probability that it is
annihilated at the later time $t$ reproducing the unperturbed ground
state $\Phi_{0}$. In the second line of this equation we have copied
the result for the contraction from (\ref{eq:contr1}) that such a
propagation of a particle on top of the unperturbed state is only
possible if $j=k$ refers to a state above the Fermi energy, in order
not to violate the Pauli principle.

In a similar way one can convince oneself that a line with an arrow
pointing down represents the propagation of a hole state, i.e., a
particle must be removed first from a state $h$ below the Fermi energy
before it is put back at a later time. With this interpretation of the
contractions visualized in the diagrams one can easily interpret the
Feynman diagrams in terms of time-dependent processes.

The single-particle Green's function can be considered as a special
example of an expectation value for the time-ordered product of two
operators calculated for the exact ground-state in
(\ref{eq:timatel}). It is defined by
\be
ig(\alpha t,\beta t') = < \Psi_{0}\vert {\cal T}\left(a_{{\rm H}\alpha}(t)
a^\dagger_{{\rm H}\beta}(t')\right) \vert \Psi_{0}> \, .
\label{eq:greend}
\ee
Note that here and in the following we have dropped the denominator
$< \Psi_{0}\vert\Psi_{0}>$, assuming that the exact ground
state is properly normalized. The creation and annihilation operators
in the Heisenberg representation are defined in an appropriate basis,
characterized by quantum numbers $\alpha$ and $\beta$. If we assume that
$\alpha$ refers to a position $\vec {r'}$ and $\beta$ to a position
$\vec{r}$ of the considered fermion in $r$ space, the single-particle Green's
function $i g(\vec{r'}t',\vec{r} t)$ describes the propagation of this
fermion from the space-time point $(\vec{r} t)$ to $(\vec{r'} t')$. In
contrast to the discussion of the contractions in the previous section,
 in this case the propagation is
with respect to the exact ground
state and the complete hamiltonian.

For a system that is invariant under translation, such as the infinite
nuclear or neutron matter, which we want to consider, the appropriate
set of basis states is that of the momentum eigenstates; the Green's
function is diagonal in this representation. Rewriting the effect of
the chronological operator ${\cal T}$ defined in (\ref{eq:chrono})
in terms of  step functions $\Theta(t-t')$, we find that
the Green's function is
given as
\bea
i g( k,t-t') &=  &\Theta(t-t') 
< \Psi_{0}\vert a_{{\rm H}k}(t)
a^\dagger_{{\rm H}k}(t')\vert \Psi_{0}> -
\Theta(t'-t) < \Psi_{0}\vert a^\dagger_{{\rm H}k}(t')
a_{{\rm H}k}(t) \vert \Psi_{0}> \nn \\
& = & \Theta(t-t') \sum_{\gamma}
e^{-i (E_{\gamma}^{(A+1)}-E_{0}^A)(t-t')} \left|
<\Psi_{\gamma}^{(A+1)} \vert a^\dagger_{k}\vert \Psi_{0}>
\right|^2  \nn\\
&& \quad - \Theta(t'-t) \sum_{\delta} e^{-i (E_{0}^A -E_{\delta}^{(A-1)})(t-t')}
\left| <\Psi_{\delta}^{(A-1)} \vert a_{k}\vert \Psi_{0}> .
\right|^2   \label{eq:greent}
\eea
To arrive at the second part of this equation we have inserted a
complete set of eigenstates for the system with $A+1$ particles
($\Psi_{\gamma}^{(A+1)}$) and $A-1$ particles ($\Psi_{\delta}^{(A-1)}$),
as appropriate, and replaced the hamiltonian in the exponential
functions of the definition for the Heisenberg operators 
 by the corresponding eigenvalues. This means that
the energies $E_{0}^A$, $E_{\gamma}^{(A+1)}$, and $E_{\delta}^{(A-1)}$
refer to the exact energies for the ground state of our reference
system, and the exact energies of eigenstates with $A+1$ and $A-1$
particles, respectively. Note that the step function can be
represented by its integral form:
\be
\Theta(t) = -\lim_{\eta\to 0} \frac{1}{2\pi i}\int_{-\infty}^{\infty}
d \omega \frac{e^{-i\omega t}}{\omega + i\eta}\, .
\ee
Thus the Fourier transformed Green's function, transforming the time
difference $t-t'$ to an energy variable $\omega$, can be written as
\bea
g(k, \omega) =& \displaystyle{\lim_{\eta\to 0}} \Biggl
(& \sum_{\gamma}\frac{ \left|
<\Psi_{\gamma}^{(A+1)} \vert a^\dagger_{k}\vert \Psi_{0}>
\right|^2 }{\omega - \left( E^{(A+1)}_{\gamma} - E_{0}^A \right) +
i\eta} + \sum_{\delta} \frac{\left|
<\Psi_{\delta}^{(A-1)} \vert a_{k}\vert \Psi_{0}>
\right|^2 }{\omega - \left( E_{0}^A - E^{(A-1)}_{\delta} \right) -
i\eta}\Biggr)\, . \label{eq:lehm}
\eea
This is the so-called spectral or Lehmann representation of the
single-particle
Green's function\cite{lehm}. Inspecting this equation one finds
that the single-particle Green's function is represented in terms of
quantities that are measurable. It shows poles at energies that
correspond to energies of the system with one particle added $(A+1)$ and one
particle removed $(A-1)$ relative to the energy of the ground state for
the reference system. The residua of these poles are given by the
spectroscopic factors, i.e., the probabilities of adding and removing
one particle with momentum $k$ to produce the specific state $\gamma$
($\delta$) of the residual system. The infinitesimal quantity $\eta$
shifts those poles below the Fermi energy (the states of the
$A-1$ system) to slightly above the real axis and those above the Fermi
energy (the states of the $A+1$ system) to slightly below the
real axis.

In our spectral representation of the single-particle Green's function
(\ref{eq:lehm}) we assumed that the spectra of the many-body
system are defined in terms of discrete energies $E^{(A+1)}_{\gamma}$.
This is true of course only for systems confined to a finite space. 
For
infinite systems, 
 the energy spectra are
continuous and it is more appropriate to introduce the so-called hole
and particle spectral functions defined by
\bea
S_{\rm h}(k,\omega ) & = &\frac {1}{\pi} \mbox{Im} g(k, \omega ), \quad \mbox{
for $\omega \leq \epsilon_{\rm F}$} \nn \\
& = & \sum_{\gamma}  {\left|
<\Psi_{\gamma}^{(A-1)} \vert a_{k}\vert \Psi_{0}>
\right|^2 } \delta \left(\omega - ( E_{0}^A - E^{(A-1)}_{\gamma})
\right)\, ,\nn \\
S_{\rm p}(k,\omega ) & = &\frac {1}{\pi} \mbox{Im} g(k, \omega ), \quad\mbox{
for $\omega > \epsilon_{\rm F}$} \nn \\
& = & \sum_{\gamma}  {\left|
<\Psi_{\gamma}^{(A+1)} \vert a^\dagger_{k}\vert \Psi_{0}>
\right|^2 } \delta \left(\omega - ( E^{(A+1)}_{\gamma} - E_{0}^A )\right)
\label{eq:specf}
\eea
where we have made reference to the case of discrete spectra in the
second and fourth lines.
These definitions imply that the single-particle
Green's function is given in terms of the spectral functions by
\be
g(k, \omega) = {\lim_{\eta\to 0}} \left( \int_{-\infty}^{\epsilon_{\rm F}}
d\omega' \frac{S_{\rm h}(k,\omega')}{\omega -\omega' - i\eta} +
\int_{\epsilon_{\rm F}}^{\infty}
d\omega' \frac{S_{\rm p}(k,\omega')}{\omega -\omega' + i\eta}
\right)\, .
\label{eq:specf2}
\ee
The single-particle Green's function or the spectral functions defining
the single-particle Green's function can be used to evaluate
observables of the system. As a first example we mention the momentum
distribution or momentum density
\be
n(k)\,  = \, < \Psi_{0}\vert a^\dagger_{k}a_{k} \vert
\Psi_{0}> \, = \,
\int_{-\infty}^{\epsilon_{\rm F}} d\omega \, S_{\rm h}(k,\omega ) \, .
\label{eq:momen}
\ee

The single-particle Green's function allows the evaluation of the
expectation value for any
single-particle operator $O$. If we return to the non-diagonal nomenclature
used, e.g., in (\ref{eq:greend}) and generalize the definition of
spectral functions in a corresponding way, such a calculation is
performed via
\be
< \Psi_{0}\vert O \vert \Psi_{0}> = \sum_{\alpha,\beta}
\int_{-\infty}^{\epsilon_{\rm F}} d\omega \, S_{\rm h}(\alpha\beta,\omega )
< \alpha \vert O \vert \beta> \, ,
\label{eq:expec}
\ee
with $< \alpha \vert O \vert \beta>$ denoting the
single-particle matrix element calculated in the basis $\alpha$, $\beta
\dots$. Furthermore, if the interaction between the fermions is a
two-body interaction, one can even calculate
 the energy of the
correlated ground state via
\bea
E_{0}^A & = &  < \Psi_{0}\vert H \vert \Psi_{0}> \nn \\
& = & \frac{1}{2} \sum_{\alpha,\beta}
\int_{-\infty}^{\epsilon_{\rm F}} d\omega \, S_{\rm h}(\alpha\beta,\omega )
\left( < \alpha \vert T_{\rm kin} \vert \beta> +
\omega\delta_{\alpha,\beta}\right) \, ,\label{eq:koltun}
\eea
with $T_{\rm kin}$ representing the single-particle operator for the kinetic
energy.

\begin{figure}[tb]
\begin{center}
\begin{minipage}[t]{10 cm}
\epsfig{file=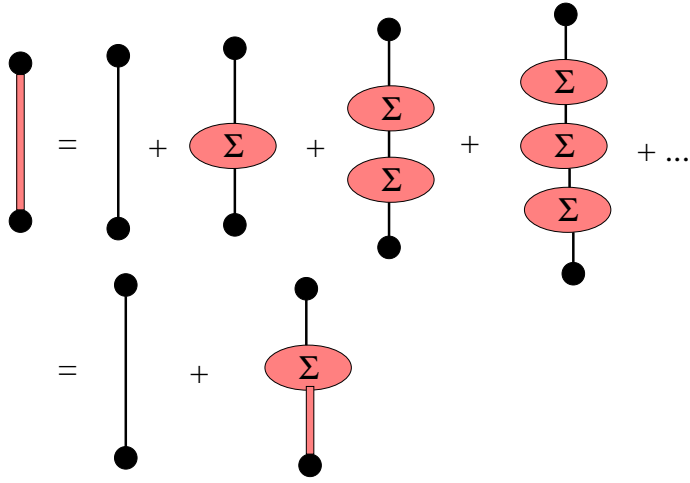,scale=0.5}
\end{minipage}
\begin{minipage}[t]{16.5 cm}
\caption{Classifying the diagrams for the single-particle Green's
function in terms of the mass operator $\Sigma$ and graphical
representation of the Dyson equation(\protect{\ref{eq:dyson1}}). 
The thick line represents the
``dressed'' single-particle Green's function while the thin lines stand for the
unperturbed Green's function $g_0$. \label{fig10}}
\end{minipage}
\end{center}
\end{figure}  

What remains to be discussed are the tools and approximations used
to determine the single-particle Green's function.
For that purpose we consider the single-particle Green's function for the
unperturbed hamiltonian
\be
g_0(\alpha\beta, \omega) = \delta_{\alpha\beta} \left\{
\frac{\Theta (\epsilon_\alpha - \epsilon_{\rm F})}{\omega
- \epsilon_{\alpha} + i\eta } +  \frac{\Theta (\epsilon_{\rm F} - 
\epsilon_\alpha
)}{\omega - \epsilon_{\alpha}^{\rm HF} - i\eta }\right\}\,. \label{eq:green0}
\ee
and classify all linked diagrams contributing to the expansion for the final,
``dressed'' single-particle Green's function following the diagrammatic
representation in the upper part of Fig.~\ref{fig10}. In this figure we group
all Feynman diagrams into parts, which are irreducible with respect to the
single-particle Green's function (represented by the ellipses $\Sigma$) and
single-particle Green's function $g_0$ connecting these irreducible parts. 
 The irreducible self-energy or mass operator $\Sigma$  contains
all diagrams of any order in the interaction $H_1$ or $V$, without an 
intermediate state, which is just a single-particle Green's function. Examples
of such contributions are displayed in Fig.~\ref{fig11}.
Note that here we do not distinguish between particle- and hole-propagation,
since the unperturbed Green's function $g_0$ (see Eq.~(\ref{eq:green0})) as well
as the resulting Green's function contains both contributions. 

The series displayed in the upper part of Fig.~\ref{fig10} represents the Dyson
equation and can be written
\bea
g(\alpha \beta, \omega) & = & g_{0}(\alpha \beta, \omega ) +
 \sum_{\gamma\delta} g_{0}(\alpha \gamma, \omega) {\Sigma}_{\gamma
\delta}(\omega)  \Bigl[ g_{0}(\delta\beta, \omega) \nn\\
&&+ \sum_{\mu\nu}
g_{0}(\delta\mu, \omega) {\Sigma}_{\mu\nu}(\omega ) g_{0}(\nu\beta,
\omega) + \dots \Bigr] \nn \\
& = & g_{0}(\alpha \beta, \omega ) +
\sum_{\gamma\delta} g_{0}(\alpha \gamma, \omega) {\Sigma}_{\gamma
\delta}(\omega)  g(\delta\beta, \omega)\, . \label{eq:dyson1}
\eea

\begin{figure}[tb]
\begin{center}
\begin{minipage}[t]{10 cm}
\epsfig{file=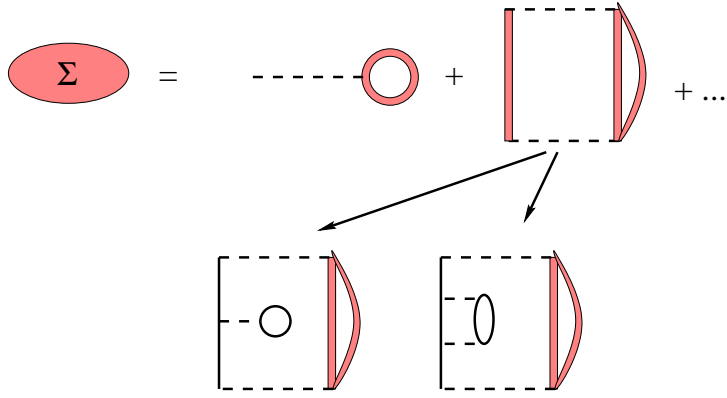,scale=0.5}
\end{minipage}
\begin{minipage}[t]{16.5 cm}
\caption{Diagrams contributing to the self-energy $\Sigma$. Note that the
thick connecting lines represent ``dressed'' single-particle Green's functions.
This means that expanding e.g. the rightmost diagram in the upper line yields
among other the terms displayed in the lower part of the figure.\label{fig11}}
\end{minipage}
\end{center}
\end{figure}

Examples of diagrams to be included in the definition of the self-energy are
displayed in Fig.~\ref{fig11}. The diagram of first order in the interaction $V$
is again the Hartree-Fock contribution. Note that it is defined in terms of the
dressed Green's function, which is expressed by the fact that the loop connected
to the $V$ interaction line is drawn as a thick line representing the dressed
Green's function. This reflects the problem of finding a self-consistent
solution of the Hartree-Fock equations. The structure of the single-particle
Green's function is identical to the unperturbed one, defined in
(\ref{eq:green0}). The only difference is that Hartree-Fock wave functions and
energies should be used.

If, however, one goes beyond the Hartree-Fock approximation by including
diagrams of higher order in the definition of the self-energy $\Sigma$, like
e.g.~the second term displayed in Fig.~\ref{fig11}, the problem of a
self-consistent determination of the Green's function gets much more involved.
To demonstrate this feature let us consider as a first step this diagram of
second order in $V$, replacing the dressed single-particle Green's functions by
the corresponding Hartree-Fock Green's function $g_0$. This means that the
intermediate states in that diagram are given in terms of intermediate 
2-particle 1-hole (2p1h) and  2-hole 1-particle (2h1p) states. The corresponding
expressions for the self-energy are given by
\bea
\Sigma_{\alpha\beta}^{(2p1h)}\omega ) & =
&  \frac{1}{2} \enspace
\sum_{\nu<F} \enspace \sum_{\rho_{1},\rho_{2}>F} \enspace \frac
{< \alpha\nu \vert V\vert \rho_{1} \rho_{2}>
<  \rho_{1} \rho_{2} \vert {V}\vert \beta\nu >}
{\omega - \left(\epsilon_{\rho_{1}}+\epsilon_{\rho_{2}}-\epsilon_\nu
\right) + i\eta }\, ,
\label{eq:self2p1h}
\eea
and
\bea
\Sigma_{\alpha\beta}^{(2h1p)}(\omega ) & =
&  \frac{1}{2} \enspace
\sum_{\rho >F} \enspace \sum_{\nu_{1},\nu_{2}<F} \enspace \frac
{< \alpha\rho \vert {V}\vert \nu_{1} \nu_{2}>
<  \nu_{1} \nu_{2} \vert {V}\vert \beta \rho >}
{\omega  - \left(\epsilon_{\nu_{1}}+\epsilon_{\nu_{2}}-\epsilon_\rho
\right)i\eta }\, .
\label{eq:self2h1p}
\eea
If we insert the sum of these two contributions into the Dyson equation
(\ref{eq:dyson1}), we obtain a single-particle Green's function, which exhibits
a much richer pole-structure. In fact assuming a discretized space of
single-particle states (as we have done also in Eqs.~(\ref{eq:self2p1h}) and
(\ref{eq:self2h1p})) one can rewrite the Dyson equation into an eigenvalue
problem\cite{skour1}
\be
\pmatrix{\epsilon_1 &\ldots &0 & a_{11} & \ldots & a_{1P} & A_{11} &
\ldots & A_{1Q} \cr
  & \ddots &  & \vdots & & \vdots &\vdots & & \vdots \cr
 0 & \ldots &\epsilon_{N} & a_{N1} & \ldots & a_{NP} & A_{N1} &
\ldots & A_{NQ} \cr
a_{11} & \ldots & a_{N1} & e_1 & & & 0 & & \cr \vdots & &\vdots & & \ddots
& & &  & \cr
a_{1P} & \ldots & a_{NP} & 0 & & e_P & & & 0 \cr
A_{11} & \ldots & A_{N1} &  & & &  E_1 & & \cr
\vdots & &\vdots& & & & & \ddots & \cr
A_{1Q} & \ldots &A_{NQ}& 0 & \ldots & 0 & & \ldots & E_Q \cr }
\pmatrix{ X_{0,k_1}^n \cr \vdots \cr X_{0,k_N}^n \cr X_1^n \cr \vdots
\cr X_P^n \cr Y_1^n \cr \vdots \cr Y_Q^n
\cr }
= \omega_n
\pmatrix{ X_{0,k_1}^n \cr \vdots \cr X_{0,k_N}^n\cr X_1^n \cr \vdots \cr
X_P^n \cr Y_1^n \cr \vdots \cr Y_Q^n
\cr } \; ,
\label{eq:geigen}
\ee
The matrix to
be diagonalized contains the Hartree-Fock single-particle energies and 
the coupling to the $P$ different $2p1h$ configurations, which is described in
terms of
\be
a_{\alpha i} = < \alpha\nu \vert V\vert \rho_{1} \rho_{2}>
\ee
and $Q$ different $2h1p$ configurations, for which we have introduced the
abbreviation
\be 
A_{\alpha i} = < \alpha\rho \vert {V}\vert \nu_{1} \nu_{2}>\, .
\ee
As long as we are still ignoring any residual
interaction between the various $2p1h$ and $2h1p$ configurations the
corresponding parts of the matrix in (\ref{eq:geigen}) are diagonal with
elements defined by $e_i$ ($E_j$) for $2p1h$ ($2h1p$)
\bea
e_{i} & = & \epsilon_{\rho_1} + \epsilon_{\rho_2} - \epsilon_{\nu}\nn\\
E_{j} & = & \epsilon_{\nu_1} + \epsilon_{\nu_2} - \epsilon_{\rho}\;,
\eea
where as before the indices $\rho_i$ and $\nu_i$ refer to particle and hole
states, respectively.

Solving the eigenvalue problem (Eq.~(\ref{eq:geigen})) one
gets as a result the single-particle Green's function
in the Lehmann representation in the discrete basis of the box defined as in
(\ref{eq:lehm}). The eigenvalues $\omega_n$ define the
position of the poles of the Green's function
\bea
\omega_n & =  \left( E^{(A+1)}_{\gamma} - E_{0}^A \right) & \quad \mbox{for}
\quad \omega_n > E_F \nn\\
\omega_n & = \left( E_{0}^A - E^{(A-1)}_{\delta} \right) & \quad \mbox{for}
\quad \omega_n < E_F \label{eq:grepo}
\eea
with $E_F$ the Fermi energy for the $A$-nucleon system. The corresponding
spectroscopic amplitudes are given by
\bea
<\Psi_0^A \vert a_{k_i} \vert \Psi_n^{A+1} > =
X_{0,k_i}^n  & \quad & for \quad \omega_n > E_F
\nonumber \\
<\Psi_0^A \vert a_{k_i}^\dagger \vert \Psi_n^{A-1}
> =  X_{0,k_i}^n & \quad & for \quad \omega_n < E_F
\label{eq:spec0}
\eea

\begin{figure}[tb]
\begin{center}
\begin{minipage}[t]{10 cm}
\epsfig{file=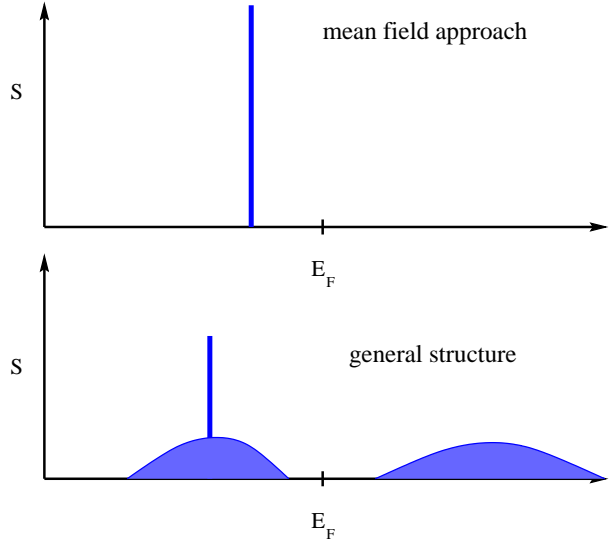,scale=0.5}
\end{minipage}
\begin{minipage}[t]{16.5 cm}
\caption{Schematic picture of the spectral function in the mean field approach
(upper part) and the general case.\label{fig12}}
\end{minipage}
\end{center}
\end{figure}

This multiplicity of the poles is a consequence of going beyond the mean-field
approximation. Within the Hartree-Fock or independent particle approach, one can
either remove a particle from the orbit $j$ (if this is an orbit below the Fermi
energy) or add a particle into this orbit (if $j$ is a state above the Fermi
energy). The spectroscopic factor is always one. This is reflected in the mean
field approximation to the single-particle Green's function by the feature, that
it exhibits for each orbit $j$ exactly one pole with a spectroscopic factor of
one. In the approach which we are discussing now this spectroscopic strength
gets redistributed as indicated in the schematic picture of Fig.~\ref{fig12}. 
The single-particle peak with strength one in the mean field approach gets
reduced to a quasi-particle peak in the spectral function with a strength $Z$,
which is smaller than one. Part of the strength gets redistributed into removal
or hole-strength, which is predominantly due to the coupling to $2h1p$
configurations. Another part, however, is shifted to energies above $E_F$ and
therefore represents a spectroscopic amplitude for adding a particle. This
implies of course that occupation numbers for the various orbits (see
(\ref{eq:momen})) will in general be different from zero and one. Also it is not
guaranteed any longer that the total particle number will be conserved,
i.e.~identical to $A$ the one of the unperturbed state 
\be
\int_0^\infty dk \, n(k) \quad \mathrel{\mathop=^{?}} A \label{eq:nconsv}
\ee
In order to obtain an approach, which is particle-number conserving, one has to
go beyond the approximation which we have discussed sofar and calculate the
Green's function in a self-consistent way. General rules for calculational
schemes of Green's functions, which conserve symmetries like the property of a
fixed particle number have been developed by Baym and Kadanov\cite{baymka}. For
our present purpose it is sufficient to say that one has to evaluate the
self-energy $\Sigma$, which enters the Dyson equation (\ref{eq:dyson1}) in a
self-consistent way in sofar that the resulting Green's function should be used
in the calculation of $\Sigma$. This self-consistency requirement is displayed
in terms of diagrams in the lower part of Fig.~\ref{fig11}.

The realization of this self-consistency requirement has to face a serious
problem: Calculating the self-energy $\Sigma$ in terms of Hartree-Fock Green's
functions led to a dressed Green's function which exhibits $L + P + Q$ poles,
where $L$ stands for the number of Hartree-Fock states of a given symmetry and
$P$ and $Q$ refer to the number of $2p1h$ and $2h1p$ states of this symmetry,
respectively. In a next step towards a self-consistent solution, one may 
consider these Green's functions with $L + P + Q$ poles already in the
calculation of the self-energy. This means that one also accounts for coupling
to $3p2h$, $3h2p$, $4p3h$ etc. configurations. This implies that the number of
poles in the Green's function resulting from this next iteration step will
dramatically increase. This ``inflation of poles'' in the single-particle
Green's function can be handled very easily by means of the so-called
``BAsis GEnerated by Lanczos'' (BAGEL) scheme\cite{skour1,skour2,taigel}.  
In this scheme on represents the Green's function in terms of a few 
``characteristic'' poles in the Lehmann representation. The number of these
poles can be kept fixed in the iteration scheme, leading to self-consistency.
These pole are determined by solving the eigenvalue Eq.~(\ref{eq:geigen}) by
means of the Lanczos scheme, starting with the single-particle states as 
appropriate initial vectors. Another possibility is to keep track of the
distribution of the spectral strength in a purely numerical way. Corresponding
calculations have been performed by van Neck et al.~\cite{neck} assuming a
model-space of finite dimension.

Up to this point we only discussed the determination of the single-particle
Green's function assuming an expansion for the self-energy, which is
self-consistent in terms of the Green's function but perturbative in terms of
the interaction $V$. From the discussion in the preceeding section we know
already that two-body correlations should be included in a non-perturbative way,
which means that one should consider all ladder diagrams, when one is calculating
e.g.~the self-energy $\Sigma$. How can this be achieved within the framework of
the Green's function approach? 

For that purpose one should solve a Dyson equation for the two-particle Green's
function. The Lehmann representation of the two-particle Green's function is
given in terms of energies and states of the systems with $A$ and $A\pm 2$
particles
\bea
g^{II}(\alpha\beta,\gamma\delta;\Omega) &=&
\sum_n {<\Psi^A_0\vert a_{\beta}a_{\alpha}\vert \Psi^{A+2}_n>
<\Psi^{A+2}_n\vert a^{\dagger}_{\gamma} a^{\dagger}_{\delta} \vert\Psi^A_0>
\over \Omega-(E^{A+2}_n-E^A_0)+i\eta} \nn\\
&&\qquad -\sum_m {<\Psi^A_0\vert a^{\dagger}_{\gamma}a^{\dagger}_{\delta}
\vert\Psi^{A-2}_m>
<\Psi^{A-2}_m\vert a_{\beta} a_{\alpha} \vert\Psi^A_0>
\over \Omega-(E^A_0-E^{A-2}_m)-i\eta}\,. \label{eq:lehmg2}
\eea
The result for the noninteracting product of dressed propagators, including
the exchange contribution reads 
\bea
g^{II}_f(\alpha\beta,\gamma\delta;\Omega)& = &
i \int {d\omega \over 2\pi}\
\{g(\alpha,\gamma;\omega) g(\beta,\delta;\Omega-\omega)
- g(\alpha,\delta;\omega) g(\beta,\gamma;\Omega-\omega) \} \nn\\
&= &\sum_{m,m'} { <\Psi^A_0\vert a_\alpha\vert\Psi^{A+1}_m>
                 <\Psi^{A+1}_m\vert a^\dagger_\gamma\vert\Psi^A_0>
                 <\Psi^A_0\vert a_\beta\vert\Psi^{A+1}_{m'}>
                 <\Psi^{A+1}_{m'}\vert a^\dagger_\delta\vert\Psi^A_0>
                 \over \Omega-\{(E^{A+1}_m-E^A_0) + (E^{A+1}_{m'}-E^A_0)\}
                 + i \eta } \nn\\
&& - \sum_{n,n'} { <\Psi^A_0\vert a^\dagger_\gamma\vert \Psi^{A-1}_n>
                 <\Psi^{A-1}_n\vert a_\alpha\vert\Psi^A_0>
                 <\Psi^A_0\vert a^\dagger_\delta\vert\Psi^{A-1}_{n'}>
                 <\Psi^{A-1}_{n'}\vert a_\beta\vert\Psi^A_0>
                 \over \Omega-\{(E^A_0-E^{A-1}_n) + (E^A_0-E^{A-1}_{n'})\}
                 + i \eta } \nn\\
&& - ( \gamma \longleftrightarrow \delta) \label{eq:2greenf}
\eea
The integration in the first line of this equation can be performed by employing the
Lehmann representation for the single-particle Green's functions.   
The ladder approximation to the two-particle propagator (\ref{eq:lehmg2}) 
is then
given by:
\bea
g^{II}_L(\alpha\beta,\gamma\delta;\Omega) &=&
g^{II}_f(\alpha\beta,\gamma\delta;\Omega) \nn\\
&& + {1\over4} \sum_{\epsilon\eta\theta\zeta}
g^{II}_f(\alpha\beta,\epsilon\eta;\Omega) <\epsilon\eta\vert V\vert\theta\zeta>
g^{II}_L(\theta\zeta,\gamma\delta;\Omega)\,.  \label{eq:dysong2}
\eea
This ladder approximation for the two-particle Green's function can then be used
to define the self-energy $\Sigma$ in a non-perturbative way\cite{wim1}. The
Dyson equations (\ref{eq:dysong2}) and (\ref{eq:dyson1}) for the two-body and
one-body Green's function have to be solved in a
self-consistent manner, this procedure has been named Self-Consistent Green
Function formalism (SCGF) and has been extensively discussed in Ref. 
\cite{wim1}. 

\subsection{\it Variational Method and Correlated Basis Function\label{sec:varia}}
An  efficient way to handle the correlations induced by the NN-interaction
is to   embody them, from the very beginning, in a trial wave function 
$\Psi_T$, which describes the system of $A$ nucleons 
\be
\Psi_T(1, ... ,A) =F(1,...,A) \Phi_{MF}(1,...,A),
\label{eq:trialwf}
\ee
where $\Phi_{MF}$ is a mean field wave function corresponding to the 
uncorrelated system  and the operator $F$ is intended to take care
 of the dynamical correlations.

Once a trial wave function is defined, the variational principle ensures
that if we are capable to calculate the expectation value of the  nuclear
 hamiltonian 
\be
\frac {\langle \Psi_T \mid H \mid \Psi_T \rangle}{\langle\Psi_T 
\mid \Psi_T\rangle} = E_T,
\label{eq:princvar}
\ee
then $E_T$ will be an upperbound to the ground state energy, $E_0$.
Parameters in the variational wave function are varied to minimize 
$E_T$ and the best $\Psi_T$ can then be used to evaluate other
observables of interest. Obviously, for the method to be 
efficient, the trial wave function must give a good representation 
of the real ground state many-body wave function. 
Although conceptually it looks very simple, the evaluation of the
 expectation value is by no means an easy task and very sophisticated
 algorithms which require
large computer capabilities have been devised during the last years.

Therefore the ingredients for a variational calculation are the hamiltonian
and 
the wave function. In addition, one requires  an efficient
 machinery to evaluate the expectation
value. By the definition of the trial wave function, 
which is given in 
configuration space, the variational method is tailored for a non-relativistic
framework where the only constituents are the nucleons,
 interacting through a local NN interaction with no energy or momentum
dependence. A realistic interaction based on the exchange of mesons between the
nucleons would be naturally given in momentum space, being non-local and energy
dependent \cite{rupr0,nijm0}. There are, however, realistic interactions which 
are specially suitable for variational calculations \cite{urbv14,argo0}.  

 Following these guidelines, one can write a realistic nuclear
hamiltonian  in the form:
\be
H=- \frac{\hbar^2}{2m} \sum_i \nabla_i^2 + \sum_{i<j} V_{ij}
\label{eq:hamil} 
\ee

where the two-body potential $V_{ij}$ has a local operatorial structure, i.e.
is given by the sum of  functions
depending on the relative distance between the nucleons and spin-isospin 
operators  build 
according to  invariance requirements. Although we devote a full section to discuss
different modern realistic potentials, it is convenient at this point to 
make some comments on the potentials usually employed in variational calculations.
For instance, the Argonne $V_{14}$\cite{argo0} 
NN potential  is given by the
sum of 14 isoscalar terms
\be
V_{ij} = \sum_{p=1,14} V_p(r_{ij}) O_{ij}^p,
\label{eq:pot1}
\ee
 with
\be
O_{ij}^{p=1,14} = \left [ 1, \vec \sigma_i \cdot \vec \sigma_j, S_{ij},
{\bf L}\cdot {\bf S}, {\bf L}^2,{\bf L}^2 \vec \sigma_i\cdot \vec \sigma_j,
({\bf L }\cdot {\bf S})^2 \right ]\otimes \left [1, \vec \tau_i \cdot 
\vec \tau_j \right ],
\label{eq:pot}
\ee
Here
\be
S_{ij}= 3(\vec \sigma_i \cdot \hat r_{ij})(\vec \sigma_j \cdot \hat r_{ij})
- \vec \sigma_i \cdot \vec \sigma_j
\ee
is the usual tensor operator, ${\bf L}$ is the relative orbital angular momentum
, and ${\bf S}$ is the total spin of the pair.

The radial components of the potential contain a long range part which
 is given by a static nonrelativistic reduction of the one pion exchange
 potential
(OPE) which contributes only to the $(\vec \sigma_i \cdot \vec \sigma_j)(\vec 
\tau_i \cdot \vec \tau_j)$ and $S_{ij} (\vec \tau_i\cdot \vec \tau_j)$.
The intermediate and short range parts are given in terms of a physically
plausible parameterization with a  reasonable number of adjustable parameters.

However, these accurate NN potentials do not satisfactorily   reproduce
the nuclear matter saturation point and underbind nuclei with  
$A > 2$ \cite{fhnc1}.   
 As discussed in the introduction, a possible 
reason for that is the existence of three body forces which can have different origins. 
We will come back to the problem of  three body forces, but here
it is important to keep in mind that the nuclear hamiltonian can certainly have
an additional term with three body forces. Usually the recent realistic 
variational calculations include also a three body force in the nuclear
hamiltonian, which is  
constructed in largely phenomenological fashion with  parameters  determined
to reproduce the saturation of nuclear matter and the 
 correct binding energy for $A=3,4$ nuclei.\cite{carlson1} 

There is also an updated version of the Argonne potential which breaks
charge independence and charge symmetry and contains some additional 
isotensor and isovector components, responsible for the breaking of
the isospin symmetries. This version of the two-body Argonne potential contains
in total 43 parameters and an operatorial structure up to 18 operators,
it is known as the Argonne $v_{18}$ NN potential \cite{Wiringa95}.
 
Once the realistic hamiltonian has been defined, which for the time being 
we suppose to have only two body forces, we need 
an ansatz for the  variational wave function. At this point, we should
 appeal to the physical
intuition and sense in order to choose an appropriate trial wave function
 for the
 system
under consideration.
 In the 
nuclear case, i.e. for nuclei and for nuclear matter around saturation,
the density is small enough to assume that two body correlations
will be the most relevant ones. A form of $F(1,2,...,A)$ that
has shown to be suitable for nuclear systems is
\be
F(1,...,A) =  S \left[ \prod_{i <j} F_2(i,j) \right ]
\label{eq:corre1}
\ee
i.e. a symmetrized product of two-body correlation operators, $F_2(i,j)$.
The natural choice is to allow the two-body correlation
 operator to have a similar structure as the  two-body NN interaction.
In  recent calculations, the ansatz for the two body correlation operator $F$ 
contains the same set of operators
as in Eq.~(\ref{eq:pot}) except the quadratic terms in the
relative angular momentum, 
\be
F_2(i,j)= \sum_{m=1,8} f^{(m)}(r_{ij}) O^{(m)}_{i,j},
\label{eq:corre2}
\ee
where the sum runs up to the spin-orbit components in Eq.~(\ref{eq:pot}).

 When the components of the correlation operator with 
$m \geq 2$  are disregarded, one recovers
 the well known Jastrow
correlation function \cite{jastrow1} , which has been largely used
 in the context of
quantum liquids \cite{feenberg} and also in nuclear physics for simple
semi-realistic interactions \cite{clark1}.

The mean field wave function $\Phi_{MF}(1,..,A)$ is a Slater determinant
of single particle wave functions, $\varphi_{\alpha}(i)$, where subscript
$\alpha$ stands for the set of quantum numbers characterizing the
single particle state, and $(i)$ indicates the spatial and spin-isospin
variables of particle $(i)$. Those single particle wave functions
are obtained by some mean field (MF) potential, which can be completely
arbitrary. 
 In the particular case of symmetric and spin and
isospin saturated nuclear matter, the mean field wave function
is a Slater determinant of plane waves:
\be
\Phi_{MF}(1,2,...,A)= Det_{ij}\left [ exp(i {\bf k}_i \cdot {\bf r}_j) 
\chi_i^S(j)
\chi_i^T(j) \right ],
\ee
with all single particle momentum states ${\bf k}_i$ occupied up to the
Fermi momentum $k_F=(6 \pi^2 \rho/d2)^{1/3}$,  where $d$ is the spin-isospin
 degeneracy of each particle level, $d=4$ for nuclear matter
and $d=2$ for neutron  matter. $\chi_i^S(j)$ and $\chi_i^T(j)$
are the spin and isospin functions. The mean field wave function
$(\Phi_{MF})$ 
 is intended
to carry the correct quantum statistics and any of the required symmetries 
of the system. As the mean field wave function has already the correct
antisymmetric character, the correlation operator requires the
presence of a symmetrizer operator {\sl S} (Eq.~(\ref{eq:corre1}))  
 to preserve  the correct symmetry behavior of the trial wave function.
Obviously in the case that the different two-body correlations
commute with each other, as it would be the case for a simple 
Jastrow correlation,  the presence of the symmetrizer is superfluous.

Besides being symmetric in all the variables, the correlation operator
possesses the cluster decomposition property, namely that upon
separating one subgroup of particles $(1,2,...,n)$ far
 from the rest $(n+1,n+2, ...,A)$, the operator $F(1,2,...,A)$
 decomposes into a product
\be
F(1,...,A)= F^{(n)}(1,...,n) F^{A-n}(n+1,...,A).
\ee
According to the clusterization property, the scalar component
 $f^{(p=1)}(r)$ in (\ref{eq:corre2}) heals to unity when r is large whereas
$f^{(p \neq 1)}(r) \rightarrow 0$.

A realistic trial wave function is expected to give an upperbound
 very near to the 
ground state energy and a wave function close to the ground state
wave function. 
The radial correlation functions $f^{(m)}(r)$ are determined by
minimizing the energy, $E_T$. However, the minimization of
the expectation value of the hamiltonian respect to arbitrary
variations of $f^{(m)}(r)$
\be
\frac {\delta}{\delta f^{(m)}} \frac {\langle \Psi_T \mid H \mid \Psi_T 
\rangle }{\langle \Psi_T \mid \Psi_T \rangle } =0
\ee
is a highly prohibitive job in the nuclear case. In practice, 
the correlations are either parameterized and the parameters
fixed by minimization or they are generated by solving 
a set of coupled differential equations obtained
by minimizing  the energy obtained in a two-body cluster
expansion.

Once the hamiltonian 
and the trial wave function are defined, one should be able to calculate the 
expectation value. The most efficient techniques are  Fermi-Hyper-Netted-
Chain (FHNC) theory and  Variational Monte Carlo (VMC) method (to be 
discussed in the next sub-section).
 FHNC is an integral equation method
that sums up series of clusters diagrams associated
 with the distribution functions of the many-body wave function. 
The method is suitable for  infinite
homogeneous systems as well as for finite, inhomogeneous systems.
In fact, the expectation value of the hamiltonian (or any other
operator) is written in terms
of $n$-body densities, 
\be
\rho_1= \frac {\langle \Psi_T\mid \sum_i \delta({\bf r}-{\bf r}_i) \mid \Psi_T\rangle}
 {\langle \Psi_T \mid \Psi_T \rangle } ,
\ee
\be
\rho_2^{(p)}(1,2)= \frac {\langle \Psi_T \mid \sum_{i \neq j} \delta( {\bf r}_i -
{\bf r}_1) \delta( {\bf r}_j -{\bf r}_2) O_{ij}^p \mid \Psi_T \rangle}
{\langle \Psi_T \mid \Psi_T \rangle }.
\ee
The densities are then cluster expanded in terms of dynamical correlations 
$h(r)=\left [f^{(1)}(r) \right ]^2 -1$,   products $f^{(1)}(r) f^{(p \geq
2)}(r)$,\ and statistical correlations,\ $\rho_0(i,j)=\sum_{\alpha}
\phi_{\alpha}^\dagger(i) \phi_{\alpha}(j)$, associated to the exchanges in the
mean field wave function. The terms of the resulting expansions are usually
called  cluster terms and are characterized by integrals containing a given
number of correlations, and exchange functions joining the  correlated
particles. In the early calculations, the expansion was stopped at low orders
\cite{jastrow1,iwamoto,clark68,guardiola1}. A great progress in the summation
of the cluster expansion was achieved by realizing that the cluster terms are
conveniently  represented by cluster diagrams \cite{ripka1,fantoni74}  and by
paying attention to the enormous amount of cancelations between the different
terms of the cluster expansion. A crucial step towards the summation was the
proper classification  of the different cluster terms, i.e. diagrams. Finally,
a close set of equations for Fermi systems represented by a wave function
containing  two-body Jastrow factor were derived \cite{fantoni75,kro75}. In
this way, the uncertainties associated to a low order expansion were removed to
a large extent. The restriction of the  FHNC equations to Bose systems is
equivalent to the earlier HNC equations used to calculate distribution
functions in the context of classical theory of liquids \cite{boer59}.

 Rules for
 constructing the different topological diagrams and derive the FHNC
 equations to sum up the different sets of diagrams have been given
several times in the literature, here
we refer the reader to two recent reviews of Fabrocini and Fantoni
 where they  give
 in a very clear and pedagogical way
the derivation of the FHNC equations for both, state independent and 
state depedent correlations,  homogeneous and inhomogeneous
systems\cite{fhnc3,fabro99}. A review which contains an extensive
analysis on  the different low order cluster expansions and a very detailed
derivation of the FHNC equations with special attention
to the different treatment of the exchange terms \cite{fantoni75,kro75}
is the one of Clark in an earlier volume of this series \cite{clark1}.
In that review one can also find results for nuclear matter with 
simple semi-phenomenological potentials that do not require the
full operatorial structure of the two-body correlations.

For brevity, here we only illustrate the main
ideas of the HNC summations by considering the case
of a system of bosons, interacting through a potential that depends only 
on the distance between the particles.
This example corresponds to a realistic situation in the context
of Condensed Matter, i.e. to liquid $^4$He at zero temperature. 
In that case the inter-atomic potential  is much simpler than
 the NN interaction, in the sense that it depends only of the distance 
between the atoms. However, as in the NN interaction, it has 
a strong repulsion at short distances and weak atraction at medium 
and large distances. A good representation of the interatomic
potential is given by the Lennard-Jones potential,
\be
V(r)= 4 \epsilon \left [ \left ( \frac {\sigma}{r} \right )^{12} - 
\left ( \frac {\sigma}{r} \right )^{6} \right ],
\label{eq:pothe}
\ee
where $\epsilon=10.22$ Kelvin (K) ( $1 eV \sim 11000 K$) and
$\sigma=2.556 $ Angstr\o m define the energy and length scales, 
respectively. In that case, the non interacting system
would be described by a wave function with all atoms
in the state of zero momentum, i.e. there is a macroscopic 
occupation of a single particle quantum state. This
phenomenon is known as Bose-Einstein condensation.
When the interactions is turned on, the correlations induced by the interaction
lead to atoms in states with momentum different from zero. However,
the liquid $^4$He still shows a condensate fraction,
i.e. a ratio between the number of atoms in the
zero momentum state and the total number of
atoms, of around $10 \%$.

A convenient  trial wave function to describe this
system is 
given by a Jastrow wave function:
\be
\Psi(1,2,...A) = F(1,2,...,A)= \prod_{i<j} f(r_{ij}),
\ee
a product of two body correlations depending only on the distance between
the particles. 
 We work in the thermodynamical
limit, allowing $N \rightarrow \infty$ and the volume $\Omega \rightarrow
\infty$ but keeping the density $\rho= N/\Omega$ constant. In this limit,
the energy per particle is given by:
\be
e(\rho)= \frac {1}{2}\rho \int d^3r g(r) \left [ V(r) - \frac {\hbar^2}{2 m_4}
\nabla^2 \ln~f(r) \right]
\ee
where $g(r)$ is the distribution function:
\be
g(r)\equiv  \frac {\rho_2^{(1)}(1,2)}{\rho^2} =
 \frac{A(A-1)}{\rho^2}\frac{\int \mid \Psi \mid^2  d^3r_3 ...d^3r_A}
{\int \mid \Psi \mid^2 d^3r_1 ...d^3r_A}
\label{eq:distri}
\ee
which gives the probability to find  two particles separated by a distance
$r$, and is normalized such that  
\be
\rho \int d^3r [g(r) -1]~ =~-1.
\label{eq:seq}
\ee
This condition can be used as an accuracy check of the 
approximations implemented in the calculation of
$g(r)$.
In this way, the problem to calculate the energy per particle is now translated in
into the calculation of the distribution function. Now, the numerator and the 
denominator of Eq.~(\ref{eq:distri}) are expanded in powers of the correlation interaction,
defined as $h(r)=[f(r)]^2-1$,
\be
g(r)= \frac {A(A-1)}{\rho^2} \frac { f^2(r_{12}) ~\int d^3r_3 ... d^3r_A  
 \left (1+\sum h(r_{ij})+ \sum
h~h+ ... \right)} {\int d^3r_1 ... d^3r_A \left (1+\sum h(r_{ij}) + ... \right )}
\label{eq:distri1}
\ee

 The terms of the resulting expansion are characterized by integrals
 containing a given number of functions $h$. Now it is  very useful to introduce
 a diagrammatic notation. This allows to classify the integrals according to
their diagrammatic representation.
The cluster diagrams are built employing the following simple rules:
 Their basic blocks are points and solid lines. Points (vertices) represent 
the coordinate ${\bf r}_i$ of a generic $i$-particle. Solid points (internal
points) imply integration over the coordinates times a factor $\rho$, circles
(external points) refer to particles labeled $1$ and $2$, and we do not
integrate over their coordinates. Solid lines are correlation
 factors and can not be superimposed. Notice that in the diagrams of the
denominator, all vertices are solid points.

\begin{figure}[tb]
\begin{center}
\begin{minipage}[t]{13 cm}
\epsfig{file=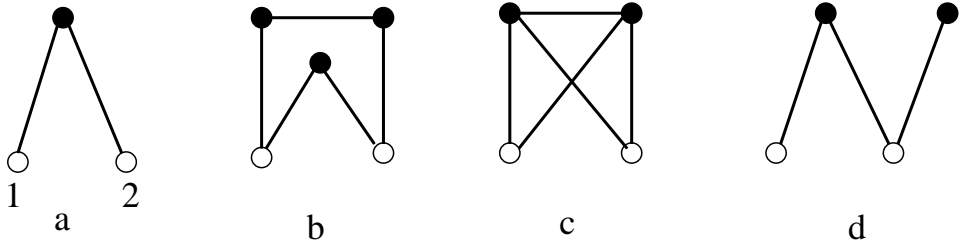,scale=0.6}
\end{minipage}
\begin{minipage}[t]{16.5 cm}
\caption{ Different HNC diagrams which appear in the calculation of the 
distribution function, $g(r_{12})$}
\label{fig:diagram}
\end{minipage}
\end{center}
\end{figure}

In Fig.~\ref{fig:diagram} we give several examples of diagrams appearing 
in the numerator of Eq.~(\ref{eq:distri1}), 
for which the expressions are given by
\be
Diagram (a) = \rho \int d^3r_3 h(r_{13}) h(r_{23})~,
\ee
\be
Diagram (b) = \rho \int d^3 r_3 d^3 r_4 d^3 r_5 h(r_{13})h(r_{14}) h(r_{23}) h(r_{24}),
\ee
and
\be
Diagram (c) = \rho^2 \int d^3r_3 d^3r_4 h(r_{13})h(r_{14})h(r_{34})h(r_{24})
h(r_{23}).
\ee

Cluster diagrams may be linked or unlinked. Unlinked diagrams have at least two
parts with no common points. They can not be drawn without separating the pen
from  the paper. The linked diagrams can be classified in reducible and
irreducible diagrams. The reducible diagrams are those which can be factorized
in a product of two or more irreducible diagrams such that one of them contains
the two external points. Diagram (d) of Fig.~\ref{fig:diagram} is  a
reducible diagram. In the infinite systems the  reducibility is closely
connected to the translationally invariance character of  the correlations. The
irreducible diagrams are then classified into: {\it Nodal} ($N$), {\it
Composite} ($X$) and the remaining ones called {\it elementary}. The {\it
nodal} diagrams have at least one node, that is an internal point  such that
all ways of going from one 1 to 2 (the two external points) should go through
it. The composite diagrams are those having two or more  (12)-sub-diagrams.
Where and (ij)-sub-diagram is a part of the diagram  connected to the rest only
through the points $i$ and $j$. Diagrams a,b,c and d of Fig.~\ref{fig:diagram}
are examples of nodal, composite, elementary  and reducible, respectively.
There are two mathematical operations related to the construction of diagrams:
the  convolution product linked to  the construction of nodal diagrams
\be
(a(r_{1i}) \mid b(r_{i2})) =\rho \int d^3r_i a(r_{1i}) b(r_{i2})
\ee
and the algebraic product $(a(r_{ij})b(r_{ij}))$ for the composite diagrams.

In evaluating the distribution function Eq.~(\ref{eq:distri1}) one can take
advantage of a lot of cancellations between the different terms. The expansion
is linked and in addition in the thermodynamical limit, for the specific case
of bosons,  is irreducible up to terms of the order $1/N$. For homogeneous
Fermi systems with Jastrow correlations, the expansion is  fully irreducible.
However, for state dependent  correlations the cluster expansion is not
irreducible, i.e. one should consider vertex corrections. In case of finite
systems, both for fermions and bosons  the cluster expansion is not irreducible
\cite{fabro99}.

The irreducible diagrams are summed up by means of the HNC equation, which
allows for  an iterative process to sum the nodal ($N(r)$) and composite
 ($X(r)$) diagrams
once the sum of the elementary diagrams ($E(r)$) is given. The HNC equation is
a non-linear integral equation relating the function $N(r)$ and $X(r)$
\be
(X(r_{1i})\mid N(r_{i2}))= N(r_{12}) - (X(r_{1i}) \mid X(r_{i2})),
\label{eq:HNC}
\ee
i.e., doing the convolution product ( associated to the construction of
nodal diagrams) of X(r) and N(r) we get all nodal diagrams except
the ones corresponding to $(X(r_{1i}) \mid X(r_{i2}))$. 
This equation  can be written also in momentum space as
\be
\tilde N(k) = \frac{\tilde X(k)^2}{1 -\tilde X(k)}
\label{eq:HNCK}
\ee
where the Fourier transform of a given function $a(r)$ is defined as
\be
\tilde a(r) =\rho \int d^3 r e^{i \vec k \cdot \vec r} a(r).
\ee

The composite diagrams are given by
\be
X(r)= f^2(r) e^{N(r)+E(r)}-1-N(r),
\label{eq:compos}
\ee
while the distribution function is expressed as
\be
g(r)= f^2(r) e^{N(r)+E(r)} = 1+X(r)+N(r)
\ee
The HNC integral equation (Eq.~(\ref{eq:HNCK})) suggests an iterative
 process to calculate
$g(r)$. At the first iteration , $X(r)=h(r)$, then 
the HNC equation is used to construct the first chain of 
diagrams. After a Fourier transformation to $r$-space, the resulting
function $N(r)$ is used to define the new $X(r)$ (Eq.~(\ref{eq:compos})).
 The number of 
diagrams summed in this way grows  tremendously and in a few 
iterations one reaches convergence. However, there is a problem with the
elementary diagrams, entering in the definition of $X(r)$, i.e. the HNC
iterative process sums up all nodal and composite diagrams ,
 once the sum of the elementary diagrams is known.
 In this sense, the function $E(r)$ is an
input for solving the HNC equation. There is no exact method to compute 
this function and approximations are necessary. The simplest option,
is to take $E(r)=0$, known as HNC/0. Actually, this approximation 
is appropriate for nuclear systems, where the density is not very high.

The Fourier transform of the radial distribution function
defines the static structure function $S(k)$,
\be
S(k)= 1~+~\rho \int d^3 r \left [ g(r) -1 \right ]
\ee
which in the case of quantum liquids is experimentally accessible
by means of elastic neutron scattering against the liquids 
\cite{lovesey}. The condition of Eq.~(\ref{eq:seq}), implies
$S(0^+)=0$ at zero temperature. When $k\rightarrow \infty $
then $S(k)\rightarrow 1$. 

\begin{figure}[tb]
\begin{center}
\begin{minipage}[t]{12 cm}
\epsfig{file=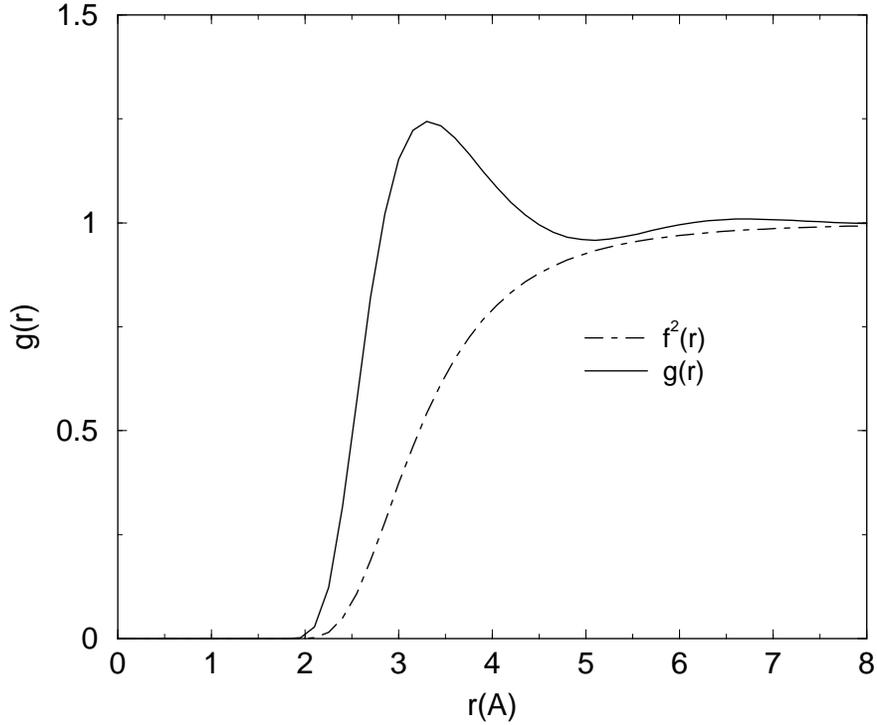,scale=0.7}
\end{minipage}
\begin{minipage}[t]{16.5 cm}
\caption{ Two body correlation function (dot-dashed line) and the corresponding
$g(r)$ calculated in HNC/0 at the experimental saturation density of liquid 
$^4$He \label{distri}}
\end{minipage}
\end{center}
\end{figure}

A two body correlation appropriate for the Lennard-Jones
potential which has been largely used in the 
literature of quantum liquids is the McMillan form
\be
f(r)= e^{- \frac{1}{2}\left [ \frac {b \sigma }{r} \right ]^5 }.
\label{eq:correh}
\ee
This correlation function together with the corresponding
distribution function are shown in Fig.~\ref{distri}
at the $^4$He saturation density $\rho_0 = 0.365 \sigma^{-3}$
in the HNC/0 approximation (i.e. by neglecting the elementary diagrams).
The correlation function (Eq.~(\ref{eq:correh})) avoids the configurations where
the two particles are close enough to feel the strong repulsion
of the potential. Finally, the distribution function has a hole 
at the origin, and the probability to find two particles
at short distances is strongly depleted. The system minimizes its energy 
by favoring distances
between the particles (maximum of g(r)) near to the   
 maximum depth of the potential. At the same time, 
the ``holes'' in the wave function yield an
increment of the kinetic energy, which was zero in the
uncorrelated system. At the end, there is a delicate balance
between the potential and the kinetic energy which 
results in the binding energy of the $^4$He atoms to be 
$-7.17 K$ at the saturation density, while the kinetic 
energy and the potential energy are $T\sim 14 K$ and
$V \sim 21 K$. These arguments of the short range correlations
apply equally well to the nuclear case, and as we 
will discuss below, the final binding energy is 
also a delicate  balance between potential and kinetic energy.
The short range correlations are  responsible for an
increment of the kinetic energy with respect to the kinetic energy of the
uncorrelated system,
which in the case of nuclear matter is given by
$3 \hbar^2 k_F^2 /10  m$, the average kinetic energy of the underlying
  free Fermi sea. 
 Before leaving this 
example completely it is  worth to mention that the HNC theory can 
treat at the same time short- and long-range
 correlations consistently \cite{ripka}, when instead of just
using   
a short-range correlation function, one considers 
the optimal correlation, solution of 
\be
\frac {\delta}{\delta f} \frac {\langle \Psi_T \mid H \mid \Psi_T \rangle}
{\langle \Psi_T \mid \Psi_T \rangle } = 0
\ee
This optimal correlation has a  long range behavior , that  translates 
in the right limit of $S(k)$  at low $k$, $S(k) \sim \hbar k/ 2 m v_s$,
being $m$ the mass of the atoms and $v_s$ the speed of sound. 
This behavior reflects the phonon nature of the low  
 energy spectrum of 
an interacting boson system.

In the case of Fermi systems, the anti-symmetrization implied by the Pauli
principle, incorporates statistical correlations to the cluster
expansions, i.e. to the diagrams. In the case of homogeneous systems,
the statistical correlations are accounted for by the Slater function 
$l(k_F r_{ij})$,
\be
\rho_0(i,j)=l(x=k_F r_{ij})=\frac {3}{x^3} \left [ \sin(x)-x~\cos(x) \right ]
\ee
This Slater function, defines the distribution function for a free Fermi 
sea,  
\be
g(r) = 1 - \frac{1}{d} l^2(k_F r).
\ee

 To take care of all topological different diagrams,  we need to introduce new
nodal and composite functions, which are related by a system of coupled 
non-linear integral equations (FHNC) \cite{fantoni75,kro75}. Once the FHNC
equations are solved one can easily calculate the energy. The presence of the
symmetrizer in the correlation operator and the non-commutativity of the
operators involved in the correlations makes it very difficult to sum  up  all
the contributions in the case of state dependent correlations. Actually, no
scheme has been devised so far, which keeps track of all possible  terms. A
complete   FHNC summation is possible  only for the Jastrow component. 
However, partial classes of diagrams  containing operatorial correlations may
exactly be summed by the Single Operator Chain approximation
(FHNC/SOC)\cite{wiringa79}. The FHNC/SOC integral equations sum all the nodal
diagrams containing only one operatorial correlation per internal side, in
addition to all the Jastrow correlated clusters. By computing its leading 
corrections, the estimated accuracy of the FHNC/SOC approximation has been set
to $\sim 1$ MeV/A for nuclear matter at saturation density \cite{wiringa80}.
The FHNC/SOC has been applied to study the equation of state of nuclear and
neutron matter \cite{fhnc1}, including three body terms in the
hamiltonian.  It is very difficult to apply the variational formalism with
state dependent correlations to study asymmetric nuclear matter with an
arbitrary asymmetry $x=\rho_p/\rho$ and  all the calculations, which are then
later on used to study $\beta-$stable matter, are based in quadratic
parameterizations of the correlation energy
\be
E(\rho,x)= T_F(\rho,x)+V_0(\rho)+(1-2 x)^2 V_2(\rho)
\ee
where
\be
T_F(\rho,x)= \frac {3}{5} \frac {\hbar^2}{2 m} (3 \pi^2 \rho)^{2/3} \left
[x^{5/3} +(1-x)^{5/3} \right ]
\ee
is the kinetic energy of the asymmetric Fermi sea and $V_0$ and $V_2$ are
related to  the correlation energy of neutron and nuclear matter. In this
parameterization, the symmetry energy of nuclear matter is
\be
E_s(\rho)= \frac {1}{8} \frac {\partial^2 E(\rho,x)}{\partial x} \mid_{x=1/2} =
\frac {5}{9} T_F(\rho,1/2) + V_2(\rho).
\ee

The extension of the FHNC/SOC to finite nuclei is an important issue in order
to have a systematic study with the same basic tools for nuclei and nuclear
matter.  After the  first efforts based on  low order cluster expansions
\cite{clarkgrana,guarditubi}, the FHNC formalism was extended to doubly closed
shell nuclei, both in $ls$ and $jj$ coupling schemes, using Jastrow correlated
wave functions \cite{co1,co2,co3} and semi-realistic interactions \cite{afnan}.
Very recently, the FHNC/SOC has been generalized to deal with realistic
potentials and state dependent correlations, with non-central components for
double closed shell nuclei as $^{16}$O and  $^{40}$Ca \cite{arias1,arias2}. The
most complete calculations \cite{arias2} deal with the Argonne
$V_{14}$,\cite{argo0} supplemented with the Urbana VII model of the three-body
interaction\cite{threeforce}. The two-body correlation function includes
operators up to
tensor components in Eq.~(\ref{eq:corre2}) and the single particle wave
functions were taken from a Woods-Saxon mean field potential. The results are
rather promising and one can conclude that now a days the calculations in
finite nuclei with realistic interactions have achieved the same degree of
accuracy than in nuclear matter. 

As we have already mentioned, a full minimization of the variational energy
respect  to arbitrary variations of $f^{(p)}(r)$ is impossible.  The
correlations, in the most sophisticated calculations, both in nuclear matter
and finite nuclei, are generated  by solving a set of coupled Euler-Lagrange
equations obtained by minimizing the second order cluster expansion of the
energy. The differential equations are solved with  the boundary conditions:
\be
f^{(p=1)}(r \geq d^{(p=1)}) =1 ,
\ee
and 
\be
f^{(p > 1)}(r \geq d^{(p>1)})=0,
\ee
where $d^{(p)}$ are the healing distances in each correlation channel. To ensure
the continuity of the first derivative of the correlations
\be
\frac {\partial f^{(p)}}{\partial r} \mid_{r=d^{(p)}} = 0
\ee
one introduces Lagrange multipliers $\lambda^{(p)}$ which are varied to satisfy the 
condition of the above equation.

Some additional variational parameters are introduced because the second order 
minimization is actually performed by using a "quenched" potential
\be
 \bar v_{ij} = \sum \alpha^{(p)} v^{(p)}(r_{ij}) \hat O^{(p)}.
\ee
These quenching parameters $\alpha^{(p)}$ are interpreted as
representing medium effects. In this way the two body correlations
contain several variational parameters
\be
F_{ij} = \sum_{p=1,8} f^{(p)}(r_{ij};d^{(p)},\alpha^{(p)}) O_{ij}^{(p)}
\ee
which will be determined by energy minimization. The number
of parameters is reduced in a drastic way by taking
 all the healing distances
equal $d^{(p)}=d_c$, except those associated to the tensor channels $d_t$. The
quenching factors are typically assumed to be 1 for the scalar channels and
those containing $L^2$, while all other channels are quenched by the same amount
  $\alpha^{(p)}= \alpha$ \cite{fhnc1,akmal98}.

A main drawback of a variational calculation is that even if the
 expectation value of the hamiltonian is evaluated exactly for a
 given trial wave function, 
one has only un upper-bound on the exact energy and an approximation
 to the exact many body wave function. The way to proceed further 
is provided by  the Correlated Basis Function (CBF) Method, which was
introduced nearly forty years ago by Feenberg and his collaborators
\cite{feenberg,clark59}. Several people have contributed later on
to make the method efficient and competitive, pushing its applications
to very different physical systems \cite{fhnc3,fhnc4}. The CBF method
provides 
 the general frame to systematically improve on a trial wave function
by introducing a complete set of correlated basis functions.
The idea is to perform a perturbation theory built on such correlated basis,
with the hope that the correlated basis states are close enough
 to the eigenstates of the hamiltonian that a lower order in the perturbation
theory will be sufficient to take care of these small differences.

The set of correlated basis states $ \Psi_n$, is built by applying the
many-body correlation operator, $F(1,2,...,A)$, usually determined in the
variational calculation, to model basis functions, $\phi_n$,
\be
\Psi_n(1,2,...,A) = F(1,2,...,A) \phi_n(1,2,...,A).
\ee

The zero order of the CBF theory, i.e. taking only one basis function, reduces to
the variational procedure. Therefore it is crucial to have a good variational
description of the ground state in order to have a fast convergence of the 
perturbative series.  The model function $\phi_0$ is also used to define the
Fermi level and as before particle states above and hole states below this Fermi
level. The model states $\phi_n$ refer then to some particle-hole configuration
with respect to $\phi_0$.

In order to perform  perturbation theory we need to decompose the
hamiltonian in an unperturbed hamiltonian $H_0$ and a perturbation
$H_I$, 
\be
 H~=~H_0~+~H_I\, .
\ee
Here they are defined in terms of matrix elements of $H$ in the correlated basis
$\mid i \rangle$. The diagonal matrix elements
\be
E_i^v~=~\frac{\langle i \mid ~H~\mid i \rangle}{\langle i \mid i\rangle}
\ee
define the unperturbed
hamiltonian:
\be
\langle i \mid H_0 \mid j \rangle = \delta_{i,j} E_i^v
\ee
while  $H_I$ is given by
\be
\langle i \mid H_I \mid j \rangle =(1 - \delta_{i,j}) \frac {\langle i 
\mid H \mid j \rangle }{\langle i\mid j \rangle}.
\ee
By construction the diagonal matrix elements of $H_I$ are zero
and therefore there are no first order terms in the perturbation expansion.

A main characteristic of the CBF theory is the non-orthogonality of the 
basis,
\be
N_{ij}=\frac {\langle i \mid j \rangle}{(\langle i \mid i \rangle \langle j \mid
j \rangle)^{1/2}} \neq \delta_{i,j}.
\ee
Therefore, it is necessary to use non-orthogonal perturbation theory \cite{fesbach}. The perturbative  correction to $E_0^v$, in a non-orthogonal basis
is given by \cite{fesbach}
\bea
\Delta E_0 = E_0 - E_0^v &=& \sum_{p \ne 0} \frac {W_{0p}(0) W_{p0}(0)}{E_0 - E_p^v}
\nn \\
&+& \sum_{q \ne p \ne 0} \frac {W_{0p}(0) W_{pq}(0) W_{q0}(0) }{(E_0-E_p^v)
(E_0-E_q^v) } + ... .
\label{eq:seriesper}
\eea
Due to the non-orthogonality of the basis,  the interaction $W_{ij}(0)$
 contains the overlap matrix element $N_{ij}$,
\be
W_{ij}(0) = H_{ij} - E_0 N_{ij}
\ee
and carries an additional energy dependence, besides the one coming from 
the denominators.

To eliminate the unknown $E_0$ from the perturbative series, one expands
Eq.~(\ref{eq:seriesper}) around $E_0=E_0^v$. In this way,
\be
W_{ij}(0)=(H_{ij} - E_0^v N_{ij}) - \Delta E_0 N_{ij} \equiv W_{ij}^v(0)
- \Delta E_0 N_{ij}.
\ee
Notice that, $W_{ij}^v(0)$ depends  on the variational
energy $E_0^v$. The expansion for the energy denominators gives
\be
\frac {1}{E_0 -E_p^v} = \frac {1}{E_0^v -E_p^v + \Delta E_0} = 
\frac {1}{E_0^v -E_p^v} \sum_{i=0}^{\infty} \left ( -
\frac {\Delta E_0}{E_0^v -E_p^v} \right )^i.
\ee
Actually, the analysis of the series is  easier 
by expressing,
\be 
W_{ij}^v(0) = W_{ij}^v(i) + (E_i^v -E_0^v) N_{ij},
\ee
with $W_{ij}^v(i)= H_{ij} - E_i^v N_{ij}$. The series can then be
written in terms of $W_{ij}^v(i)$ 
\bea
\Delta E_0 &=& \sum_{p\ne 0} (W_{0p}^v(0) - \Delta E_0 N_{0p}) 
\left ( \frac {W_{p0}^v(p)}{E_0 - E_p^v} - N_{p0} \right ) 
\nn \\
&+& \sum_{q \ne p \ne 0} (W_{0p}^v(0)- \Delta E_0 N_{0p} ) 
\left ( \frac {W_{pq}^v(p)}{E_0 -E_p^v} - N_{pq} \right ) 
\left ( \frac {W_{q0}^v(q)}{E_0 -E_q^v} - N_{q0} \right )+ ...
\label{eq:seriesper1}
\eea 
Due to the correlations introduced in the wave functions, the calculation of 
matrix elements
is much more involved than in normal perturbation theory. The 
CB matrix elements are constructed from the matrix elements of $F^+ F$ and
$F^+ H F$ on Fermi gas states, characterized by the number of particles and holes.
The matrix elements are calculated by cluster expansions. 
After expanding the energy denominators of the series
 (Eq.~(\ref{eq:seriesper1})),   a careful analysis of the cluster expansion
of the matrix elements and of the energy dependence of the different terms
leads to the conclusion that $\Delta E_0$ is given by the sum of all
linked Goldstone type diagrams (see section \ref{sec:holeline}) 
built up with $W_{ij}^v(i)$ and $N_{ij}$
interaction boxes. The interaction boxes (or equivalently the matrix elements)
are the sums of only linked cluster diagrams. Therefore, the CBF 
perturbation series is well behaved in the thermodynamic limit \cite{fantoni84}.

In dealing with  CBF perturbation theory we are confronted 
with two types of diagrams,
the Goldstone type diagrams representing the terms of the series and the 
 diagrams corresponding to the cluster expansion of the 
matrix elements, which are summed up by means of FHNC techniques.

Perturbative CBF calculations for the binding energy of nuclear matter
have been performed up to second order. In this case, the
perturbative correction to the ground state energy is expressed as:
\bea
\Delta E_0 &=& \sum_{p \ne 0} W_{0p}^v(0) \left ( \frac {W_{p0}^v (p)}
{E_0^v -E_p^v} - N_{p0} \right ) \nonumber \\
&=& \sum_{p \ne 0} \frac {W_{0p}^v(0) W_{p0}^v(0)}{E_0^v - E_p^v} =
\sum_{p \ne 0} \frac {\mid H_{0p} - E_0^v N_{p0} \mid ^2 }
{E_0^v - E_p^v},
\eea
and has been calculated within the set of $2p-2h$ correlated basis states. 
In the thermodynamical limit, the energy denominators 
 can be expressed in terms of the variational estimate of the 
single particle energies,$e^v(p)$,
\be
E_0^v -E^v({\bf p}_1, {\bf p}_2, {\bf h}_1, {\bf h}_2)=e^v(h_1)+e^v(h_2)-e^v(p_1)
-e^v(p_2).
\ee
As expected, the second order  correction to the ground state energy, increases
the binding. In the first calculations, the correlation operator
 was taken of Jastrow type and in this case the perturbative second order
correction was large \cite{krot81}. On the other hand, the corrections
are much smaller, when the correlation with the full operatorial structure is
used.Therefore, a significant improvement in the convergence
of the series is expected. Typical values of second order corrections to the 
binding energy at saturation density of nuclear matter range from 
$\sim -1$ to $\sim -3$ MeV. A very recent estimation for the 
Argonne $V_{18}$ and the Urbana IX model for the three-body force, is $-1.89$ MeV
at saturation density \cite{fhnc2}.

The CBF perturbation theory has also been implemented to study other
observables,  besides the hamiltonian, like the self-energy, the momentum
distribution, the single particle spectral function,  the different nuclear
response functions, etc. In the previous section \ref{subsec:green} we have
considered the self-energy of a nucleon as the  key quantity to calculate the
single-particle Green's function. Here one can calculate the imaginary part and
split it similar to (\ref{eq:self2p1h}) and (\ref{eq:self2h1p})  into two
pieces, the correlation term, i.e. coupling to $2h1p$ correlated configurations
and the  polarization term corresponding to the coupling to $2p1h$
configurations. The imaginary part of $\Sigma^{2h1p}(k,\omega)$ for a particle
state, in second order perturbation theory is defined as
\be
Im \Sigma^{2h1p}(k,\omega < e^v(k_F))= \frac {\pi}{2} \sum \mid \langle {\bf k}
 \mid
H -E_0^v \mid {\bf p}_1; {\bf h}_1,{\bf h}_2 \rangle \mid^2  
\delta(\omega+ e^v(p_1)-e^v(h_1)-e^v(h_2))
\ee
while
\be
Im \Sigma^{2p1h}(k,\omega >e^v(k_F))= \frac {\pi}{2} \sum \mid \langle {\bf k}
\mid H - E_0^v \mid {\bf p}_1, {\bf p}_2; {\bf h}_2 \rangle \mid^2
\delta(\omega + e^v(h_2)-e^v(p_1)-e^v(p_2)),
\ee
Similar expressions hold for the hole states. Using dispersion relations
one can calculate the corrections to the variational estimate of the 
single particle potential
\be 
 \Delta e(k) = e(k)-e^v(k)=\frac {P}{\pi} \int_{-\infty}^{\infty} 
\frac {Im \Sigma(k,\omega)}{e^v(k)- \omega} d \omega .
\ee

In the case of the single particle momentum distributions, the second order
perturbative correction to $n^v(k)$ has been obtained by calculating the
expectation value of the occupation number operator $a^{\dagger}_{\bf k} a_{\bf k}$
on the perturbed ground state $\mid 0_{per} \rangle $ containing two particle 
two-hole correlated states:
\be
\mid 0_{per} \rangle = \mid 0_v \rangle + \frac {1}{4} \sum \alpha(
{\bf p}_1,{\bf p}_2;{\bf h}_1,{\bf h}_2) \mid {\bf p}_1 {\bf p}_2; {\bf h}_1
{\bf h}_2 \rangle ,
\ee
where
\be
\alpha({\bf p}_1,{\bf p}_2;{\bf h}_1,{\bf h}_2) =
\frac {\langle 0_v \mid H -E_0^v \mid {\bf p}_1,{\bf p}_2;{\bf h}_1,{\bf h}_2 
\rangle }{(e^v(h_1)+e^v(h_2)-e^v(p_1)-e^v(p_2))},
\ee
and keeping the terms of order $\alpha^2$ \cite{fantopa84}.

However, there are still some problems with orthogonality corrections
when the series is truncated at a given order. Moreover, it is not straightforward
to extract the orthogonalized eigenvectors in a non-orthogonal CBF theory.
Therefore, even if the applications of the non-orthogonal perturbation theory
are rather successful, it is desirable to have also a method available
in a orthogonal scheme. In this way,  the $N_{ij}$ boxes 
would be eliminated from the diagrams and the parallelism with normal
 perturbation theory becomes more  
 transparent. To this end, one can try a Schmidt orthogonalization process,
however it happens that the variational estimates of the hamiltonian on the 
Schmid orthogonalized states are worse than the ones calculated with the
non-orthogonal basis. Very recently, a new orthogonalization procedure has been 
devised \cite{fantoni88}, in which  the diagonal matrix elements of
the hamiltonian and of the identity operator are preserved.  It is a two
step process which combines the Schmid orthogonalization with the
L\"owdin transformation \cite{lowdin}. Standard perturbation theory may be used
in this new set of  states. Although both schemes should be equivalent, in practice orthogonal
CBF theory is more efficient particularly in calculating the spectral functions
\cite{benhar90,benhar92} or the nuclear response functions \cite{fantopa87,
fantofa89,fabro97}.

The one-body Green's function has also been studied in the framework of CBF.
 Instead of computing the Green's function from a Dyson equation,
 as in the perturbative scheme discussed in the previous section, one 
directly calculates the single particle spectral functions. 
The starting point is the variational estimate,
which is equivalent to approximate the true ground state by the variational
ground state wave function and take 1 particle and 2 particle-1 hole 
basis correlated states for  A+1 particles  
 in the case of 
$S_p(k,\omega)$ and the   
1 hole and 2 hole -1 particle correlated states for A-1 particles 
 for $S_h(k,\omega)$. 
The projection into   1 hole (1 particle) state is the dominant
one and is given by
\be
S_{1h(1p)}^{(v)}(k,\omega) = \mid \varphi_{\bf k}^{h(p)} \mid^2
 \delta(\omega-e^v(k)) 
\Theta \left [\pm (k_F-k) \right ],
\ee
where the upper (lower) sign in the argument of the $\Theta$-function applies
to the hole (particle) spectral function, $e^v(k)$ is the variational single 
particle energy and the overlap matrix element $\varphi_{{\bf k}}^{h(p)}$
is defined by
\be
\varphi_{{\bf k}}^{h}= \langle 0_v \mid a_{{\bf k}}^{\dagger} \mid h \rangle .
\ee
Where $\mid h \rangle$ is the basis correlated state of A-1 particle build on  a
 Slater determinant
where the state with momentum $k$ is missing.  In a similar way,
\be
\varphi_{{\bf k}}^{p} = \langle 0_v \mid a_{{\bf k}} \mid p \rangle.
\ee
where in this case the state $\mid p \rangle$ represents the correlated state
of A+1 particles build on a Slater determinant that besides having all 
the states occupied up to the Fermi level it has an additional particle
on the state ${\bf k}$ above the Fermi level. Actually these type of 
intermediate states are the only non-vanishing contributions in any 
uncorrelated ground-state. In such a case, the spectral functions $S_{unc,h(p)}
(k,\omega)$
are just $\delta$-functions with strength unity:
\be
S_{unc,h(p)}= \delta (\omega -\frac {\hbar k^2}{2m}) \Theta \left [ \pm 
(k_F-k) \right ].
\ee
As  correlations are present in the CBF states also other intermediate
states contribute to the zero order approximation to the spectral
function.  The contribution from 2h-1p (2p-1h) correlated states is spread
out in energy and can be interpreted as a background contribution to be added
to the quasi-particle part ($\delta$-function) of the spectral function. 
This spreading produces a considerable quench of the quasi-particle peak. 
On top of this variational estimate one can build perturbation corrections,
 which take into account the admixture of more complicated configurations in the
states considered in the definition of the spectral function. For instance,
the 
2p-2h admixture in the ground state, and the 2h-1p (2p-1h) in the
one-hole (particle) have been studied \cite{benhar90,benhar92}.
As a consequence of these corrections, the $\delta$-peak of the quasiparticle,
acquires a width (except for $k=k_F$), which is related to the imaginary part 
of the self-energy at the quasiparticle-energy. For $k=k_F$, the on-shell
imaginary part of the self-energy is zero and the quasiparticle peak is a
 $\delta$-function which strength ($Z_{k_F}$) which defines the discontinuity
 of the momentum
distribution at $k=k_F$.

We can conclude that CBF has reach its maturity and
systematic studies of different observables, for both nuclear matter
and nuclei are available. The starting wave function is realistic enough that
the perturbative corrections to the different quantities can be safely
calculated up to second order.  The formal properties of the perturbative series 
have been established and the efforts now can be addressed to improve the 
evaluation of the matrix elements or to study
 also higher orders in the perturbative expansion.

\subsection{\it Variational and Quantum Monte Carlo Techniques}

There are various possibilities to apply the techniques of Monte Carlo (MC)
sampling in calculations of quantum many-body systems. For example
 one can use MC based algorithms to sample
the multidimensional integral occurring in the evaluation of the
expectation value of the hamiltonian,
 or any other operator, on a given
trial function (Eq.~(\ref{eq:princvar})). In fact for simple interactions
 and central correlations the
procedure is well established and it has been applied both for bosons and 
fermions, for systems with finite and infinite particle numbers. This type 
of calculations are known as Variational Monte Carlo (VMC) 
methods. However, there are also methods based on stochastic algorithms, which 
aim at an exact integration of the many-body Schr\"odinger equation. Those are 
generically known as Quantum Monte Carlo (QMC) methods.
VMC and QMC calculations in systems of liquid helium, both  $^4$He and 
$^3$He, which follow Bose and Fermi statistics respectively, have motivated an 
enormous progress in Monte Carlo Methods.  
However, the nuclear many-body hamiltonian is much more complicated and
requires to introduce strong state dependence correlations in the nuclear 
wave function. 
Therefore the progress in accurate nuclear ground state calculations
based on MC techniques has been rather slow. 
It is only recently, that the drastic improvement
of the computational resources, i.e. parallel computers, allows to increase
the number of nucleons involved in the calculations.

Once the trial wave function is chosen, the variational Monte Carlo
method  provides an exact - at least in the statistical sense - evaluation of 
the expectation value, which is free of the approximations involved
in the integral equation methods discussed in the previous section. However,
we should not consider these methods as competing ones but rather as a 
complementary 
ones. For instance, it is clear that Monte Carlo  methods have difficulties
in considering the long-range behavior of the distribution function of an
infinite homogeneous system because of the finite size of the simulating
box (see below). On the other hand one can very well determine this long-range 
behavior in the context of HNC methods. In the case of finite nuclei, 
Monte Carlo methods have enormous problems in increasing the number of 
nucleons of the nucleus under consideration. On the other hand recent 
calculations with FHNC 
techniques can easily be applied to nuclei with A=40. For the systems,
where both methods can be applied, they can help each other in looking
for the appropriate correlation, in understanding the physical requirements
of the wave function and in evaluating the accuracy of the approximations
involved in the FHNC method.

Recent reviews on Quantum Monte Carlo
 methods can be found in Refs.\cite{monc1,kalos1,schmid1}. Another
 review-like book with 
presentation of the methodology and physical applications is \cite{fabromon}. 
Two introductions into this field and rather comprehensive reviews
by R. Guardiola
 on both the
Variational and the Quantum Monte Carlo method   
 have been published in Refs.\cite{guarmc1,guarmc2}. More oriented to nuclear 
physics is the paper of Carlson and Wiringa  on variational Monte Carlo
in  finite nuclei \cite{carlswi1}. A recent overview including also Green 
Function Monte Carlo for
nuclear systems
is due to S. Pieper \cite{piepermc1}. Finally, the generalization to finite 
temperature, focusing
mainly to applications in condensed matter physics 
 has recently been reviewed by D. Ceperley \cite{monc2}.

Here we will mainly discuss VMC calculations. In order to explain the main 
ideas we will, as in the previous sub-section, use a simple system of
bosons, obeying the hamiltonian of Eq.~(\ref{eq:hamil}) with the scalar
interaction of Eq.~(\ref{eq:pothe}). First we consider a droplet
of $A$ $^4$He atoms. A trial function satisfying the symmetry requirements
is given by,
\be
\Psi({\bf r}_1,{\bf r}_2,...,{\bf r}_A)= \prod_{i<j}^A f(r_{ij}) \prod_{i=1}^A
\phi(r_i)
\label{drop1}
\ee
i.e., the product of a Jastrow factor times a model wave function that assigns
the same single particle wave function to all the atoms. In this case, for
spinless particles and local interactions, the expectation value of the
hamiltonian  can easily be written in a form appropriate for
Monte Carlo  calculations. Let ${\bf R}$ represent the set of the $3A$ spacial
coordinates of the atoms in the droplet. We define a local energy $E_L({\bf R})$
by
\be
E_L({\bf R})= \frac {1}{\Psi({\bf R})} H \Psi({\bf R})
\ee
and introduce a probability distribution function
\be
p({\bf R})= \frac {\mid \Psi({\bf R}) \mid ^2 }{\int d{\bf R'} \mid 
\Psi({\bf R}') \mid^2 },
\ee
which by definition is normalized to 1. Notice also that for local potentials,
the contribution of the potential to the local energy is just the value of the
potential for the given configuration ${\bf R}$. The energy is
given by
\be
E_v= \frac {\langle \Psi \mid H  \mid \Psi \rangle}{\langle \Psi \mid \Psi 
\rangle }  =  \frac {\int d{\bf R} \frac {\mid \Psi ({\bf R}) \mid ^2 H 
\Psi(R)}{\Psi(R)}}{\int d {\bf R'} \mid \Psi({\bf R'}) \mid^2}  
 = \int d {\bf R}\, p({\bf R})\,E_L({\bf R}).
\label{eq:monte3}
\ee
The Monte Carlo way of evaluating this integral is to consider $p({\bf R})$ as the
importance sampling distribution and $E_L({\bf R})$ as the function to be sampled.
The essence of the Monte Carlo method is to generate a set of statistically 
independent configurations (${\bf R}_i$) which are distributed according to the 
 probability distribution
$p({\bf R})$. Once this set of points $({\bf R}_i)$ is obtained, the expectation
value may be calculated through
\be
E_v = \lim_{N \rightarrow \infty} \frac {1}{N} \sum_{i=1}^N E_L({\bf R}_i).
\ee
Actually, this way to organize the integral of Eq.~(\ref{eq:monte3})
is a special example of the more general
expression
\be
E_v=\frac {\int d {\bf R} \frac {\Psi^{\dagger}({\bf R}) H \Psi({\bf R})}
{W({\bf R})} W({\bf R})}{\int d {\bf R'} \frac {\Psi^{\dagger}({\bf R'}) 
\Psi({\bf R'})}{W({\bf R'})} W({\bf R'}) },
\ee
where $W({\bf R})$ is a  probability distribution function. 
When $W({\bf R})= \Psi^{\dagger}({\bf R}) \Psi({\bf R}) $ we recover
 Eq.~(\ref{eq:monte3}). Once we have a set of configurations
in the $3A$ dimensional space proportional to $W({\bf R})$ then the calculation
of $E_v$ or the expectation value of any other operator $\hat O$
\be
           O_v = \frac {\langle \Psi \mid \hat{O} \mid \Psi \rangle }
{\langle \Psi \mid \Psi \rangle }
\ee
is straightforward
\be
          O_v = \frac {\sum \frac {\Psi^{\dagger}({\bf R}) \hat{O} \Psi({\bf R})}
{W({\bf R})} }{\sum \frac { \Psi^{\dagger}({\bf R}) \Psi({\bf R})}
{W({\bf R})}},
\ee
where the sum runs over all the configurations (${\bf R}_i$).
 
In order to get an estimate for the accuracy of the MC evaluation of the
integrals, the samplings are grouped in blocks, each block with 
a sufficiently large number of points $({\bf R}_j\ j=1,{\rm nmov})$, ${\rm nmov}$
being the number of configurations in each block. The central limit 
theorem guarantees that the average values of the energy obtained in each
block $i$,
\be
E_v^i = \frac{1}{{\rm nmov}} \sum_{j=1}^{{\rm nmov}} E_L({\bf R}_j)
\ee
 are distributed as a Gaussian centered around the true average value $E_v$
\be
E_v = \frac {1}{{\rm nblock}} \sum_{i=1}^{{\rm nblock}} E_v^i
\ee
where ${\rm nblock}$ is the number of blocks and the  statistical
 error is given by
\be
\sigma = \frac {1}{\sqrt{{\rm nblock}-1}} \left ( \frac{1}{{\rm nblock}}
 \sum_{i=1}^{{\rm nblock}}
(E_v^i)^2 - E_v^2 \right )^{1/2}.
\ee
 
The configurations $({\bf R}_i)$ following the importance sampling function
can be generated through the so-called Metropolis algorithm\cite{MRRTT}, 
which yields a sequence of random numbers, a Markov chain, distributed 
according to the required distribution probability.  The way to generate 
a sequence of configurations  $({\bf R}_i)$ by means of the Metropolis 
algorithm  is rather simple and may be described as follows:
Assume that we are at a given configuration ${\bf R}_{old}$, and the
value of the wave function for this configuration is $w_{old} = 
\Psi({\bf R}_{old})$
 then one tries
to generate a new configuration by defining $3A$ new coordinates 
\be 
{\bf R}_{new} = {\bf R}_{old} + ({\rm RAND}() - 0.5)* step.
\ee
where ${\rm RAND}()$ produces random numbers uniformly distributed in 
$[0,1]$ and  $step$ defines the scale of the Monte Carlo moves. 
The value of the wave function at ${\bf R}_{new}$ is $w_{new}$.

Now we perform the Metropolis question to see if we are going to accept this new
configuration ${\bf R}_{new}$ in our Markov-chain: 
If the probability of the new configuration $p({\bf R}_{new})$ is larger than 
the older,
i.e. if $w_{new}^2 > w_{old}^2$, then ${\bf R}_{new}$ is accepted. If the 
probability is smaller, then ${\bf R}_{new}$ is accepted with probability
$w_{new}^2/w_{old}^2$. Both cases are included in the following condition:
\be
\mbox{if} ( w_{new}^2/w_{old}^2  < RAND() ) \label{eq:MRRTT}
\ee
 then ${\bf R}_{new}$ is accepted and is the starting configuration for the 
next move. If the condition (\ref{eq:MRRTT}) is not fulfilled, the move is
rejected and we try to generate another configuration from ${\bf R}_{old}$.
 One keeps track of
the number of acceptances, and a efficient calculation should 
keep the percentage of acceptances between $40\%$ and $60\%$. Also one can 
allow for 
several moves between the different blocks in order to reduce the correlations
between the different calculations of the energy. It is also convenient
to perform several moves at the beginning of the process, in order to reach
 the region where the Metropolis 
algorithm works properly. In the Monte Carlo language this is known
as thermalization. 

In the case of infinite homogenous systems one can of course not perform a 
simulation with an infinite
 number of particles. The usual way of dealing with these systems consists in 
representing the system in a simulation cell with periodic boundary conditions.
If the size of the cells is $L^3$, the density of the system determines
 the number of particles $A$ that we should consider for each cell, 
$\rho  = A/L^3$. Metropolis moves
are always recast into the main simulation cell, but in implementing the
algorithm one must consider the particles in the periodic images as well. 
At a given density, the number of particles which can be treated in a
calculation defines a clear limitation
to the size of the simulation cell. Typical number of particles in the
simulation box range between 50 and  100. 

There are no essential differences in the case of fermion systems 
with respect to what we have discussed for boson systems so far. 
There are, however,
some technical problems, associated with the calculation of determinants
(the model wave function should be antisymmetric) and the
manipulation of spin and isospin degrees of freedom.
When the interaction and the correlations do not depend on the discrete
(spin and isospin) degrees of freedom,
 the trial wave function can be considered
as the product of a Jastrow factor and as many Slater determinants as fermion
species, i.e. one corresponding to spin up and the other to spin down. Once
a configuration is defined, each determinant is a number and the acceptance
criterium for a Metropolis move contains  the squared quotient of
 two determinants times the contributions of the
correlation factors.  The number of
operations involved in the computation of a determinant is proportional to $A^2$
as a consequence, moving $A$ particles involves $A^3$ operations.  There
is an algorithm, based on the calculation  of a determinant by using the matrix
of cofactors
 which reduces the number of operations
to $A^2$, which is of course important for calculation with larger number of 
fermions\cite{ceper77}.

In the variational method one aims to a minimization of the energy. This implies
that one should perform  VMC calculations for several variational parameters.
 This
usually requires huge amounts of computing time. Besides,  the statistical
fluctuations of the results may give an incorrect localization of the minimum. 
To avoid this problem, a method  called   reweighting of configurations has been
introduced. This method tries to use configurations corresponding to the
same importance sampling function to calculate the energy 
for  wave functions which have  small changes in their variational parameters
\cite{ceper77}.  

In the nuclear case, the strong state dependence of the interaction
and the correlation, produce  several complications that limit the application
of the method to a small number of nucleons. Actually all the applications
with realistic potentials refer to light nuclei and only very recently 
attempts have been made to study homogeneous systems \cite{schmid99}.
 
For a finite number of nucleons,  the trial wave function
is a symmetrized product of non-commuting two-body operatorial correlations
$F_2(i,j)$ (Eq.~(\ref{eq:corre2})) times a Slater determinant of 
single particle functions 
(Eq.~(\ref{eq:trialwf})). The trial wave function must be  translationally
 invariant, and the correlation function fulfils this condition automatically .
Also the implementation of the translational invariant to the model wave 
function is rather  
 straightforward 
in the Monte Carlo method. For instance, if the model function has some dependence
on the single particle coordinates ${\bf r}_i$ , one just substitutes that set
of coordinates by $\tilde {\bf r}_i$ with $\tilde{\bf r}_i = {\bf r}_i - 
\bf R_{cm}$ and ${\bf R}_{cm}= 1/A \sum {\bf r}_i$. The fact that the trial 
wave function describes a localized bound state, is either reflected in the 
correlation operator
\be 
F_{ij}(r_{ij}) \rightarrow 0 ~~~~~~, r_{ij} \rightarrow \infty
\ee
or in the model wave function, by requiring that the model wave function
tends to zero when one of the particles is moved far away.

For very light nuclei one can take the model function ($\phi$) as a pure 
spin-isospin 
function with no spatial dependence, for instance for triton and the $^4$He
nucleus the following functions have been used \cite{carlswi1,lomnitz}
\be
\mid \phi(^3H) \rangle = {\cal A} \mid \downarrow n \uparrow n \downarrow p
 \rangle 
\ee
and 
\be
\mid \Phi(^4He) \rangle = {\cal A} \mid \downarrow n \uparrow n \downarrow p
\uparrow n \rangle 
\label{eq:hel}
\ee
where $\cal A$ is the antisymmetrizer and the up- and down-arrows indicate the
spin of the different nucleons. In this case the localization is impose on
the correlation functions. One of the problems is that the size of the
spin-isospin space of A nucleons increases very rapidly. There are $N_s=2^A$
 possible spin states ranging from all particles having their spins pointing down
to all having their spins pointing up.  If one fixes the charge of the 
system, i.e. the number of protons, then there are  
\be
                 N_{char}= \frac { A !}{Z ! (A-Z) ! }
\ee
states in the isospin space with the same total charge.
 The total dimension is therefore $N_{dim}=
N_s \times N_{char}$
For instance in the case of the $^4$He, $N_s= 12$ and $N_{char}=6$ with 
$N_{dim}=96$. While for the $^3$H, $N_s=8$ and $N_{char}=3$ while 
$N_{dim}=24$. However for the $^{16}O$, $N_s=65536$ and $N_{char}=12870$ 
while $N_{dim}= 843448320$. Each correlation operator $F_{ij}$ is a very
sparse matrix in the spin-isospin space. In evaluating the
expectation value, this matrix acts on a vector  in spin-isospin space to produce
another vector. For instance the spin-isospin vector corresponding to
the model wave function of $^4$He (Eq.~(\ref{eq:hel})) has 24 non zero
 components, as 
a result of the antisymmetrization. When the correlations act on those
initial vectors other states are generated. One can try a straightforward
 generalization of the VMC methods discussed above by writing
\be
\Psi = \sum_{p=1}^M \Psi_p
\ee
where $\Psi_p$ corresponds to the term in $\Psi$ in which the A(A-1)/2 correlation
operators $F_{ij}$ operate in a specific order labeled by the index $p$.
The energy expectation value is given by
\be
E_v = \frac {\sum_{p,q=1}^M  \int \Psi_p^{\dagger} H \Psi_q d {\bf R} }
{\sum_{p,q=1,M} \int \Psi_p^{\dagger} \Psi_q d {\bf R}}\, 
\label{eq:order}
\ee
where the spin isospin degrees of freedom are treated explicitly.
 However, to sample 
the coordinates using the Metropolis algorithm would still be not practical, 
because it would be necessary to take into account all the $(A(A-1)/2)^2$
possible terms, related to the ordering of the correlation operators in
 each side of
the expectation value. Instead, one can sample randomly the coordinates ${\bf R}$
and the order labels $p$ and $q$  using for the importance sampling function
\be
W_{pq}({\bf R}) = \mid Re[\Psi_p^{\dagger}({\bf R})  \Psi_q({\bf R})\mid 
\ee
where the absolute magnitude is required in order to be sure that the weight 
function is never less than zero. The statistical error per sample in 
configuration space is increased by sampling the order of the operators.
 However,
the time saved by evaluating each time only one term in the sum 
(Eq.~(\ref{eq:order}))
increases the efficiency of the calculation. The way to proceed is the following:
From an initial configuration $X =({\bf R},p\& q)$ a trial configuration
$X'=({\bf R}',p'\& q')$ is generated. 
The Metropolis condition (\ref{eq:MRRTT}) is now applied to the sampling
functions $W_{pq}({\bf R})$ and $W_{p'q'}({\bf R'})$ to see if $X'$ is accepted
or rejected. 

Special codifications
of the action of the operators on the spin-isospin vectors, based on 
binary representations of those vectors are useful.  
However, calculations performing the full spin-isospin summations,
as the ones just described, are feasible with present computers only up to A=8. 

To study heavier nuclei one needs to sample also the spin-isospin states
in a random walk which takes place in a combined  spin, isospin, and
coordinate space \cite{carlson85}. The main problem in sampling over spin-isospin
states is the increase in the statistical error if 
the sampling is done blindly. In order to reduce the variance,
 the spin-isospin states must be treated in a manner similar to that commonly 
used for the spatial
coordinates. In order to devise the proper strategy, we write the expectation
value of the hamiltonian as
\be
E_v = \frac {\int d{\bf R} \sum_{a,b} \Psi_a^{\dagger}({\bf R}) H_{ab}
\Psi_b({\bf R})}{\int d{\bf R} \sum_a \Psi_a^{\dagger}({\bf R}) \Psi_a({\bf R})},
\ee
in this case $a$ and $b$ represent spin-isospin states. The basis of spin-isospin
states is just the one discussed above, i.e. the states where each nucleon 
has a definite third component of spin and isospin. 
The Monte Carlo strategy consists in sampling values of ${\bf R}$ and $a$,
 while explicitly summing over all 
spin-isospin states $b$, for each choice of ${\bf R}$ and $a$. The full summation
over $b$, although it is a lengthy  calculation, reduces the statistical 
error of the calculation significantly, since for the exact wave function
\be 
\sum_b H_{ab} \Psi_b = E \Psi_a
\ee
for each configuration (${\bf R}, a$).

The generalization of the standard Metropolis algorithm in order to sample
${\bf R}$ and $a$ in the combined coordinate-spin-isospin space, must allow
not only to move the particles in coordinate space but also
for changes in the orientation of the spins and isospins of the nucleons. 
The  way to do that is by flipping the third component of the spin and/or
 by exchanging the spin-isospin of two nucleons, chosen in a random way.
Of course these are not the only possibilities but these movements are easy to
implement and enough to sample
the full spin-isospin space. As before, the proposed move is accepted 
or rejected according to the ratio of the squares of the wave function evaluated
at $({\bf R},a)$ and $({\bf R}',a')$, as in standard variational Monte Carlo.
In this way the Monte Carlo method has been applied to $^{16}O$, to 
calculate the energy, form factor and charge distribution, using 
the Reid V6 interaction, i.e. including up to the tensor channels in the
general expression of the force (Eqs.~(\ref{eq:pot1},\ref{eq:pot})).

Another alternative is to perform  a cluster expansion for the non-central
correlations while the central correlations and the antisymmetry are treated
to all orders. 
The different terms are exactly evaluated  with  Monte Carlo techniques.
 This procedure has allowed to study $^{16}O$ with 
the full Argonne $V_{14}$ two-nucleon interaction, supplemented with the
Urbana VII three-nucleon potentials, using  a trial wave function which
contains pair- and triplet-correlation operators.  Not only the 
binding energy but also other quantities like density distribution, momentum
distributions, charge form factors and response functions have been 
studied \cite{pieper90,pieper92}.

To go beyond VMC one needs to invent an algorithm to 
generate the exact ground state wave function.
 These new algorithms are generally called
as Quantum Monte Carlo and they try to integrate the many-body
Schr\"odinger equation by means of stochastic procedures. The simplest 
of these methods is the so called Diffusion Monte Carlo (DMC) which approximates
the many-body time dependent Green's function by calculating the time evolution
operator  
 in small time steps. The starting point is to consider the many-body
 Schr\"odinger equation
in imaginary time,
\be
 - \frac {\partial \Psi({\bf R},t)}{\partial t} = (H - E) \Psi({\bf R},t)
\label{eq:time1}
\ee
The formal solution of Eq.~(\ref{eq:time1}) is
\be
 \Psi({\bf R},t)= \exp{\left (- \left[ H - E \right ]t\right )} \Psi({\bf R},0)
\ee
where the propagator $\exp{\left  [-(H-E) t\right ]}$ is called Green's function,
and $E$ is a convenient energy shift. The imaginary time evolution of an arbitrary
starting state $\Psi({\bf R},0)$, once expanded in  the basis of stationary states
 ($\phi_i({\bf R})$) of the hamiltonian, is given by
\be
\Psi({\bf R},t)= \sum_i \exp {\left (- \left [E_i -E \right ] t\right )}
 c_i \phi_i({\bf R}).
\ee
The amplitudes of the different states change with time, increasing or 
decreasing depending on the sign of $(E_i -E)$. Independently of the
value of the energy shift $E$, the most important amplitude
after a long time will be the one corresponding to the state with the lowest
energy $E_0$.  
In other words the DMC method projects out the  ground state wave function from the
trial wave function using
\be
\Psi_0 = \lim_{t \rightarrow \infty }~~ \exp {\left [-(H-E_0)t \right ]} \Psi_T ~,
\ee
provided that the overlap between the starting trial function and the true
ground state is different than zero ($c_0 \neq 0$).
The eigenvalue $E_0$ is  exactly calculated during the process. 
The starting trial wave function is represented by a set of random 
vectors or walkers  $\left ( {\bf R}_1,
{\bf R}_2,.... {\bf R}_{Nw} \right )$ distributed with probability 
proportional to the trial wave function, in such a form that the time
evolution of the wave function is represented by the evolution of the set 
of walkers. Notice that the quantity interpreted as a probability distribution
function is the wave function of the system, and not its squared as in VMC.
Along the iteration process, going from a wave function to another wave function,
means to change the random numbers representing the wave function.
 Each walker generates a number of descendants (none, one or more than one) which
will finally accommodate to the exact wave function. The method is iterative and
the ground state is asymptotically approached. In its simplest form, the
DMC is only appropriate to describe the ground state of boson systems. In this
case the wave function is positive and can be used as a probability distribution
function. In the case of fermions, the ground state wave function is not 
positive definite, and one should be careful to interpret the wave function
 as a probability
distribution function, giving rise to the so called "sign problem".
 Importance sampling combined with the so called fixed
node approximation allows to build a method also in this case.  

The actual computation of the time evolution is done in small time steps $\tau$,
by writing  the Green's function  as a product over short time intervals
\be
\exp{\left (-\left [H-E\right ] t\right )} = \prod_{i=1}^n
 \exp{\left (- \left [H-E \right ] \tau\right )},
\ee
where $\tau=t/n$.  These short time propagators are then approximated as 
the product of free particle propagators and a propagator involving the
potential, 
\be
\exp{(-H \tau)}\approx \exp {(-V \tau)} \exp{(-K \tau)} 
\ee
where $V$ is the interaction and $K$ the kinetic energy operator. Their
coordinate representation is rather simple,
\be
G_K({\bf R}',{\bf R},t) =\langle {\bf R}' \mid \exp{-K t} \mid {\bf R}\rangle
= \frac {1}{(4 \pi D t)^{3A/2}} \exp{\left (-({\bf R}' -{\bf R})^2 /4Dt\right )}.
\ee
Where $D = \hbar^2/2m$ is called the diffusion constant. If the potential is local
the Green's function related to the potential is,
\be
G_V({\bf R}',{\bf R},t)=\langle {\bf R}' \mid \exp{(-V t)}\mid {\bf R}\rangle=
\exp{(-V({\bf R})t)} \delta({\bf R}- {\bf R}').
\ee

The coordinate representation of the total Green function for short times 
including the energy shift is given by
\be
G({\bf R}',{\bf R}, \tau)=G_K({\bf R}',{\bf R},\tau)
 exp{\left [E-V({\bf R})\right]}+ O(\tau^2).
\ee
Using the above expressions,  the propagation in coordinate space reads 
\be
\langle {\bf R} \mid \Psi(\tau+ \delta \tau )\rangle = \int d {\bf R}' \langle 
{\bf R} \mid G(\delta \tau) \mid {\bf R}' \rangle  \langle {\bf R}' \mid 
\Psi(\tau)\rangle .
\ee

This integral equation is solved by using Monte Carlo methods to sample the free
particle propagator and including branching  to take
the potential into account, thus allowing for the process of replication
of the walkers. The equation is iterated until convergence. 
In conclusion the scheme is well defined once one has found an appropriate 
approximation for the short time Green's function, and a collection of walkers 
describing
the starting state has been chosen. These walkers are evolved in time, 
obtaining a new collection of walkers. If the number of time steps is large
enough the final walkers
will represent the ground state wave function.  An important improvement is
obtained by introducing an importance sampling function to increase
the statistical accuracy of the method.

The method is well established for systems of bosons interacting through central
scalar interactions. For fermions one must deal with the so called 
sign problem, on which a lot of progress has been achieved in the last years,
see \cite{boronat99,moroni95}.
Realistic applications are in the context of quantum liquids to study
$^4$He and $^3$He. In the nuclear case, using realistic interactions
there are only applications for a few number of nucleons. Recent 
calculations are able to arrive up to 7 nucleons, calculating the ground 
state and low-lying excited states \cite{pudliner97}. Also new algorithms are 
being
designed to treat the nuclear  or neutron matter case with realistic interactions
\cite{schmid99,koonin99}.

\newcommand{\slasm}[1]{\mbox{$#1\hskip-0.5em{/}$}}
\section{Effects of Correlations derived from Realistic Interactions}
\subsection{\it Models for the NN Interaction\label{sec:nninter}}

In our days there is a general agreement between physicists working on this
field, that quantum chromo dynamics (QCD) provides the basic theory of 
the strong 
interaction. Therefore also the roots of the strong interaction between two
nucleons must be hidden in QCD. For nuclear structure calculations, however, one
needs to determine the NN interaction at low energies and momenta, a region in
which one cannot treat QCD by means of perturbation theory. On the other hand, 
the
system of two interacting nucleons is by far too complicate to be treated by
means of lattice QCD calculations. Therefore one has to consider
phenomenological models for the NN interaction.

Attempts have been made to develop models for the NN interaction, which consider
the QCD degrees of freedom, quarks and gluons explicitly. As an example we
mention the non-relativistic constituent quark models, which are very 
successful in describing the properties of baryons\cite{isgur}. Based on such
constituent quark models, so-called quark cluster models have been developed for
the NN interaction\cite{faes0,yap,bra}. These models
 include the effects of
one-gluon-exchange terms and account for the exchange of mesons
 between the
quarks. This exchange of mesons within the quark model is schematically 
displayed in the left part of Fig.~\ref{fig13}. The exchange of a
quark-antiquark pair is substituted by the exchange of mesons of various kinds. 
This interpretation is in line with the arguments of t'Hooft and
Witten\cite{thooft,witten}. They demonstrated that in the low-energy regime, the
relevant degrees of freedom of QCD should be well described in terms of a meson
theory. This means that also the quark-antiquark exchange processes displayed in
that figure should be dominated by the exchange of the collective
quark-antiquark modes, i.e.~the exchange of mesons. Such constituent quark
cluster models are very successful in describing the main features of the
baryon-baryon interaction, however, the accuracy of the fits of these
interactions to the empirical NN scattering phase shifts is not sufficient to
use such interactions based on a quark model for nuclear structure calculations.  

\begin{figure}[tb]
\begin{center}
\begin{minipage}[t]{14 cm}
\epsfig{file=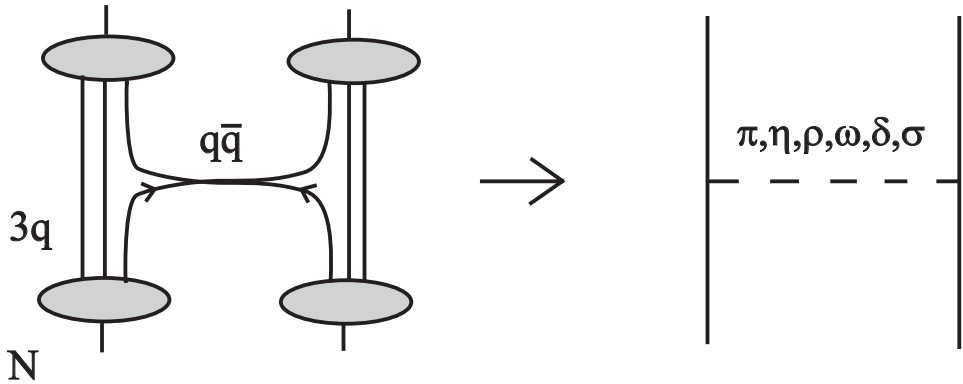,scale=1.0}
\end{minipage}
\begin{minipage}[t]{16.5 cm}
\caption{Nucleon-Nucleon interaction in the quark- and meson-exchange picture.
The lines in the left part of the figure refer to propagating quarks, whereas
those in the right part refer to nucleons and mesons. \label{fig13}}
\end{minipage}
\end{center}
\end{figure}

If one considers the meson degrees of freedom as the most important degrees of
freedom for QCD at low energies, it is quite natural to describe the NN
interaction in terms of mesons, which are exchanged between nucleons. The
internal structure of the nucleons can then be parameterized in terms of
meson-nucleon form-factors. Detailed discussions of these meson exchange models 
have been presented, e.g. in
 \cite{rupr0,erkelenz,carl,broja}. The basic idea
of this meson exchange model dates back to 1935, when Yukawa suggested that a
new particle with ``intermediate'' mass, eventually called a meson, should be
responsible for the strong interaction between nucleons\cite{yukawa}.

In modern one-meson or one-boson-exchange (OBE) model of the interaction one
assumes that the basic NN interaction is described by the exchange of all
mesons displayed in the right-hand part of Fig.~\ref{fig13}. The diagram of
that figure can be interpreted as a Feynman diagram representing the simplest
non-trivial contribution to the two-nucleon Green's function. Such
contributions can be evaluated by using essentially the same rules as discussed
in subsection \ref{subsec:green}. The basic building blocks for evaluating
these contributions are the operators for the meson-nucleon vertices, which are
given by
\bea
\Gamma_{\rm s}\  = & \sqrt{4\pi } g_{\rm s} &\quad \mbox{for a scalar meson,}
\nn\\[2mm]
\Gamma_{\rm ps}\  = & i\sqrt{4\pi } g_{\rm ps} \gamma^5 &\quad \mbox{for a
pseudoscalar meson,} \nn \\
\Gamma_{\rm v}\  = & \sqrt{4\pi }
\left[g_{\rm v}\gamma^{\mu}+\frac{f_{\rm v}}{2m}i\sigma^{\mu\nu}k_{\nu}\right]
&\quad \mbox{for a vector meson.}\label{eq:mesnucvert}
\eea
The $\gamma^{\mu}$ are the usual matrices using the conventions
of Bjorken and Drell\cite{bjorkdre}, the tensor operator is defined by the 
commutator $\sigma^{\mu\nu}=i [\gamma^\mu,\gamma^\nu]$ and $k_{\mu}$ refers 
to the 4-momentum of the exchanged meson. Matrix elements of these operators
have to be calculated between Dirac spinors for plane waves. In the helicity
representation these Dirac spinors are given as
\be
u(\vec q, \lambda ) = \sqrt{\frac{E_q+m}{2m}}\left( \matrix{ 1\cr \frac{2\lambda
q}{E_q+m}\cr}\right) \vert \lambda > \label{eq:diracspinor}
\ee
where $\lambda$ refers to the helicity, i.e.~the projection of the nucleon spin
on the direction of the momentum $\vec q$ and 
\be
E_q = \sqrt{\vec q^2 + m^2}
\ee 
is the relativistic energy of the free nucleon. Note that the Dirac spinors of
Eq.~(\ref{eq:diracspinor}) are normalized due to
\be
\bar u(\vec q, \lambda ) u(\vec q, \lambda ) =  u^\dagger(\vec q, \lambda )
\gamma^0 u(\vec q, \lambda ) = 1\, .
\ee
Instead of using the pseudoscalar coupling $\Gamma_{\rm ps}$ in
(\ref{eq:mesnucvert}) one also often employs the pseudovector coupling
\be
\Gamma_{\rm pv} =i\sqrt{4\pi }\frac{f_{\rm pv}}{m_{\rm mes}}\gamma^\mu \gamma^5
k_{\mu} \, .
\label{eq:psvecc}
\ee
Both couplings yield equivalent results for on-shell nucleons, if one
identifies $m_{\rm mes}g_{\rm ps}=2mf_{\rm pv}$. For nucleons described by 
Dirac spinors different from those of a free particle like in
(\ref{eq:diracspinor}), however, the
pseudovector coupling suppresses the enhancements due to the
antiparticle admixture as compared to pseudoscalar coupling.

If now one considers the interaction of two nucleons in the center-of-mass
frame with momenta $\vec{q}$ and $\vec{-q}$ before and the momenta
$\vec{q'}$ and $\vec{-q'}$ after the interaction, the matrix element for
the exchange of a meson of the kind $\alpha$ is given by
\be
{\cal V}_{\alpha}(\vec{q'},\vec{q}) =
\bigl(\bar{u}(-q') \Gamma_{\alpha} u(-q)\bigr)\, P_{\alpha}(k)\,
\bigl(\bar{u}(q') \Gamma_{\alpha} u(q)\bigr)\, ,
\label{eq:vborn}
\ee
with the Dirac spinors $u$ as defined in (\ref{eq:diracspinor}). 
Momentum conservation requires that
the 4-momentum of the exchanged meson is $k=q-q'$ and the meson
propagators are given by
\bea
P_{\rm s} & = & \frac{1}{k^2-m_{\rm s}^2} \quad \mbox{ for scalar and
pseudoscalar mesons,}\nn\\
P_{\rm v} & = & \frac{-g_{\mu\nu}+k_{\mu}k_{\nu}/m_{\rm v}^2}{k^2-m_{\rm
v}^2}\quad
\mbox{for vector mesons.}
\label{eq:mesprop}
\eea
A very efficient way for the evaluation of the OBE matrix elements in
(\ref{eq:vborn}) using the helicity representation for the Dirac
spinors has been presented in \cite{erkelenz}. The two-particle states
can be expanded in terms of eigenstates with respect to the total
angular momentum $J$. Since the OBE amplitudes are invariant under rotation
the angular momentum is a good quantum number and one obtains matrix
elements
\be
\langle \lambda_{1}'\lambda_{2}'q' \vert{\cal V} \vert\lambda_{1}\lambda_{2}q
\rangle_{J} = 2\pi \int_{0}^\pi d\theta\,
\sin{\theta}\,\,d^J_{\lambda\lambda'}(\theta ) \,\,
\langle \lambda_{1}'\lambda_{2}'
{\bf q}' \vert{\cal V} \vert\lambda_{1}\lambda_{2} {\bf q} \rangle\, ,
\label{eq:helici}
\ee
where $\lambda_{i}$ denotes the helicity of nucleon $i$, $\theta$ is the angle
between the momenta $\vec{q}$ and $\vec{q'}$, and the $d^J_{\lambda\lambda'}$
are the reduced rotation matrices with $\lambda = \lambda_{1}-\lambda_{2}$ and
$\lambda' = \lambda_{1}'-\lambda_{2}'$. Inspection of the symmetries for the
matrix elements in (\ref{eq:helici}) shows that there are six independent
matrix elements between the various helicity states for each $J$. These matrix
elements in the helicity representation can easily be transformed into the
conventional basis of partial waves for two nucleons
\be
\vert \lambda_{1}\lambda_{2}q \rangle_{J}\,  \Longrightarrow \, \vert
^{2S+1}L_{J} q \rangle\, ,
\ee
where $S$ identifies the total spin of the nucleons, $L$ is the orbital
angular momentum of the relative motion, which is usually labeled by
the letter $S$, $P$, $D\dots$ for $L=0,1,2\dots$, and $J=L+S$ is the total
angular momentum. The Pauli principle for the interacting nucleons
requires that the total isospin $T_{\rm iso}$
is related to these quantum numbers by
the requirement that the sum $L+S+T_{\rm iso}$ is an odd number.

As discussed before, one can account for the composite structure of the mesons 
and baryons by introducing form factors. This means that  one may consider the 
coupling constants in (\ref{eq:mesnucvert}) not as universal constants but as
depending on the 4-momenta of the interacting hadrons. A simple choice for
such a form factor, which is commonly used, is to assume that it depends
only on the momentum transfer $k$, the momentum of the meson, and takes
the form
\be
g_{\alpha}(k) = g_{\alpha}\left( \frac{\Lambda_{\alpha}^2-m_{\alpha}^2}
{\Lambda_{\alpha}^2 - k^2} \right)^\nu\, ,
\label{eq:formf}
\ee
with a cut-off parameter $\Lambda$ and an exponent $\nu$, which is
1 for the so-called monopole form factor. 

The operators defining the meson nucleon vertices in (\ref{eq:mesnucvert}) as
well as the meson propagators in (\ref{eq:mesprop}) refer to a relativistic
description. This means that the two-body amplitudes which we have defined so
far should be considered as an irreducible interaction kernel ${\cal V}$ to be 
used in a
relativistic equation like Bethe-Salpeter equation, which can schematically
be written 
\be
{\cal T}_{\rm BS}(q',q;P) = {\cal V}(q',q;P) + \int d^4k\frac{i}{(2\pi )^4}
{\cal V}(q',k;P) 
{\cal G}_{\rm BS}(k;P) {\cal T}_{\rm BS}(k,q;P)\, .
\label{eq:bethesal}
\ee
Note that this is an integral equation in the 4-dimensional space of momentum
vectors. The  momenta of the
interacting nucleons are defined in terms of the center-of-mass
momentum $P$, which is conserved, and the relative momenta $q,q'$, and $k$ in
such a way that the momenta of the particles are, e.g., $p_{i}=1/2 P \pm  k$.
In the center of mass frame the total momentum $P$ has a time like
component, which is identical to the total energy $\sqrt{s}$, with $s$
referring to the corresponding Mandelstam variable (see e.g. \cite{itzyk}), 
whereas the space component of $P$ is equal to zero. The
uncorrelated two-particle Green's function ${\cal G}_{\rm BS}$ occurring in
(\ref{eq:bethesal}) can be written as a product of relativistic single-particle
Green's functions
\be
{\cal G}_{\rm BS}(k;P) =
\left(\frac{1}{\frac{1}{2}\slasm{P}+\slasm{k}-m+i\eta}\right)^{(1)}
\left(\frac{1}{\frac{1}{2}\slasm{P}-\slasm{k}-m+i\eta}\right)^{(2)}
\, ,
\label{eq:gbethes}
\ee
where the superscript $(1)$ or $(2)$
refers to the corresponding nucleon. It is
common practice to ignore the propagation of the solutions of negative
energy and furthermore reduce the 4-dimensional Bethe--Salpeter equation
(\ref{eq:bethesal}) to an integral equation in three dimensions by
fixing the time component of $k$ in a covariant way. One of the
possible choices is the approach suggested by \cite{bbs}:
\be 
{\cal G}_{\rm BBS}(k;P) = \delta(k_{0})\frac{i}{2\pi}
\frac{m^2}{E_{k}}\frac{\Lambda^{+(1)}(k)\Lambda^{+(2)}(-k)}{\frac{1}{4}s
-E_{k}^2+i\eta}\, ,
\ee
with $\Lambda^{+(i)}$ referring to the projector on Dirac states with 
positive energy for particle $i$
\bea
\Lambda^{+(i)}(k) & = & \left(\frac {\gamma^0 E_k - \vec \gamma \cdot \vec k +
m}{2m}\right)^{(i)} \nn \\
& = & \sum_{\lambda_i} \vert u(\vec k, \lambda_i )><\bar u (\vec k, \lambda_i
)\vert 
\eea
with $u$ the Dirac spinors of (\ref{eq:diracspinor}). 
The assumption of the Blankenbecler--Sugar propagator, 
that the time-like component $k_{0}$ vanishes, means that
for the propagation of the intermediate states both nucleons are
considered to be off-shell by the same amount and the irreducible
interaction terms do not transfer energy between the interacting
nucleons. This implies that no energy transfer should be assumed for
the meson propagators if one considers the contributions of OBE in the
Blankenbecler--Sugar approximation. Replacing the Bethe--Salpeter
propagator ${\cal G}_{\rm BS}$ in (\ref{eq:bethesal}) by the
Blankenbecler--Sugar approximation one obtains
\bea
{\cal T}(q',q) &= &{\cal V}(q',q) + \int \frac{d^3k}{(2\pi )^3}
{\cal V}(q',k) \frac{m^2}{E_{k}}\frac{\Lambda^{+(1)}(k)\Lambda^{+(2)}(-k)}
{q^2-k^2+i\eta} {\cal T}(k,q)\, ,
\label{eq:bbs}
\eea
where we have replaced $s$ by $4E_{q}^2$, using $E_{q}=\sqrt{q^2+m^2}$.
Assuming that the matrix elements of ${\cal V}$ are calculated between
spinors, which correspond to the solution of the Dirac equation with
positive energy (to account for the projectors $\Lambda^+$), we may define
\bea
V({\bf q',q}) & = & \sqrt{\frac{m}{E_{q'}}} {\cal V}(q',q;P)
\sqrt{\frac{m}{E_{q}}}\, ,\nn\\[1mm]
T_{\rm scat}({\bf q',q}) & = & \sqrt{\frac{m}{E_{q'}}} {\cal T}(q',q;P)
\sqrt{\frac{m}{E_{q}}}\,
.
\label{eq:minrel}
\eea 
This allows us to rewrite the Blankenbecler--Sugar equation (\ref{eq:bbs})
as
\be
T_{\rm scat}({\vec q',\vec q}) = V({\vec q',\vec q}) + \int \frac{d^3k}
{(2\pi )^3} V({\vec
q',\vec k}) \frac{m}{q^2-k^2+i\eta}T_{\rm scat}({\vec k,\vec q})\, ,
\label{eq:lipschwn}
\ee
which has the form of the non-relativistic Lippmann--Schwinger equation
(\ref{eq:lipschw}) for the scattering $T_{\rm scat}$ matrix. 
This means that if we evaluate the
relativistic matrix elements of ${\cal V}$ and apply the so-called
``minimal relativity'' factors of (\ref{eq:minrel}), the
(relativistic) Blankenbecler--Sugar equation (\ref{eq:bbs}) becomes identical
to the non-relativistic scattering equation. The states of this
scattering equation can be rewritten in the usual partial-wave basis
and the integral (\ref{eq:lipschwn})  can be solved with the
techniques as described by \cite{haftab}.

Alternatives to the Blankenbecler--Sugar approach to reduce the Bethe--Salpeter
equation to a three-dimensional integral equation have been developed. As
examples we mention the approaches introduced by Kadychevsky\cite{kady}, 
Gross\cite{gross}, Thompson\cite{thompson}, Schierholz\cite{schier},
and Erkelenz\cite{erkelenz}. A detailed discussion of the various approaches 
has been presented by Brown and Jackson\cite{broja}.

With the OBE ansatz one can now solve the Blankenbecler--Sugar or a
corresponding scattering equation and adjust the parameter of the OBE model to
reproduce the empirical NN scattering phase shifts as well as binding energy and
other observables for the deuteron. Typical sets of parameters resulting from
such fits are listed in table~\ref{tab:obe}.

\begin{table}
\begin{center}
\begin{minipage}[t]{16.5 cm}
\caption{Parameters of the realistic OBE potentials Bonn $A$, $B$ and $C$ (see
table A.1 of \protect{\cite{rupr0}}).
The second column displays the type of
meson: pseudoscalar (ps), vector (v) and scalar (s) and the third its
isospin $T_{\rm iso}$.}
\label{tab:obe}
\end{minipage}
\begin{tabular}{rrrr|rr|rr|rr}
\hline
&&&&&&&&&\\[-2mm]
&&&&\multicolumn{2}{c}{Bonn A}&\multicolumn{2}{c}{Bonn
B}&\multicolumn{2}{c}{Bonn C}\\
Meson &&$T_{\rm iso}$&$m_{\alpha}$&$g^2_{\alpha}/4\pi$&$\Lambda_{\alpha}$
&$g^2_{\alpha}/4\pi$&$\Lambda_{\alpha}$&$g^2_{\alpha}/4\pi$&$\Lambda_{\alpha}$\\
&&&[MeV]&&[MeV]&[MeV]&[MeV]\\
&&&&&&&&&\\[-2mm]
\hline
&&&&&&&&&\\[-2mm]
$\pi$ & ps & 1 & 138.03 & 14.7 & 1300 & 14.4 & 1700 & 14.2 & 3000\\[2mm]
$\eta$ & ps & 0 & 548.8 & 4 & 1500 & 3 & 1500 & 0 & -\\[2mm]
$\rho$ & v & 1 & 769 & 0.86$^{\rm a}$ & 1950 & 0.9$^{\rm a}$ & 1850 & 
1.0$^{\rm a}$ & 1700 \\[2mm]
$\omega$ & v & 0 & 782.6 & 25$^{\rm a}$ & 1350 & 24.5$^{\rm a}$ & 1850 &
24$^{\rm a}$ & 1400\\[2mm]
$\delta$ & s & 1 & 983 & 1.3 & 2000 & 2.488 & 2000 & 4.722 & 2000\\[2mm]
$\sigma^{\rm b}$ & s & 0 & 550$^{\rm b}$ & 8.8 & 2200 & 8.9437 & 1900 & 8.6289 &
1700\\
&&&(710-720)$^{\rm b}$ & 17.194 & 2000 & 18.3773 & 2000 & 17.5667 & 2000\\
&&&&&&&&&\\[-2mm]\hline
\end{tabular}
\begin{minipage}[t]{16.5 cm}
\vskip 0.5cm
\noindent
$^{\rm a}$ The tensor coupling constants are $f_{\rho}$=6.1 $g_{\rho}$
and $f_{\omega}$ = 0. \\
$^{\rm b}$ The $\sigma$ parameters in the first line apply for NN channels 
with isospin 1, while those in the second line refer to isospin 0 channels. In
this case the masses for the $\sigma$ meson of 710 (Bonn A) and 720 MeV (Bonn B
and C) were considered.
\end{minipage}
\end{center}
\end{table}

Some of the OBE parameters, such as the masses of the $\pi$, $\eta$,
$\rho$, $\omega$, and $\delta$ mesons, are not free parameters but are taken
from the mass table \cite{padat}. Other parameters, such as the
cut-off parameters $\Lambda$ and the contributions from $\eta$ and
$\delta$ exchange, do not effect the fit very much but are used as a
fine tuning. The coupling
constant for the $\pi$ is very well constrained by the $\pi N$
scattering data.

Also the coupling constants for the $\rho$ meson, in
particular the large tensor coupling $f_{\rho}$ are deduced from a
dispersion analysis of $\pi N$ data in
 \cite{hoehl}. As we will see
below, this strong coupling for the $\rho$ is of some significance for
the nuclear structure calculation. A non-relativistic reduction for the
$\rho$ exchange, similar to the one performed for the $\pi$ in the
preceding subsection, yields a tensor component for the NN interaction
with a sign opposite to the one deduced from one-pion-exchange in
(\ref{eq:vpi2}). Therefore a strong $\rho$ exchange reduces the
tensor force originating from the $\pi$ exchange significantly.

The $\omega$ coupling constant used in the OBE potential displayed here
but also in other realistic OBE models is rather large. A simple quark
model with SU$(3)$ flavor predicts the $\omega$ coupling to be nine times
as large as that for the corresponding isovector vector meson, the $\rho$.
The strong $\omega$ exchange contribution, however, is required to
obtain sufficient repulsion for the NN interaction at short distances.
A possible reason for this discrepancy may be that the $\omega$
exchange in the OBE model contains an effective parameterization of
short-range repulsion originating from quark effects\cite{faes0}. Another
explanation would be that  
this strong $\omega$ exchange simulates
also
repulsive $\pi-\rho$ exchange terms with intermediate isobar
excitations\cite{anast1} (see also discussion below).

The only part of the OBE model, that has a purely phenomenological
origin is the $\sigma$ exchange, which describes the medium-range
attraction of the NN interaction. This exchange of the scalar $\sigma$ meson 
is used to describe various two-$\pi$ exchange processes, which are irreducible
with respect to intermediate NN states and therefore not accounted for by
summing ladder diagrams of one-pion-exchange in the Lippmann-Schwinger equation.
Such two-$\pi$ exchange processes can be studied in a systematic way by means 
of dispersion relations
 using experimental information from $\pi$N scattering
processes. Such studies of the medium-range attraction employing dispersion
relations have been performed e.g.~by Vinh Mau and collaborators\cite{paris} and
the group in Stony Brook\cite{durso}. They are the basis of the medium range
attraction, which is used in the so-called Paris potential\cite{pari2}. 

\begin{figure}[tb]
\begin{center}
\begin{minipage}[t]{11 cm}
\epsfig{file=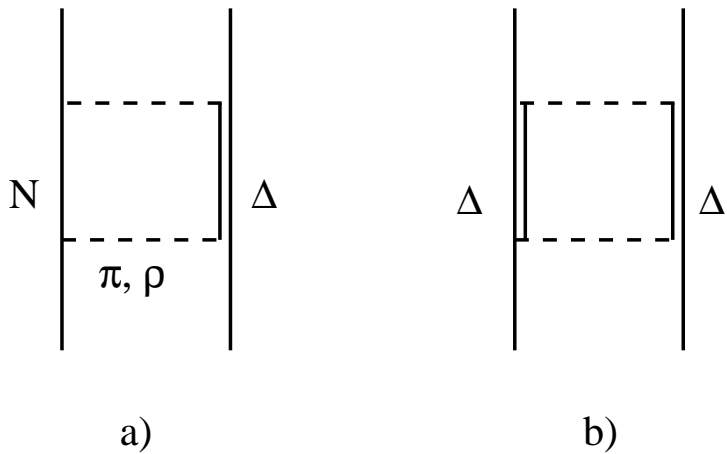,scale=0.7}
\end{minipage}
\begin{minipage}[t]{16.5 cm}
\caption{Contributions to the NN interaction beyond the OBE approach due to 
intermediate $N\Delta$ (a) and $\Delta\Delta$ excitations (b)\label{fig14}}
\end{minipage}
\end{center}
\end{figure}

Another way to determine the effects of these irreducible two-$\pi$ exchange 
processes is to evaluate such contributions explicitly. A rather important
contribution of this kind are the processes displayed in
Fig.~\ref{fig14}, in which the interacting nucleons are excited to intermediate
$N\Delta$ or $\Delta\Delta$ excitations (with $\Delta$ representing the isobar
excitation of the nucleon at 1232 MeV with spin and isospin 3/2). Since the 
$N\rightarrow \Delta$ excitation requires a change of the spin and isospin of the
baryon, such excitations could only be formed by the exchange of non-scalar,
isovector mesons like the $\pi$ and $\rho$ meson. The dominant contribution will
be the iterated two-$\pi$ exchange which yields attraction. In a simple OBE
model this attraction would be described in terms of $\sigma$ meson exchange.
Note that in $NN$ channels with isospin $T=0$ only $\Delta\Delta$ intermediate
states can be reached because of isospin conservation. In $T=1$ channels one
obtains contributions from intermediate $N\Delta$ and $\Delta\Delta$ states.
This is a plausible explanation for the feature that the masses of the $\sigma$
mesons listed in table~\ref{tab:obe} are larger for $T=0$ than for $T=1$.

Corresponding terms with $\pi$ and $\rho$ exchange are repulsive and of 
shorter range\cite{anast1}. They might be simulated by a strong $\omega$
exchange as discussed above\cite{oldrep}.

During the past few years, considerable progress has been made in constructing
realistic models for the NN interaction. In 1993, the Nijmegen group published a
new phase-shift analysis including selected proton-proton and proton-neutron
scattering data below a laboratory energy of 350 MeV with a $\chi^2$ per datum
of 0.99 for 4301 data\cite{nijmp}. Based on these data charge-dependent $NN$
potentials have been constructed by the Nijmegen group\cite{nijm1}, the Argonne
group\cite{Wiringa95} (Argonne $V_{18}$) and Machleidt et al.\cite{cdb} (CD-Bonn)
which reproduce the $NN$ data with a $\chi^2$ of 1.03, 1.09 and 1.03,
respectively.  In order to achieve fits with
such a high accuracy one has to go beyond the OBE ansatz discussed so far.
Machleidt et al.\cite{cdb} obtain such an accurate fit in the so called CD-Bonn 
potential\ by adjusting the parameters of the $\sigma$ meson exchange
in each partial wave, separately.  

One may say that these charge-dependent $NN$ interactions are essentially
phase-shift equivalent, the on-shell matrix elements of the NN transition 
matrix $T$ are almost identical. This
does, however, not imply that the models for the NN interaction underlying
these descriptions are identical. Moreover, the off-shell properties of each
potential may be rather different.  All models for the NN interaction $V$
include a one-pion exchange (OPE) term, using essentially  the same $\pi NN$
coupling constant. However, even this long range
part of the NN interaction, which is believed to be well understood, is treated
quite differently in these models. 

The CD-Bonn potential is based on the relativistic meson field theory, which
has been outlined above. Including the ``minimal relativity'' factors of 
(\ref{eq:minrel}) one obtains an expression for the One-Pion-Exchange
contribution to the NN interaction of two nucleons in the $^3S_1$ channel, in
plane wave states with momenta $k$ and $k'$ for the initial and final state,
respectively, which is of the form
\be
        < k'\vert V_{SS}^\pi \vert k > =
         -\frac{g_\pi^2}{4\pi}\frac{1}{2\pi m^2}
         \sqrt{\frac{m^2}{E_kE_{k'}}} \int_{-1}^1 d \cos \theta
          \left( \frac{\Lambda^2 -m_\pi^2}{\Lambda^2 + q^2}\right)^2
         \frac{k'k\cos\theta - (E_kE_{k'}-m^2)}{q^2 + m_\pi^2},
\label{eq:vpi1}
\ee
where $q^2$ denotes the momentum transfer, 
\be
 q^2  =  ({\vec k}-{\vec k'})^2 = k^2 +
                {k'}^2 - 2 k k' \cos\theta\, .
\ee
Note that the expression (\ref{eq:vpi1}) contains a dependence on the momenta
$k$ and $k'$, which cannot be reduced to a dependence on the momentum transfer.
This demonstrates that this expression for the One-Pion-Exchange yields a
non-local interaction. Results for such matrix elements as a
function of $k$, keeping $k'$ = 95 MeV/c fixed, are displayed in Fig.\
\ref{fig:vpi}\cite{artu971}.
 
\begin{figure}[tb]
\begin{center}
\begin{minipage}[t]{14 cm}
\epsfig{file=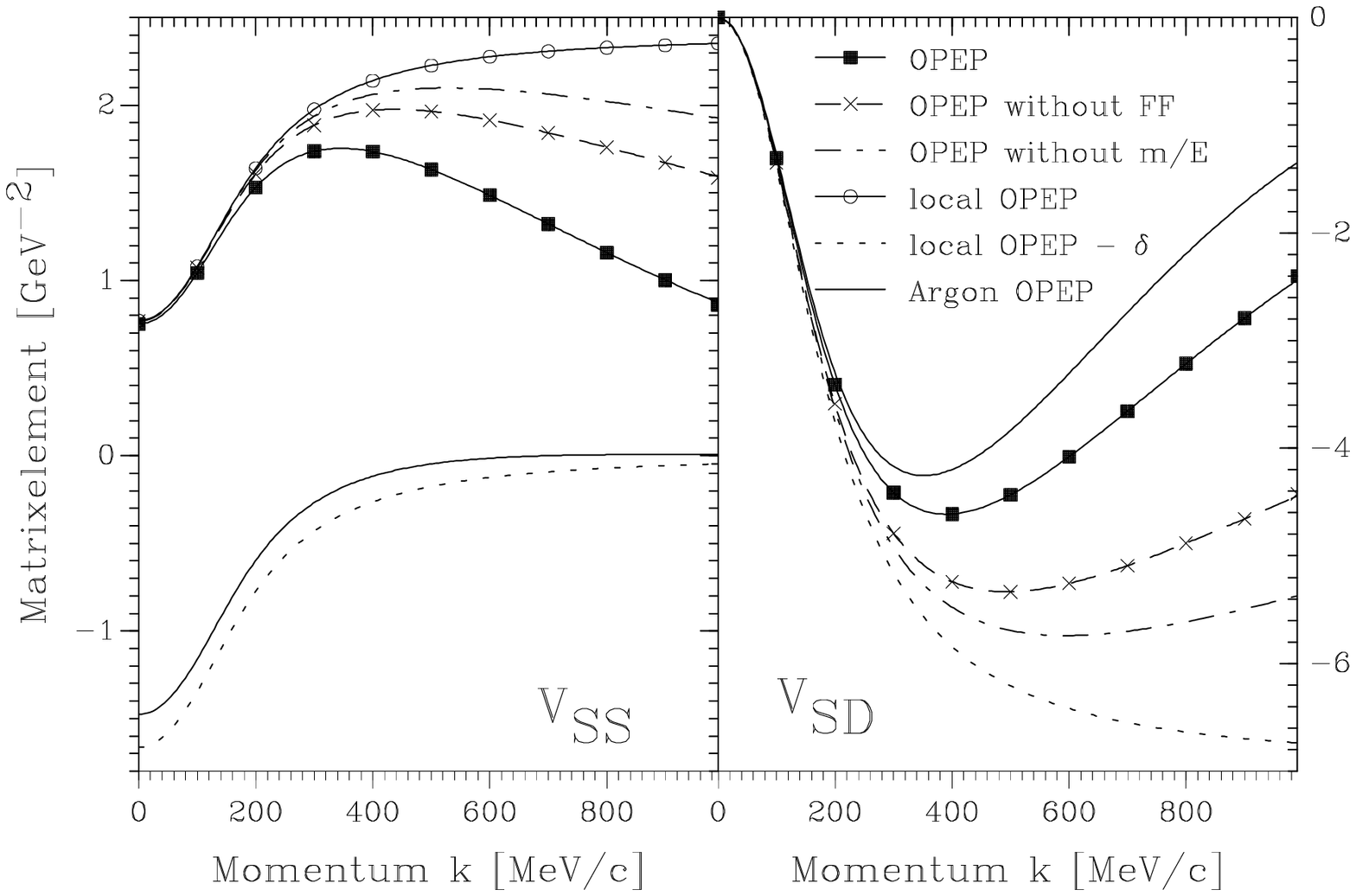,scale=0.75}
\end{minipage}
\begin{minipage}[t]{16.5 cm}
\caption{Plane-wave matrix elements of the one-pion-exchange potential (OPEP)
using various approximations. As an example, the matrix elements in momentum
space $\langle k'\vert V\vert k\rangle$ are shown as functions of $k$ for a
fixed value of $k'$ = 95 MeV/c. The left part of the figure exhibits matrix
elements for the partial waves $^3S_1$--$^3S_1$, while the right part shows the
tensor component in the $^3S_1$--$^3D_1$ channel with $k$ referring to the
momentum in the $^3D_1$ partial wave.\label{fig:vpi}}
\end{minipage}
\end{center}
\end{figure}  

If now we introduce in (\ref{eq:vpi1}) the nonrelativistic approximation for
\be 
     E_kE_k' - m^2 \approx \frac{1}{2}k^2 + \frac{1}{2} {k'}^2,
\label{eq:rel}
\ee
ignore the form-factor $(\Lambda^2 -m_\pi^2)/(\Lambda^2 + q^2)$ and the 
``minimal relativity'' factors, we obtain
\be
      \langle k'\vert V_{SS}^\pi \vert k \rangle_{\mbox{local}} =
      - \frac{g_\pi^2}{4\pi}\frac{1}{2\pi M^2}\int_{-1}^1 d \cos \theta
        \left(\frac{m_\pi^2}{2(q^2+m_\pi^2)} -
        \frac{1}{2}\right).
\label{eq:vpi2}
\ee
This can be viewed as the local approximation to the OPE since the matrix
element depends on the momentum transfer $q^2$, only. It can easily be
transformed into the configuration-space representation, resulting in a Yukawa
term plus a $\delta$ function, which originates from the Fourier transform of
the constant $1/2$ in Eq.\ (\ref{eq:vpi2}). The various steps leading to this
result are shown in Fig.\ \ref{fig:vpi}. It is obvious from this figure that
all of the steps leading to the local expression (\ref{eq:vpi2}) are not really
justified for momenta $k$ around and above 200 MeV, a region of relative
momenta which is of importance in the deuteron wave function. This is true for
the matrix elements $V_{SS}$ as well as $V_{SD}$.

Potentials like the Argonne but also the Nijmegen potentials contain the OPE
contribution in the local approximation regularizing the limit of small $r$.
In the case of the Argonne potentials the regularization is made in terms of a 
Gaussian function\cite{Wiringa95}.  It is remarkable that this regularization
leads to matrix elements in the $SD$-channel
which are close to those derived from the relativistic expression of the Bonn
potential. This is not the case in the $SS$ channel, where the removal of the
$\delta$ function term is very significant. This comparison of the various
approximations to the OPE part of the NN interaction demonstrates that even
this long range part of the NN interaction is by no means settled. The local
approximation and the regularization by form factors have a significant effect.

The description of the short-range part is also different in these models.  The
NN potential Nijm-II \cite{nijm1} is a  purely local potential in the sense that
it uses the local form of the OPE potential for the long-range part and
parameterizes the contributions of medium and short-range in terms of local
functions (depending only on the relative displacement between the  two
interacting nucleons) multiplied by a set of spin-isospin operators.  The same
is true for the Argonne $V_{18}$ potential \cite{Wiringa95}. The NN potential
denoted
by Nijm-I \cite{nijm1} uses also the local form of OPE but includes a $\bf p^2$
term in the medium- and short-range  central-force (see Eq.\ (13) of Ref.\
\cite{nijm1})   which may be interpreted as a non-local contribution to the
central force. The CD-Bonn potential is based consistently upon relativistic
meson field theory \cite{rupr0}. Meson-exchange Feynman diagrams are typically
nonlocal expressions that are represented in momentum-space in analytic form.
It has been shown \cite{cdb} that  ignoring the non-localities in the  OPE
part leads  to a larger tensor component in the bare potential.

\begin{table}
\begin{center}
\begin{tabular}{| c | r r r r r r |}
\hline
Pot. & $T_S$ & $T_D$ & $V_{SS}$ &  $V_{DD}$ & $V_{SD}$ & $P_D$ \\
& [MeV] & [MeV] &[MeV] &[MeV] &[MeV] & [ \% ] \\
\hline
&&&&&&\\
CD-Bonn & 9.79 & 5.69 & -4.77 & 1.34 & -14.27 & 4.83 \\
Argon $V_{18}$ & 11.31 & 8.57 & -3.96 & 0.77 & -18.94 & 5.78 \\
Nijm I & 9.66 & 7.91 & -1.35 & 2.37 & -20.82 & 5.66 \\
Nijm II & 12.11 & 8.10 & -5.40 & 0.59 & -17.63 & 5.64 \\
&&&&&&\\
\hline
\end{tabular}
\begin{minipage}[t]{16.5 cm}
\caption{Contributions to the kinetic and potential energy  of the deuteron
originating from the  $^3S_1$ and $^3D_1$ parts of the wave function as
defined  in Eq.\ \protect\ref{eq:split}.  Results are listed for the
charge-dependent Bonn potential (CD-Bonn \protect\cite{cdb}), the Argonne
$V_{18}$ \protect\cite{Wiringa95}, the  Nijmegen potentials Nijm I and Nijm II
\protect\cite{nijm1}.  The last column of this Table shows the calculated
$D$-state probabilities of the deuteron.}
\label{tab:tab1}
\end{minipage}
\end{center}
\end{table}

By construction,
all realistic NN potentials reproduce the experimental value for the
energy of the deuteron of --2.224 MeV.
However, the various contributions
to the total deuteron energy
originating from kinetic energy and potential energy in the
$^3S_1$ and $^3D_1$ partial waves of relative motion,
\begin{eqnarray}
   E & = & \langle \Psi_S \vert T \vert \Psi_S\rangle +
            \langle \Psi_D \vert T \vert \Psi_D\rangle +
           \langle \Psi_S\vert V \vert \Psi_S\rangle +
           \langle \Psi_D\vert V \vert \Psi_D\rangle + 2
           \langle \Psi_S\vert V \vert \Psi_D\rangle \nonumber \\
     & = & T_S + T_D + V_{SS} + V_{DD} + V_{SD}, \label{eq:split}
\end{eqnarray}
exhibit quite different results. This can be seen from the numbers listed in
Table \ref{tab:tab1}. In this table, we
display the various contributions to the deuteron binding
energy employing the four
potentials introduced above.

The kinetic energies are significantly larger for the local potentials
$V_{18}$ and Nijm II than for the two interaction models CD-Bonn and Nijm I
which contain non-local terms.
The corresponding differences in the $S$-wave functions
can be seen in Fig.\ \ref{fig:deuwav}. The local potentials yield a stronger
suppression of the $^3S_1$ wave function for small relative distances. This
reflects stronger repulsive short-range components of the local interactions.
These stronger short-range components are accompanied by larger high momentum
components in the momentum distribution, which yields larger kinetic energies. 

\begin{figure}[tb]
\begin{center}
\begin{minipage}[t]{12 cm}
\epsfig{file=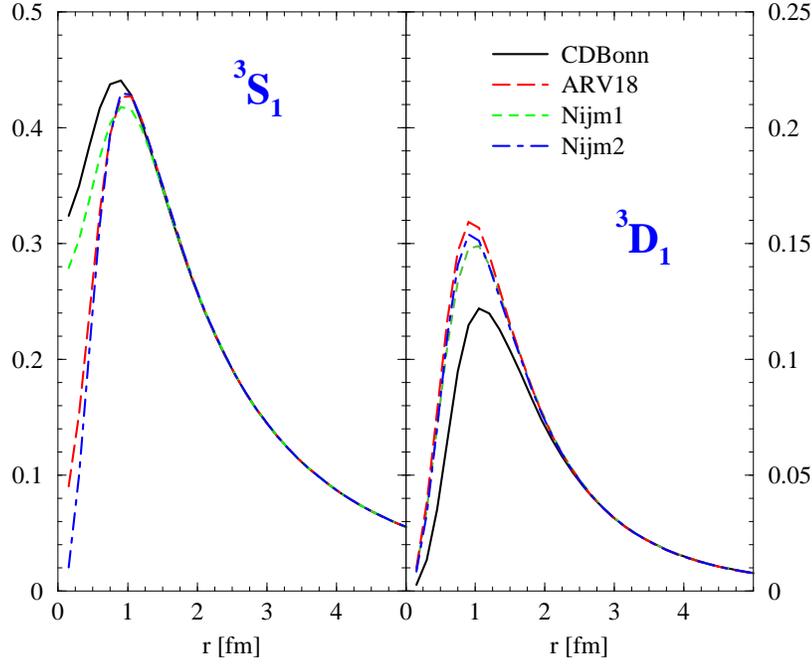,scale=0.65}
\end{minipage}
\begin{minipage}[t]{16.5 cm}
\caption{Wave function for the deuteron ($^3S_1$ and $^3D_1$) calculated for
different realistic interactions. \label{fig:deuwav}}
\end{minipage}
\end{center}
\end{figure}

Comparing the contributions to the potential energy, displayed in
Table \ref{tab:tab1}, one finds large differences
particularly for the tensor contribution $V_{SD}$. The dominant part of this
tensor contribution should originate from the tensor component of the
one-pion-exchange potential which we discussed above.

Although the modern NN potentials yield essentially the same NN scattering phase
shifts and the same
binding energy for the deuteron, there are significant differences
in the contributions to both the kinetic and potential energy of the deuteron
in the various partial waves. Speaking in general terms, these
differences can be traced back to
off-shell differences between the potentials. In particular it is the
inclusion of non-local contributions in the long-range ($\pi$ exchange) as well
as short-range part of the NN interaction, which is responsible for these
differences.

While the definition of a realistic two-body interaction between nucleons is a
rather well defined subject with only little differences between the various
models, the situation is much less clear for three-body and other many-body
forces. As an example let us consider the process displayed in
Fig.~\ref{fig15}a) with one of the three nucleons being excited to the $\Delta$
resonance in the intermediate state. In a many-body theory which does not 
consider isobar
degrees of freedom explicitly, this process should be included as two-meson
exchange three-nucleon interaction. If one calculates the expectation value of
this three-nucleon force with the uncorrelated ground state of the hole line
expansion, the Goldstone diagrams displayed in Fig.~\ref{fig15}b) and c) occur.
For a nuclear system with total isospin $T=0$ the contribution in b) vanishes
since it represents a $\Delta$ - hole (isospin T=1 or 2) admixture to the 
ground state. The diagram of Fig.~\ref{fig15}c) is closely related to the ground
state expectation value of the two-body interaction term shown in
Fig.~\ref{fig14}a). Fig.~\ref{fig15}c corresponds to a correction of that 
two-body diagram which is due to the fact that the intermediate nucleon states
below the Fermi level are blocked by the Pauli principle. Such corrections have
been included in nuclear structure calculations, which employed NN interactions
of the sort shown in Fig.~\ref{fig14}\cite{oldrep,anast2}. This demonstrates the
model dependence of such three-nucleon terms.

\begin{figure}[tb]
\begin{center}
\begin{minipage}[t]{12 cm}
\epsfig{file=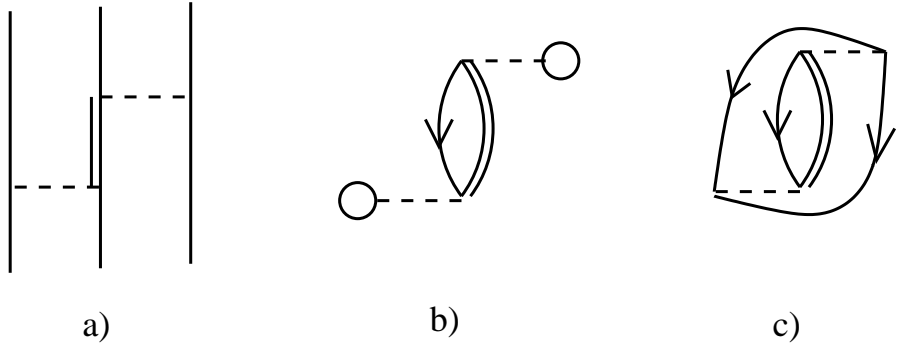,scale=0.6}
\end{minipage}
\begin{minipage}[t]{16.5 cm}
\caption{Contributions to the three-body forces as discussed in the 
text\label{fig15}}
\end{minipage}
\end{center}
\end{figure}

The models for the three-nucleon interactions which are frequently used in
nuclear many-body calculations are based on rather phenomenological grounds.
Typically they include a local parameterization of the two-pion exchange terms
with intermediate $\Delta$ excitations of the form\cite{fuji}
\be 
V_{ijk} = A_{2\pi}\sum_{cycl} \left( \left\{ X_{ij},X_{ik}\right\} \left\{
\vec{\tau}_i\cdot\vec{\tau}_j,\vec{\tau}_i\cdot\vec{\tau}_k \right\}  +
\frac{1}{4}\left[ X_{ij},X_{ik}\right] \left[
\vec{\tau}_i\cdot\vec{\tau}_j,\vec{\tau}_i\cdot\vec{\tau}_k \right] \right)
\ee
where 
\be
X_{ij} = \frac{e^{-m_\pi r_{ij}}}{m_\pi r_{ij}}\vec \sigma_i \cdot \vec \sigma_j
+ \frac{e^{-m_\pi r_{ij}}}{m_\pi r_{ij}} \left[1 + \frac{3}{m_\pi r_{ij}} +
\frac{3}{(m_\pi r_{ij})^2}\right] S_{ij}
\ee
and the symbols $[,]$ and $\{,\}$ denote commutator and anti-commutator,
respectively. Furthermore one introduces a repulsive short range
term\cite{carlson1}. The parameters are then adjusted to obtain a good fit to
the binding energies of few-body nuclei and nuclear matter. Various three-body
forces, labeled e.g. UVII, UIX etc.  
have been introduced in this way by the Urbana group\cite{threeforce}.

\subsection{\it Ground state Properties of Nuclear Matter and Finite Nuclei}

In the first part of this section we would like to discuss the convergence of
the many-body approaches and compare results for nuclear matter as obtained from
various calculation schemes presented in section 2. 
The convergence of the hole-line expansion for nuclear matter has been 
investigated during the last few years in particular by the group in
Catania\cite{song1,song2}. Continuing the earlier work of Day\cite{day81} they
investigated the effects of the three-hole-line contributions for various
choices of the auxiliary potential $U$ (see Eq.~\ref{eq:ubhf}). In particular
they considered the standard or conventional choice, which assumes a
single-particle potential $U=0$ for single-particle states above the Fermi level,
and the so-called ``continuous choice'', which has been advocated by the Liege
group\cite{bruek3}. This continuous choice supplements the definition of the
auxiliary potential of the hole states in Eq.~(\ref{eq:ubhf}) with a 
corresponding definition (real part of the BHF self-energy) also for the
particle states with momenta above the Fermi momentum, $k >k_F$. In this way
one does not have any gap in the single-particle spectrum at $k=k_F$.

Fig.~\ref{fig:baldo} shows the results of BHF calculations for the Argonne
$V_{14}$\cite{argo0} interaction, assuming the standard choice (labeled BHF-s 
for standard) and the continuous choice (BHF-g) for the auxiliary potential.
At $k_F = 1.4$ fm$^{-1}$, which corresponds to a density close to the empirical
saturation density,  the standard choice yields an energy of $-10.9$ MeV per
nucleon while the continuous choice leads to $-17.1$ MeV. At first sight this
dependence on the auxiliary potential, a difference of 6 MeV seems very large.
One should keep in mind, however, that a Hartree-Fock calculation at the same
density for the same interaction yields +41.8 MeV per nucleon. This means that
the inclusion of two-hole line contributions yields an attractive contribution
of 52.7 MeV and 58.9 MeV for the standard and continuous choice, respectively.
This means that the uncertainty, how to choose the auxiliary potential, leads to
a 10 percent effect in this two-hole line contribution.

\begin{figure}[tb]
\begin{center}
\begin{minipage}[t]{12 cm}
\epsfig{file=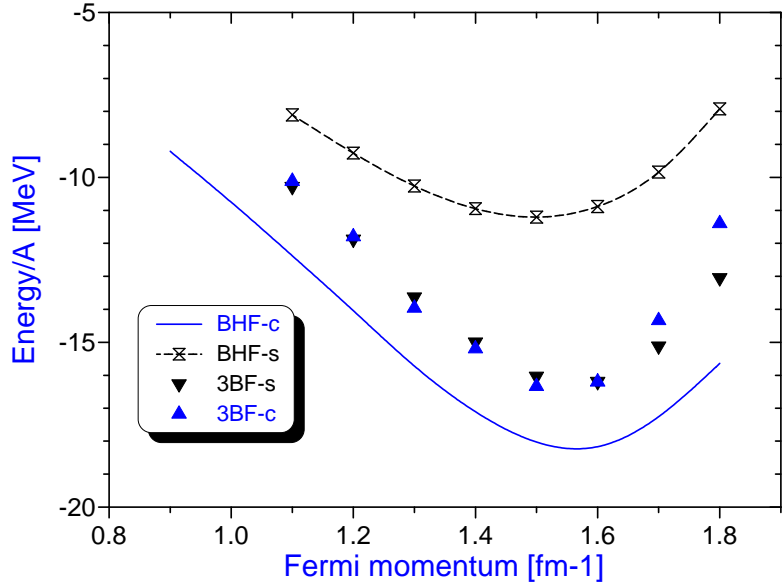,scale=0.6}
\end{minipage}
\begin{minipage}[t]{16.5cm}
\caption{Energy per nucleon calculated for symmetric nuclear matter as a
function of the Fermi momentum. Continuous lines represent the results of BHF
calculations assuming the standard choice (BHF-s) and the continuous choice
(BHF-c) for the auxiliary potential. The symbols (3BF) show results of
calculations  \protect{\cite{song2}} with inclusion of the three-hole line
contributions. The results were obtained using the Argonne $V_{14}$
interaction\protect{\cite{argo0}}.\label{fig:baldo}}
\end{minipage}
\end{center}
\end{figure}

Looking at the Bethe-Goldstone equation (\ref{eq:betheg}) it is easy to
understand that the continuous choice leads to more binding energy. The absolute
values for the attractive energy denominators in the Bethe-Goldstone equation,
which correspond to the excitation energies of two-particle two-hole excitations
are smaller if one accounts for an attractive single-particle potential also in
the case of the particle states. This enhances the effects of correlations,
leading to more attractive matrix elements for the G-matrix, and provides more
binding energy.

The Catania group also evaluated the three-hole line contributions to the energy
of nuclear matter for both choices of the auxiliary potential $U$. Results of
such calculations, including the sum of all Bethe-Fadeev ladders, are displayed
by symbols in Fig.~\ref{fig:baldo}. The contributions of the three-hole line
terms are -3.1 MeV and +1.9 MeV per nucleon at $k_F$ = 1.4 fm$^{-1}$ for the
standard and continuous choice, respectively. This is about a factor 20 smaller
than the two-hole line contributions discussed above. This indicates a very good
convergence of the hole line expansion for nuclear matter around saturation
density. This conclusion is supported by the fact that the results are rather
independent on the choice of the auxiliary potential $U$ after the effects of
three-hole lines are included. Differences can be observed only for densities, 
which correspond to $k_F$ around 1.7 fm$^{-1}$ and larger. At those higher
densities the effects of  four-hole line and higher order terms may get
important.
Also it is worth noting that the BHF results using the continuous choice are
closer to the results with inclusion of three-hole line terms. 

How do such results obtained from the hole line expansion compare with those
derived from variational calculations? In order to discuss this question we
consider as a first step again the results obtained for the Argonne $V_{14}$
interaction at a density of 0.185 fm$^{-3}$, which corresponds to a Fermi 
momentum $k_F$ of 1.4 fm$^{-1}$. As we have seen from the discussion above, the
hole-line expansion yields -15. to -15.2 MeV per nucleon if the effects of 
three-hole line contributions are taken into account. The variational
calculation of Wiringa et al.\cite{fhnc1} predict an upper bound of -13.4 MeV
using the FHNC/SOC approximation at this density. The perturbative corrections
within the framework of the CBF theory (see Eq.~(\ref{eq:seriesper1}) and
subsequent discussion) may provide another -1.5 to -2 MeV\cite{akmal98} so that
these very different approaches yield very similar results. 

In comparing such numbers one must keep in mind that the results of both 
approaches should be considered with a kind of ``error bar''. This error bar
should account for conceptual problems: On one side we do not know about the 
effects of higher order terms in the hole line expansion, on the other side we
don't know about the relevance of e.g.~elementary diagrams left out in the
FHNC/SOC approach. However one should also be aware of uncertainties which are
due to technical problems. For example, in performing self-consistent BHF 
calculations one typically solves the Bethe-Goldstone equation assuming an
angle-averaged approximation for the Pauli operator\cite{haftab,gammel}. Recent
investigations show that an exact treatment of the Pauli operator may can
modify the calculated energy up to around 0.5 MeV per
nucleon\cite{ksuzuk,schiller}. Also details of the parameterization of the
single-particle energies can lead to differences of this order of magnitude.
Uncertainties due to technical reasons must also be considered for the
variational calculations. As an example we mention the restriction to a
specific variational ansatz for the wave  function. As an indication for
uncertainties of variational calculations we mention the calculation of the
kinetic energy. The kinetic energy can be calculated using different
expression. If all many-body clusters are calculated completely these
expressions yield the same results. So the difference in energies obtained
using the so-called Jackson-Feenberg and the Pandharipande-Bethe expression is
a measure of the error\cite{zabodif}. 
This estimates yields error-bars of the order of 1 MeV
per nucleon\cite{fhnc2} at densities around 0.16 fm$^{-1}$,  the empirical
saturation density of nuclear matter.     

\begin{figure}[tb]
\begin{center}
\begin{minipage}[t]{12 cm}
\epsfig{file=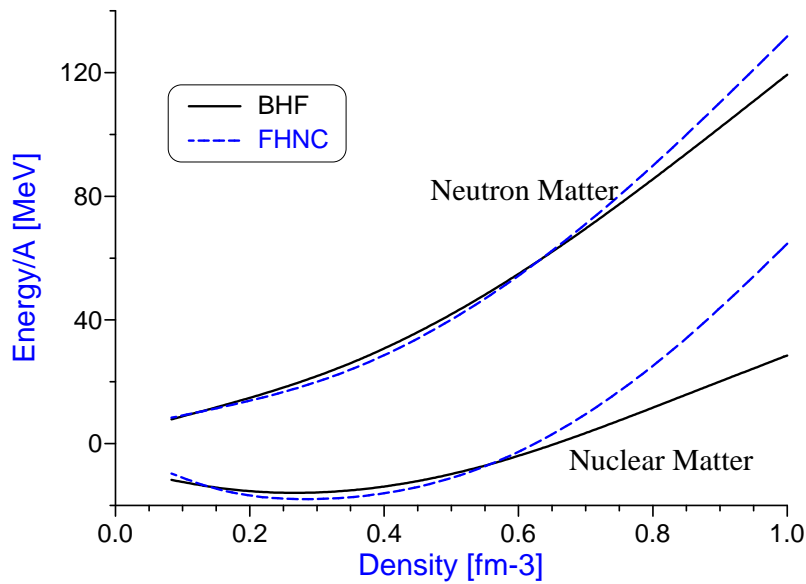,scale=0.6}
\end{minipage}
\begin{minipage}[t]{16.5cm}
\caption{Energy per nucleon calculated for symmetric nuclear matter and pure
neutron matter as a function of the Fermi momentum.  The results were obtained
from BHF\protect{\cite{engvik}} and FHNC\protect{\cite{akmal98}}using the
Argonne $V_{18}$ interaction\protect{\cite{argo0}}.\label{fig:engv}}
\end{minipage}
\end{center}
\end{figure}

A comparison between the energies for nuclear matter and neutrons at various
densities calculated in the framework of the variational approach\cite{akmal98} 
and the BHF approximation\cite{engvik} is displayed in Fig.~\ref{fig:engv}.
For both calculations the charge-dependent version of the Argonne potential
\cite{Wiringa95}, $V_{18}$ has been used. One finds a rather good agreement
between the two approaches at densities up to 0.6 fm$^{-3}$, which is about 4
times the saturation density of nuclear matter. The predictions of the two 
approaches deviate from each other at higher densities. This differences at
higher densities are to be expected since the convergence of the hole-line 
expansion should be good at low densities only. Also the variational approach
should be less accurate at high densities, genuine n-body correlations should
become important. The comparison also shows, however, that quite reliable
estimates for the equation of state for symmetric nuclear matter as well as
asymmetric nuclear systems can be deduced from such microscopic many-body
calculations up to the densities shown in  Fig.~\ref{fig:engv}. 

While the techniques developed for the variational calculations impose a
restriction to the NN interaction to local potentials, more general
interactions can be considered in the Brueckner hole-line approximation.
Therefore in comparing the features in the predictions derived for various
two-body interactions we will restrict ourselves to the discussion of results
obtained within the framework of BHF.  Results for the saturation points of
nuclear matter, i.e.~the minima of the energy per nucleon versus density
curves, are displayed in Fig.~\ref{fig:coester}. The results presented include
predictions from rather old versions of the NN interaction\cite{oldrep} like
the Reid  soft-core potential\cite{reid} (E/A=-10.3 MeV, $k_F$=1.4 fm$^{-1}$),
but also modern versions like the charge-dependent Bonn potential\cite{cdb}
(E/A=-19.7 MeV, $k_F$=1.68 fm$^{-1}$) and Argonne $V_{18}$\cite{Wiringa95} 
(E/A=-16.8 MeV, $k_F$=1.59 fm$^{-1}$). All these saturation points form a band,
the well known Coester band, an observation which has been made already in
1970\cite{coestba}: Some interactions, like the Reid soft-core, produce a 
minimum around the empirical saturation density, $k_F=1.36$, but  predict an
energy of only -10 to -11 MeV per nucleon, which is too small as compared to the
value -16 MeV taken from the 
volume term of the empirical mass formula. Other
interactions predict an energy, which is close to the empirical value but
saturate at a density, which is around 60 percent too high (or even higher). 

\begin{figure}[tb]
\begin{center}
\begin{minipage}[t]{8 cm}
\epsfig{file=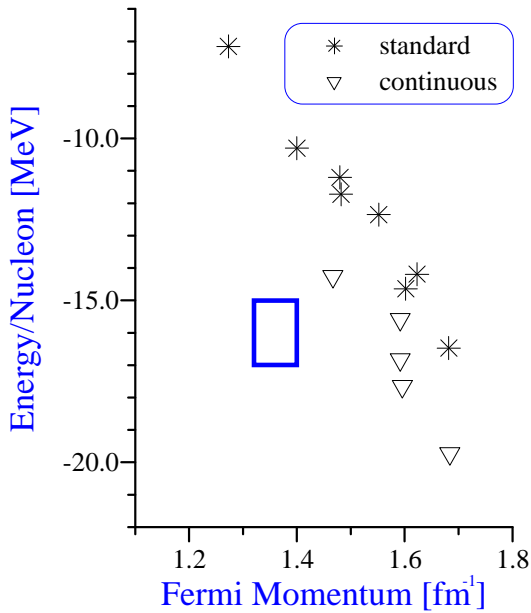,scale=0.8}
\end{minipage}
\begin{minipage}[t]{16.5cm}
\caption{Saturation points of nuclear matter calculated for various realistic NN
interactions. Some of those results are taken from \protect{\cite{oldrep}}.
Results were obtained using BHF with standard choice for the auxiliary
potential, as well as the continuous choice.\label{fig:coester}}
\end{minipage}
\end{center}
\end{figure}

Traditionally, the saturation properties of the different interactions have been
related to the $d$-state probability in the deuteron predicted by this
interaction\cite{oldrep} using the following argument: All interactions produce
essentially the same phase-shifts for NN scattering, i.e.~they yield the same
$T$-matrix. One may distinguish ``soft'' interactions, in which the attraction
of $T$ originates predominantly from the bare potential $V$. The attractive
terms of higher order, reflecting the importance of correlations are of little
importance. Such soft interactions show a small amount of correlations, which is
connected to a small $d$-state probability, a direct indication for weak tensor
correlations. The key-quantity of the BHF approximation is the $G$-matrix, which
as we have discussed already in section \ref{sec:holeline}, corresponds to the
$T$-matrix except that the terms of higher order in $V$ are quenched because of
the Pauli operator and dispersive corrections (the absolute values of the
energy denominator are larger in the Bethe-Goldstone equation for the nuclear 
medium than in the Lippmann-Schwinger equation). For a ``soft'' interaction this
quenching of the attraction in the higher order terms shows only a little
effect: the $G$-matrix is almost as attractive as $T$ therefore one obtains
large binding energies at large densities. In contrast, a ``stiff'' interaction
produces more correlations, the attractive contributions of the higher order terms
in $T$ are more important, the quenching mechanism is more important and one
obtains saturation points for nuclear matter at smaller densities and smaller
energies.

In order to explore these saturation features a little bit more in detail, we
will focus our attention to those NN interactions which have recently been
fitted with  high precision to NN scattering phase shifts as we already
discussed in the preceeding section. In particular we would like to explore the
contributions of various components of these interactions to the calculated
binding energy.  
The BHF approach yields the total energy of the system including effects of
correlations. Since, however, it does not provide the correlated many-body
wave function, one does not obtain any information about e.g.~the expectation
value for the kinetic energy using this correlated many-body state. To obtain
such information one can use the Hellmann-Feynman theorem, which may be
formulated as follows: Assume that one splits the total hamiltonian into
\begin{equation}
H = H_0 + \Delta V
\end{equation}
and defines a hamiltonian depending on a parameter $\lambda$ by
\begin{equation}
H(\lambda ) = H_0 + \lambda \Delta V\,.
\end{equation}
If $E_\lambda$ defines the eigenvalue of
\begin{equation}
H(\lambda ) \vert \Psi_\lambda > = E_\lambda\vert \Psi_\lambda >
\end{equation}
the expectation value of $\Delta V$ calculated for the eigenstates of the
original hamiltonian $H=H(1)$ is given as
\begin{equation}
<\Psi \vert \Delta V \vert \Psi > = \left. \frac{\partial E_\lambda}{\partial
\lambda} \right|_{\lambda=1}\, . \label{eq:helfey}
\end{equation}
The BHF approximation can be used to evaluate the energies $E_\lambda$, which
also leads to the expectation value $<\Psi \vert \Delta V \vert \Psi >$
employing this Eq.~(\ref{eq:helfey}). In the present work we are going to apply
the Hellmann-Feynman theorem to determine the expectation value of the kinetic
energy and of the one-pion-exchange term $\Delta V = V_\pi$ contained in the
different interactions.

\begin{table}[t]
\begin{center}
\begin{tabular}{c|rrrr|rrrr}
\hline 
&&&&&&&&\\[-2mm]
& CDBonn & ArV18 & Nijm1 & Nijm2 & A & B & C & Reid\\
&&&&&&&&\\[-2mm]
\hline
&&&&&&&&\\[-2mm]
$<E>$ & -17.11 & -15.85 & -15.82 &  -13.93 & -16.32 & -15.32 & -14.40
& -12.47\\
$<V>$ & -53.34 & -62.92 & -55.08 & -61.94 & -52.44 & -53.03 & -54.95 &
 -61.51\\
$<T>$ & 36.23 & 47.07 & 39.26 & 48.01 & 36.12 & 37.71 & 40.55 &
 49.04 \\
$<V_{\pi}>$ & -22.30 & -40.35 & -28.98 & -28.97 & -12.48 & -26.87 & -45.74 &
 -27.37 \\
$<E>_{\mbox{HF}}$ & 4.64 & 30.34 & 12.08 & 36.871  & 7.02 & 10.07 & 29.56 &
 176.25\\
&&&&&&&&\\[-2mm]
\hline
&&&&&&&&\\[-2mm]
$P_D$ [\% ] & 4.83 & 5.78 & 5.66 & 5.64 & 4.38 & 4.99 & 5.62 & 6.47\\
&&&&&&&&\\[-2mm]
\hline
\end{tabular}
\begin{minipage}[t]{16.5 cm}
\caption{Energies calculated for nuclear matter with Fermi momentum $k_F$ = 1.36
fm$^{-1}$. Results are listed for the energy per nucleon calculated in BHF
($<E>$) and Hartree-Fock ($<E>_{HF}$) approximation.  Furthermore the
expectation value for the NN interaction $<V>$, the kinetic energy
$<T_{\mbox{Kin}}>$ and the one-pion-exchange term $<V_{\pi}>$ are listed. For
completeness we also give the D-state probability calculated for the deuteron
$P_D$. Results are presented for the charge-dependent Bonn  (CDBonn)
\protect\cite{cdb}, the Argonne $V_{18}$ (ArV18) \protect\cite{Wiringa95} and 
two
Nijmegen (Nijm1, Nijm2) \protect\cite{nijm1} interactions. For a comparison
results are also given for three older versions of the Bonn interaction (A,B,C)
\protect\cite{rupr0} and the Reid soft core potential \protect\cite{reid},
which is supplemented in partial waves in which it is not defined by the OBE C
potential. All energies are given in MeV per nucleon. \label{tabhelf}}
\end{minipage}
\end{center}
\end{table}

First differences in the prediction of nuclear properties\cite{artu99} 
obtained from the modern interactions
 are displayed in table~\ref{tabhelf} which contains various
expectation values calculated for nuclear matter at the empirical saturation
density, which corresponds to a Fermi momentum $k_F$ of 1.36 fm$^{-1}$. The most
striking indication for the importance of nuclear correlations beyond the mean
field approximation may be obtained from the comparison of the energy per
nucleon calculated in the mean-field or Hartree-Fock (HF) approximation. All
energies per nucleon calculated in the (HF) approximation are positive.
therefore far away from the empirical value of -16 MeV. Only after inclusion of
NN correlations in the BHF approximation results are obtained which are close to
the experiment. While the HF energies range from 4.6 MeV in the case of CDBonn
to 36.9 MeV for Nijm2, rather similar results are obtained in the BHF
approximations. This demonstrates that the effect of correlations is quite
different for the different interactions considered. However it is worth noting
that all these modern interactions are much ``softer'' than e.g.~the old Reid
soft-core potential\cite{reid} in the sense that the HF result obtained for the
Reid potential (176 MeV) is much more repulsive.

Another measure for the correlations is the enhancement of the kinetic energy
calculated for the correlated wave function as compared to the mean field
result which is identical to $T_{FG}$, the energy per particle of the free Fermi
gas. At the empirical density this value for $T_{FG}$ is 23 MeV per nucleon.
One finds that correlations yield an enhancement for this by a factor which
ranges from 1.57 in the case of CDBonn to 2.09 for Nijm1. It is remarkable that
the effects of correlations, measured in terms of the enhancement of the kinetic
energy or looking at the difference between the HF and BHF energies, are
significantly smaller for the interactions CDBonn and Nijm1, which contain
non-local terms. The non-local interactions tend to be ``softer'' in the sense
discussed above and therefore lead to more binding energy in nuclear matter.

\begin{figure}[tb]
\begin{center}
\begin{minipage}[t]{12 cm}
\epsfig{file=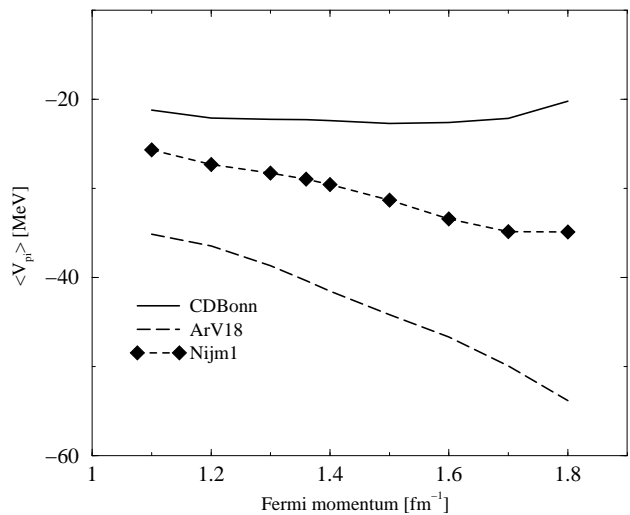,scale=0.5}
\end{minipage}
\begin{minipage}[t]{16.5 cm}
\caption{Expectation value of the $\pi$-exchange contribution to the energy of
nuclear matter as a function of density   \label{fig:pionc} }
\end{minipage}
\end{center}
\end{figure}

The table~\ref{tabhelf} also lists the expectation value for the pion-exchange
contribution $V_\pi$ to the two-body interaction. Here one should note that the
expectation value of $V_\pi$ calculated in the HF approximation is about 15 MeV
almost independent of the interaction considered. So it is repulsive and
completely due to the Fock exchange term. If, however, the expectation value for
$V_\pi$ is evaluated for the correlated wave function, one obtains rather
attractive contributions ranging from -22.30 MeV per nucleon (CDBonn) to -40.35
MeV (ArV18). This expectation value is correlated to the strength
of the tensor force or the D-state probability $P_D$ calculated for the
deuteron (see table~\ref{tabhelf} as well). Interactions with larger $P_D$, like
the $ArV18$, yield larger values for $<V_{\pi}>$. For a further support of this
argument we also give the results for three different version of
charge-independent Bonn potentials A, B and C, defined in \cite{rupr0}.
All this demonstrates that pionic and tensor correlations are very
important to describe the binding properties of nuclei. In fact, the gain in
binding energy due to correlations from $V_\pi$ alone is almost
sufficient to explain the difference between the HF and BHF energies.

The importance of pionic correlations has been emphasized by Akmal and
Pandharipande\cite{fhnc2}. They observe an enhancement of the pionic
correlations in FHNC calculations for nuclear matter at high densities and 
interprete this change in the wavefunction as an indicator for a phase
transition to pion condensation\cite{pioco1,pioco2}. We do not intend to go into
a detailed discussion of pion condensation\cite{prevco}, but simply show the 
expectation 
 values for $<V_{\pi}>$ as a function of density in Fig.~\ref{fig:pionc}. 
Indeed one finds a smooth
enhancement of this expectation value with density if one considers the local
representation of the $\pi$-exchange as contained in the Argonne $V_{18}$
interaction. It may be questionable if this should be called an indication for a
phase transition. Quite a different behavior is obtained if the non-local 
components are taken into account as they appear e.g. in the CDBonn potential.   

Inspecting the expectation values for the kinetic energies we observe a feature
very similar to the one observed for the deuteron (see section
\ref{sec:nninter}): the local interactions,
ArV18 and Nijm2, yield larger kinetic energies than CDBonn and Nijm1, which
contain nonlocal terms. This is independent of the density considered.

\begin{figure}[tb]
\begin{center}
\begin{minipage}[t]{12 cm}
\epsfig{file=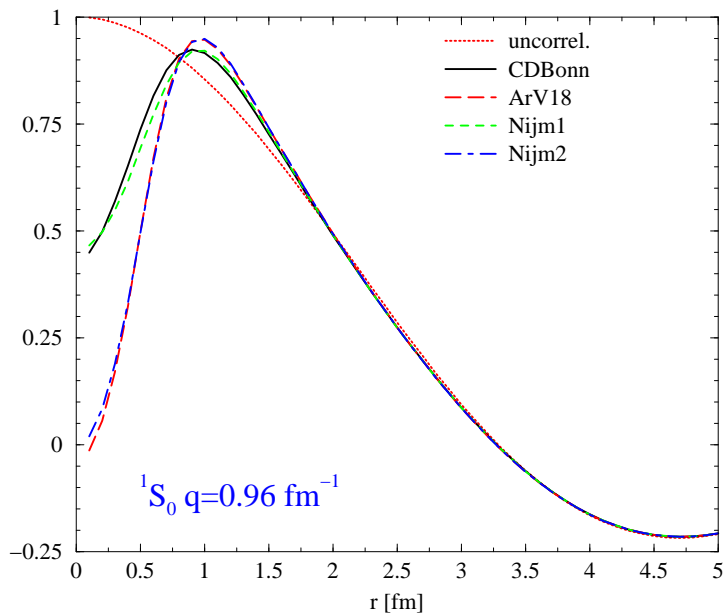,scale=0.6}
\end{minipage}
\begin{minipage}[t]{16.5 cm}
\caption{Correlated wave functions  $\vert \psi_{ij} >$ as defined in
(\protect\ref{eq:gcorel2}) as a function of the relative distance for the $^1S_0$
partial wave. Results are shown for a pair of nucleons in nuclear matter at
empirical saturation density, which heal to an uncorrelated two-nucleon wave
function with momentum $q$ = 0.96 fm${-1}$ at larger distances. The curves are
labeled by the interactions, which were considered. \label{chfig2} }
\end{minipage}
\end{center}
\end{figure}

A different point of view on nuclear correlations may be obtained from
inspecting the  relative wave functions for a correlated pair $\vert
\psi_{ij}>$ defined in (\ref{eq:gcorel2}). Results for such correlated wave
functions for a pair of nucleons in nuclear matter at empirical saturation
density are displayed in Figs.~\ref{chfig2} and \ref{chfig3}. As an example we
consider wave functions which ``heal'' at large relative distances to an
uncorrelated two-nucleon wave function with momentum $q$ = 0.96 fm$^{-1}$
calculated at a corresponding average value for the starting energy.

Fig.~\ref{chfig2} shows relative wave functions for the partial wave $^1S_0$.
One observes the typical features: a reduction of the amplitude as compared to
the uncorrelated wave function for relative distances smaller than 0.5 fm,
 reflecting the
repulsive core of the NN interaction, an enhancement for distances between
$\approx$ 0.7 fm and 1.7 fm, which is due to the attractive components at
medium range, and the healing to the uncorrelated wave function at large $r$.
One finds that the reduction at  short distances is much weaker for the
interactions CDBonn and Nijm1 than for the other two. This is in agreement with
the discussion of the kinetic energies  and the difference between HF and BHF
energies (see table~\ref{tabhelf}). The nonlocal interactions CDBonn and Nijm1
are able to fit the NN scattering phase shifts with a softer central core than
the local interactions.

Very similar features are also observed in the $^3S_1$ partial wave displayed in
the left half of Fig.~\ref{chfig3}. For the $^3D_1$ partial wave, shown in the
right part of Fig.~\ref{chfig3}, one observes a different behavior: All NN
interactions yield an enhancement of the correlated wave function at $r\
\approx$ 1 fm. This enhancement is due to the tensor correlations, which couples
the partial waves $^3S_1$ and $^3D_1$. This enhancement is stronger for the
interactions ArV18, Nijm1 and Nijm2 than for the CDBonn potential. Note that the
former potential contains a pure nonrelativistic, local one-pion-exchange term,
while the CDBonn contains a relativistic, nonlocal pion-exchange contribution.
See also the discussion of the wave function for the deuteron in the preceeding
section \ref{sec:nninter} and Fig.~\ref{fig:deuwav}.

\begin{figure}[tb]
\begin{center}
\begin{minipage}[t]{12 cm}
\epsfig{file=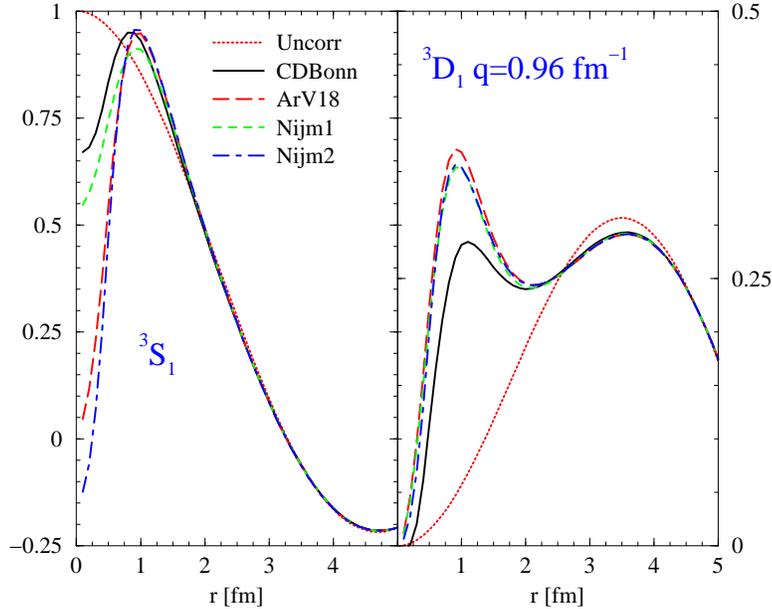,scale=0.6}
\end{minipage}
\begin{minipage}[t]{16.5 cm}
\caption{Correlated wave functions as a function of the relative distance for
the $^3S_1$ and $^3D_1$ partial waves. Further details see
Fig.~\protect\ref{chfig2}. \label{chfig3} }
\end{minipage}
\end{center} 
\end{figure}

These most recent parameterizations of the nucleon-nucleon interaction  not
only predict different saturation properties of symmetric nuclear matter.  Also
one can observe differences in the calculated symmetry energy at high
densities\cite{engvik}. The differences between the predictions derived from
these modern potentials, however, are significantly smaller than those derived
from older models of the NN interaction. All NN interactions now predict a
steady increase of the symmetry energy with density. This may be considered as
an improvement in the parameterization of the NN interactions. Nevertheless the
remaining differences lead to non-negligible differences in nuclear
astrophysics, in the studies of neutron star matter\cite{prakash,heisel}. The
CD Bonn potential predicts a slightly larger symmetry energy at high densities.
This leads to a larger proton fraction in $\beta$-stable matter. As a
consequence one finds that the so-called direct URCA process, a very efficient
cooling mechanism for neutron stars by neutrino emission, may occur at densities
around 0.88 fm$^{-3}$, 1.05 fm$^{-3}$, or 1.25 fm$^{-3}$ according to the
predictions of the CD Bonn, Argonne $V_{18}$, or Nijmegen I potential.

These modern NN interactions account for a breaking of isospin symmetry to
reproduce pp and pn phase shifts accurately. Even after the electromagnetic
effects have been removed, the strong interaction between two protons is in
general less attractive than the pn interaction. If included in the calculation,
these isospin symmetry breaking effects lead to differences in the predicted
energy of the order of 0.3 MeV per nucleon\cite{ispos1}. Isospin symmetry
breaking effects as well as charge-symmetry breaking effects (i.e.~differences
between proton-proton and neutron-neutron interactions which are partly
due to the different masses of the nucleons) of these interactions have also 
been considered to explore their impact on the so-called Nolen-Schiffer
anomaly\cite{ispos2}.

Up to this point, we were mainly concerned with bulk properties of nuclear
matter and correlations which are relevant for all nucleons in the Fermi sea.
Special attention has also been paid to correlations around the Fermi level.
This includes pairing correlations, which have been studied in the framework of
the BCS approach by various groups\cite{baldbcs,mortbcs}. Traditionally, such
studies concentrate on the $S=0$, $T=1$ pairing effects. Correlations,
however, are even stronger for the $S=1, T=0$ channel and corresponding pairing
effects have been studied within the framework of the Green's function
approach\cite{vonderf,schnell}.
Instabilities of normal nuclear matter with respect to such pairing correlations
are of particular importance at smaller densities, below the saturation density
of nuclear matter.

We are now going to discuss the situation of many-body calculations based on
realistic interactions for finite nuclei. On one hand such calculations are 
in general much more difficult to perform since one has to determine also the
appropriate single-particle wave functions within the self-consistent
calculations. In nuclear matter the basis of single-particle wave functions, the
plane waves, is determined by the symmetry of the problem. This additional
complication shows up in all the many-body approaches, which we consider. For
the hole-line or Green' function approach this implies that one has to determine
the basis, in which the single-particle self energy is diagonal. The same basis
must the also be used to define quantities like the Pauli operator in the
Bethe-Goldstone equation. One meets a very similar problem in the coupled
cluster method, for which one has to determine also the amplitude $S_1$ in a
self-consistent way. For variational calculations the additional complication
shows up in the fact that one also has to determine the long-range part of the
wave function, e.g.~in terms of an appropriate mean field wave function 
$\Phi_{MF}$ in the trial  wave function of Eq.~(\ref{eq:trialwf}) from the
variational calculation.

On the other hand, however, the calculations for finite nuclei might lead to
more reliable results as the average density is lower than for nuclear matter at
saturation density. All the expansion techniques which we discussed should
converge better for small densities, which means the results obtained at a given
order might be more reliable for the finite system.

Some features discussed for the infinite systems should also hold for finite
nuclei. This should be true in particular for short-range correlations. Such
short range correlations should not be affected by the long-range behavior of
the wave functions. To repeat the argument from a different point of view: 
Short-range correlations are described in terms of particle-hole excitations of
such high energies that the shell structure of the single-particle spectrum of
finite nuclei is not relevant. The information on such short-range correlations 
in finite nuclei might therefore be deduced from the corresponding information
in nuclear matter by means of a local density
approximation\cite{lda1,lda2,lda3,lda4}. The situation is different for the
long-range correlations, which are built in terms of low energy configurations
for which the shell structure is relevant. Such features of the correlations
may also vary, depending on the specific nucleus under consideration.

Many-body calculations using the hole-line expansion, the coupled cluster
method, the Green's function approach or variational calculations have been done
for various different nuclei. In the following discussion we will mainly focus
our attention to results on $^{16}$O, since for this double magic nucleus
results from the different approaches are available. 

The complication in solving the BHF equations for finite nuclei have been
discussed e.g.~in \cite{sauer}, where also a computer program for solving the
Bethe-Goldstone equations has been made available. Different techniques to solve
the Bethe-Goldstone equation have been described by Barrett et al.\cite{bruce}
and Kuo and coworkers\cite{tomg}. Also in BHF calculations for finite nuclei one
has to choose the spectrum of the intermediate particle states in the
Bethe-Goldstone equation. The standard choice is the same as for nuclear matter,
use pure kinetic energies. Old studies of three-body terms in the coupled
cluster method led to the suggestion that a shift of the kinetic energy spectrum
by a value around -8 MeV, would minimize the effects of three-body
correlations\cite{zabo8}. Such results indicate that the effects of 
three-body
correlations may even be smaller than for nuclear matter (see the convergence
argument given above), however, a systematic study of three-hole line
contributions for various interactions should be performed and is still 
missing.

\begin{figure}[tb]
\begin{center}
\begin{minipage}[t]{12 cm}
\epsfig{file=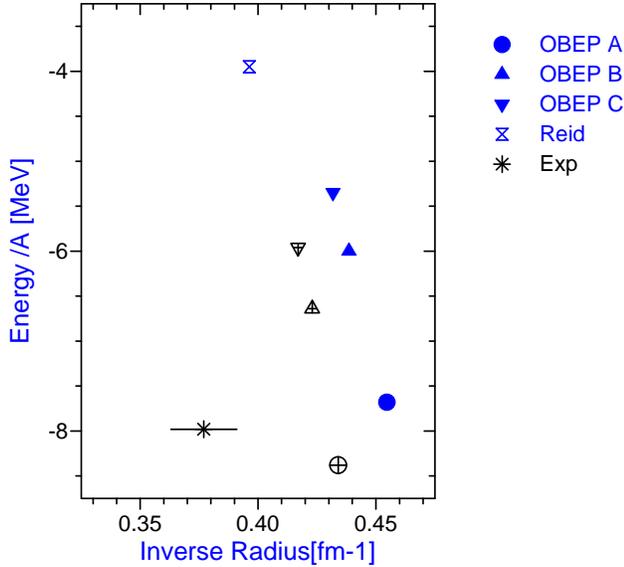,scale=0.6}
\end{minipage}
\begin{minipage}[t]{16.5 cm}
\caption{Ground state properties for $^{16}$O as predicted in BHF calculations
using different OBE potentials from \protect{\cite{rupr0}}. In terms of open
symbols we also present the results after corrections due to general ring
diagrams are included.  \label{fig:cos16} }
\end{minipage}
\end{center} 
\end{figure}

Results for the energy and the radius of the charge distribution of $^{16}$O
obtained from BHF calculations~\cite{carlo} using various meson exchange 
potentials are displayed in Fig.~\ref{fig:cos16}. The plot of the results in
terms of the energy per nucleon versus the inverse of the radius of the charge
distribution has been chosen to have a presentation of results which is similar
to the plot of saturation point for nuclear matter in Fig.~\ref{fig:coester}.
The results for nuclear matter and finite nuclei are very similar in so far
that also the calculated ground state properties for $^{16}$O show a behavior
like the Coester band discussed for nuclear matter. There are interactions like
the OBEP A (defined in Table A.1 of \cite{rupr0}), which yield about the correct
binding energy per nucleon of -7.98 MeV but predict a radius of the charge
distribution, which is much smaller than the experimental one. Therefore we also
have in this case a density which is too large. Other OBE interaction, which
show a larger D-state probability in the deuteron, like OBEP B and OBEB C, are
``stiffer'' and lead to slightly larger radii but less binding energy. Note that
all these modern OBE potential yield more binding than the old Reid soft-core 
potential\cite{reid}. 

The BHF approximation can be considered as a very efficient tool to account for
the effects of two-nucleon short range correlations. This approach, however,
might be too simple to account for long-range correlations corresponding to low
energy configurations in the nuclear shell-model. Therefore attempts have been
made to consider a model space, in the sense as we introduced this term in the
discussion around Eq.~(\ref{eq:heffmod}), in which these long range are treated
explicitly, and use the lowest order hole-line expansion to account for the
short-range correlations outside this model space. This is similar to the
concept of model space for nuclear matter as it has been discussed by Kuo et
al.~\cite{tomnm,song,jiang}. One possible way along this line is to consider a
solution of the Bethe-Goldstone equation (\ref{eq:betheg}) using a Pauli
operator which restricts intermediate two-nucleon states to configurations which
are outside the model-space considered. One may then use the resulting
$G$-matrix in shell-model configuration mixing calculation, in which all
nucleons are considered as active particles which are allowed to form all
configurations within the model space. Such kind of no-core shell-model 
calculations have recently been done by Barrett and
coworkers\cite{zheng1,navra1}. Due to the dramatic increase of configurations
with increasing number of active nucleons, however, such studies are restricted
to very light nuclei.

For heavier nuclei one can consider a model space defined in terms of
configurations in e.g.~an oscillator basis and consider in a systematic way
the
effects of correlations defined within this model space. One important class of
such long-range configurations are the RPA correlations represented by the
so-called particle-hole ring diagrams. They correspond to an iteration of the
particle-hole ladders to any order. Within such a limited model space the 
number of particle-particle configurations is of the same order as the number of
hole-hole configurations. Therefore one may like to treat particle-particle and
hole-hole ladders on the same footing as it is done in the Green's function
formalism. This is achieved by summing the contribution of all particle-particle
hole-hole ring diagrams. A technique has been developed which allows the
consistent summation of all particle-hole and particle-particle hole-hole ring
diagrams leading to a so-called Super RPA (SRPA)\cite{ellis}. It is remarkable
that the resulting SRPA equations yield a stable solution only if the
self-energy, used to calculate the single-particle Green's function, is
determined in a self-consistent way\cite{heinz}. 

Such long-range correlations lead to an additional binding energy for closed
shell nuclei like $^{16}$O of around 1 MeV per nucleon. They might be
characterized by the depletion of the hole-state occupation. Again for the
example of $^{16}$O, the SRPA correlations in a model space which includes the
$1p0f$ oscillator states leads to an occupation probability for the $0p$ hole
states which is of the order of 0.85. Note that this depletion of the hole state
occupation due to the long-range correlations should be added to the depletion
which is due to the short-range correlations (see discussion in the next
subsection). The long-range correlations also have an effect on the calculated
radii. The effects of these SRPA correlations on the ground-state properties of
$^{16}$O are also shown in Fig.~\ref{fig:cos16} for the interactions OBEP A to
C. The SRPA correlations yield a small improvement of the results obtained
within the BHF approach. This improvement, however, is not sufficient to meet
the experimental data.

Calculations for finite nuclei using the ``Exp(S)'' or Coupled Cluster Method
have been performed already 25 years ago by Zabolitzky\cite{zab1,zab2}. Using
the Reid soft-core potential he finds that the SUB2 approximation yields results
which are very similar to those obtained in the BHF approximation (see result
for the Reid potential in Fig.~\ref{fig:cos16}) and that the effects of
three-body correlations, included in the SUB3 approach, leads to an additional 
binding energy of around 0.5 MeV per nucleon connected with an enhancement of
the wound integral or depletion of the hole-state occupation by around 0.05
percent. This indicates a very good convergence of the Coupled Cluster Method as
well as the hole-line expansion. 

More recent calculations using the language of the Coupled Cluster Approach 
have been reported by Heisenberg and Mihaila\cite{heis1} for the Argonne
$V_{18}$
interaction\cite{Wiringa95}. The calculations are performed in a configuration
space defined in terms of harmonic oscillator functions. They use a $G$-matrix
approximation for the two-body interaction and calculate a mean field which they 
correct to account for 3p3h and 4p4h correlations.  They obtain for the
case of $^{16}$O a binding energy of -5.9 MeV per nucleon and a radius of the
charge distribution of 2.81 which is even larger than the experimental value.
Unfortunately, there are no BHF calculations available for the same
interaction. However, looking at the results obtained for nuclear matter,
one would expect that BHF calculations for the $V_{18}$ interaction might produce
results with a binding energy around -6 MeV but a smaller radius. It is not
clear whether these possible differences are due to the 3p3h and 4p4h
corrections included in \cite{heis1}, due to the restricted oscillator space 
or due to a lack of self-consistency in solving the Bethe-Goldstone equation.

Variational calculations using the FHNC summation techniques which we discussed
in section \ref{sec:varia} have recently been performed for our reference
nucleus $^{16}$O as well\cite{arias2}. As we have discussed before the 
techniques of these  variational calculation restrict their applications to 
local interactions. In the work of Fabrocini et al.~\cite{arias2} various
versions of the Argonne potentials $V_{14}$ and $V_8$
 have been used supplemented by
three-nucleon potentials of the Urbana group. One of the main problems for such
variational calculations for finite systems is to obtain a reliable description
for the long range structure of the wave function, i.e.~the shape of the
single-particle wave functions which form the mean field part of the trial wave
function (\ref{eq:trialwf}). In ref.~\cite{arias2} two different models for
these single-particle wave functions are considered: a harmonic oscillator model
and a Woods Saxon parameterization. The optimal parameters are determined from
the variational calculation. It turns out, however, that the functional exhibits
a rather flat minimum as a function of the parameters characterizing the
single-particle waves. This is mainly due to the balance between kinetic energy
and potential energy. In particular for wave functions with small radii one
observes a cancellation between these contributions, which is rather sensitive
to the details of the variational form of the wave function.

In fact, one finds minima of the variational calculations, which are quite
different for the two parameterization of the single-particle wave functions
considered. Therefore also the expectation values of single-particle operators
like the radius are rather model-dependent. The Woods Saxon parameterization 
yields a radius for $^{16}O$, which is 0.2 fm smaller than the one derived from
the harmonic oscillator parameterization. This certainly a drawback of
variational calculations for finite systems. On the other hand, however, the
variational calculation yield upper bounds for the energy of the order of -5.2
MeV per nucleon in the case of $^{16}$O which is a very good benchmark for other
many-body calculations. This is supported by the fact that the FHNC/SOC
calculations yield results very close to corresponding Variational Monte-Carlo
calculation in cases for which such a comparison is available\cite{arias2}.
   
Very extensive extensive Variational Monte Carlo (VMC) as well as Green's
function Monte Carlo (GFMC) calculations have recently been performed for nuclei
with particle number up to $A=8$\cite{piepermc1,pudliner97}. Results of such
calculations for the ground state of these nuclei are presented in
table~\ref{tabmoca}. The calculations leading to the results displayed in that
figure used the Argonne $V_{18}$ potential for the two-body interaction supplemented 
by the Urbana IX three-nucleon force. This version IX of the Urbana three
nucleon forces has been adjusted to reproduce together with the $V_{18}$ NN
interaction the binding energy of $^3$H and to give a reasonable saturation 
point for nuclear matter. Because of these adjustments the calculated energies
cannot directly be compared to the results which we discussed before in which no
three-nucleon force has been employed. The expectation value of the three
nucleon force alone yields a contribution to the energy of $^7$Li of -8.9 MeV.

\begin{table}
\begin{center}

\begin{tabular}{c|rrr|rr}
\hline
&&&&&\\[-2mm]
&\multicolumn{3}{c|}{Energy [MeV]}&\multicolumn{2}{c}{Radius [fm]}\\
$^AZ$ & VMC & GFMC & Exp. & VMC & Exp. \\
&&&&&\\[-2mm] 
\hline
&&&&&\\[-2mm]
$^3$H & -8.32 & -8.47 & -8.48 & 1.59 & 1.60 \\
$^4$He & -27.78 & -28.30 & -28.30 & 1.47 & 1.47 \\
$^6$He & -24.87 & -27.64 & -29.27 & 1.95 &\\
$^6$Li & -28.09 & -31.25 & -31.99 & 2.46 & 2.43 \\
$^7$He & -24.43 & -25.16 & -28.82 & &\\
$^7$Li & -32.78 & -37.44 & -39.24 & 2.26 & 2.27 \\
$^8$He & -19.71 & -25.77 & -31.41 & &\\
$^8$Li & -29.70 & -38.26 & -41.28 & &\\
$^8$Be & -48.06 & -54.66 & -56.50 & &\\
&&&&&\\ 
\hline
\end{tabular}
\begin{minipage}[t]{16.5 cm}
\caption{Results for ground state properties of nuclei with nucleon number up to
$A=8$ as obtained in VMC and GFMC calculations. These results for the total
energy and rms radii for the protons have been obtained using the Argonne
$V_{18}$
for the two-body interaction supplemented by the Urbana IX three-nucleon force.
The results displayed in this table have been copied from 
\protect{\cite{piepermc1}} and \protect{\cite{pudliner97}}.
}
\label{tabmoca}
\end{minipage}
\end{center}
\end{table}

Despite the use of this adjusted three-nucleon force it is quite remarkable to
see how well the GFMC calculations reproduce the experimental values for the
binding energies. It is also very satisfactory to see that he variational
calculations (VMC) yield predictions for the energies which are above those 
obtained in GFMC by values which are typically less than 1 MeV per nucleon.
Here one must keep in mind, however, that for these very light nuclei more
sophisticated shapes could be considered for the trial wave function than it was
possible for heavier nuclei. In particular one does not start assuming a set of
single-particle wave function in the mean field part of the trial function, but
determines also the long range part of the radial shape in terms of correlation
functions.

Also the results for the radii are in very good agreement with the empirical
data. Again one should keep in mind, however, that this success is at least
partly due to the use of the adjusted three-nucleon force, which has been
determined to remove the problem of the Coester band in nuclear matter. This
indicates that sophisticated variational calculations using a hamiltonian which 
has been adjusted to reproduce the saturation point in nuclear matter tend to
give proper saturation mechanisms also for finite nuclei. This leads to the
expectation that hamiltonians, which lead to a saturation of nuclear matter at a
too high density shall predict radii for finite nuclei, which tend to be too
small. 

\subsection{\it Single Particle Properties in Nuclear Matter and Finite Nuclei} 
\label{subsec:single}

Single particle properties are most conveniently described in terms of the
Green's function which has been introduced in section \ref{subsec:green}.
All single particle properties can be calculated from the single-particle
Green's function, in the sense that the  expectation value of any 
 one-body operator  can be obtained by using the hole spectral function (Eq. 
(\ref{eq:expec})). In this section we will mainly discuss the properties
of the hole spectral function in finite nuclei. However,  we will also
present some  results in nuclear matter which show some features
of the spectral functions. A recent review on the effects 
of correlations in the independent particle motion in fermion
systems,  with special emphasis in the nuclear case,
 can be found in Ref. \cite{sick97}. 

The  physical meaning of the hole 
spectral function can be read from Eq. (\ref{eq:specf}): 
it gives the probability of removing a particle with momentum ${\bf k}$ from 
the system of A particles leaving the resulting (A-1) system with an 
energy $E^{A-1}=E_0^A -\omega$, where $E_0^A$ is the ground state energy 
of the $A$ particle system. Analogously, the particle spectral function
$S_p(k,\omega)$ is the probability of adding a particle with momentum $k$ and 
leaving the resulting $(A+1)$-system with an energy $E^{A+1} = \omega + E_0^A$.
(see also Fig.~\ref{fig12} and discussion in section \ref{subsec:green}}

In the Self-Consistent Green's Function (SCGF) formalism, the self-energy is 
the key quantity to determine the one-body Green's function. The 
self-energy  takes into account the strong interactions that a 
nucleon in the nuclear medium has with the other nucleons.
Since the self-energy is determined from the two-body effective
interaction between two dressed particles (Fig.(\ref{fig11})), and this
 requires
in turn the knowledge of the propagator, one needs to deal with 
a coupled problem which must be solved self-consistently \cite{wim1}. 
In 
the case of an infinite system, the self-energy is diagonal in $k$
and the formal solution of the Dyson's equation (Eq.(\ref{eq:dyson1})) is 
particularly simple
\be
g(k,\omega)=\frac {1}{\omega - \frac {k^2}{2m} -\Sigma(k,\omega)} .
\ee

\begin{figure}[t]
\begin{center}
\begin{minipage}[t]{12 cm}
\begin{center}
\epsfig{file=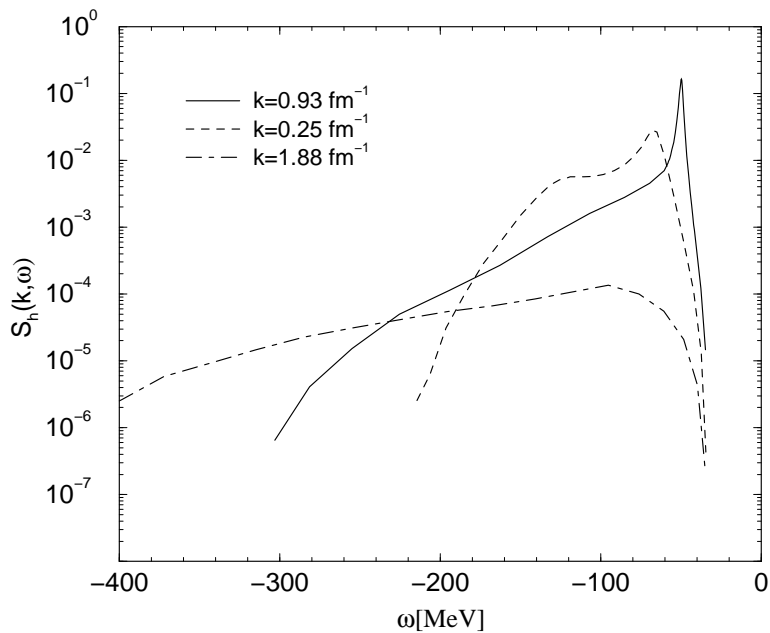,scale=0.6}
\end{center}
\end{minipage}
\begin{minipage}[t]{16.5 cm}
\caption{ Hole spectral function (in MeV$^{-1}$) for three 
different momenta in nuclear matter at $k_F=1.36$ fm$^{-1}$.  
\label{fig:figsingle1}}
\end{minipage}
\end{center}
\end{figure} 
By combining the Lehmann representation (\ref{eq:lehm}),
 which expresses the single particle
propagator in terms of the spectral functions, with the Dyson equation 
(\ref{eq:dyson1}),
which relates the propagator to the self-energy, the following
expressions for the spectral functions are obtained :
\be
S_{h(p)}(k,\omega)= \pm \frac {1}{\pi} \frac {Im \Sigma(k,\omega)} 
{(\omega -k^2/2m - Re \Sigma(k,\omega))^2 + (Im \Sigma (k, \omega))^2} ~,~ 
\mbox{for} \ \omega < \epsilon_F ~ (\omega > \epsilon_F). 
\ee

Due to the short range repulsion present in any realistic interaction, a 
meaningful approximation to the effective interaction, used 
in the calculation of the self-energy, should sum up ladder
diagrams which in this approach include also hole-hole propagation,
either to all orders \cite{vonderf,wim1} or  to a second order in the
propagation of holes \cite{baldo92,lda4}. 
The inclusion of  hole-hole propagation is required since at the single
particle level it yields the coupling to the excited states of the 
(A-1) particle system. The  self-energy is used in the Dyson equation
and the resulting single particle propagator is plugged again in the
calculation of the effective interaction, until self-consistency is achieved. 
In the calculations presented here, 
self-consistency has been established
only for the quasi-particle energies, i.e. at the level of the 
real on-shell part of the self-energy
\be
\epsilon_{qp} (k) = \frac {k^2}{2 m} + Re \Sigma (k,\epsilon_{qp}(k)).
\ee
When this self-consistency is achieved, the complete energy
dependence of the self-energy can be studied  \cite{vonderf}. 
The imaginary part of the self-energy is different from zero 
over a wide range of energies, being positive below $\epsilon_F$ and
negative above. This yields a spreading of the single-particle strength, but 
for $k< k_F$ $S_h$ will still contain a peak at the quasi-particle energy. 
Since $S_p$  contains some fraction of the total strength the 
occupation probability (Eq. (\ref{eq:momen}) ) for $k< k_F$ will be depleted to
a value below one. 

The energy dependence of the hole spectral function is 
shown in Fig. \ref{fig:figsingle1} for several momenta. These spectral
functions have been obtained for the Reid soft core
potential at $k_F=1.36$ fm$^{-1}$ in the framework of
SCGF\cite{vonderf}. The hole spectral function in nuclear
matter has been also calculated for the Urbana V$_{14}$ interaction
 \cite{urbv14}
in the framework of  CBF theory, by including
one hole and 2h1p intermediate correlated states\cite{benhar89}.
 The agreement
between both methods is rather satisfactory.
For $k < k_F$ the hole
spectral function shows a sharp peak (quasi-hole peak) located
at the quasi-particle energy that concentrates most of the strength. 
When $k$ approaches $k_F$ the quasi-hole peak becomes sharper. 
In the limit $k=k_F$, the peak is just a $\delta$-function, the 
strength of which defines  the discontinuity of the momentum 
distribution at $k_F$. The typical occupation probability for $k<k_F$
is about $80 \%$. Since the number of particles must be conserved,
it is necessary to have partial occupation of states $k>k_F$, which 
were unoccupied in the mean field description. A smooth distribution
of strength is observed for $k > k_F$ indicating the possibility
to find a nucleon in a momentum that would be empty in absence
of correlations.  

\begin{figure}[t]
\begin{center}
\epsfig{file=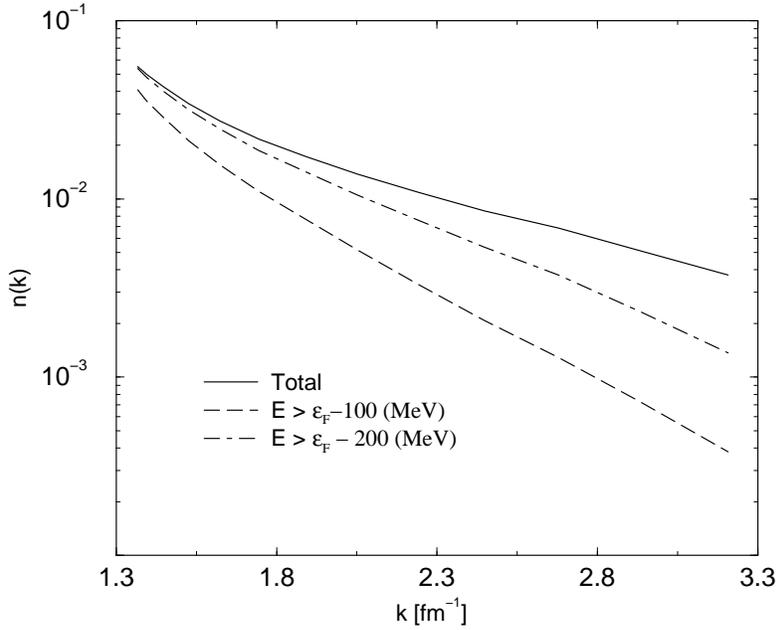,scale=0.6}
\begin{minipage}[t]{16.5 cm}
\caption{ Momentum distribution for $k>k_F$ obtained with various 
energy cut-offs in the integration of the spectral functions.
\label{fig:figsingle2}}
\end{minipage}
\end{center}
\end{figure}
 
A very important question concerns the location of the
high momentum components in the nucleus
as a function of the energy. To illustrate this point,
Fig.\ref{fig:figsingle2} shows the occupations for $k>k_F$ obtained by 
integrating $S_h(k,\omega)$ over excitation energies up 
to 100 Mev, 200 Mev 
and up to infinity. One concludes that it is necessary
 to integrate
over high excitation energies of the $A-1$ system in order to
catch the high-momentum components. As a result of correlations there 
are particles promoted above the Fermi sea and  $n(k)$
is characterized by a depletion below $k_F$ and a long tail
(occupation above $k_F$) which gives $\approx$ a $62\%$ of 
the total kinetic energy. The distribution, when
the nuclear matter is considered as a normal Fermi fluid,
is discontinuous at $k_F$ and the discontinuity is 
given by $Z_{k_F}$. Basically all calculations,
independently of the method and of the interaction used,
provide rather similar results for the depletion.
The result for the occupation at $k=0$ using the  Reid soft core
at $k_F=1.36$ fm$^{-1}$ within SCGF is $n(0)=0.83$ 
\cite{vonderf}. CBF gives
the same result for Urbana $V_{14}$, however the calculation
was performed at slightly smaller density $k_F=1.33$ fm$^{-1}$ 
\cite{benhar90}. 
The calculations in the framework of Green function formalism
but using a self-energy with  up to second order in the 
propagation of holes, give also very similar results
\cite{lda4,baldo92}.
However, the discontinuity at $k_F$ is more sensitive
to both the interaction and the method. Under the same 
conditions as before, $Z_{k_F}=0.72$ for the Reid 
and 0.7 for the Urbana $V_{14}$. We will have further
discussions on $n(k)$ in connection with the momentum
distribution in $^{16}$O.

The spectral function fulfils energy weighted sum-rules,
which besides providing useful information on the
spreading of the strength can also be used as a way to
control the approximations employed in the calculation.
The two lowest sum-rules have a simple expression. $m_0(k)$
is obtained by using the completeness relation in the energy integration 
of the spectral functions,   
\bea
m_0(k) = &\int _{-\infty}^{\epsilon_F} d \omega S_h(k,\omega) + 
\int _{\epsilon}^{\infty} d \omega S_p(k,\omega) \nn \\ 
 = & \langle \Psi_0^A \mid \left \{ a_{{\bf k}}, a_{{\bf k}}^{\dagger}\right \}
\mid \Psi_0^A \rangle = 1 ,
\eea
due to the fermion character of the nucleons, the anti-commutator 
$\left  \{ a_{{\bf k}}, a_{{\bf k}}^{\dagger}\right \}$ is equal
to the unit operator.

 A similar procedure, leads to the first-order
energy weighted sum rule $m_1(k)$:
\bea
m_1(k) = &\int_{-\infty}^{\epsilon_F} \omega S_h(k,\omega) d \omega 
+ \int _{\epsilon_F}^{\infty} \omega S_p(k,\omega) d \omega \nn \\
 & = \langle \Psi_0^A \mid \left \{ \left [ a_{{\bf k}}, H \right ], 
a_{{\bf k}}^{\dagger} \right \} \mid \Psi_0^A \rangle  .
\eea

In order to evaluate the right hand side of $m_1(k)$ it is necessary to assume a 
Hamiltonian, which in the present case is taken to be nonrelativistic and with 
only two-body forces. Under these assumptions,
\be
 \langle \Psi_0^A \mid \left \{ \left [ a_{{\bf k}}, H \right ],
a_{{\bf k}}^{\dagger} \right \} \mid \Psi_0^A \rangle =
\frac {k^2}{2m} + \frac {1}{(2 \pi)^3} \int d^3 k' n(k') \langle {\bf k},
{\bf k'} \mid V \mid {\bf k}, {\bf k'} \rangle _a,
\label{eq:summ1}
\ee 
where $ \langle {\bf k},{\bf k'} \mid V \mid {\bf k},{\bf k'} \rangle_a$ is the
anti-symmetrized two-body matrix element of the bare nucleon-nucleon 
interaction. The second term on the right hand side of Eq. (\ref{eq:summ1})
can be identified with the energy independent part of the self-energy which is  
obtained as the high frequency limit of $\Sigma(k,\omega)$ \cite{baldo92}. 
These sum-rules have been investigated in Ref. \cite{polls94}. There,  it was
 found
 that
the SCGF approach  yields spectral functions that 
fulfil both sum-rules  rather accurately.
 Due
to the short range repulsion of the potential and to the positive character
of the kinetic energy contribution, the right hand side of Eq.(\ref{eq:summ1})
is usually positive definite and  rather large. For instance using the
 Reid potential it is around 300 MeV at saturation density for $k=0$.  
For  
momenta below $k_F$ , most of the strength ($\approx 80 \%$) is below 
$\epsilon_F$
which is a negative quantity. Therefore the contribution of the hole part is 
negative and it must be  the positive contribution of the
high energy tail of the particle part of
the spectral function which, besides compensating the negative contribution
of the hole part, brings the left hand side to fulfil the equality. This
confirms the need for the appearance of single-particle strength at very
high energies.

\begin{figure}[t]
\begin{center}
\epsfig{file=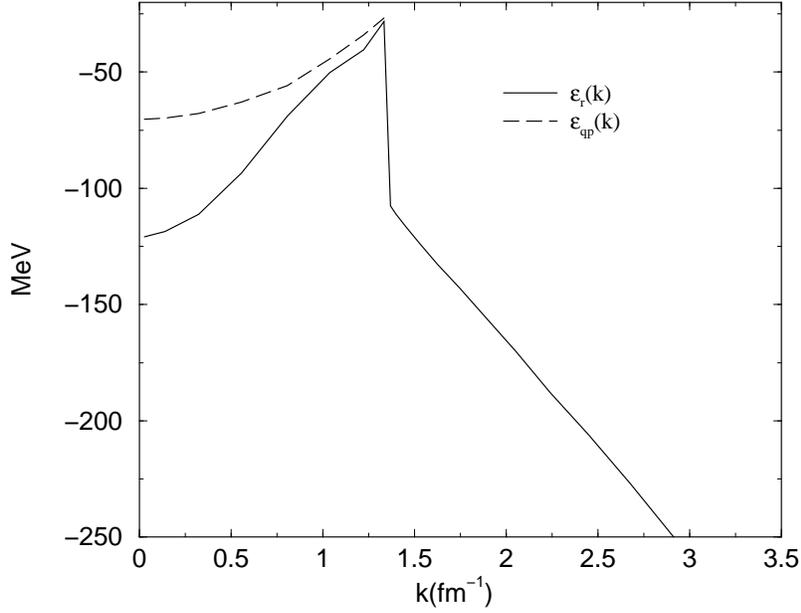,scale=0.6}
\begin{minipage}[t]{16.5 cm}
\caption{ Momentum dependence of the removal energy and the quasi-particle 
energies, in nuclear matter at $k_F=1.36$ fm$^{-1}$.
\label{fig:figsingle3}}
\end{minipage}
\end{center}
\end{figure} 

Another useful check for the spectral functions is the density sum-rule 
\be
\rho = \frac {\deg}{(2 \pi )^3} \int d^3 k ~n(k) = \frac {\deg}{(2 \pi)^3}
\int d^3 k \int_{-\infty}^{\epsilon_F} S_h(k,\omega) d \omega \, , 
\ee
with $\deg$ denoting the degeneracy of the single-particle level, which is 4
for nuclear matter. 
This sum-rule is well respected by the different approaches. In the case 
of the SCGF formalism it is fulfilled within $1 \%$. The integration up to 
$k_F$ provides $81 \%$ of the sum-rule and one gets a $97 \%$ if the
integration is carried up to $3 k_F$.

In the case of two body interactions, the hole spectral function gives access, 
through the Koltun sum-rule (\ref{eq:koltun}),
to the binding energy per particle,
\be
e(\rho)= \frac {deg}{\rho (2 \pi)^3}\int d^3 k \int_{-\infty}^{\epsilon_F} 
\frac {1}{2} \left ( \frac {k^2}{2m} + \omega \right ) S_h(k,\omega) d \omega ,
\ee 
to the kinetic energy per particle,
\be
t(\rho) = \frac {deg}{\rho (2 \pi )^3 } \int d^3 k \frac {k^2}{2m}  \int_{-\infty}
^{\epsilon_F} d \omega S_h(k,\omega) = 
\frac  {deg}{\rho (2 \pi )^3 } \int d^3 k \frac {k^2}{2m} n(k)
\ee
and to the potential energy per particle $v(\rho)=e(\rho)-t(\rho)$. Introducing
the removal energy $\epsilon_r(k)$,
\be
\epsilon_r(k) = \frac {\int_{-\infty}^{\epsilon_F} d \omega~
 \omega ~S_h(k,\omega) }
{\int_{-\infty}^{\epsilon_F} d \omega S_h(k,\omega) }\label{eq:removal}
\ee
one can express the binding energy as
\be
e(\rho)= \frac {deg}{\rho (2 \pi)^3} \int d^3 k \frac {1}{2} \left ( \frac {k^2}
{2m} + \epsilon_r(k) \right ) n(k).
\ee

In Brueckner-Hartree-Fock, $\epsilon_r(k) =k^2/2m + u(k)$, i.e. it coincides
with the quasiparticle 
energy, and $n(k)=\theta (k_F-k)$, therefore
\be
e_{BHF} = \frac{deg}{\rho (2 \pi )^3} \int d^3 k \left ( \frac  {k^2}{2m} +
\frac {1}{2} u(k) \right ).
\ee
Due to  correlations, the spreading of the spectral function implies
that $\epsilon_r(k) \ne \epsilon_{qp}(k)$. Obviously this comparison between 
$\epsilon_r(k)$ and $\epsilon_{QP}$ has sense only for $k \leq k_F$.

  The momentum dependence of the removal energy is shown in Fig. 
\ref{fig:figsingle3} 
together with the quasi-particle energies (below $k_F$) for the 
Reid soft core within SCGF. $\epsilon_r(k)$
is  an increasing function of $k$ for $k< k_F$ and  decreases quite fast
for 
$k>k_F$ having a large discontinuity at $k_F$.
For $k=0.027$ , $\epsilon_r=-121 $ MeV, while $\epsilon_{QP}=-70.38$ MeV, and the
occupation probability  
$n(0.027)= 0.83$.

Averaging (\ref{eq:removal}) over all momenta one can define a mean removal 
energy, $\tilde \epsilon_r$, by
\be 
\tilde \epsilon_r = \frac {\int d^3 k \int_{-\infty}^{\epsilon_F} d \omega \omega
S_h(k,\omega)}{\int d^3 k \int S_h(k,\omega) d\omega }= t(\rho)+ 2 v(\rho)
\ee
In the case that we are considering as an example, 
i.e. nuclear matter with the Reid soft core, at $k_F=1.36$ fm$^{-1}$ in the
framework of SCGF,  $\tilde \epsilon_r \approx -86$ MeV. On the other hand,
the binding energy is $\approx -18$MeV, which is too large.
However,
what we want to show here is the balance of kinetic and potential energy
and the contribution to the Koltun sum-rule from  momenta above $k_F$. 

This binding energy is the result of a cancellation between a
kinetic energy,  
$t(\rho)= 48.4$ MeV, and a potential energy ,$v(\rho)=-66.7$ MeV.
 The contributions to $t(\rho)$
and $v(\rho)$ of $k'$s below $k_F$ are the 38$\%$ and the 47$\%$ respectively. 
This supports the crucial role of the high momentum components also in 
the energetics of the system, when the kinetic and the potential energy 
are calculated separately. One should keep in mind that the contribution
 of the low
momenta  ($k < k_F$) to the density sum-rule was $\approx 80 \%$ and their
contribution to the total energy is around $70\%$.  The details depend on the
interaction used. Note that the expectation value for the kinetic energy
obtained in SCGF agrees rather well with the result listed in table
\ref{tabhelf} for the Reid potential. The comparison in that table also shows
that the non-local OBE interactions are softer than the Reid potential leading
to smaller kinetic energies and smaller contributions from high momenta.    

To carry on the full self-consistent scheme, and calculate the 
effective interaction with dressed particles, i.e. considering the
spectral functions in the intermediate states in calculating the
G-matrix is a very difficult task. There are however, recent attempts
in that direction  \cite{dickhoff96,bozek99} and also efforts
 to analyze the meaning of
using dressed particles when one considers the scattering in the medium
\cite{dickhoff98}.

The calculation of the single particle spectral function for finite nuclei
is much more complicated. Both the self-energy and the Green's function
are not anymore diagonal in  momentum and furthermore, one should deal with
discrete (bound) and continuum single particle basis states.  
Microscopic calculations have mainly been performed for very light nuclei
\cite{carlson98}. 
For heavier nuclei, several procedures based on local 
density approximation have been devised \cite{lda1,lda2,lda3}.  Here we will
pay more attention to the explicit calculations of the spectral function
directly in finite nuclei.  The calculations have been mainly dealing  with
the hole part of the spectral function for 
$^{16}$O nucleus for which also systematic exclusive electron 
scattering  measurements exist, which will be discussed in the next section. 
A convenient way to take advantage of the spherical symmetry is to
introduce a partial wave decomposition of the spectral function and work
in the single-particle basis characterized by the orbital angular momentum $l$, total 
angular momentum $j$, isospin third component $\tau$ and momentum $k$,
\be
S_{lj\tau}(k,\omega)= \sum_n \mid \langle \Psi_n^{A-1} \mid a_{kl\tau}
 \mid \Psi_0^A 
\rangle \mid ^2 \delta(\omega - (E_0^A - E_n^{A-1}))
\ee
  The momentum distribution
\be
n_{lj\tau}(k) = \langle \Psi_0^A \mid a_{klj\tau }^{\dagger} a_{klj\tau}\mid 
\Psi_0^A \rangle = \sum_n \mid \langle \Psi_n^{A-1} \mid a_{klj\tau} \mid
\Psi_0^A \rangle \mid ^2,
\label{eq:single1}
\ee
is obtained by integrating $S_{lj\tau}(k,\omega)$ over  the excitation
 energies of the
$A-1$ system. In the independent particle model (IPM), the sum in this equation is 
typically reduced to one term, if ($l,j,\tau$) refer to a single-particle orbit
occupied in $\Psi_0^A$. In that case, Eq.(\ref{eq:single1}) yields the square
of the momentum space wave function for this single particle state. The contribution
$n_{lj\tau}(k)$ vanishes in the IPM if no state with this quantum numbers is
 occupied. If correlations are present beyond the IPM approach this
 simple picture is no longer 
true and one can study the effects of correlations on $n(k)$.
Taking into account the degeneracy factors of each orbital
one obtains the total momentum distribution
\be
n(k) = \sum_{l,j,\tau} (2 j + 1) n_{lj\tau}(k).
\ee

The spectral function for the various partial waves, $S_{lj\tau}(k,\omega)$ can be 
obtained from the imaginary part of the corresponding single-particle Green's 
function 
$g_{lj \tau}(k,\omega)$ which can be evaluated by solving the Dyson equation 
(Eq. (\ref{eq:dyson1}))
\be
g_{lj}(k_1,k_2;\omega) = g_{lj}^{(0)}(k_1,k_2;\omega) +
 \int dk_3 \int dk_4 g_{lj}^{(0)}(k_1,
k_3;\omega) \Delta \Sigma_{lj}(k_3,k_4;\omega) g_{lj}(k_4,k_2;\omega),
\label{eq:single2}
\ee
where $g^{(0)}$ refers to a Hartree-Fock propagator and $\Delta \Sigma_{lj}$
 represents contributions to the real and imaginary part of the irreducible
self-energy , which go beyond the Hartree-Fock approximation of the nucleon 
self-energy used to derive $g^{(0)}$. Notice that here and in the following we
 have dropped the isospin quantum number $\tau$.
 Ignoring the Coulomb interaction between the 
protons the Green function are identical for $N=Z$ nuclei and therefore independent
of the quantum number $\tau$.

As in the nuclear matter case,  the key point of this approach is the
 calculation
of the self-energy, its calculation as well as the solution of the Dyson 
equation has
been discussed in detail in Ref. \cite{artur1}. However we include here a brief
summary of the relevant aspects of the method.

The self-energy is evaluated in terms of a $G$ matrix which is obtained
 as a solution of the Bethe-Goldstone equation for nuclear-matter. 
The Bethe-Goldstone equation is solved at a given density and starting
 energy.
In the calculations reported here, the chosen interaction is the one-boson-
exchange (OBE) potential B \cite{rupr0} for $k_F=1.4 fm^{-1}$ and  
starting energy -10 MeV. 
These choices are rather arbitrary. It turns out, however, that the final result
is not sensitive to this choice. The self-energy contains a Hartree-Fock
contribution plus terms of second order in $G$ with intermediate
two-particle-one-hole (2p1h) and two-hole-one-particle states, assuming harmonic
oscillator states for the hole states and plane waves appropriately
orthogonalized for the particle states. The techniques, which are needed to
evaluate matrix elements in such a mixed representation are described in
\cite{artur1}. 

Using such matrix elements one can determine the imaginary part of the 2p1h and
2h1p contributions, $W_{lj}^{2p1h}$ and $W_{lj}^{2h1p}$ in a straightforward
way.

\begin{figure}[t]
\begin{center}
\epsfig{file=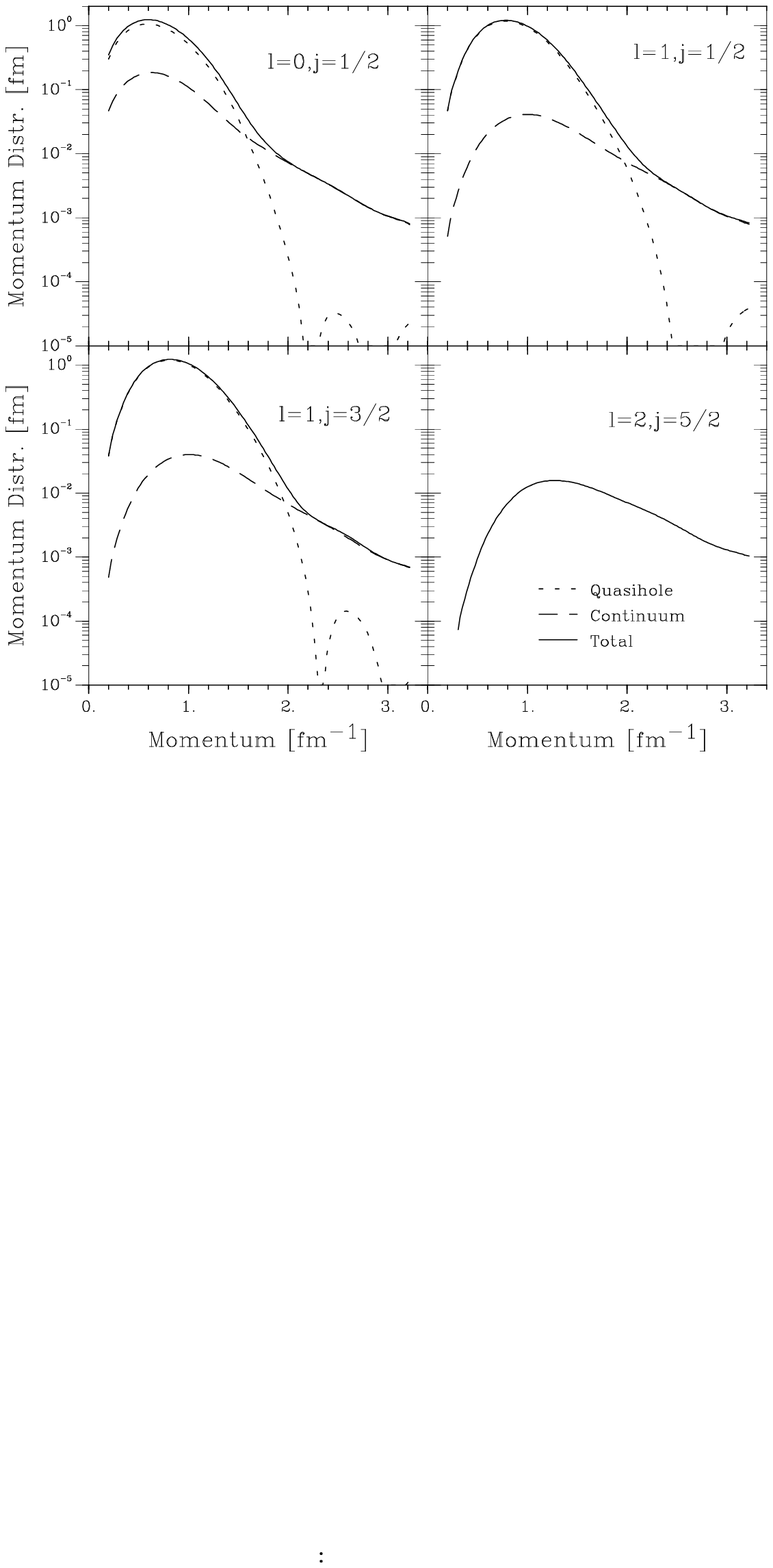,scale=0.8}
\begin{minipage}[t]{16.5 cm}
\caption{ Momentum distribution (solid line) for different partial waves in $^{16}$O. The momentum 
distribution is the sum of the quasihole contribution (short-dashed line) and
the continuum contribution (long-dashed line). \label{fig:figsingle5}}
\end{minipage}
\end{center}
\end{figure}

The real parts of the self-energy are calculated by means of dispersion
relations like,
\be
V_{lj}^{ 2p1h}(k,k';\omega) =\frac {{\cal P}}{\pi}
 \int_{-\infty}^{\infty}
\frac {W_{lj}^{ 2p1h}(k,k';\omega')}{\omega'-\omega} d\omega',
\label{eq:disper}
\ee
where ${\cal P}$ indicates a principal value integral and $k$, $k'$ refer to 
the modulus of the single-particle momentum $k$ for the nucleon under
consideration with angular momentum quantum numbers $l$ and $j$. 
The imaginary part  $W^{ 2p1h}$ is different from zero only for
 positive energies. Since the diagonal matrix elements of $W^{ 2p1h}$
are negative, the dispersion relation (Eq.(\ref{eq:disper}))
implies that the diagonal elements of $V^{ 2p1h}$ will be attractive for
negative energies. They will decrease and change sign only for large 
positive values for the energy of the interacting nucleon. 

Since the Hartree-Fock contribution $\Sigma^{HF}$ is calculated in
 terms of a nuclear G-matrix, it already contains $ 2p1h$ terms.  
In order to avoid such an over-counting of particle-particle
ladder terms, a correction term ($V_c$), which contains the $2p1h$ contribution
 calculated in nuclear matter with the same starting energy and Pauli
 operator as used in the  Bethe-Goldstone equation,
 is subtracted from $V^{ 2p1h}$.
A  dispersion relation similar to Eq.(\ref{eq:disper}) holds for $V^{2h1p}$ and 
$W^{2h1p}$.

Summing up the different contributions, the self-energy is expressed
as
\be
\Sigma = \Sigma^{HF} + \Delta \Sigma = \Sigma^{HF} + (V^{ 2p1h}-V_c + 
V^{2h1p})+ i (W^{ 2p1h}+ W^{ 2h1p}).
\ee

The resulting self-energy is non-local and energy dependent.  It 
represents a microscopic derivation of the optical potential which
can be used to analyze the data of nucleon-nucleus scattering.  
Careful studies of the non-locality of this self-energy have been
presented in Ref. \cite{borromeo}. In this reference, a parameterization of
the self-energy, for both the real and the imaginary part, was given
in terms of local energy dependent potential of the Woods-Saxon 
form.

Once the self-energy is calculated one must solve the Dyson equation (Eq. 
(\ref{eq:single2}))
for the single particle propagator. To this aim, it is useful to discretize
the integrals involved in this equation by considering a complete basis within 
a spherical box of a radius $R_{box}$. The calculated observables are independent of
the choice of $R_{box}$, if it is chosen to be around 15 fm or larger. 
A complete and orthonormal set of regular basis functions within this box is given
by
\be
\Phi_{iljm}({\bf r})= \langle {\bf r} \mid k_i ljm \rangle = N_{il} j_l(k_ir)
{\cal Y}_{ljm}(\theta,\phi).
\ee
In this equation ${\cal Y}_{ljm}$ represent the spherical harmonics including
the spin degrees of freedom, $N_{il}$ is a normalization constant and $j_l$ denote
the spherical Bessel functions for the discrete momenta $k_i$ which fulfil
\be
   j_l(k_i R_{box})=0.
\label{eq:momcre}
\ee

These basis functions, defined for discrete values of the momentum $k_i$ within
the box, differ from the plane wave states defined in the continuum with the
same momenta just by the normalization constant, which is $\sqrt{2/\pi}$ for
the latter. 

As a first step, one constructs the Hartree-Fock approximation for the
single-particle Green's function in the "box basis". To this end, the
Hartree-Fock Hamiltonian is diagonalized,
\be
\sum_{n=1}^{N_{max}} \langle k_i \mid \frac {k_i^2}{2m}\delta_{in} + 
\Sigma_{lj}^{HF} \mid k_n \rangle \langle k_n \mid \alpha \rangle_{lj} 
= \epsilon_{\alpha l j}^{HF} \langle k_i \mid \alpha \rangle _{lj}.
\ee
The set of basis states in the box is truncated by assuming an appropriate
$N_{max}$ (Usually $N_{max} =20$ is enough).
 In the basis of Hartree-Fock states $\mid \alpha \rangle $, 
the Hartree-Fock propagator is diagonal and given by
\be
g_{lj}^{(0)}(\alpha, \omega)= \frac {1}{\omega - \epsilon_{lj}^{HF} \pm i\eta},
\ee
where the sign in front of the infinitesimal imaginary quantity $i\eta$ is 
positive (negative) if $\epsilon_{\alpha l j}^{HF}$ is above (below) the 
Fermi energy. With these ingredients one can solve the Dyson equation 
(Eq. (\ref{eq:single2})).
An efficient procedure is to determine first the reducible self-energy,
\be
\langle \alpha \mid \Sigma_{lj}^{red}(\omega) \mid \beta \rangle =  
\langle \alpha \mid \Delta \Sigma_{lj}(\omega) \mid \beta \rangle + \sum_{\gamma}
\langle \alpha \mid \Delta \Sigma_{lj}(\omega) \mid \gamma \rangle 
 \times g_{lj}^{(0)}(\gamma;\omega) \langle \gamma \mid \Sigma_{lj}^{red}(\omega)
 \mid
\beta \rangle 
\ee
and obtain the propagator from
\be
g_{lj}(\alpha,\beta;\omega)= \delta_{\alpha \beta} g_{lj}^{(0)}(\alpha;\omega) +
g_{lj}^{(0)}(\alpha;\omega) \langle \alpha \mid \Sigma_{lj}^{red}(\omega)
 \mid \beta 
\rangle g_{lj}^{(0)}(\beta;\omega).
\ee

\begin{figure}[t]
\begin{center}
\epsfig{file=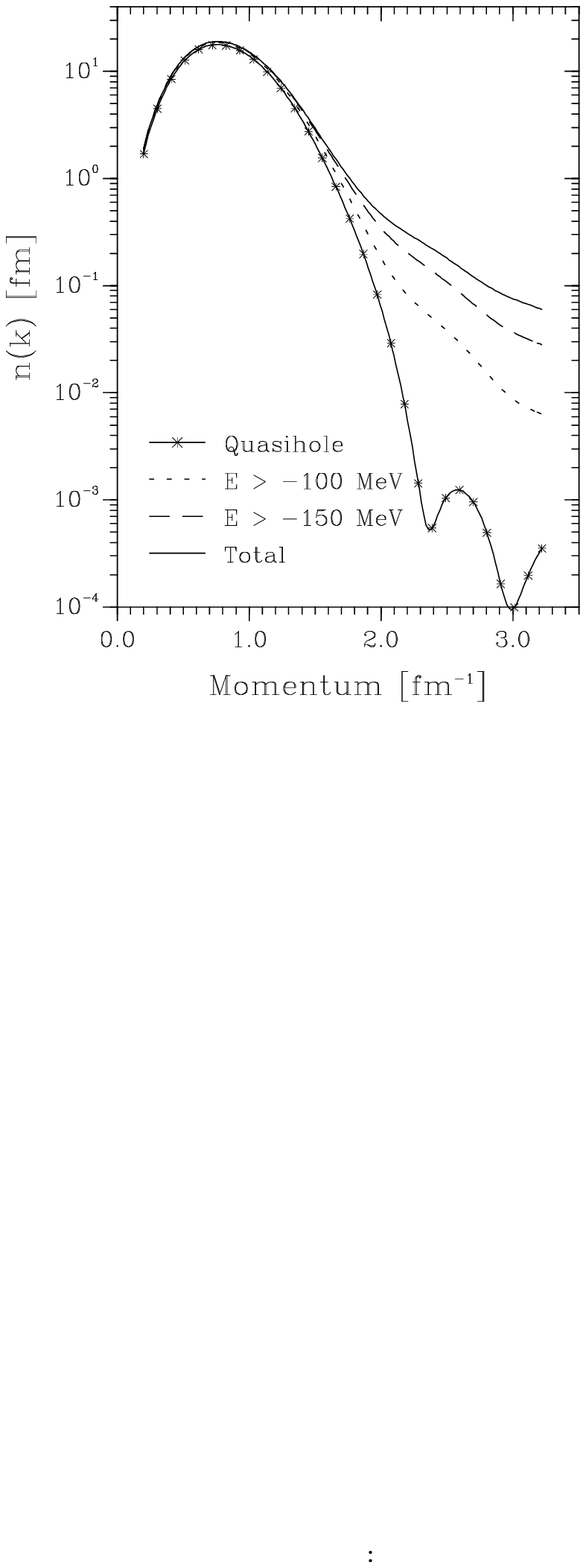,scale=0.7}
\begin{minipage}[t]{16.5 cm}
\caption{ The total momentum distribution of $^{16}$O. The quasihole
contributions also shown together with the results obtained with various 
energy cutoffs in the integration of the spectral functions. 
\label{fig:figsingle6}} 
\end{minipage}
\end{center}
\end{figure}

Using this representation of the Green's function one can calculate the
 spectral function in the "box basis" from
\be
\tilde S_{lj}^c(k_m,k_n;\omega)=\frac{1}{\pi} Im \left ( \sum_{\alpha, \beta}
\langle k_m \mid \alpha \rangle _{lj} g_{lj}(\alpha,\beta;\omega) \langle
 \beta \mid k_n \rangle _{lj} \right ).
\label{eq:conti}
\ee
For energies  below the lowest sp energy of a given Hartree-Fock state
 (with $lj$)  this spectral function is different from zero only due to the 
imaginary part of $\Sigma^{red}$. This contribution involves 
the coupling to the continuum of 2h1p states and is therefore
\be
\sum_{n=1}^{N_{max}} \langle k_i \mid \frac {k_i^2}{2m} \delta_{in} + \Sigma_{lj}
^{HF} + \Delta \Sigma_{lj}(\omega=\epsilon_{\Upsilon l j}^{qh}) \mid  k_n \rangle 
\langle k_n \mid \Upsilon \rangle _{lj} = \epsilon_{\Upsilon l j}^{qh} 
\langle k_i \mid \Upsilon \rangle _{lj}.
\ee
Since, in this approach $\Delta \Sigma$ contains a sizable imaginary part only 
for energies  below $\epsilon_{\Upsilon}^{qh}$, the energies of the
quasihole states are real and the continuum contribution is separated in energy
from the  quasihole contribution. The quasihole contribution to the hole
spectral function is given by
\be
\tilde S_{\Upsilon lj}^{qh}(k_m,k_n;\omega)= Z_{\Upsilon l j} 
\langle k_m \mid \Upsilon
\rangle _{lj} \langle \Upsilon \mid k_n \rangle _{lj} 
\delta(\omega-\epsilon_{\Upsilon l j}^{qh} ),
\ee 
with the spectroscopic factor for the quasihole state given by
\be
Z_{\Upsilon l j } = \left ( 1 - \frac {\partial \langle \Upsilon \mid \Delta 
\Sigma_{lj}(\omega) \mid \Upsilon \rangle }{\partial \omega} \mid 
\epsilon_{\Upsilon l j}^{qh}
\right ) ^{-1}.
\label{eq:discret}
\ee
Finally, the continuum contribution of Eq. (\ref{eq:conti})
 and the quasihole parts of Eq. (\ref{eq:discret})
can be added and renormalized to obtain the spectral function in the continuum 
representation at the momenta defined by Eq. (\ref{eq:momcre}):
\be
S_{lj}(k_i,\omega)=\frac {2}{\pi} \frac {1}{N_{il}^2}
 \left ( \tilde S_{lj}^c(k_i,\omega) +
\sum_{\Upsilon} \tilde S_{\Upsilon l j}^{qh}(k_i,\omega) \right ) .
\ee

\begin{figure}[t]
\begin{center}
\begin{minipage}[t]{16 cm}
\begin{center}
\epsfig{file=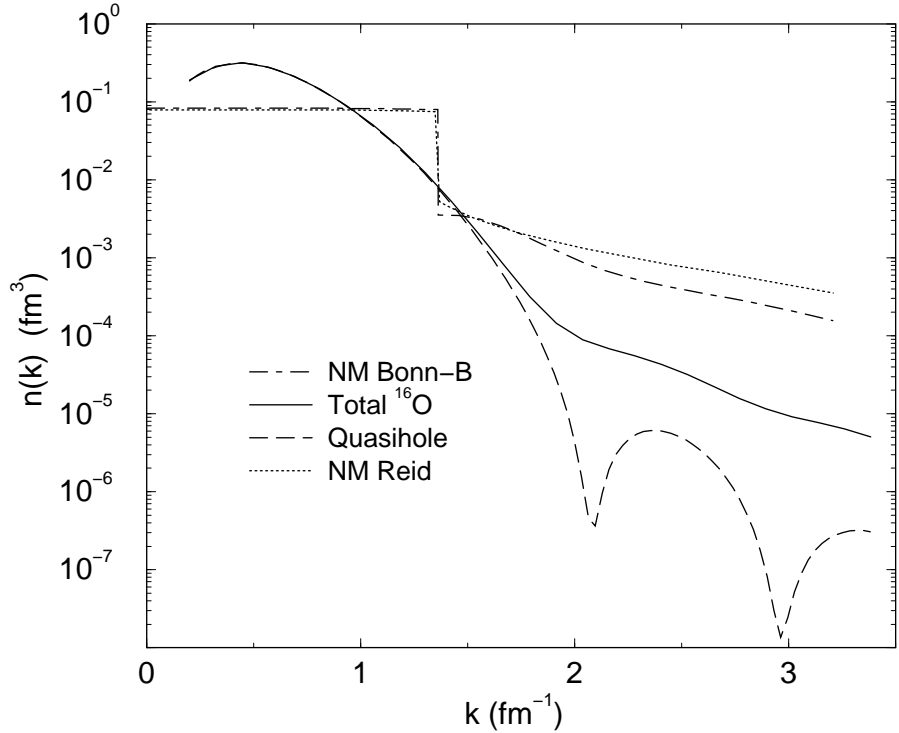,scale=0.7}
\end{center}
\end{minipage}
\begin{minipage}[t]{16.5 cm}
\caption{ Momentum distribution of $^{16}$O, obtained in the Green 
function approach employing the OBE potential, compared with $n(k)$
obtained in nuclear matter. Also reported are the $n(k)$ of nuclear
matter within the SCGF approach for the Reid potential.
\label{fig:figsingle7}} 
\end{minipage}
\end{center}
\end{figure}

The quasi-hole states with a probability defined by 
the spectroscopic factor $Z_{lj}$ (in this example 
0.78, 0.91 and 0.90 in the case of $s_{1/2}$,$p_{3/2}$ and $p_{1/2}$) are 
located at the corresponding quasi-hole energies -34.30 MeV, -17.90 MeV and
-14.14 MeV, respectively. Some strength has been moved to more negative
values of $\omega$ in the (2h1p) continuum.

The momentum distribution $n_{lj}(k)$ is given by 
the energy-integrated spectral function. Here it is also convenient to separate
the contribution in two pieces: the continuum and the quasi-hole part
\be
n_{lj}(k) = \int_{-\infty}^{\epsilon_F} d\omega \left [
 \tilde S_{lj}^c (k,\omega) +
\tilde S_{lj}^{qh}
(k,\omega) \right ].
\ee
This separation into the two parts is displayed in Fig. \ref{fig:figsingle5}
 for various 
partial waves. This figure shows quite clearly that the momentum distribution at
 small momenta is dominated by the quasihole contribution (for those partial waves
for which exists) whereas the high momentum components are given by 
the continuum part. In order to show the importance of the continuum part of
the spectral functions as compared  to the quasihole contribution, the 
particle numbers for each partial wave including the degeneracy of the 
states,
\be
\tilde n_{ij} = 2 (2j+1) \int_{-\infty}^{\epsilon_F} d\omega
 \int_0^{\infty} dk k^2 
S_{lj}(k,\omega),
\ee
separating the quasihole and the continuum contributions are reported
in Table \ref{table:single1}. In this  approach, there are
only 14.025 out of 16 nucleons in the quasi-hole states. Another
1.13 "nucleons" are found in the $2h1p$ continuum with 
partial wave quantum numbers of the $s$ and $p$ shells, 
while an additional 0.79 "nucleons" are obtained from the continuum
with orbital quantum numbers of the $d$,$f$ and $g$ shells. The
quasi-hole strength can be identified with the
experimental spectroscopic factor. In this  approach, 
for the $p_{1/2}$ it turns to be $90 \%$ which is too 
large compared with the experimentally determined
value, $63 \%$. This discrepancy , can be associated to the
fact that this approach is treating mainly the short-range
correlations. Long-range (low-energy) correlations, 
will typically yield another $10\%$ reduction of
the quasihole strength \cite{skour2,nili1,geurts96}. 
Similar results are obtained  with
variational Monte Carlo techniques, calculating the overlap
between a correlated ground state wave function
for $^{16}$O and the quasi-hole states 
of the ($A-1$) residual system ($^{15}N$) \cite{radici94}.
However, 
an appropriate treatment of center of mass motion can enhance the spectroscopic
factor by up to  $7 \%$\cite{neck98} increasing
again the discrepancy with the experimental data. The absolute value for the
spectroscopic factors seems not to be understood yet.

\begin{table}
\begin{center}
\begin{tabular}{c|rr|r}
&&&\\
$lj$&\multicolumn{1}{c}{$\hat n^{qh}$}
&
\multicolumn{1}{c|}{$\hat n^c$}&\multicolumn{1}{c}{$\hat n$}
 \\
&&&\\ \hline
&&&\\
$s_{1/2}$ & 3.120 & 0.624 & 3.744 \\
$p_{3/2}$ & 7.314 & 0.332 & 7.646 \\
$p_{1/2}$ & 3.592 & 0.173 & 3.764 \\
&&&\\
$d_{5/2}$ & 0.0 &  0.234 & 0.234 \\
$d_{3/2}$ & 0.0 & 0.196 & 0.196 \\
$f_{7/2}$ & 0.0 & 0.117 & 0.117 \\
$f_{5/2}$ & 0.0 & 0.140 & 0.140 \\
$g_{9/2}$ & 0.0 & 0.040 & 0.040 \\
$g_{7/2}$ & 0.0 & 0.064 & 0.064 \\
&&&\\ \hline
$\sum$ & 14.025 & 1.920 & 15.945 \\
&&&\\ \hline
\end{tabular}
\begin{minipage}[t]{16.5 cm}
\caption{ Distribution of nucleons in $^{16}$O. Listed are the total occupation
number $\tilde n$ for various partial waves  and  also the 
contributions from the quasihole ($\tilde n^{qh}$) and the 
continuum part ($\tilde n^c$) of the spectral function separately. The last
line gives the sum of particle numbers for all partial wave listed. \label{table:single1}}
\end{minipage}
\end{center}
\end{table}

The sum of the particle numbers listed in Table \ref{table:single1}
 is slightly smaller (15.945)
than the particle number corresponding to $^{16}$O. There are  several 
possible sources for this discrepancy: First of all the analysis only accounts for
momenta below 3.3 fm$^{-1}$ and the partial waves with $l>4$ have not been 
considered. The restriction in $k$ is determined by the choice of $N_{max}$ in
truncating the "box basis". Inspecting the decrease of the occupation numbers
listed in Table \ref{table:single1} with increasing $l$ one can expect
 that the "missing" nucleons
may be found in partial waves with $l>4$.
 Furthermore, however, one must also keep
in mind that this approach to the single-particle Green function is not 
number conserving, as the Green functions used to evaluate the self-energy
 are not determined in a self-consistent way \cite{wim1}.
 
Similarly to Fig.\ref{fig:figsingle2} for nuclear matter,
 the momentum distribution including
the quasi-hole states obtained with various energy cutoffs is shown in 
Fig. \ref{fig:figsingle6}.  The quasihole part reflects the momentum components that
one can detect  in knock-out reactions with small energy transfer.  
 As in the
nuclear matter case, the high momentum components due to 
short-range correlations can be detected only for large excitation energies.

The total momentum distribution is displayed again in Fig.\ref{fig:figsingle7}
  and compared
to predictions for nuclear matter. In order to enable the comparison  with
nuclear matter results, the momentum distributions have been normalized such 
that $\int d^3 k n(k)$ yields 1. In order to show the sensitivity of the
calculated momentum distribution  to the nucleon-nucleon interaction, 
the results obtained for Reid soft-core in SCGF are also shown\cite{vonderf}. The
nuclear matter results for the OBE-B potential have been obtained by
using a self-energy which contains only second order terms on the
propagation of holes \cite{lda4}. The OBE-B is consider to be softer than
the older Reid potential. This is reflected by the fact that the momentum distribution
obtained for the Reid potential yields larger occupation values for 
$k > k_F$ than those  obtained for the OBE potential. The comparison between nuclear 
matter and finite nuclei seems to indicate  that
the enhancement of the momentum distribution predicted in this approach 
for high momenta is below the corresponding prediction derived for nuclear matter.
These problem has been clarified in Ref.\cite{lda4} were an LDA estimation 
of the contribution to $n(k)$ of the partial wave not included in the calculation
brought the tail of $n(k)$  in $^{16}$O in close agreement with the 
nuclear matter results, when calculated with the same interaction.

\begin{figure}[t]
\begin{center}
\begin{minipage}[t]{18 cm}
\begin{center}
\epsfig{file=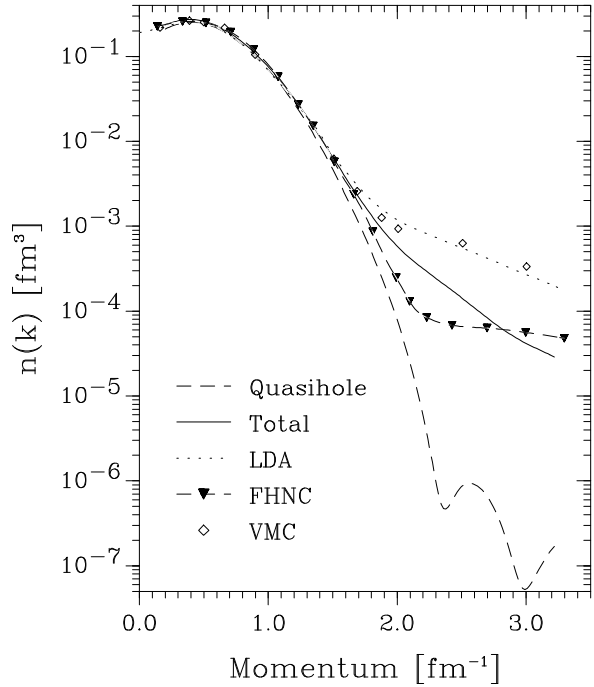,scale=0.6}
\end{center}
\end{minipage}
\begin{minipage}[t]{16.5 cm}
\caption{ Comparison of different approaches to calculate the momentum 
distribution in $^{16}$O. The results obtained from the spectral function are given 
separately including
both the quasihole and the continuum part (total). The curve labeled LDA corresponds
to  the results
from  the local density  approximation (LDA) obtained in Ref. \cite{lda3}, while the FHNC
belongs to the
Fermi hypernetted chain approach  of Ref. \cite{co2} and VMC stands for  the variational
Monte Carlo calculation of Ref. \cite{pieper92}. 
 \label{fig:figsingle8}} 
\end{minipage}
\end{center}
\end{figure}

In Fig. \ref{fig:figsingle8} we compare the $n(k)$ in $^{16}$O obtained in different 
approaches. The results obtained in the Green function approach (total) 
are compared with those obtained in LDA from Ref. \cite{lda3}, which were
obtained using the Urbana V$_{14}$ interaction. Also shown are the results, 
obtained by a direct FHNC calculation of the expectation value of the number
occupation operator in a correlated wave function of the ground state of $^{16}$O,
where the correlations were optimize by minimizing the ground state energy
calculated with the semi-realistic Afnan and Tang interaction \cite{co2} and
finally the variational Monte Carlo calculations of Ref.\cite{pieper92} obtained 
with a ground state wave function containing two- and three- operatorial correlations
which minimize the ground state energy for the Argonne V$_{14}$ interaction.

Now we want to illustrate the effects of correlations, which are taken 
into account in the Green's function approach  beyond the BHF approximation, 
on the ground state
properties of $^{16}$O . To calculate these properties one needs also 
the non-diagonal part of the density matrix  which is given by
\be
\tilde n_{lj}(k_i,k_n)=\int_{-\infty}^{\epsilon_F} d\omega \frac {1}{\pi} 
Im \left ( \sum_{\alpha, \beta} \langle k_i \mid \alpha \rangle_{lj} 
g_{lj}(\alpha,\beta;\omega) \langle \beta \mid k_n \rangle_{lj} \right ).
\label{eq:densma}
\ee
and contains as in the case of the spectral function, a continuous
contribution and a part originating from the quasihole states,
\be
\tilde n_{lj}^{qh}(k_i,k_n) = \sum_{\Upsilon} Z_{\Upsilon l j} \langle k_i 
\mid \Upsilon \rangle_{lj} \langle \Upsilon \mid k_n \rangle_{lj} .
\ee
With this density matrix, the expectation value for the square 
of the radius can be calculated according to 
\be
\langle \Psi_0^A \mid r^2 \mid \Psi_0^A \rangle = \sum_{lj} 2 (2j+1)
\sum_{i,n=1}^{N_{max}} \langle k_i \mid r^2 \mid k_n \rangle_l n_{lj}
(k_i,k_n).
\ee
The matrix elements of $r^2$ in the basis states are given by
\be
\langle k_i \mid r^2 \mid k_n \rangle_l = N_{il} N_{nl} \int_0 ^{R_{max}}
dr r^4 j_l(k_i r) j_l(k_n r).
\ee

The total energy of the ground state is obtained from the 
Koltun sum rule (\ref{eq:koltun})
\be
E_0^A = \sum_{lj} 2 (2j+1) \sum_{i=1}^{N_{max}} \int_{-\infty}^{\epsilon_F}
d\omega \frac {1}{2} \left ( \frac {k_i^2}{2 m} + \omega \right ) 
\left ( \tilde S_{lj}^c(k_i,\omega) + \sum_{\Upsilon} \tilde S_{\Upsilon l j}^{qh}
(k_i,\omega) \right ).
\label{eq:koltun1}
\ee

Using the previous equations we calculate the contributions
of the different partial waves to the binding energy, the results
are reported in Table \ref{table:single2}. 
As a first step we consider the Hartree-Fock (HF) approximation. 
The resulting binding energy per
nucleon (-1.93 MeV) is quite small. This is probably due to the
use of a G matrix calculated at the saturation density of nuclear
matter, which overestimates the Pauli effects as compared 
to a BHF calculation directly for $^{16}$O. 

The treatment of the Pauli operator is improved by adding the 
$2p1h$ part ,with the corresponding correction term, to the self-energy.
This approximation can be considered as an approximation to a 
full BHF and we will labeled "BHF". This correction increases
the binding energy to -4.01 MeV. This number is in reasonable
agreement with self-consistent BHF calculations performed for
$^{16}$O using the same interaction \cite{carlo}.
However, as the single particle states are more bound than 
the single particle states obtained in the HF calculations,
the gain in binding energy is accompanied by a reduction of
 the calculated radius of the nucleon distribution.

The inclusion of the 2h1p contributions to the self-energy reduces the
absolute values of the quasi-hole energies. Despite of this
reduction, the total binding energy is increased compared to BHF
calculations. This increase of the binding energy
is mainly due to the continuum part of the spectral function. 
Comparing the various contributions, one finds that 
only the 37 $\%$ of the total energy is due to the 
quasi-hole part in Eq. (\ref{eq:koltun1}). The dominant
part (63 $\%$) results from the continuum part of the spectral functions
although this continuum part only represents 11 $\%$ of the nucleons. 
The inclusion of $2h1p$ terms increases also the radius, moving 
the results for the ground state off the Coester band. 

\begin{table}
\begin{center}
\begin{tabular}{c|rrr|rrr|rrr}
&&&&&&&&&\\
&\multicolumn{3}{c|}{HF} &\multicolumn{3}{c|}{BHF} &\multicolumn{3}{c}
{Total} \\
$lj$&\multicolumn{1}{c}{$\epsilon$}&\multicolumn{1}{c}{$t$}&
\multicolumn{1}{c|}{$\Delta E$}
&\multicolumn{1}{c}{$\epsilon$}&\multicolumn{1}{c}{$t$}&
\multicolumn{1}{c|}{$\Delta E$}
&\multicolumn{1}{c}{$\epsilon$}&\multicolumn{1}{c}{$t$}&
\multicolumn{1}{c}{$\Delta E$}\\
&&&&&&&&&\\
\hline
&&&&&&&&&\\
$s_{1/2}$ qh & -36.91 & 11.77 & -50.28 & -42.56 & 11.91 & -61.30 & -34.30
& 11.23 & -35.98 \\
$s_{1/2}$ c & &&&&&& -90.36 & 17.09 & -22.89 \\
$p_{3/2}$ qh & -15.35 & 17.62 & 9.08 & -20.34 & 18.95 & -5.59 & -17.90 &
18.06 & 0.37 \\
$p_{3/2}$ c  & & & & & & & -95.19 &
35.19 & -9.96\\
$p_{1/2}$ qh & -11.46 & 16.63 & 10.34& -17.07 & 18.46 &  2.76 & -14.14 &
17.19 & 5.47 \\
$p_{1/2}$ c  & & & & & & & -103.62&
35.94 & -5.84\\
$l>1$ c  & & & & & & & -98.87 &
63.17 & -12.27 \\
&&&&&&&&&\\
\hline
&&&&&&&&&\\
$E/A$&\multicolumn{3}{c|}{-1.93} &\multicolumn{3}{c|}{-4.01}
&\multicolumn{3}{c}
{-5.12} \\
$\left\langle r \right\rangle$&\multicolumn{3}{c|}{2.59}
&\multicolumn{3}{c|}{2.49} &\multicolumn{3}{c}
{2.55} \\
&&&&&&&&&\\
\end{tabular}
\begin{minipage}[t]{16.5 cm}
\caption{Groundstate properties of $^{16}$O. Listed are the energies
$\epsilon$ and kinetic energies $t$ of the quasihole states (qh) and the
corresponding mean values for the continuum contribution (c), normalized to 1,
for the various partial waves. Multiplying the sum: $1/2(t+\epsilon )$ of these
mean values with the corresponding particle numbers of Tab.I, one obtains
the contribution $\Delta E$ to the energy of the ground state.
 Results are presented for
the Hartree-Fock (HF), Brueckner-Hartree-Fock (BHF) and the Green function
approach (Total). 
 The particle numbers for the qh states in HF and BHF
are equal to the degeneracy of the states, all other occupation numbers
are zero. The results for the radii are given in fm, all other entries in MeV.
\label{table:single2}}
\end{minipage}
\end{center}
\end{table}

\begin{figure}[t]
\begin{center}
\epsfig{file=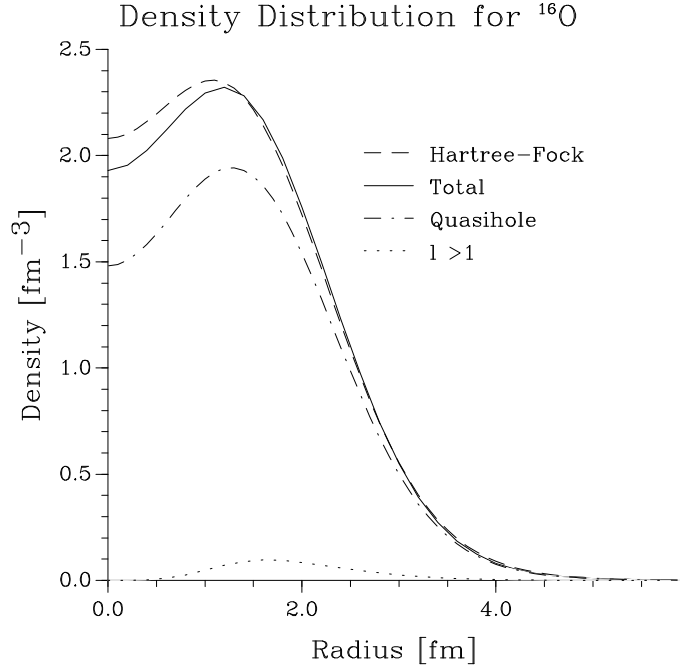,scale=0.7}
\begin{minipage}[t]{16.5 cm}
\caption{Density distribution of nucleons in $^{16}$O as a function
of the distance from the center of the nucleus.  
\label{fig:figsingle10}} 
\end{minipage}
\end{center}
\end{figure}

The relative importance of the various contribution to the single-particle
strength can also be seen from inspecting Fig.~\ref{fig:figsingle10} which shows
the density profile of $^{16}$O and some of its components. This figure also
demonstrates that the strength located in single-particle states with $l>1$,
which would not be occupied in the mean field approach, provide a small but
non-negligible contribution.

Another way of describing correlation effects in a single-particle basis is 
the representation of the single-particle density matrix in terms of natural
orbits\cite{polls95,lowdin1,antonov,neck93}. 

\subsection{\it Correlations in Nucleon Knock-Out Experiments}

The uncorrelated Hartree-Fock state of nuclear matter is given as a Slater
determinant of plane waves, in which all states with momenta $k$ smaller than
the Fermi momentum $k_F$ are occupied, while all others are completely
unoccupied. Correlations in the wave function beyond the mean field approach will
lead to occupation of states with $k$ larger than $k_F$. Therefore correlations
should be reflected in an enhancement of the momentum distribution at high 
momenta. Indeed, as we have already discussed above,   microscopic
calculations exhibit such an enhancement for nuclear matter as well as for
finite nuclei\cite{lda4,artur1}. At first sight one feels encouraged to measure 
this momentum distribution
by means of exclusive $(e,e'p)$ reactions at low missing energies, such that
residual nucleus remains in the ground state or other well defined bound state.
From the momentum transfer $q$ of the scattered electron and the momentum $p$ of
the outgoing nucleon one can calculate the momentum of the nucleus before the
absorption of the photon and therefore obtain direct information on the momentum
distribution of the nucleons inside the nucleus.

This idea, however, suffers from a little inaccuracy. In such exclusive
$(e,e'p)$ experiments one does not measure the whole momentum distribution but
rather the spectral function, at the energy which correspond to the specific
final state. As we have seen in the discussion of the preceeding section the
spectral function at low energies does not show this enhancement of the high
momentum components, which shows up only if one integrates the spectral function
up to high excitation energies (see Fig.~\ref{fig:figsingle6} and discussion 
there). A
general review on nucleon knock-out by means of electromagnetic probes is
presented in the book of Boffi et al.\cite{bofbook} 

\begin{figure}[tb]
\begin{center}
\begin{minipage}[t]{12 cm}
\epsfig{file=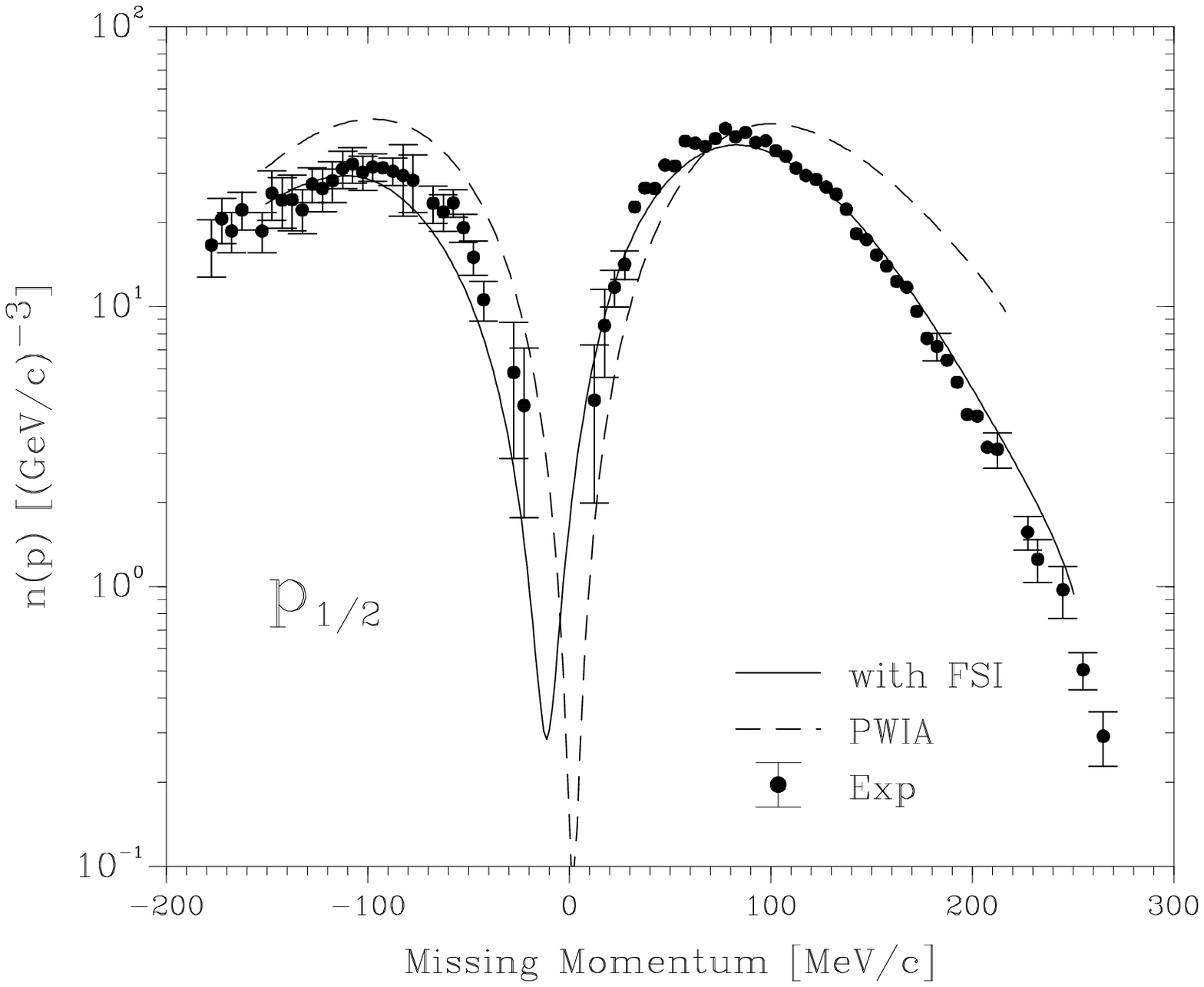,scale=0.6}
\end{minipage}
\begin{minipage}[t]{16.5cm}
\caption{ Reduced cross section for the $^{16}O(e,e'p)^{15}N_{gs}$ reaction 
in parallel kinematics. Results from \protect{\cite{poll1}} with and without
(plane wave impulse approximation, PWIA) inclusion of FSI are compared to
experimental data taken by Leuschner et al.\protect{\cite{leusch}}
\label{fig:marco}}
\end{minipage}
\end{center}
\end{figure}

A second problem is related to the fact that the nucleon, which is knocked out
in such an $(e,e'p)$ process, feels the interaction with the remaining nucleus.
It will be retarded and might also get reabsorbed. These effects of the final
state interaction (FSI) can be taken into account by means of an optical 
potential. As an example for the importance of FSI effects we show in
Fig.~\ref{fig:marco} the reduced cross section for the
$^{16}O(e,e'p)^{15}N_{gs}$ reaction with and without inclusion of FSI effects.
The reduced cross section is defined as the cross section divided by the
elementary electron - nucleon cross section times a kinematical factor. 
Data were taken \cite{leusch} in so-called parallel kinematics. This means 
that the missing momentum of the proton, the momentum of emerging proton minus
the momentum transfer is parallel or antiparallel (negative values) to the
momentum transfer. Note that the missing momentum roughly corresponds to the 
momentum of the nucleon before absorption of the virtual photon. The retardation
effects yield a reduction of the momentum for the outgoing proton. Therefore FSI
effects lead to a shift of the cross section to smaller missing momenta, which
can nicely be seen from Fig.~\ref{fig:marco}.

\begin{figure}[t]
\begin{center}
\begin{minipage}[t]{8 cm}
\epsfig{file=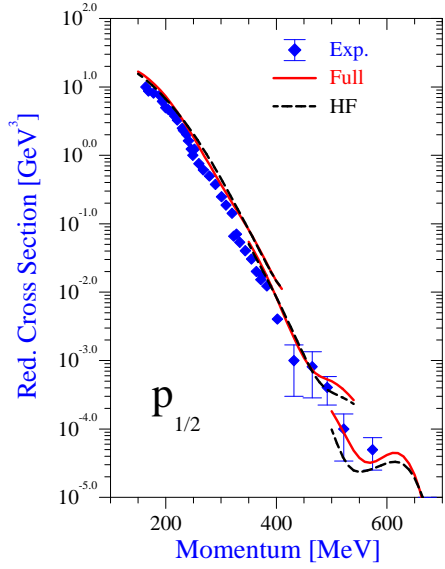,scale=0.5,angle=270}
\end{minipage}
\begin{minipage}[t]{16.5cm}
\caption{Reduced cross section for the $^{16}$O($e,e'p$)$^{15}$N
reaction leading to the ground state ($1/2^-$) of $^{15}$N in the
kinematical conditions considered in the
experiment at MAMI (Mainz) \protect\cite{mainz}. Results for the
mean-field description (HF) and the correlated spectral function
(Full) are presented.
\label{fig:nili}}
\end{minipage}
\end{center}
\end{figure}

The data displayed in Fig.~\ref{fig:marco} exhibit results only up to moderate
values of the missing momenta. Experiments trying to explore high missing
momenta were performed at MAMI in Mainz\cite{mainz1} and NIKHEF in
Amsterdam \cite{adam1}. These data can be well reproduced over a range of missing
momenta going up to 600 MeV/c by calculations which account for the FSI using 
a relativistic model for the optical potential\cite{nili}. The shape of the
calculated cross section, however, is rather insensitive to the use of spectral
functions, which are derived either from mean field or more general
single-particle Green's function (see Fig.~\ref{fig:nili}). The only difference
is the global spectroscopic factor which is adjusted to reproduce the total
cross section. 

This demonstrates that exclusive one-nucleon knock-out experiments only yield 
limited information on NN correlations. Therefore one tries to investigate
exclusive $(e,e'NN)$ reactions, i.e.\ triple coincidence experiments in which
the energies of the two outgoing nucleons and the energy of the scattered
electron guarantee that the rest of the target nucleus remains in the ground
state or a well defined excited state. The idea is that processes in which the
virtual photon, produced by the scattered electron, is absorbed by a pair of
nucleons should be sensitive to the correlations between these two nucleons.

\begin{figure}[th]
\begin{center}
\epsfig{file=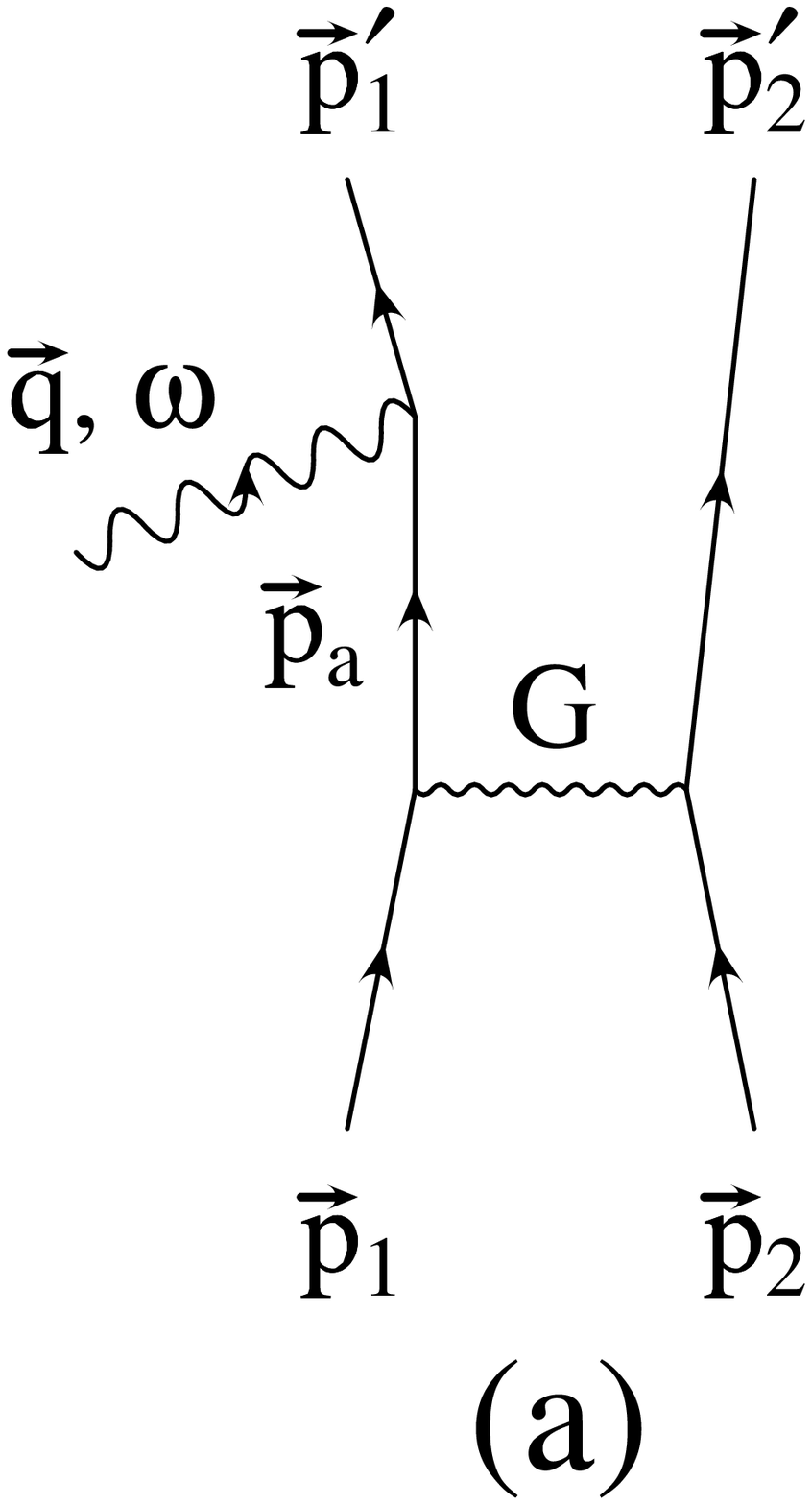,scale=0.18}\hspace{.5cm}
\epsfig{file=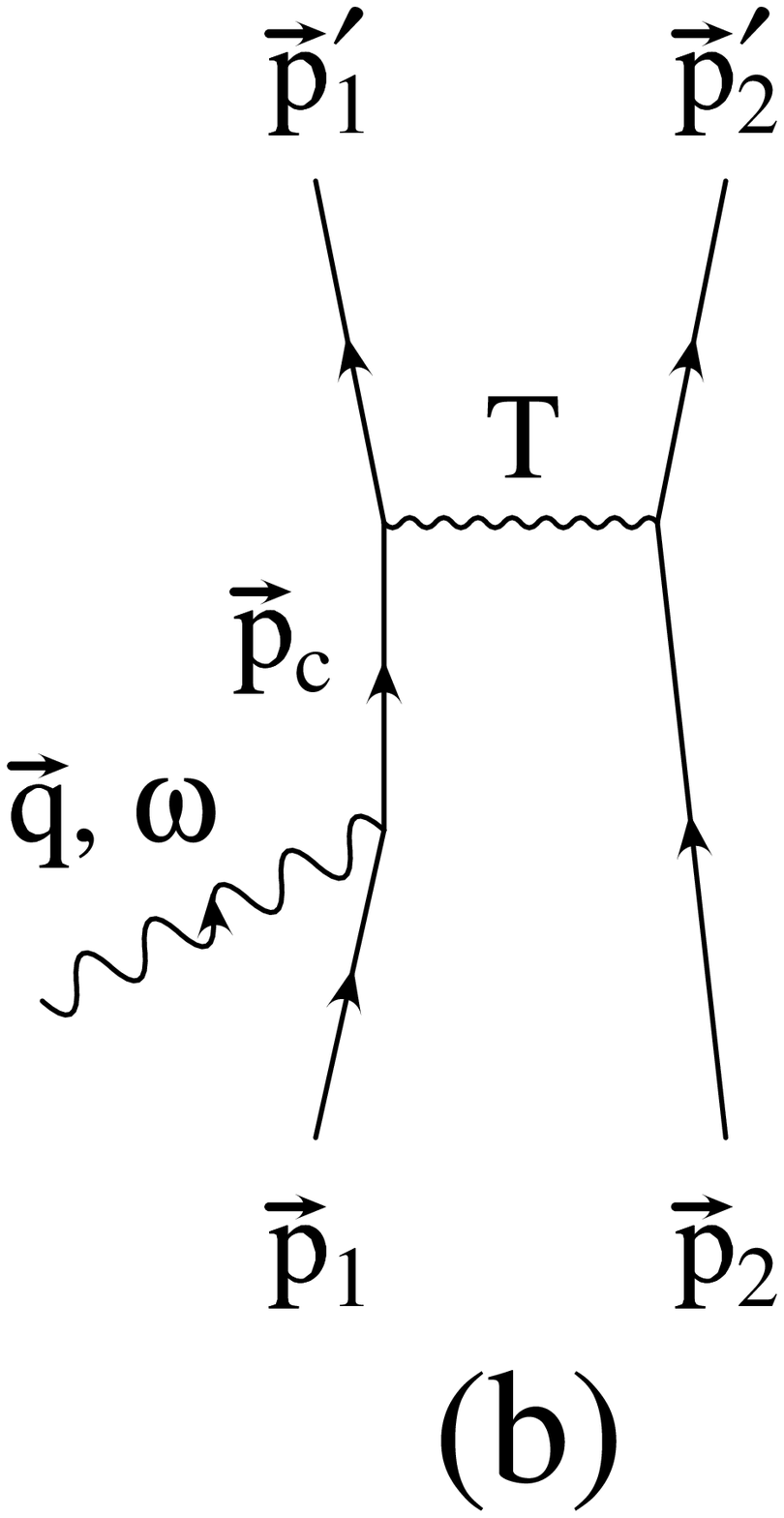,scale=0.18}\hspace{.5cm}
\epsfig{file=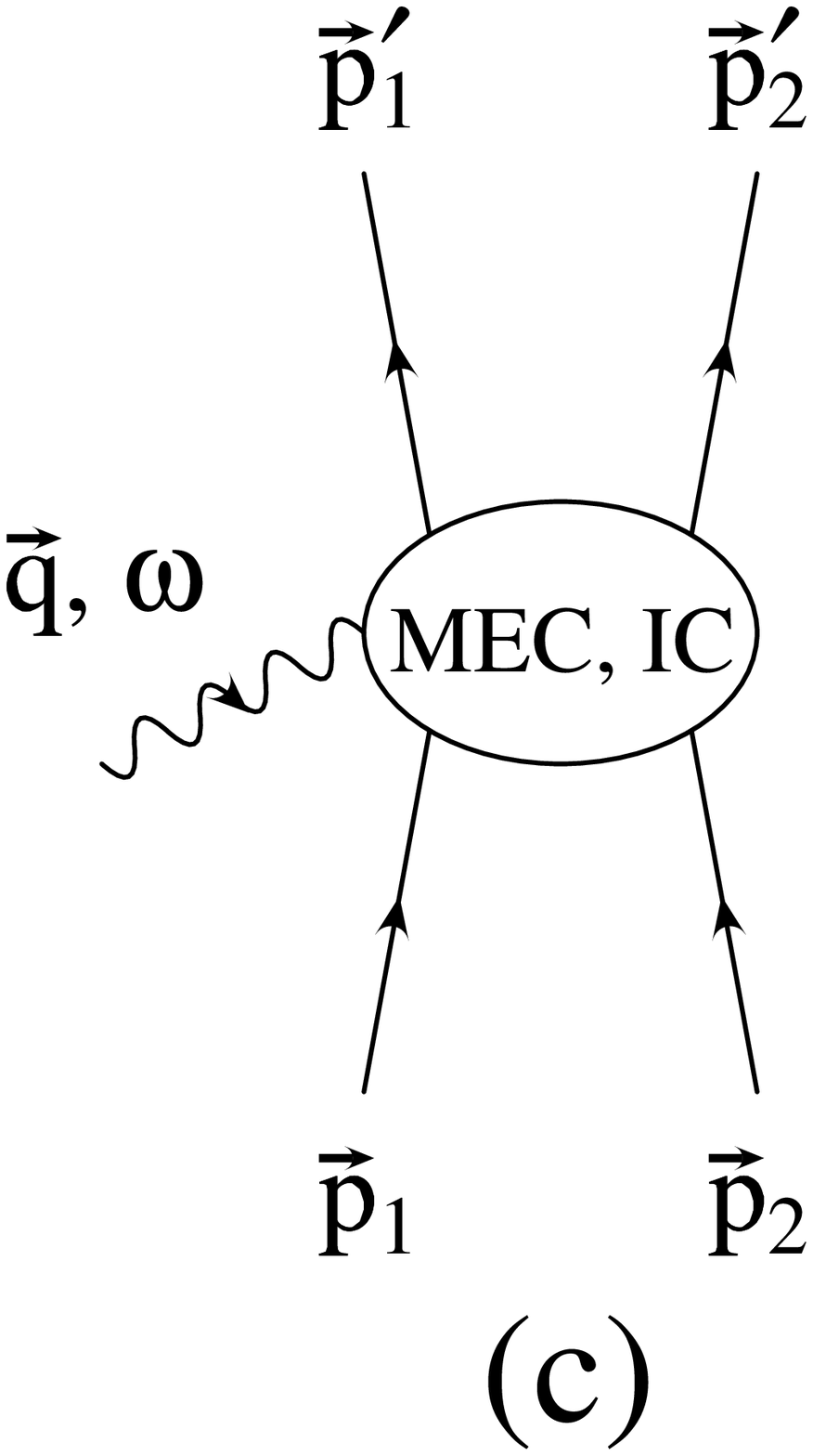,scale=0.18}
\begin{minipage}[t]{16.5cm}
\caption{\label{gfig3}
Diagrams for the different processes contributing to the $(e,e'2N)$
reaction. Diagram (a) and (b) show the absorption of the photon by a
single nucleon. The nucleon-nucleon correlations are described by the $G$
matrix. Diagram (c) depicts photon absorption via meson exchange (MEC) or
isobaric currents (IC)}
\end{minipage}
\end{center}
\end{figure}

Unfortunately, however, this process which is represented by the diagram
in Fig.~\ref{gfig3}a, competes with the other processes described by the
diagrams of Fig.~\ref{gfig3}b and c. These last two diagrams refer to the
effects of final-state-interaction (FSI) and contributions of two-body currents.
Here we denote by final state interaction not just the feature that each of the
outgoing nucleons feels the remaining nucleus in terms of an optical potential.
Here we call FSI the effect, that one of the nucleons absorbs the photon,
propagates (on or off-shell) and then shares the momentum end energy of the
photon by interacting with the second nucleon which is also knocked out of the
target. The processes described in Figs.~\ref{gfig3}a and \ref{gfig3}b,
correlations and FSI, are rather similar, they differ only by the time ordering
of NN interaction and photon absorption. Therefore it seems evident that one
must consider both effects in an equivalent way. Nevertheless, most studies up 
to now have ignored this equivalency but just included the correlation effects 
in terms of a correlated two-body wave function. For the sake of consistency
one should assume the same interaction to be responsible for the correlations 
and this two-body FSI. Correlations can be evaluated in terms of the Brueckner 
G-matrix while the T-matrix derived from the very same interaction should be
used to determine FSI.

The nine-fold differential cross section of such an $(e,e'2N)$ reaction can be
written as a product of a matrix element of the leptonic current $j_\mu$ times
the matrix elements of the corresponding hadronic current $< J^\mu >$ calculated
between the initial and final nuclear state
\be
\frac{{\rm d}^9\sigma}{{\rm d}\tilde{E_1}{\rm d}\tilde{\Omega_1} {\rm d}
\tilde{E_2}{\rm d}\tilde{\Omega_2} {\rm d}E_e'{\rm d}\Omega_e'} = 
K \vert j_\mu < J^\mu > \vert ^2 \label{eq:cross0}
\ee
with $K$ a kinematical factor. This expression can be rewritten into the form
\cite{eepwq1,eepwq2}
\bea
\frac{{\rm d}^9\sigma}{{\rm d}\tilde{E_1}{\rm d}\tilde{\Omega_1} {\rm d}
\tilde{E_2}{\rm d}\tilde{\Omega_2} {\rm d}E_e'{\rm d}\Omega_e'}
&= &\frac{1}{4}\,\frac{1}{(2\pi)^9}\, \tilde{p_1}\, \tilde{p_2}\, \tilde{E_1}\,
\tilde{E_2}\,
\sigma_{\rm Mott} \,
\Big\{ v_C W_L + v_T W_T +v_S W_{TT} + v_I W_{LT} \Big\}\,
\nonumber \\
&& \times (2\pi)\, \delta (E_f-E_i)
\eea
where $\tilde{E_1},\tilde{E_2}$ and $\tilde{p_1},\tilde{p_2}$ denote the
energies and momenta of the outgoing nucleons, respectively.  The
virtual photon created in the electron scattering process carries a transfered
momentum $\vec{q}$ and energy $\omega$. The leptonic structure functions
$v_i$ ($i=C,T,S,I$) are defined by 
\bea
v_C&=& \Big( \frac{q_{\mu}q^{\mu}}{\vec{q}\,^2} \Big)^2 \nonumber \\
v_T&=& \tan^2 \frac{\theta_e}{2} - \frac{1}{2} \Big(
\frac{q_{\mu}q^{\mu}}{\vec{q}^{\,2}} \Big) \nonumber \\
v_I&=& \frac{q_{\mu}q^{\mu}}{\sqrt{2}|\vec{q}\,|^3}\,(E_e+E_e')\,
\tan\frac{\theta_e}{2}\nonumber \\
v_S&=& \frac{q_{\mu}q^{\mu}}{2\vec{q}\,^2}
\eea
Here, $\theta_e$ is the angle of the scattered electron with respect to the
incident electron beam and $E_e,E_e'$
are the energies of the incident and the scattered electron, respectively.
 The nuclear structure functions $W_i$ ($i=L,T,TT,LT$ stands for longitudinal,
transverse, etc.) contain the matrix
elements of the nuclear current operator for a given photon polarization
$\lambda$. These matrix elements have to be calculated for the various processes
displayed in Fig.~\ref{gfig3}.

\begin{figure}[tb]
\begin{center}
\epsfig{file=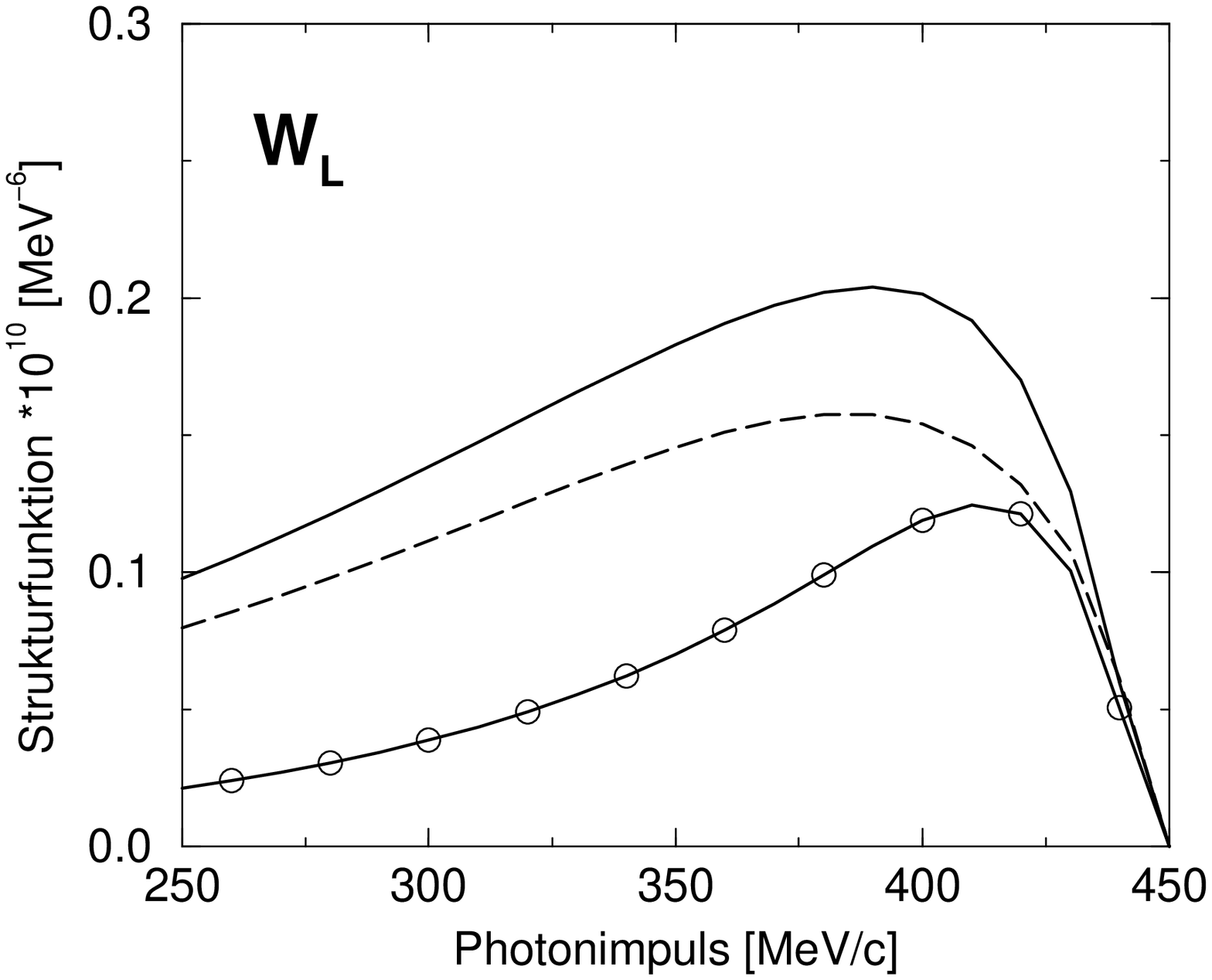,scale=0.4}\hspace{.5cm}
\epsfig{file=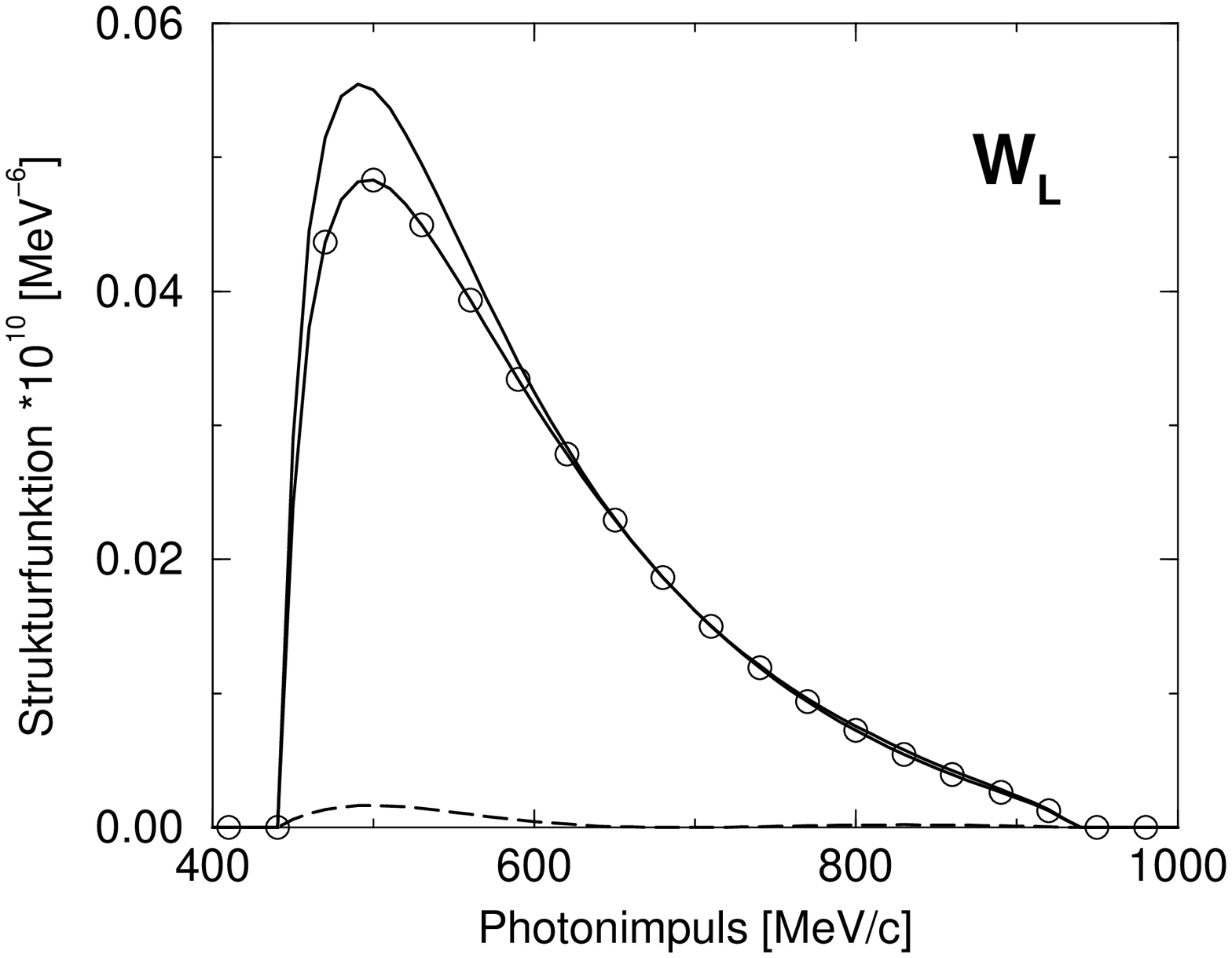,scale=0.4}
\begin{minipage}[t]{16.5cm}
\caption{\label{fig:knoe1}
 Longitudinal structure function for the knockout of a proton-proton pair
as a function of the photon momentum $q$, keeping the photon energy constant 
at $\omega=215\,{\rm MeV}$. The left figure shows results for a 'super parallel' 
kinematical situation with angles $\theta_{p,1}'=0^{\rm o}$ and
$\theta_{p,2}'=180^{\rm o}$ of the two protons with respect to the
direction of the photon momentum, while the figure on the right-hand side
displays results assuming $\theta_{p_i}=\pm 30^{\rm o}$. The dashed line
represents the contribution from correlations, the line with circles those of
the FSI and the solid line the coherent sum of these two}
\end{minipage}
\end{center}
\end{figure}

The relative importance of the various contributions under different 
kinematical setups, can be explored in calculations of the structure function
$W_i$ for nuclear matter at saturation density\cite{knoed}. 
As a first example we consider the longitudinal structure function
for the knockout of a proton-proton pair. In this case one can ignore the
effects of meson-exchange currents (MEC) and isobar currents (IC) displayed in
Fig.~\ref{gfig3}, since two protons do not exchange charged mesons. 
One of the protons is assumed to be emitted parallel
to the momentum of the virtual photon with an energy of $T_{p,1}=156\,{\rm
MeV}$, while the second is emitted antiparallel to the photon momentum with an
energy of $T_{p,2}=33\,{\rm MeV}$ (see left part of Fig.~\ref{fig:knoe1}).
This is called the `super-parallel kinematic',
which should be appropriate for a separation of longitudinal and transverse
structure functions. In this situation the dominant contribution to the
longitudinal response function is due to correlation effects (dashed line).
But also the FSI effects contribute in a non-negligible way to the cross section
(curve with circle symbols), although the two protons are emitted in 
opposite directions.
The effects of FSI are much more important, if we request that the two protons
are emitted in a more symmetric way. As an example we show the longitudinal
structure function for ($e,e'pp$), requesting that each of the protons carries
away an energy of 70 MeV and is emitted with an angle of $30^{\rm o}$ or
$-30^{\rm o}$ with respect to the momentum transfer $q$ of the virtual photon.
Corresponding results are displayed in in the right part of
Fig.~\ref{fig:knoe1}.  For this kinematical
situation the FSI contribution is much more important than the correlation
effect.

\begin{figure}[tb]
\begin{center}
\epsfig{file=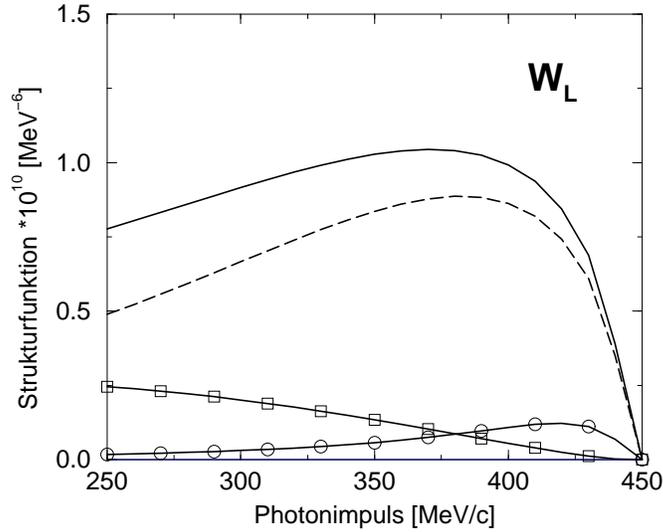,scale=0.5}
\begin{minipage}[t]{16.5cm}
\caption{\label{fig:knoe2}
Longitudinal structure function for ($e,e'pn$) for the kinematical condition as
displayed in the right part of Fig.~\protect{\ref{fig:knoe1}}. The contribution
of MEC is shown by the solid line with box symbol}
\end{minipage}
\end{center}
\end{figure}

The situation is even more complicated in the case of the ($e,e'pn$) reaction,
since in this case also MEC effects need to be considered. Results for the
longitudinal structure function for ($e,e'pn$) assuming the same kinematical
setup as in the right part of Fig.~\ref{fig:knoe1} are displayed in
Fig.~\ref{fig:knoe2}. The resulting structure function for ($e,e'pn$) 
is almost an order of magnitude larger than for the corresponding ($e,e'pp$) 
case.  The dominating
contribution to the longitudinal response is again the correlation part.
Comparison with Fig.~\ref{fig4} demonstrates that the $pn$ correlations are
significantly larger than those for the $pp$ pairs. This supports our
conclusion from discussing the results of table \ref{tabhelf} that the pionic or
tensor correlations which are different for isospin $T=0$ and $T=1$ pairs play
an important role and are even more important than the central correlations,
which are independent of the isospin. Note that the MEC contribution for this
`superparallel kinematic' is smaller than the correlation effect. 
This is due to a strong cancellation between the pion
seagull and the pion in flight contributions to the MEC. In fact, the dominant
MEC contribution in this kinematical setup is due to the coupling of the photon
to the $\rho$ meson. This does not hold for other situations, in which the pion
contributions to MEC are dominant.

The calculation of $(e,e'NN)$ reactions for finite nuclei typically account for
the FSI effects only on the level of the mean field approximation. This means
one 
considers an optical potential for the outgoing nucleons but ignores the FSI
effects which are due to the residual interaction between the two ejected
nucleons. As a typical example we will consider again the closed shell nucleus
$^{16}$O as the target nucleus. For an exclusive experiment leading to a discrete
final state of the daughter nucleus with well defined angular momentum $J$ and
isospin $T$ the initial state $\Phi_i$ , for which the matrix elements 
$< J^\mu >$ (see Eq.~(\ref{eq:cross0}) of the current operator have to be
calculated can be expanded in terms of correlated two-hole wave functions
\be
\Phi_i^{JT} (\vec r_1, \vec r_2) = \sum_{\nu_1 \nu_2} a_{\nu_1\nu_2}^{JT}
< \vec r_1, \vec r_2 \vert \Psi_2 \vert \nu_1\nu_2 JT > \label{eq:stauf0}
\ee  
where we have used the nomenclature of the coupled cluster method introduced in
subsection 2.2. The expansion coefficients $a^{JT}_{\nu_1\nu_2}$ are determined
from a configuration mixing calculation of the two-hole states in $^{16}$O,
which can be coupled to the angular momentum and parity of the requested state
within the model space considered for the treatment of long range
correlations\cite{giussta}.

\begin{figure}[tb]
\begin{center}
\epsfig{file=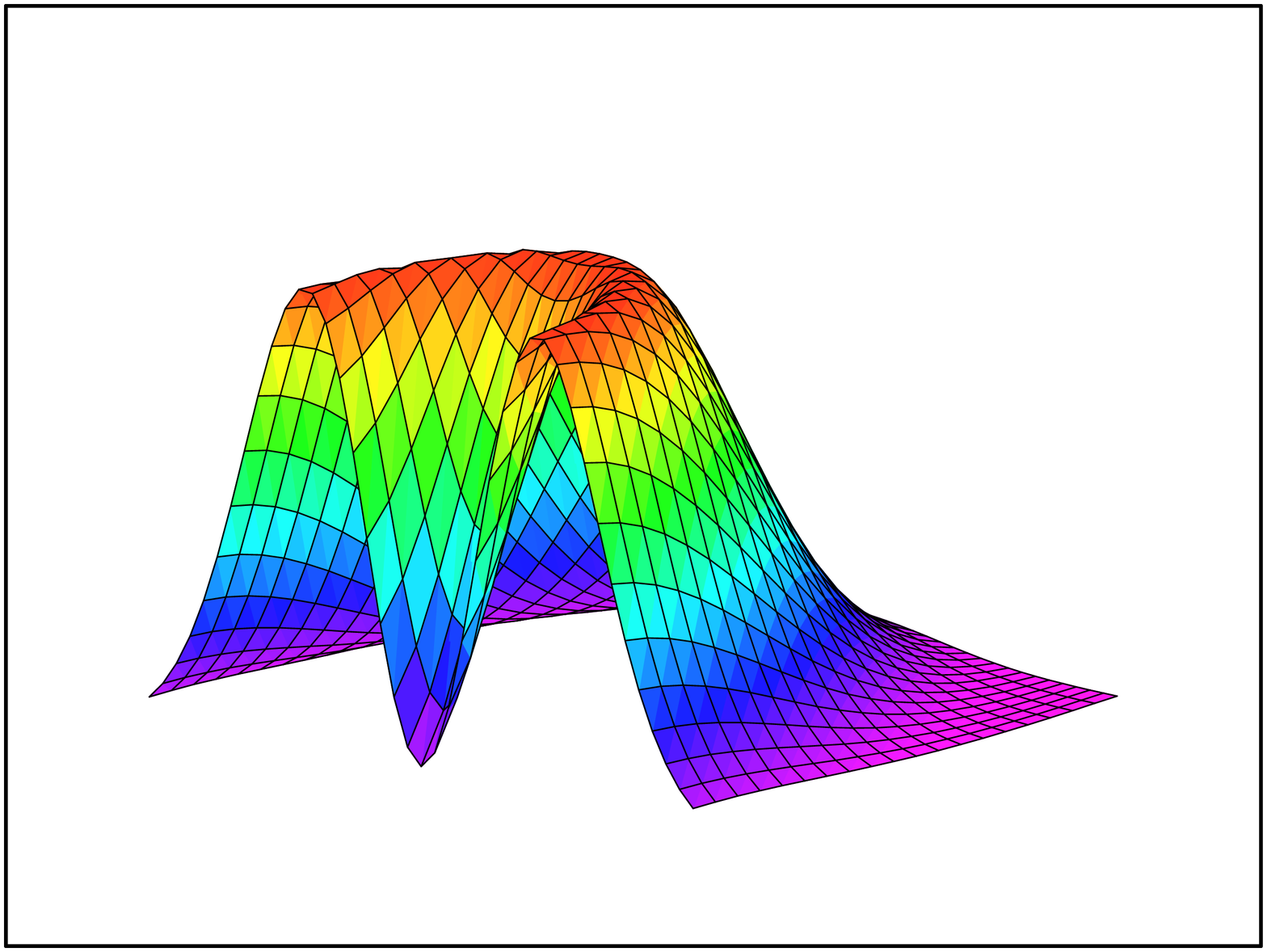,scale=0.25,angle=270}\hspace{.5cm}
\epsfig{file=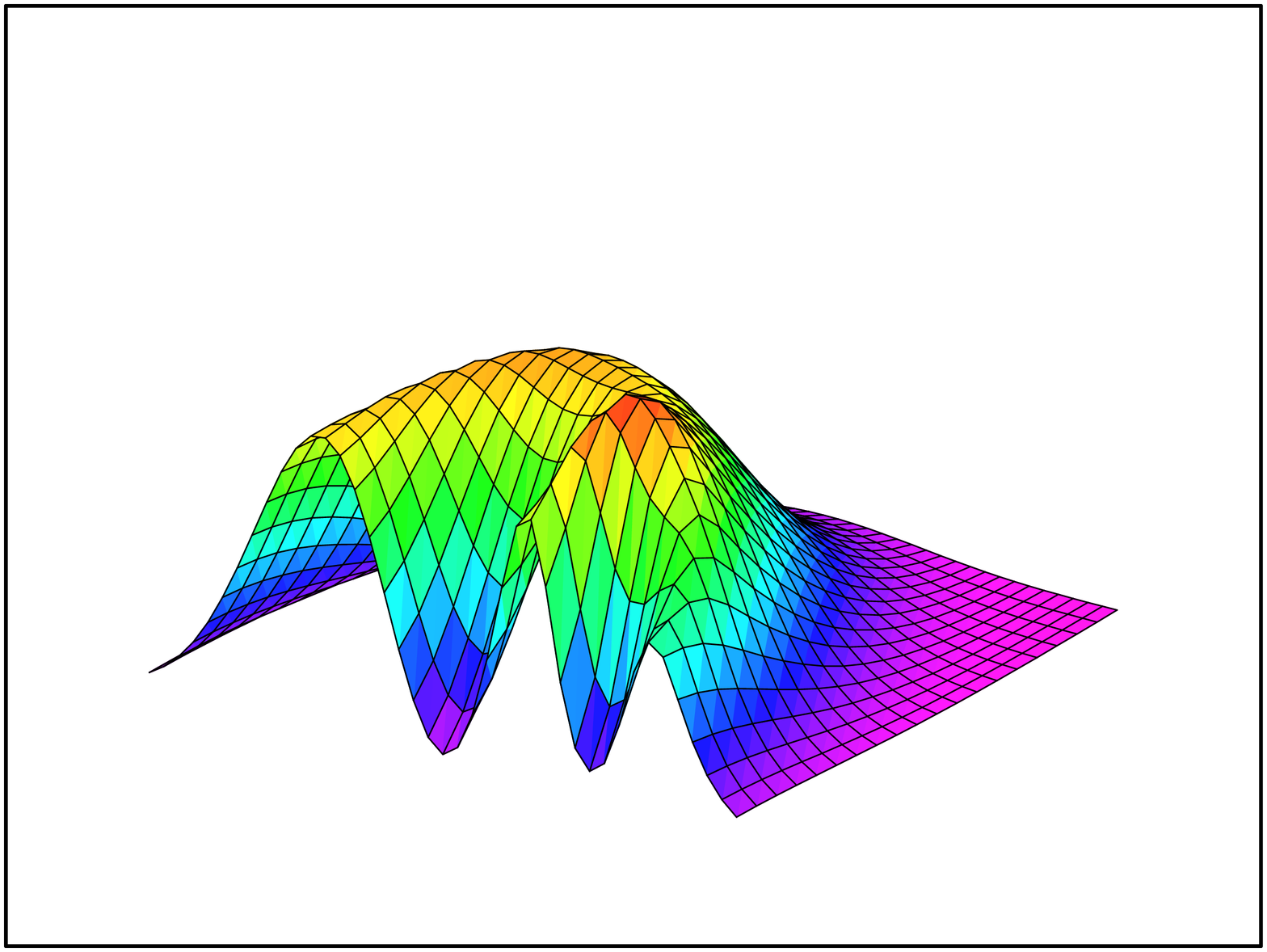,scale=0.25,angle=270}

\epsfig{file=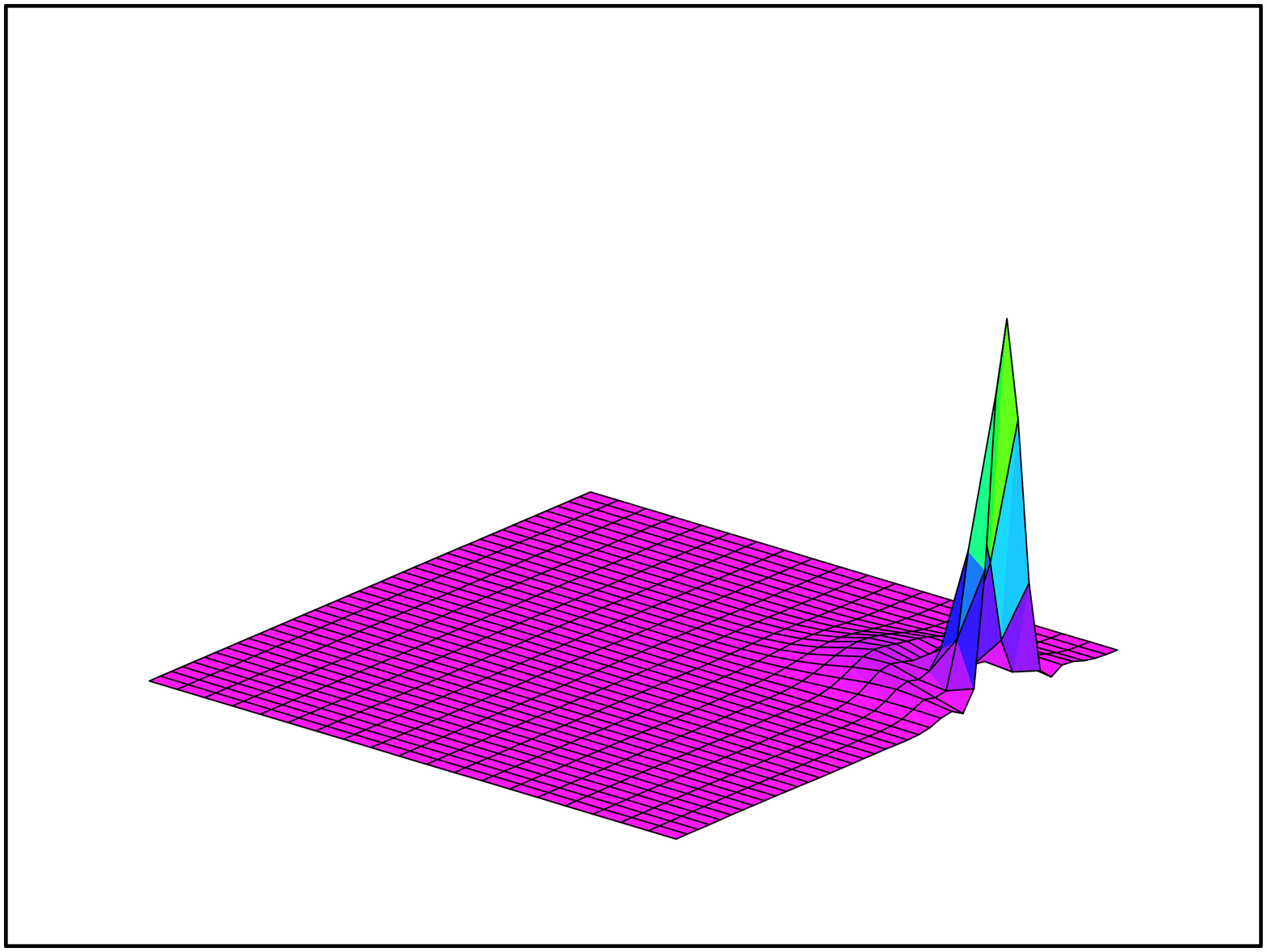,scale=0.25,angle=270}\hspace{.5cm}
\epsfig{file=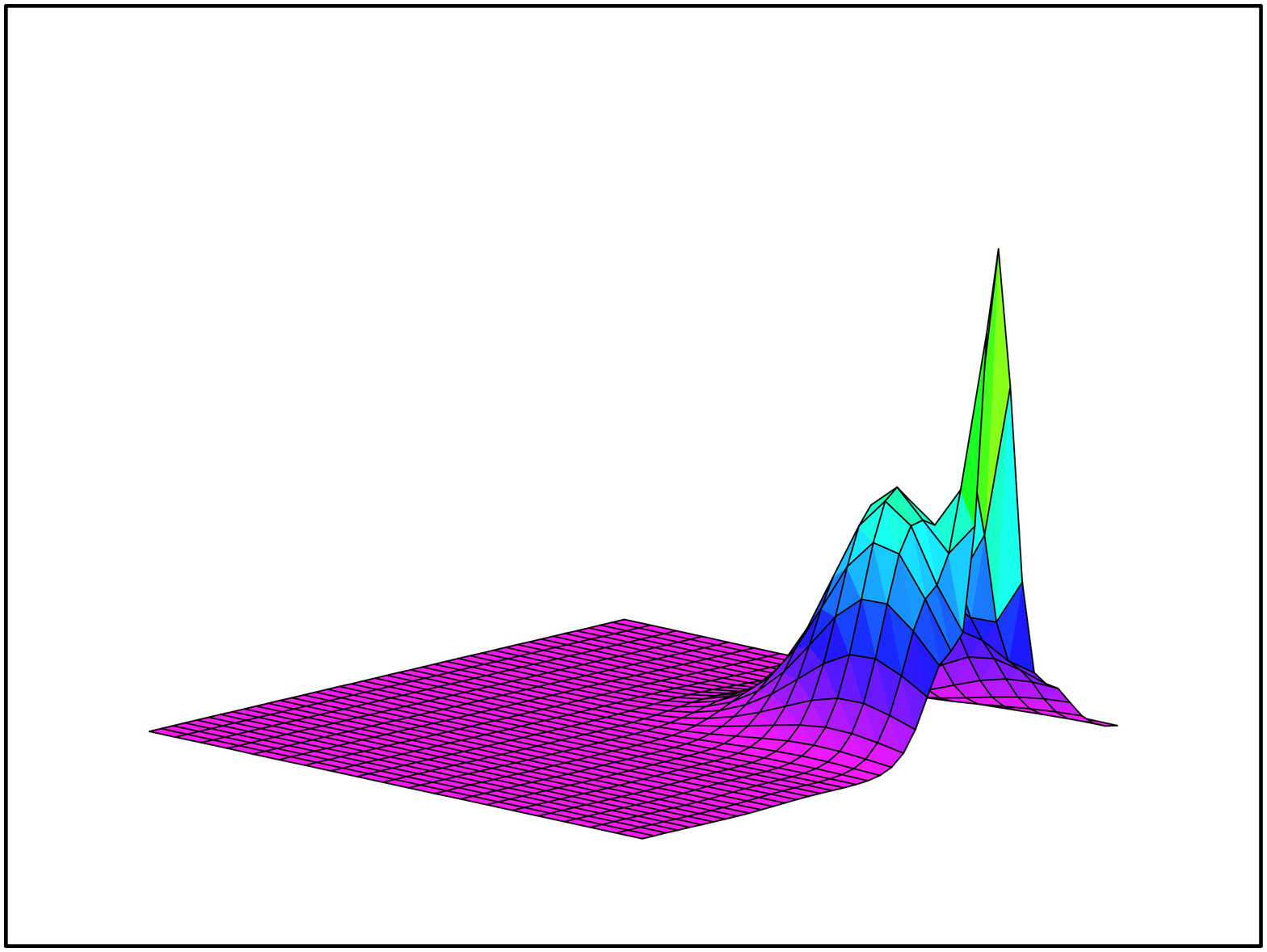,scale=0.25,angle=270}
\begin{minipage}[t]{16.5cm}
\caption{\label{fig:cors2} 
Two-body density for the removal of two nucleons from
$p_{1/2}$ shell in $^{16}$O. The upper part displays the results for the total
density $\Psi_2^2$ without (left) and with inclusion (right) of NN correlations.
The lower part shows the defect function squared for $pp$ (left) and $pn$
knock-out (right). See text for further description.}
\end{minipage}
\end{center}
\end{figure}

As an example, exhibiting the effects of correlations, we display in
Fig.~\ref{fig:cors2} the two-body densities, resulting from the knock-out of
two nucleons from the $p_{1/2}$ shell in $^{16}$O. In Fig.~\ref{fig:cors2} such
two-body densities are displayed for  a fixed $\vec r_1 = (x_1=0, y_1=0, z_1=2\
{\rm fm})$ as a function of $\vec r_2$, restricting the presentation to the
$x_2,z_2$ half-plane with ($x_2 >0, y_2 =0$). The upper left part of this
figure displays the two-body density without correlations ($\hat S_2 = 0$). One
observes that the two-body density, displayed  as a function of the position of
the second particle $\vec r_2$ is not affected by the position of the first one
$\vec r_1$. Actually, the two-body density displayed is equivalent to the
one-body density. This just reflects the feature of independent particle
motion. If correlation effects are included, as it is done in the upper right
part of Fig.~\ref{fig:cors2}, one finds a drastic reduction of the two-body
density at $\vec r_2 = \vec r_1$ accompanied by a slight enhancement at medium
separation between $\vec r_1$ and $\vec r_2$.

In order to amplify the effect of correlations, the lower part of
Fig.~\ref{fig:cors2} displays the corresponding correlation densities (i.e.~the
corresponding amplitudes  $\hat S_2$ squared). While the left part shows the
correlation density for the removal of a proton-proton pair, the corresponding
density for a proton-neutron pair is displayed in the right part. Comparing
these figures one sees that that the $pn$ correlations are significantly
stronger than the $pp$ correlations. This is mainly due to the presence of
pionic or tensor correlations in the case of the $pn$ pair. This part of
Fig.~\ref{fig:cors2}  also exhibits quite nicely the range of the correlations.
This range is short compared to the size of the nucleus even in the case of the
$pn$ correlations. All results displayed in this figure have been obtained
using the Argonne V14 potential for the NN interaction\cite{argo0}.

\begin{figure}[tb]
\begin{center}
\epsfig{file=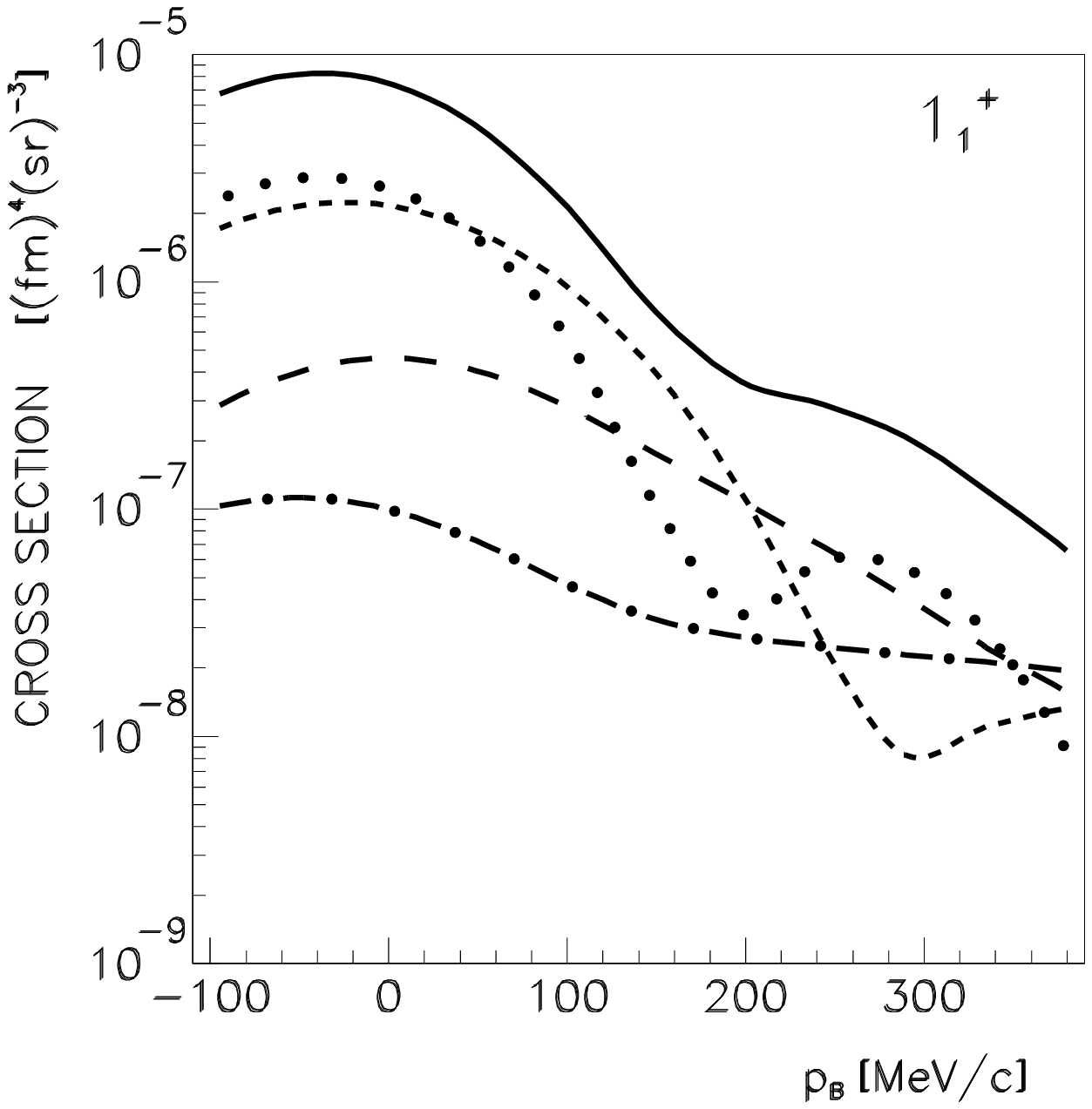,scale=0.50}\hspace{.5cm}
\epsfig{file=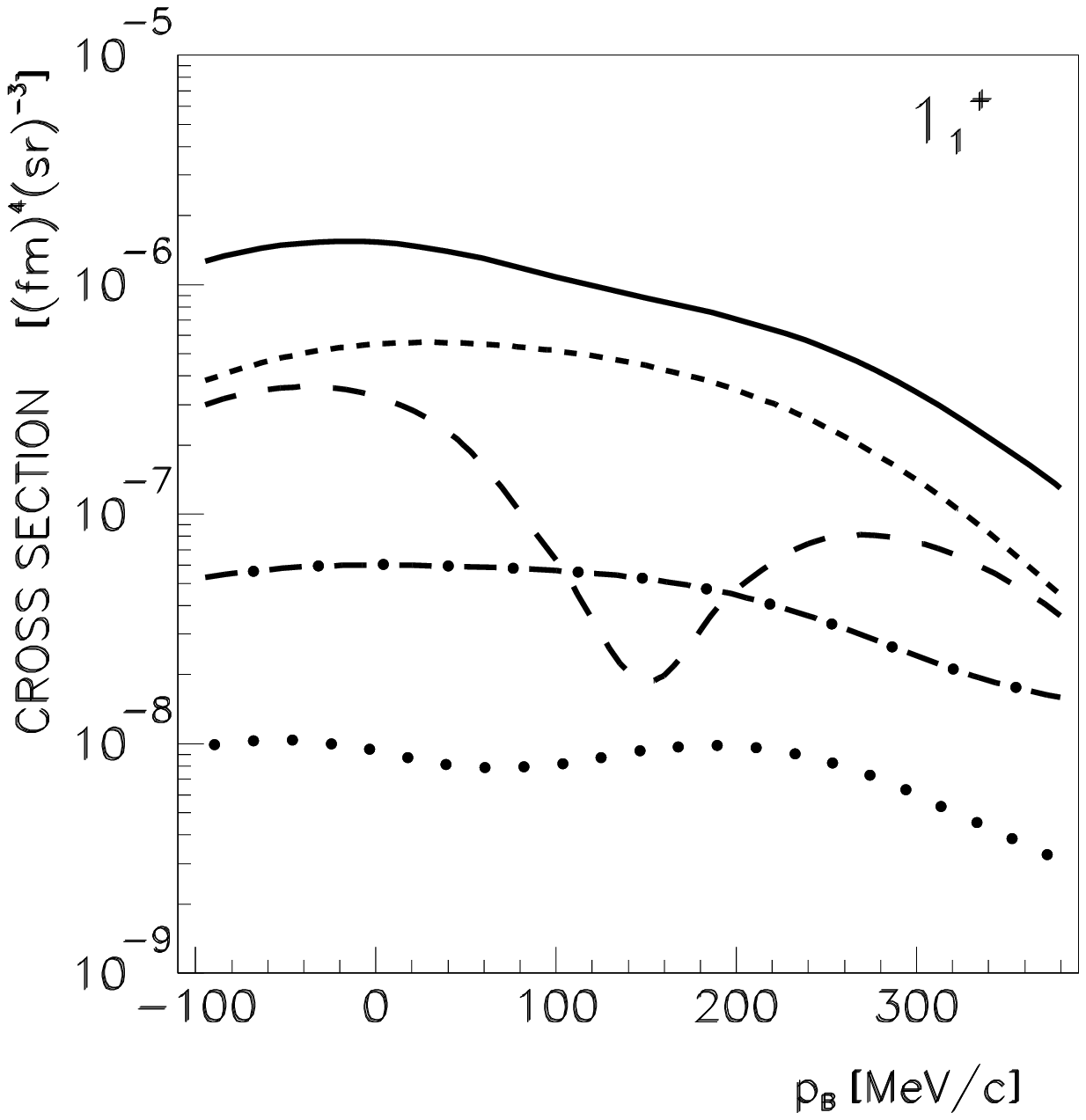,scale=0.50}
\begin{minipage}[t]{16.5 cm}
\caption{\label{fig:eenn}
The differential cross section  of the $^{16}$O(e,e$'$pn)
reaction as a function of the recoil momentum $p_{\mathrm{B}}$ for the
transition to the $1^+_1$ ground state of $^{14}$N ($E_{2\mathrm{m}} = 22.96$
MeV), in the super-parallel kinematics with $E_{0} = 855$ MeV, and
$\omega = 215$ MeV $q = 316$ MeV/$c$. The recoil-momentum distribution is
obtained changing the kinetic energies of the outgoing nucleons. Separate
contributions of the one-body, seagull, pion-in-flight and $\Delta$-current are
shown by the dotted, short-dashed, dot-dashed and long-dashed lines,
respectively.  Positive (negative) values of
$p_{\mathrm{B}}$ refer to situations where
${\mbox{\boldmath $p$}}_{\mathrm{B}}$ is parallel (antiparallel) to
${\mbox{\boldmath $q$}}$.
The calculations leading to the left part used correlated wave functions calculated in the frameworkof the CCM, while for the right part a simple Jastrow correlation function has been considered.
}
\end{minipage}
\end{center} 
\end{figure}

Results for the cross section of exclusive ($e,e'pn$) reactions on $^{16}$O
leading to the ground state of $^{14}$N are displayed in Fig.~\ref{fig:eenn}.
The calculations have been performed in the super-parallel kinematic, which we
already introduced before. The kinematical parameters correspond to those
adopted in a recent $^{16}$O(e,e$'$pp)$^{14}$C experiment at
MAMI~\cite{rosnere}. In order to allow a direct comparison of ($e,e'pp$) with
($e,e'pn$) experiments, the same setup has been proposed for
the first experimental study of
the $^{16}$O(e,e$'$pn)$^{14}$N reaction~\cite{MAMI}. This means that we assume
an energy of the incoming electron $E_0 =855$ MeV, electron
scattering angle $\theta=
18^{\mathrm{o}}$, $\omega = 215$ MeV and $q= 316$ MeV/$c$. The proton is
emitted parallel and the neutron antiparallel to the momentum transfer $\q$. 

Separate contributions of the different terms of
the nuclear current are shown in the figure and compared with the total cross
section\cite{giussta}. The contribution of the one-body current, entirely due to
correlations, is large. It is of the same size as that of the pion seagull
current. The contribution of the $\Delta$-current is much smaller at lower
values of $p_{\mathrm{B}}$, whereas for values of $p_{\mathrm{B}}$ larger than
100 MeV/$c$ it becomes comparable with that of the other components.
It is worth noting the the total cross section is about an order of magnitude
larger than the one evaluated for the corresponding ($e,e'pp$)
experiment\cite{gius1}. This confirms our finding about the relative cross
sections for $pp$ and $pn$ knock out, which we have discussed above for the
study in nuclear matter.

The right-hand part of Fig.~\ref{fig:eenn} shows the quantities as the left part
but calculated with the simpler prescription
of correlations, i.e. by the product of the pair function of the shell
model, described for $1^+_1$ as a pure ($p_{1/2}$)$^{-2}$ hole, and of a
Jastrow type central and state independent correlation function.
The large differences between the cross sections in the two parts of
Fig.~\ref{fig:eenn} indicate that a refined description of the two-nucleon
overlap, involving a careful treatment of both aspects related to nuclear
structure and NN correlations, is needed to give reliable predictions of the
size and the shape of the ($e,e'pn$) cross section. 

\section{Conclusion}

The main aim of this review has been to demonstrate that nuclear systems
are very intriguing many-body systems.  They are non-trivial
systems in the sense that they require the treatment of correlations beyond the
mean field or Hartree-Fock approximation. Therefore, from the point of view of 
many-body
theory, they can be compared to other quantum many-body systems like liquid He,
electron gas, clusters of atoms etc. A huge amount of experimental data is
available for real nuclei with finite number of particles as well as for the
infinite limit of nuclear matter or the matter of a neutron star. 

It is a challenge for theoretical physics to develop  many-body 
theories for a description of these data, which are based on a realistic
model for the NN interaction and yield predictions of the many-body data which
are free of any parameters. A lot of progress has been made during the last
years to develop such tools of many-body theory 
as the Brueckner hole-line expansion, 
the coupled cluster or ``exponential S'' method, the self-consistent calculation
of Green's function and variational calculations, which are based on cluster
expansion techniques as well as Monte Carlo methods. These techniques are very
useful for the study of nuclear structure but are also used in investigationis 
of other quantum-many-body sytems.

The different techniques yield predictions for the bulk properties of nuclear
systems at normal densities which tend to agree with each other. This is
certainly true for the treatment of short-range correlations. Additional effort
may be required for a reliable treatment of long-range correlations in finite
nuclei. Those correlations are
related to the low energy configurations within a shell-model basis. Shell-model
configuration mixing calculations using Monte-Carlo techniques\cite{koonin} or 
other sophisticated methods to deal with nuclear structure calculations in large
model spaces\cite{vampir} might be appropriate and should be combined with the 
formalisms to handle short-range correlations.

Many-body theories have
reached a level of sophistication that their predictions can be considered as a
reliable test of models for the NN interaction.
Characteristic differences can be observed if either a complete meson exchange
model is used, which leads to a non-local NN interaction, or if the local
approximation is considered. In any case, however, one finds that a two-nucleon
interaction alone does not lead to the empirical saturation point of nuclear
matter and also fails to reproduce binding energy and density of finite nuclei.
The calculations tend to predict binding energies which are too small and/or
densities which are too large. 

This can be corrected by introducing empirical three-nucleon forces. It is
not clear, however, whether such three-nucleon forces simulate sub-nucleonic
degrees of freedom ($\Delta$ excitations of the nucleons) or relativistic
features as they are contained in the Walecka model\cite{serot}. Further studies
are needed to clarify this point. One must try to inspect special observables
like e.g.~the spin-orbit splitting in the single-particle potential, the
energy dependence of the optical potential or the response functions for 
($e,e'p)$ experiments\cite{ulrich,bill}, which are sensitive to the relativistic
features.

There is also a challenge for a cooperation between
experimental and theoretical physics to search for observables,
which test the significance of the short-range NN correlations and thereby the
short-range structure of the underlying NN interaction. The study of exclusive
$(e,e'NN)$ experiments seems to be an appropriate tool to explore the
proton-proton and proton-neutron correlations.

\bigskip
\noindent{\bf Acknowledgments}\\[1ex]
A large part of the work which has been presented here, has been obtained in 
collaborations with many colleagues. In particular we would like to thank
K.~Allaart, K.~Amir-Azimi-Nili, P.~Czerski, W.H.~Dickhoff,  A. Fabrocini,
R.~Fritz, C.~Giusti, M.~Hjorth-Jensen, M.~Kleinmann, D.~Kn\"odler,
R.~Machleidt, F.D.~Pacati, A.~Ramos, E.~Schiller, L.D. Skouras, and J.~Udias.
This work has been supported by grants from  the DFG (SFB 382,  GRK 135 and  
Wa728/3), DGICYT (Spain) grant No. PB95-1249 and the program SGR98-11
from Generalitat de Catalunya.


\begin{thebibliography}{99}
\itemsep -2pt 
\bibitem{faes0} A. Valcarce, A. Buchmann, F. Fern\'andez, and Amand Faessler,
\Journal{\PRC} {51}{1480} {1995} 
\bibitem{rupr0} R. Machleidt, \Journal{\em Adv. Nucl. Phys.}{19}{189}{1989}
\bibitem{nijm0} M.M. Nagels, T.A. Rijken, and J.J. de Swart, \Journal{\PRD} 
{17} {768} {1978}
\bibitem{argo0} R.B. Wiringa, R.A. Smith, and T.L. Ainsworth, \Journal{\PRC}
{29} {1207} {1984}
\bibitem{urbv14} I.E. Lagaris and V.R. Pandharipande, \Journal{\NPA}{359} 
{331} {1981}
\bibitem{skyrme} M. Brack, C. Guet, and H.-B, Hakansson, \Journal {\PREP}{123}
{275}{1985}
\bibitem{art99} H. M\"uther and A. Polls, \Journal{\PRC} {}{in press}{1999}, 
preprint nucl-th/9908002.
\bibitem{hamada} T. Hamada and I.D. Johnston, \Journal{\em Nucl. Phys.} {34}
{382} {1962}
\bibitem{bruek1} K.A.~Brueckner, \Journal{\em Phys. Rev.} {97} {1353}{1955}
\bibitem{bruek2} H.A.~Bethe, \Journal{\em Ann. Rev. Nucl. Sci.}{21}{93}{1971}
\bibitem{bruek3} J.P. Jeukenne, A. Legeunne, and C. Mahaux, \Journal{\PREP}{25}
{83} {1976}
\bibitem{kuem} H.~K\"ummel, K.~H.~L\"uhrmann, and J.~G.~Zabolitzky,
\Journal{\PREP} { 36} {1} {1978}
\bibitem{bish} R.~F.~Bishop,
 in {\em Microscopic Quantum Many-Body Theories and Their
Applications}, eds. J.~Navarro and A.~Polls (Springer 1998) 
\bibitem{wim1} W.H. Dickhoff and H. M\"uther, \Journal{\em Reports on Progress 
in Physics} {11} {1947} {1992}
\bibitem{fhnc1} R.B. Wiringa, V. Fiks, and A. Fabrocini, \Journal{\PRC} {38}
{1010} {1988}
\bibitem{fhnc2} A. Akmal and V.R. Pandharipande, \Journal{\PRC} {56} {2261}
{1997}
\bibitem{fhnc3} S. Fantoni and A. Fabrocini, 
in {\em Microscopic Quantum Many-Body Theories and Their
Applications}, eds. J.~Navarro and A.~Polls (Springer 1998)
\bibitem{monc1} K.E. Schmidt and D.M. Ceperley, in {\em Monte Carlo Methods
III}, ed. K. Binder (Springer 1991)
\bibitem{monc2} D.M. Ceperley, \Journal{\em Rev. Mod. Phys.} {67} {279} {1995}
\bibitem{nikhef} I. Bobeldijk et al., \Journal{\em Phys. Rev. Lett.} {73} 
{2684}{1994}
\bibitem{mainz} K.I. Blomqvist {\it et al.}, \Journal{\PLB} {344} {85}{1995}
\bibitem{artur1} H. M\"uther, A. Polls, and W.H. Dickhoff, \Journal{\PRC}
{51} {3040}{1995}
\bibitem{wim2} H. M\"uther and W.H. Dickhoff, \Journal{\PRC}{ 49}
{R17}{1994}
\bibitem{poll1} A. Polls, M. Radici, S. Boffi, W.H. Dickhoff, and H. M\"uther,
\Journal{\PRC} {55} {810}{1997}
\bibitem{onder} C.\ J.\ G.\ Onderwater {\it et al.}, \Journal{\PRL}{81} {2213} 
{1998}
\bibitem{rosner} G.\ Rosner, Proc. on {\em ``Perspectives in Hadron Physics''},
eds. S.~Boffi, C.~Cioffi degli Atti and M.~Giannini, (World Scientific
1998).
\bibitem{paris} R. Vinh Mau in {\it Mesons in Nuclei} Vol. I (North-Holland,
Amsterdam, 1979)  
\bibitem{durso} J.W. Durso, M. Saarela, G.E. Brown, and A.D. Jackson
\Journal{\NPA} {278} {445}{1977}
\bibitem{wal1} J.D.\ Walecka, \Journal{\ANNP} { 83} {491} {1974}
\bibitem{serot} B.D.\ Serot and J.D.\ Walecka, \Journal{\em Adv.\ Nucl.\ Phys.}
{16} {1}{1986}
\bibitem{anast} M.R.\ Anastasio, L.S.\ Celenza, W.S.\ Pong, and C.M.\ Shakin,
\Journal{\PREP}{ 100} {327} {1983}
\bibitem{brock} R. Brockmann and R. Machleidt, \Journal{\PRC}{42}{1965}{1990}
\bibitem{malf1} B.\ Ter Haar and R.\ Malfliet, \Journal{\PREP}{ 149} {207}
{1987}
\bibitem{weigel} H.\ Huber, F.\ Weber, and M.K.\ Weigel, \Journal{\PL}{
B317} {485} {1993}
\bibitem{broc1} H. M\"uther, R. Machleidt, and R. Brockmann, \Journal{\PLB}{202} 
{483}{1988}
\bibitem{fri93} R. Fritz, H. M\"uther, and R. Machleidt, \Journal{\PRL} {71}{46}
{1993} 
\bibitem{fri94} R. Fritz and H. M\"uther, \Journal{\PRC}{49}
{633}{1994}
\bibitem{boersma} F. Boersma and R. Malfliet, \Journal{\PRC} {49}
{233}{1994}
\bibitem{elster} R. Machleidt, K. Holinde, and Ch. Elster, \Journal{\PREP}
{149}{1}{1987}
\bibitem{oldrep} H. M\"uther, \Journal{\em Prog. Part. Nucl. Phys.}
{14}{123}{1985}
\bibitem{franz} F. Osterfeld, \Journal{\em Rev. Mod. Phys.}{64}{491}{1992}
\bibitem{rhom} R. Rapp, R. Machleidt, J.W. Durso, and G.E. Brown, \Journal{\PRL}
{82}{1827}{1999}
\bibitem{pimas1} R. Rapp, J.W. Durso, Z. Aouissat, G. Chanfray, O. Krehl, P.
Schuck, J. Speth, J. Wambach, \Journal{\PRC} {59}{R1237}{1999}
\bibitem{pimas2} T. Hatsuda, T. Kunihiro, H. Shimzu,
\Journal{\PRL}{82}{2840}{1999}
\bibitem{v28} R.B. Wiringa, R.A. Smith, and T.L. Ainsworth, \Journal{\PRC}
{29}{1207}{1987}
\bibitem{ruder} H. Riffert, H. M\"uther, H. Herold, and H. Ruder: 
{\it Matter at High Densities in Astrophysics}, Springer Tracts in Modern 
Physics 133, (Springer Verlag, 1996)
\bibitem{glaude} P.J. Brussard and P.W.M. Glaudemans: {\it Shell model
applications in nuclear spectroscopy} (North-Holland, Amsterdam, 1977) 
\bibitem{tomnm} T.T.S. Kuo, Z.Y. Ma, and R. Vinh Mau, \Journal{\PRC} {33} {717}
{1986}
\bibitem{fesh}  H. Feshbach, \Journal{\ANNP} {19} {287} {1962}
\bibitem{bloho} C. Bloch and H. Horowitz, \Journal {\em Nucl. Phys.} {8} 
{51}{1958}
\bibitem{bran} B.H. Brandow, \Journal {\em Rev. Mod. Phys.} {39} {771}{1967}
\bibitem{kulera} T.T.S. Kuo, S.Y. Lee, and K.F. Ratcliff, \Journal
{\NPA}{176}{65}{1971}  
\bibitem{MKO} M. Hjorth-Jensen, T.T.S. Kuo, and E. Osnes, \Journal{\PREP} {261}
{125}{1995}
\bibitem{leesuz} S.Y. Lee and K. Suzuki, \Journal{\PLB}{91}{79}{1980}
\bibitem{artufold} H. M\"uther, A. Polls and T.T.S. Kuo, \Journal{\NPA}{435}
{548}{1985}   
\bibitem{BBP} H.A. Bethe, B.H. Brandow and A.G. Petschek, \Journal{\PREV} {129}
{225}{1962}
\bibitem{maha1} J.D. Jeukenne, A. Lejeune and C. Mahaux, \Journal{\PREP}
{25}{83}{1971}
\bibitem{rajar} R. Rajarman and H.A. Bethe, \Journal{\em Rev. Mod. Phys.}
{39}{745}{1967}
\bibitem{day81} B.D. Day, \Journal{\PRC} {24}{1203}{1981}
\bibitem{song1} H.Q. Song, M. Baldo, U. Lombardo, and G. Giansiracusa,
\Journal{\PLB} {411}{237}{1997}
\bibitem{song2} H.Q. Song, M. Baldo, G. Giansiracusa, annd U. Lombardo, 
\Journal{\PRL} {81} {1584}{1998}
\bibitem{haftab} M.I. Haftel and F. Tabakin, \Journal{\NPA}{158}{1}{1970}
\bibitem{sauer} H. M\"uther and P. U. Sauer, {\it Computational Nuclear
Physics 2}, eds. K. Langanke, J. A. Maruhn and S. E. Koonin (Springer
Verlag 1992)
\bibitem{coes0} F. Coester, \Journal {\em Nucl. Phys.} {7} {421}{1958}
\bibitem{coes1} F. Coester and H. K\"ummel, \Journal {\em Nucl. Phys.} {17}
{477}{1960}
\bibitem{thoules} D.J. Thouless, \Journal {\em Nucl. Phys.} {21} {225}{1960}
\bibitem{zab1} J.G. Zabolitzky, \Journal{\NPA} {228}{272}{1974}
\bibitem{song} H.Q. Song, S.D. Yang, and T.T.S. Kuo \Journal{\NPA} {462} {491}
{1987}
\bibitem{jiang} M.F. Jiang, T.T.S. Kuo, and H. M\"uther, \Journal{\PRC} {38}
{2408} {1988}
\bibitem{ellis} H.A. Mavromatis, P. Ellis, and H. M\"uther, \Journal{\NPA}
{530} {251}{1991}
\bibitem{zab2} J.G. Zabolitzky, \Journal{\NPA} {228}{285}{1974}
\bibitem{stauf} M. Stauf, ``Diplomarbeit'' (University T"ubingen, 1998)
\bibitem{heis1} J.H. Heisenberg and B. Mihaila, \Journal{\PRC} {59} {1440}
{1999}
\bibitem{emrich} K. Emrich, J.G. Zabolitzky, and K.H. L\"uhrmann, \Journal{\PRC} 
{16} { 1650}{1977}
\bibitem{fetwal} A.L. Fetter and J.D. Walecka, {\it Quantum Theory of Many 
Particle Systems} (McGraw-Hill New York, 1971)
\bibitem{negor} J. Negele and H. Orland {\it Quantum Many-Particle Systems}
(Addison-Wesley Redwood City. 1988)
\bibitem{mattuk} R.D. Mattuck, {\it A Guide to
Feynman Diagrams in the Many-Body Problem} (McGraw-Hill New York, 1976)
\bibitem{mahau1} C. Mahaux and R. Sartor, \Journal{\em Adv.\ Nucl.\ Phys.}
{20}{1}{1991}
\bibitem{gellow} M. Gell-Mann and F.E. Low, \Journal{\PREV}{84}{350}{1951}
\bibitem{wick} G.C. Wick,\Journal{\PREV}{80}{268}{1950} 
\bibitem{lehm} H. Lehmann, \Journal{\NCA}{11}{342}{1954}
\bibitem{skour1} H. M\"uther and L.D. Skouras, \Journal{\NPA}{555}{541}{1993}
\bibitem{baymka} G. Baym and L.P. Kadanov, \Journal{\PREV}{124}{287}{1961}
\bibitem{skour2} H. M\"uther and L.D. Skouras, \Journal{\NPA}{581}{247} {1995}
\bibitem{taigel}  H. M\"uther, T. Taigel and T.T.S. Kuo, \Journal{\NPA} {482}
{601} {1988}
\bibitem{neck} D. Van Neck, M. Waroquier, and J. Ryckebusch, \Journal{\NPA}
{530} {347} {1991}
\bibitem{carlson1} J. Carlson, V. R. Pandharipande, and R.B. Wiringa, \Journal{\NPA}
{401}{59}{1983}
\bibitem{threeforce} R. Schiavilla, V. R. Pandharipande and R. B. Wiringa,
\Journal{\NPA}{449}{219}{1986}
\bibitem{Wiringa95} R.B. Wiringa, V. G. J. Stoks, and R. Schiavilla,
\Journal{\PRC} {51}{38}{1995}
\bibitem{jastrow1} R. Jastrow, \Journal{\PREV}{98}{1479}{1955}
\bibitem{feenberg} E. Feenberg, {it Theory of Quantum Fluids}
 (Academic, New York,1969)
\bibitem{clark1} J.W. Clark, \Journal{\em Prog. Part. Nucl. Phys.}{2}{89}{1979}
\bibitem{fabro99} A. Fabrocini and S. Fantoni,
 in {\it Advances in Quantum Many-Body Theories}, Vol. 2,  eds. R.F. Bishop 
and N.R. Wilets, (World Scientific, Singapore ,1999).
\bibitem{iwamoto} F. Iwamoto and M. Yamada, \Journal{\PRO}{17}{543}{1957}
\bibitem{clark68} J. W. Clark and P. Westhaus, \Journal{\MATH}{9}{131}{1968}
\bibitem{guardiola1} R. Guardiola and A. Polls, \Journal{\NPA}{342}{385}{1980}
\bibitem{ripka1} M. Gaudin, J. Gillespie, and G. Ripka, \Journal{\NPA}{176}{237}
{1971}
\bibitem{fantoni74} S. Fantoni and S. Rosati, \Journal{\NCA}{20}{179}{1974}
\bibitem{fantoni75} S. Fantoni and S. Rosati, \Journal{\NCA}{25}{593}{1975}
\bibitem{kro75} E. Krotscheck and M.L. Ristig, \Journal{\NPA}{242}{389}{1975}
\bibitem{boer59} J.M.J. van Leeuwen, J. Groeneveld, and J. de Boer,
\Journal{\PHYS}{25}{792}{1959}
\bibitem{lovesey} S.W. Lovesey, {\it Theory of Neutron Scattering from Condensed
Matter}, Vol. 1, (Clarendon Press, Oxford, 1986)
\bibitem{ripka} G. Ripka, \Journal{\PREP}{56}{1}{1979}
\bibitem{wiringa79} V.R. Pandharipande and R.B. Wiringa, \Journal{\RMP}{51}{821}
{1979}
\bibitem{wiringa80} R.B. Wiringa, \Journal{\NPA}{338}{57}{1980}
\bibitem{clarkgrana} J.W. Clark, in {it The Many-Body Problem, Jastrow Correlations Versus
Brueckner Theory}, eds. R. Guardiola and J. Ros, Lectures Notes in Physics, Vol. 138. 
(Springer-Verlag, Heidelberg, 1980)
\bibitem{guarditubi} R. Guardiola, A. Faessler, H. M\"uther, and A. Polls, 
\Journal{\NPA}{371}{79}{1981} 
\bibitem{co1} G. Co', A. Fabrocini, S. Fantoni, and I. Lagaris, \Journal{\NPA}{549}
{439}{1992}
\bibitem{co2} G. Co', A. Fabrocini, and S. Fantoni, \Journal{\NPA}{568}{73}{1994}
\bibitem{co3} F. Arias de Saavedra, G. Co', A. Fabrocini, and S. Fantoni, 
\Journal{\NPA}{605}{359}{1996}
\bibitem{afnan} I. R. Afnan and Y. C. Tang, \Journal{\PREV}{175}{1337}{1968}
\bibitem{arias1} A. Fabrocini, F. Arias de Saavedra, G. Co', and P. Folgarait,
\Journal{\PRC}{57}{1668}{1998}
\bibitem{arias2} A. Fabrocini, F. Arias de Saavedra, and G. Co', Preprint
IFUP-TH 57/99.
\bibitem{akmal98} A. Akmal, V.R. Pandhariapnde, and D.G. Ravenhall,
\Journal{\PRC}{58} {1804}{1998}
\bibitem{clark59} J.W. Clark and E. Feenberg, \Journal{\PREV}{113}{388}{1959}
\bibitem{fhnc4} A. Fabrocini and S. Fantoni, in 
{\it First International Course
on Condensed Matter, ACIF series}, Vol. 8, eds. D. Prosperi, S. Rosati and S.
Violini, (World Scientific, Singapore, 1987)
\bibitem{fesbach} P.M. Morse and H. Feshbach, in {\it Methods of Theoretical 
Physics},(Addison-Wesley,1991)
\bibitem{fantoni84} S. Fantoni, \Journal{\PRB}{29}{2544}{1984}
\bibitem{krot81} E. Krotscheck, R.A. Smith and A.D. Jackson, \Journal{\PLB}{104}
{421}{1981}
\bibitem{fantopa84} S. Fantoni and V.R. Pandharipande, \Journal{\NPA}{427}{473}
{1984}
\bibitem{fantoni88} S. Fantoni and V.R. Pandhariapnde, \Journal{\PRC}{37}{1697}
{1988}
\bibitem{lowdin} P.O. L\"owdin, \Journal{\CHEM}{18}{365}{1950}
\bibitem{benhar90} O. Benhar, A. Fabrocini and S. Fantoni, \Journal{\PRC}{41}
{R24}{1990}
\bibitem{benhar92} O. Benhar, A. Fabrocini and S. Fantoni, \Journal{\NPA}{550}
{201}{1992}
\bibitem{fantopa87} S. Fantoni and V.R. Pandharipande, \Journal{\NPA}{473}{234}
{1987}
\bibitem{fantofa89} A. Fabrocini and S. Fantoni, \Journal{\NPA}{503}{375}{1989}
\bibitem{fabro97} A. Fabrocini, \Journal{\PRC}{55}{338}{1997}
\bibitem{kalos1} D. M. Ceperley and M.H. Kalos, in {\it Monte Carlo Methods 
in Statistical Physics}, ed. K. Binder, (Springer Verlag, Berlin, 1979).
\bibitem{schmid1} K. Schmidt and M. H. Kalos, in {\it Applications of the 
Monte Carlo Methods in Statistical Physics}, ed. K. Binder, (Springer-Verlag, 
Berlin, 1984).
\bibitem{fabromon} S. Caracciolo and A. Fabrocini Eds., {\it Monte Carlo Methods in Theoretical Physics}, (ETS Editrice, Pisa, 1991).
\bibitem{guarmc1} R. Guardiola 
 in {\em Microscopic Quantum Many-Body Theories and Their
Applications}, eds. J.~Navarro and A.~Polls (Springer 1998)
\bibitem{guarmc2} R. Guardiola 
{\it First International Course
on Condensed Matter, ACIF series}, Vol. 8, eds. D. Prosperi, S. Rosati and S.
Violini, (World Scientific, Singapore, 1987)
\bibitem{carlswi1} J. A. Carlson and R. B. Wiringa in {\it Computational Nuclear
Physics 1},eds. K. Langanke, J.A. Marhun and S.E. Koonin, (Springer-Verlag, 1991).
\bibitem{piepermc1} S.C. Pieper in 
 in {\em Microscopic Quantum Many-Body Theories and Their
Applications}, eds. J.~Navarro and A.~Polls (Springer 1998)
\bibitem{MRRTT} N.~Metropolis, A.~Rosenbluth, M.~Rosenbluth, A.~Teller, and 
E.~Teller, \Journal{\em J. Chem. Phys.}{21}{1087}{1955}
\bibitem{ceper77} D. Ceperley, G.V. Chester and M.H. Kalos, \Journal{\PRB}{16}
{3081}{1977}
\bibitem{schmid99} K.E. Schmidt and S. Fantoni, \Journal{\PLB}{446}{99}{1999}
\bibitem{lomnitz} J. Lomnitz-Adler, V.R. Pandhariapnde and R.A. Smith, 
\Journal{\NPA}{361}{399}{1981}
\bibitem{carlson85} J. Carlson and M.H. Kalos, \Journal{\PRC}{32}{2105}{1999}
\bibitem{pieper90} S.C. Pieper, R.B. Wiringa and V.R. Pandharipande, 
\Journal{\PRL}{64}{364}{1990}
\bibitem{pieper92} S. C. Pieper, R.B. Wiringa and V.R. Pandharipande, 
\Journal{\PRC}{46}{1741}{1992}
\bibitem{pudliner97} B. S. Pudliner, V.R. Pandharipande, J. Carlson, S.Pieper and
R.B. Wiringa, \Journal{\PRC}{56}{1720}{1997}
\bibitem{moroni95} S. Moroni, S. Fantoni and G. Senatore, \Journal{\PRB}{52}
{13547}{1995}
\bibitem{boronat99} J. Boronat and J. Casulleras, in {\it Proceedings
of the MBX, Advances in Quantum Many-Body Theory}, Vol. 3. (World Scientific,
Singapore, 2000).
\bibitem{koonin99} H.M. M\"uller, S.E. Koonin, R. Seki, U. van Kolck, 
nucl-th/9910038.
 \bibitem{isgur} N. Isgur and G. Karl, \Journal{\PRD} {18}{4187} {1978}
\bibitem{yap} K. Yazaki, \Journal{\em Prog. Part. Nucl. Phys.} {24} {353} {1990}
\bibitem{bra} K. Br\"auer, A. Faessler, F. Fernandez, and K. Shimizu, 
\Journal{\NPA}{507} {599} {1990}
\bibitem{thooft} G. T'Hooft, \Journal{\NPB}{75}{461}{1974}
\bibitem{witten} E. Witten, \Journal{\NPB}{160}{57}{1979}
\bibitem{erkelenz} K. Erkelenz, \Journal{\PREP} {13}{191}{1974}
\bibitem{carl} K. Holinde, \Journal{\PREP} {68}{121}{1981}
\bibitem{broja} G.E. Brown and A.D. Jackson, {\it The Nucleon-Nucleon 
Interaction} (North-Holland Pub. Comp. Amsterdam, 1976)
\bibitem{yukawa} H. Yukawa, \Journal{\em Proc. Phys. Math. Soc. Jpn}
{17}{48}{1935}
\bibitem{bjorkdre} J.D. Bjorken and S.D. Drell, {\it Relativistic Quantum
Mechanics}, (McGraw-Hill New York, 1964)
\bibitem{itzyk} C. Itzykson and J.-B. Zuber, {\it Quantum Field Theory} 
(McGraw-Hill New York, 1980)
\bibitem{bbs} R. Blankenbecler and R. Sugar, \Journal{\PREV}{142}{1051}{1966}
\bibitem{kady} V.G. Kadychevsky, \Journal{\NPB}{6}{125}{1968}
\bibitem{gross} F. Gross, \Journal{\PREV}{186}{1448}{1969} 
\bibitem{thompson} R.H. Thompson, \Journal{\PRD}{1}{110}{1970}
\bibitem{schier} G. Schierholz, \Journal{\NPB}{40}{335}{1972}
\bibitem{padat} Particle Data Group, \Journal{\PRD}{50}{1173}{1994}
\bibitem{hoehl} G. H\"ohler and E. Pietarinen, \Journal{\NPB}{95}{210}{1975} 
\bibitem{anast1} K. Holinde, R. Machleidt, M. Anastasio, A. Faessler, and H.
M\"uther, \Journal{\PRC} {18}{780} {1978}
\bibitem{pari2} M. Lacombe, B. Loiseau, J.M. Richard, R. Vinh Mau, J. Cote, P.
Pires, and R. de Tourreil, \Journal{\PRC}{21}{861}{1980}
\bibitem{nijmp} V.G.J. Stoks et al. \Journal{\PRC}{48}{792}{1993}
\bibitem{nijm1} V.G.J. Stoks, R.A.M. Klomp, C.P.F. Terheggen,
and J.J. de Swart, \Journal{\PRC}{49} {2950} {1994}
\bibitem{cdb} R. Machleidt, F. Sammarruca, and Y. Song, \Journal{\PRC}{53}
{R1483} {1995} 
\bibitem{artu971} A. Polls, H. M\"uther, R. Machleidt, and M. Hjorth-Jensen,
\Journal{\PLB}{432} {1} {1998}
\bibitem{anast2} M. Anastasio, H. M\"uther, A. Faessler, K. Holinde, 
and R. Machleidt, \Journal{\PRC} {18}{2916} {1978}
\bibitem{fuji} J. Fujita and H. Miyazawa, \Journal{\PRO}{17}{360}{1957}
\bibitem{gammel} K.A.~Brueckner and J.L. Gammel,
\Journal{\PREV}{109}{1023}{1958}
\bibitem{ksuzuk} K. Suzuki, R. Okamoto, M. Kohno, and S. Nagata,
\Journal{\NPA}{}{in press}{1999}, nucl-th9907050
\bibitem{schiller} E. Schiller, H. M\"uther, and P.~Czerski,
\Journal{\PRC}{59}{2934}{1999}, Erratum \Journal{\PRC}{60}{059901}{1999}
\bibitem{zabodif} J.G. Zabolitzky, \Journal{\PRA}{16}{1258}{1977}
\bibitem{engvik} L. Engvik, M. Hjorth-Jensen, R. Machleidt, H. M\"uther, and A.
Polls, \Journal{\NPA}{627}{85}{1997}
\bibitem{reid} R. Reid, \Journal{\ANNP} {50}{411}{1968}
\bibitem{coestba} F. Coester, S. Cohen, B. D. Day and C. M. Vincent
\Journal{\PRC} {1} {769}{1970}
\bibitem{artu99} H. M\"uther and A. Polls, \Journal{\PRC}{}{in print}{2000},
preprint nucl-th/9908002
\bibitem{pioco1} A.B. Migdal, \Journal{\RMP}{50}{10}{1978}
\bibitem{pioco2} G.E. Brown and W. Weise, \Journal{\PREP}{27}{1}{1976}
\bibitem{prevco} W.H. Dickhoff, A. Faessler, J. Meyer-ter-Vehn, and H. M\"uther,
\Journal{\PRC}{23}{1154}{1981}
\bibitem{prakash} M. Prakash, \Journal{\PREP}{242}{191}{1994}
\bibitem{heisel} H. Heiselberg and M. Hjorth-Jensen, \Journal{\PREP}{to be
published}{}{1999}; preprint nucl-th/9902033
\bibitem{ispos1} H. M\"uther, A. Polls, and R. Machleidt, \Journal{\PLB}
{445}{259}{1999}
\bibitem{ispos2} C. Harzer, H. M\"uther, and R. Machleidt, \Journal{\PLB}
{459}{1}{1999}
\bibitem{baldbcs} M. Baldo, J. Cugnon, A. Lejeune, and U. Lombardo, 
\Journal{\NPA}{515}{409}{1990}
\bibitem{mortbcs} \O. Elgaroy and M. Hjorth-Jensen,
\Journal{\PRC}{57}{1174}{1998}
\bibitem{vonderf} B.E. Vonderfecht, W.H. Dickhoff, A. Polls, and A. Ramos, 
\Journal{\NPA}{555}{1}{1993}
\bibitem{schnell} A. Schnell, G. R\"opke, and P. Schuck,
\Journal{\PRL}{83}{1926}{1999}
\bibitem{lda1} S. Stringari, M. Traini, and O. Bohigas, \Journal{\NPA}
{516}{33}{1990}
\bibitem{lda2} D. van Neck, L. Dieperink, and E. Moya de Guerra, \Journal{\PRC} 
{51}{1800}{1995}
\bibitem{lda3} O. Benhar, A. Fabrocini, S. Fantoni, and I. Sick, \Journal{\NPA}
{579}{493}{1994}
\bibitem{lda4} H. M\"uther, G. Knehr, and A. Polls, \Journal{\PRC}
{52}{2955}{1995}
\bibitem{bruce} B.R. Barrett, R.G.L. Hewitt, and R.J. McCarthy, \Journal{\PRC}
{3}{1137}{1971}
\bibitem{tomg} E.W. Krenciglowa, C.L. Kung, T.T.S. Kuo, and E. Osnes, \Journal
{\ANNP} {101}{154}{1976} 
\bibitem{zabo8} J.G. Zabolitzky, \Journal{\PLB}{47}{487}{1973}
\bibitem{carlo} K.W. Schmid, H. M\"uther, and R. Machleidt, \Journal{\NPA}
{530}{1991}{14}
\bibitem{zheng1} D.C. Zheng, B.R. Barrett, J.P. Vary, W.C. Haxton, and C.L.
Song, \Journal{\PRC}{52}{2488}{1995}
\bibitem{navra1} P. Navratil and B.R. Barrett, \Journal{\PRC}{57}{3119}{1998}
\bibitem{heinz} E. Heinz, H. M\"uther, and H.A. Mavromatis, \Journal{\NPA}
{587}{77}{1995}
\bibitem{sick97} V. R. Pandharipande, I. Sick, and P. K. A. deWitt Huberts,
\Journal{\RMP} {69}{981}{1997}
\bibitem{baldo92} M. Baldo, I. Bombaci, G. Giansiracusa, U. Lombardo, C. Mahaux,
and, R. Sartor, \Journal{\NPA}{545}{741}{1992}
\bibitem{benhar89} O. Benhar, A. Fabrocini and S. Fantoni \Journal{\NPA}{505}
{267}{1989} 
\bibitem{polls94} A. Polls, A. Ramos, J. Ventura, S. Amari, and W.H. Dickhoff,
\Journal{\PRC}{49}{3050}{1994}
\bibitem{dickhoff96} C.C. Gearhart, W.H. Dickhoff, A. Polls, and A. Ramos,
\Journal{\INT}{5}{261}{1996}
\bibitem{bozek99} P. Bozek, \Journal{\PRC}{59}{2619}{1999}
\bibitem{dickhoff98} W.H. Dickhoff, \Journal{\PRC}{58}{2807}{1998}
\bibitem{carlson98} J. Carlson and R. Schiavilla, \Journal{\RMP}{70}
{743}{1998}
\bibitem{borromeo} M. Borromeo, D. Bonatsos, H. M\"uther, and A. Polls,
\Journal{\NPA}{539}{189}{1992}
\bibitem{nili1} K. Amir Azimi Nili, H. M\"uther, L.D. Skouras, and A. Polls,
\Journal{\NPA}{604}{245}{1996}
\bibitem{geurts96} W.J.W. Geurts, K. Allaart, W.H. Dickhoff, and H.
M\"uther, \Journal{\PRC}{53}{2207}{1996}
\bibitem{radici94} M. Radici, S. Boffi, S.C. Peiper, and V. R. Pandharipande,
\Journal{\PRC}{50}{3010}{1994}
\bibitem{neck98} D. Van Neck, M. Waroquier, A.E.L. Dieperink, S.C. Pieper, and
V.R. Pandharipande, \Journal{\PRC}{57}{2308}{1998}
\bibitem{polls95} A. Polls, H. M\"uther, and  W. H. Dickhoff, \Journal{\NPA}
{594}{117}{1995}
\bibitem{lowdin1} P.-O. L\"owdin,\Journal{\PREV}{97}{1474}{1955}
\bibitem{antonov} M. V. Stoitsov, A.N. Antonov and S.S. Dimitrova,
\Journal{\PRC}{48}{74}{1994}
\bibitem{neck93} D. Van Neck, M. Waroquier and K. Heyde, \Journal{\PLB}
{314}{255}{1993}
\bibitem{bofbook} S. Boffi, C. Giusti, F.D. Pacati, and M. Radici, {\em
Electromagnetic Response of Atomic Nuclei} (Oxford University Press, Oxford
1996)
\bibitem{leusch} M. Leuschner et al., \Journal{\PRC}{49}{955}{1994} 
\bibitem{mainz1}  K.I. Blomqvist et al., \Journal{\PLB}{344}{85}{1995}
\bibitem{adam1}  I. Bobeldijk et al., \Journal{\PRL}{73}{2684}{1994}
\bibitem{nili} K. Amir-Azimi-Nili, J.M. Udias, H. M\"uther, L.D. Skouras,
and A.Polls, \Journal{\NPA} {625} {633} {1997}
\bibitem {eepwq1} C. Giusti and F. D. Pacati, \Journal{\NPA}{535}{573}{1991}
\bibitem{eepwq2} J. Ryckebusch, M. Vanderhaeghen, K. Heyde, and M.
Waroquier, \Journal{\PLB}{350}{1}{1995} 
\bibitem{knoed} D. Kn\"odler, H. M\"uther, and P. Czerski, preprint
nucl-th9909051
\bibitem{giussta} C.~Giusti, H.~M\"uther, F.~D.~Pacati, and M.~Stauf,
\Journal{\PRC}{60}{054608}{1999} 
\bibitem{rosnere}
G.~Rosner, {\em Proceedings of the 10th Mini-Conference on Studies of
Few-Body Systems with High Duty-Factor Electron Beams}, NIKHEF, Amsterdam 1999,
in press
\bibitem{MAMI}
J.~R.~M.~Annand, P.~Bartsch, D.~Baumann, J.~Becker, R.~B\"ohm, D.~Branford,
S.~Derber, M.~Ding, I. ~Ewald, K.~F\"ohl, J.~Friedrich, J.~M.~Friedrich,
P.~Grabmayr (spokesperson), T.~Hehl, D.~G.~Ireland, P.~Jennewein, M.~Kahrau,
D.~Kn\"odler, K.~W.~Krygier, A.~Liesenfeld, I.~J.~D.~MacGregor, H.~Merkel,
K.~Merle, P.~Merle, U.~M\"uller, H.~M\"uther, A.~Natter, R.~Neuhausen,
Th.~Pospischil, G.~Rosner (spokesperson), H.~Schmieden, A.~Wagner,
G.~J.~Wagner, Th.~Walcher, M.~Weis, S.~Wolf,
MAMI proposal Nr: A1/5-98.
\bibitem{gius1} C.~Giusti, F.~D.~Pacati, K.~Allaart, W.~J.~W.~Geurts,
W.~H.~Dickhoff, and H.~M\"uther,
\Journal {\PRC}{57} {1691} {1998}
\bibitem{koonin} S.E. Koonin, D.J. Dean, and K. Langanke, \Journal{\PREP}
{278}{1}{1997}
\bibitem{vampir} E. Hammaren, K.W. Schmid, and A. Faessler, \Journal{\em Eur.
Phys. J. A}{2}{371}{1998}
\bibitem{ulrich} S. Ulrych and H. M\"uther, \Journal{\NPA} {641} {499} {1998}
\bibitem{bill} William Bertozzi, Contribution to the Workshop on
`Electromagnetically induced two-hadron emission', Granada 99

\end{thebibliography}
\end{document}